\documentclass[11pt]{article}
\usepackage[colorlinks = true,
            bookmarks = false,
            linkcolor = blue,
            urlcolor  = blue,
            citecolor = blue,
            anchorcolor = blue]{hyperref}
\usepackage[letterpaper, left = 1.05in, top = 1.05in, right = 1.05in, bottom = 1.05in, nohead, includefoot, verbose, ignoremp]{geometry}
\usepackage{float}
\usepackage{amsfonts}
\usepackage{bbm}
\usepackage{amsmath}
\usepackage{graphicx}
\usepackage{subcaption}
\usepackage[affil-it]{authblk}
\usepackage{natbib}
\usepackage{xy}\xyoption{all} \xyoption{poly} \xyoption{knot}
\usepackage{verbatim}
\usepackage{color,ulem}
\usepackage[medium]{titlesec}
\usepackage{wrapfig}
\usepackage{changepage}
\usepackage{algorithm2e}
\usepackage{enumitem,cleveref}

\newcommand{\subscript}[2]{$#1 _ #2$}

\newcommand{\Polya}{P\'{o}lya}

\title{Dynamics of homelessness in urban America\thanks{This work was supported by Zillow, AFOSR Grant FA9550-16-1-0038, and NSF CAREER Award IIS-1350133. The authors thank Surya Tokdar, Svenja Gudell, Cory Hopkins, and Melissa Allison for helpful discussions.}}
\author{Chris Glynn\thanks{Peter T. Paul College of Business and Economics, University of New Hampshire, and Department of Statistics, University of Washington,  \href{mailto:christopher.glynn@unh.edu}{christopher.glynn@unh.edu}}\hspace{.25cm}and\hspace{.25cm}Emily B. Fox\thanks{Departments of Statistics and Computer Science, University of Washington, \href{mailto:ebfox@uw.edu}{ebfox@uw.edu}} }
\date{\vspace{-5ex}}

\begin{document}
\maketitle
\begin{abstract}
The relationship between housing costs and homelessness has important implications for the way that city and county governments respond to increasing homeless populations.  Though many analyses in the public policy literature have examined inter-community variation in homelessness \textit{rates} to identify causal mechanisms of homelessness \citep{Culhane2013, lee2003determinants, fargo2013}, few studies have examined time-varying homeless \textit{counts} within the same community \citep{mccandless2016bayesian}.   To examine trends in homeless population counts in the 25 largest U.S. metropolitan areas, we develop a dynamic Bayesian hierarchical model for time-varying homeless count data.  Particular care is given to modeling uncertainty in the homeless count generating and measurement processes, and a critical distinction is made between the counted number of homeless and the true size of the homeless population.  For each metro under study, we investigate the relationship between increases in the Zillow Rent Index and increases in the homeless population.  Sensitivity of inference to potential improvements in the accuracy of point-in-time counts is explored, and evidence is presented that the inferred increase in the rate of homelessness from 2011-2016 depends on prior beliefs about the accuracy of homeless counts.  A main finding of the study is that the relationship between homelessness and rental costs is strongest in New York, Los Angeles, Washington, D.C., and Seattle.          
\end{abstract}

\section{Introduction}
\label{sec:Introduction}
Counts of people experiencing homelessness in cities such as Seattle, Los Angeles, and New York reveal alarming year-over-year increases in the raw numbers of enumerated individuals.  In addition to rising counts of homeless, rental costs in these cities are significantly increasing as well.  The relationship between housing costs and homelessness is a topic of great public importance and has received considerable attention \citep{hanratty2017local, fargo2013, Culhane2013, stojanovic1999, oflaherty1995, sclar1990homelessness}.

Several challenges exist in quantifying the impact of increased rental costs on the size of the homeless population.  The first challenge is that point-in-time homeless counts often occur on a single night in January and are thus subject to significant sampling variability.  The second challenge is that the accuracy of the count itself is not the same from one year to the next.  Differences in the number of volunteers, weather, and count methodologies lead to counts that are difficult to compare year-over-year.  

These facts beg the question: are homeless populations across the county increasing?  Or do the reported counts simply represent a higher fraction of the homeless population?  Changing count accuracy over time directly impacts inferred trends in the size of the homeless population.  In light of this fact, we investigate the impact of different count accuracy trajectories on the inferred change in homelessness rates from 2011-2016.

Inference on the relationship between trends in rental costs and trends in the homeless population is related to other trend analyses with data quality challenges \citep{Tokdar2011, coles2006extreme, cornulier2011bayesian, kery2010hierarchical}.  Although we observe the number of \textit{counted} homeless, we do not observe the true size of the homeless population.  Plant-capture methods \citep{laska1993plant,schwarz1999estimating} have demonstrated that homeless counts systematically understate the size of the total homeless population \citep{Hopper2008}.  One strategy to include the uncounted number of homeless in the analysis is to build a mechanism for the imperfect counting process into the statistical model, as \citet{mccandless2016bayesian} have done with plant-capture data from Edmonton, Canada.  

In this paper, the total size of the homeless population is imputed, and uncertainty in the total homeless population and count accuracy is propagated to our assessment of the relationship between rental costs and homelessness. Our goal is to \textit{jointly} model the collection of homeless count time series from the 25 largest metropolitan areas in the United States.  In contrast to \citet{mccandless2016bayesian}, who treat time-indexed counts as exchangeable, we directly model temporal dependence.  We develop a Bayesian dynamic modeling framework to investigate the relationship between the number of homeless and rent costs subject to different prior beliefs about count accuracy over time.  

The data in our analysis comes from three sources: the U.S. Census Bureau; the U.S. Department of Housing and Urban Development (HUD); and the housing website Zillow.  The data include the total population, point-in-time homeless counts, and the Zillow Rent Index (ZRI) for continuums of care that service the 25 largest metro areas from 2011-2016. 

Numerous previous studies have utilized inter-community variation in homelessness \textit{rates} to identify potential causal mechanisms of homelessness \citep{fargo2013, Culhane2013, raphael2010housing, lee2003determinants, early2002subsidized, quigley2001economics, quigley2001homeless, troutman1999public, hudson1998estimating, grimes1997assessing, honig1993causes, burt1992over, bohanon1991economic, appelbaum1991scapegoating, quigley1990does}.  Studies that model homelessness rates, defined as  $\frac{\mbox{total homeless}}{\mbox{total population}}$, assume that both the numerator and denominator are observed without error.  In practice, there is significant uncertainty in both the numerator and denominator in any such homelessness rate calculation.  To account for that uncertainty, we directly model time-varying \textit{counts} within the same community.  Working with time series of count data has two advantages.  First, statistical models of counts more aptly characterize the sampling variability in the observed data; and second, focusing on within community variation over time avoids drawing conclusions from data generated across different municipal and state governments, climates, and social structures. A major contribution of our work is the development of a statistical framework that enables researchers, policymakers, and local continuum coordinators to address five specific questions for each metro:
\begin{enumerate}[leftmargin=2cm, rightmargin=1cm,label=(\subscript{Q}{\arabic*})]
    \item When adjusting for increases in count accuracy and total population growth, is the \textit{rate} of homelessness increasing? \label{Q1}
    \item If ZRI increases by $x\%$, what are the predicted increases in the counted and total number of people experiencing homelessness? \label{Q2}
    \item Expense and logistical challenges preclude more than one homeless count per year in many metros.  If a second point-in-time count were to be conducted in a given year, what is the expected range in the number of homeless to be enumerated?\label{Q3}
    \item Given that $C$ homeless are counted and count accuracy is imperfect, what is the expected range in the total number of people experiencing homelessness at a point in time? \label{Q4}
    \item What is the one-year-ahead forecast of the total homeless population in 2017? \label{Q5}
\end{enumerate}

We identify New York, Los Angeles, Seattle, and Washington, D.C. as metros where (i) the inferred rate of homelessness significantly increased from 2011-2016 and (ii) there exists a strong relationship between housing costs and homelessness.  We find that predicted increases in counted homeless due to increased ZRI are robust to different prior beliefs about time-varying count accuracy; however, we present evidence that the inferred change in the homelessness rate from 2011-2016 is sensitive to the trajectory of count accuracy.  This point is emphasized to encourage researchers, policymakers, and continuum leaders to carefully quantify their beliefs and uncertainty about count accuracy.      

The prior beliefs that we incorporate in this analysis are informed by existing literature and discussions with count coordinators, volunteers, and homelessness experts from around the country.  Incorporating the expert opinions of count coordinators in every metro in the sample will lead to a more informed study.  Our goal in this paper is to advance the statistical methodology utilized by researchers to analyze data on homelessness.  We view this as a demonstration of a modeling framework that will benefit from a partnership between private companies with relevant data, HUD, and local continuums of care. 

In Section \ref{sec:Data}, we discuss the data used in our analysis and necessary pre-processing steps to account for geographic mismatches between counties and continuums of care.  Section \ref{sec:Model} describes the Bayesian dynamic model that hierarchically shares information across all metros under study.  Efficient information sharing, both locally in time and hierarchically across all metros, facilitates sharper inference on the relationship between rental costs and homelessness.  Our hierarchical dynamic model allows us to estimate local relationships between homelessness and rental costs whereas the cross-sectional regression model in \citet{Culhane2013} estimates a single global effect.  We discuss prior information and how that information translates to prior distributions for model parameters in Section \ref{sec:Priors}.  Model fitting with a custom Markov chain Monte Carlo algorithm is discussed in Section \ref{sec:MCMC}, and Section \ref{sec:Results} presents results and addresses questions \ref{Q1} - \ref{Q5}. Section \ref{sec:Discussion} concludes with a discussion of our findings.   

\section{Data}
\label{sec:Data}
The data in our study comes from three different sources: the U.S. Census Bureau, HUD, and the housing website Zillow.  For the continuums of care in the 25 largest metros, we observe a collection of three time series that correspond to  (i) the total number of people living in the metro, (ii) the \textit{counted} number of homeless in the continuum(s) of that metro, and (iii) the ZRI for the metro.

For the total population data, we use county-level population estimates reported by the U.S. Census Bureau \citep{Census}.  The homeless counts are the number of individuals experiencing homelessness (both sheltered and unsheltered) at a point-in-time as reported by HUD \citep{HUD}.  While the first two series of interest are fairly self-explanatory, the ZRI warrants further detail.  Below is a description from Zillow \citep{ZRI}.
\begin{adjustwidth}{2cm}{}
\textit{Similar to the Zillow Home Value Index (ZHVI), we created the Zillow Rent Index (ZRI) to track the monthly median rent in particular geographical regions. Like the ZHVI, we sought to create an index for rents that is unaffected by the mix of homes for rent at any particular time. This makes temporal comparisons of rents more valid since the index is tracking the rents for a consistent stock of inventory.}
\end{adjustwidth}
The geographic coverage of Zillow's housing data, the temporal invariance in the rental stock underlying the metric, and ease of temporal comparison make ZRI a natural choice for summarizing time-varying rental costs across all U.S. metropolitan areas. 

\begin{table}[ht!]
\centering
\caption{HUD continuums of care and state counties that correspond to the 25 largest metropolitan areas under study.  In cases where more than one continuum of care is in a county, we aggregate homeless counts to form a synthetic continuum for that county.  When a single continuum spans multiple counties, we construct a population-weighted ZRI measure and aggregate total population figures across the multiple counties.}
\label{tab:GeoKeys}
\tiny 
\begin{tabular}{rlll}
  \hline
 & Metro area & HUD continuum of care & Counties \\ 
  \hline
1 & New York, NY & NY-600 & New York, Bronx, \\ & & &  Queens, Kings, Richmond \\ 
  2 & Los Angeles-Long Beach-Anaheim, CA & CA-600, CA-606, CA-607, CA-612 & Los Angeles \\ 
  3 & Chicago, IL & IL-510, IL-511 & Cook \\ 
  4 & Dallas-Fort Worth, TX & TX-600 & Dallas \\ 
  5 & Philadelphia, PA & PA-500 & Philadelphia \\ 
  6 & Houston, TX & TX-700 & Harris, Fort Bend \\ 
  7 & Washington, DC & DC-500 & District of Columbia \\ 
  8 & Miami-Fort Lauderdale, FL & FL-600 & Miami-Dade \\ 
  9 & Atlanta, GA & GA-500, GA-502 & Fulton \\ 
  10 & Boston, MA & MA-500 & Suffolk \\ 
  11 & San Francisco, CA & CA-501 & San Francisco \\ 
  12 & Detroit, MI & MI-501, MI-502 & Wayne \\ 
  13 & Riverside, CA & CA-608 & Riverside \\ 
  14 & Phoenix, AZ & AZ-502 & Maricopa \\ 
  15 & Seattle, WA & WA-500 & King \\ 
  16 & Minneapolis-St Paul, MN & MN-500 & Hennepin \\ 
  17 & San Diego, CA & CA-601 & San Diego \\ 
  18 & St. Louis, MO & MO-500, MO-501 & St. Louis \\ 
  19 & Tampa, FL & FL-501 & Hillsborough \\ 
  20 & Baltimore, MD & MD-501, MD-505 & Baltimore \\ 
  21 & Denver, CO & CO-503 & Adams, Arapahoe, Boulder, \\ & & &  Broomfield, Denver, Douglas, \\ & & &  Jefferson \\ 
  22 & Pittsburgh, PA & PA-600 & Allegheny \\ 
  23 & Portland, OR & OR-501 & Multnomah \\ 
  24 & Charlotte, NC & NC-505 & Mecklenburg \\ 
  25 & Sacramento, CA & CA-502 & Sacramento \\ 
   \hline
\end{tabular}
\end{table}

One of the challenges in working with the HUD point-in-time data is that the jurisdiction of the HUD-defined continuums of care do not always agree with the boundaries of cities or counties.  Often, each county will have a single continuum; however, cases exist where this is not true.  In some counties, there may be more than one continuum (e.g, Cook County, IL and Fulton County, GA have two).  Other times, there may be multiple counties in a single continuum (e.g., the Denver, CO continuum spans seven different counties, and the New York City continuum spans five).  Table \ref{tab:GeoKeys} maps the 25 metros under study to the underlying HUD continuum(s) of care.  In each metro, if the continuum does not match up with a single county, we construct a synthetic unit of analysis by following one of two approaches.  If a county includes multiple continuums, we aggregate homeless totals reported by each continuum in the county.  If a continuum includes multiple counties, we aggregate population totals reported by the different counties that make up a single continuum and construct a population-weighted ZRI metric. 

\begin{figure}[ht!]
\centering
\begin{subfigure}{.3\textwidth}
  \centering
\includegraphics[width=1\textwidth]{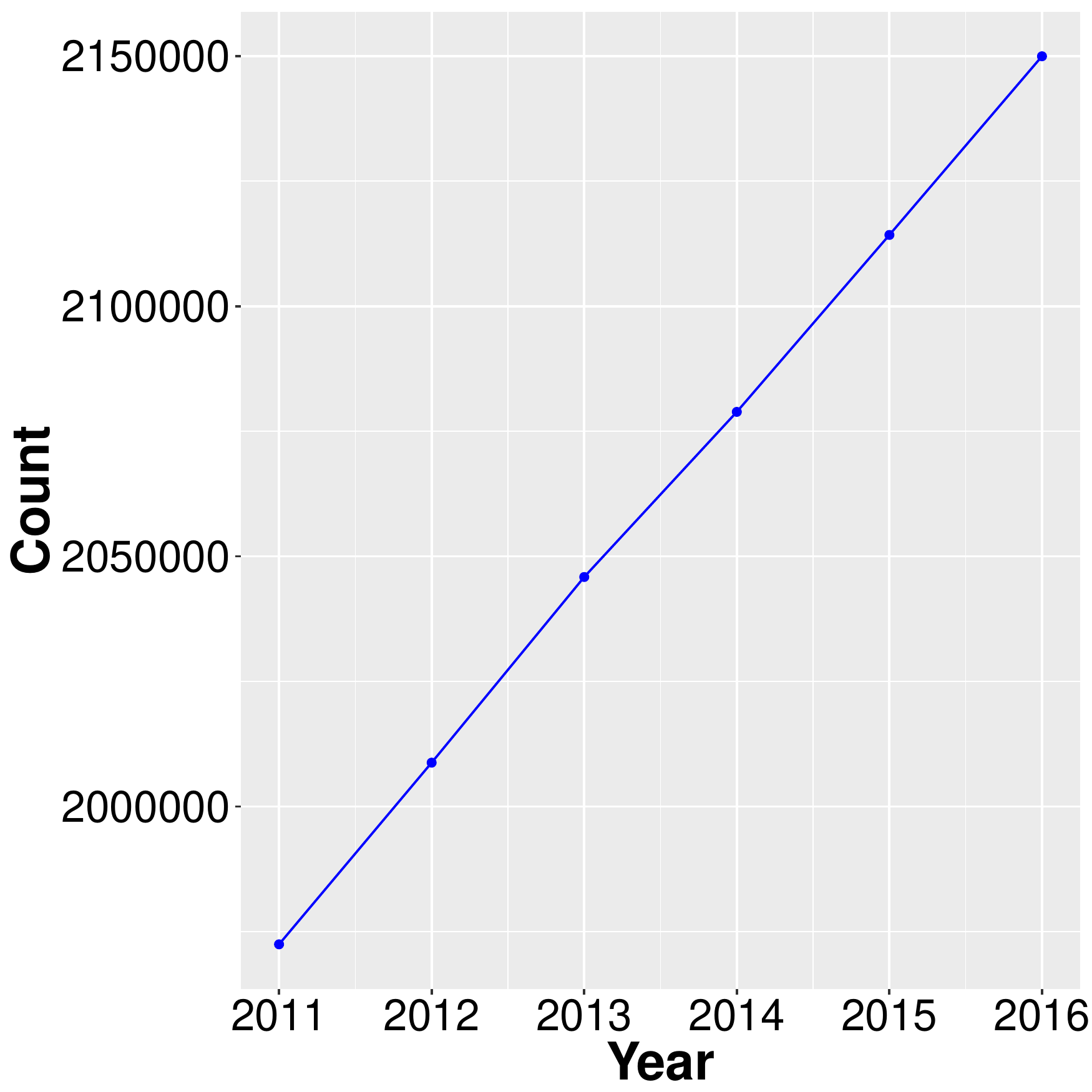}
\caption{Total population}
\label{subfig:Total_pop}
\end{subfigure}
\begin{subfigure}{.3\textwidth}
  \centering
\includegraphics[width=1\textwidth]{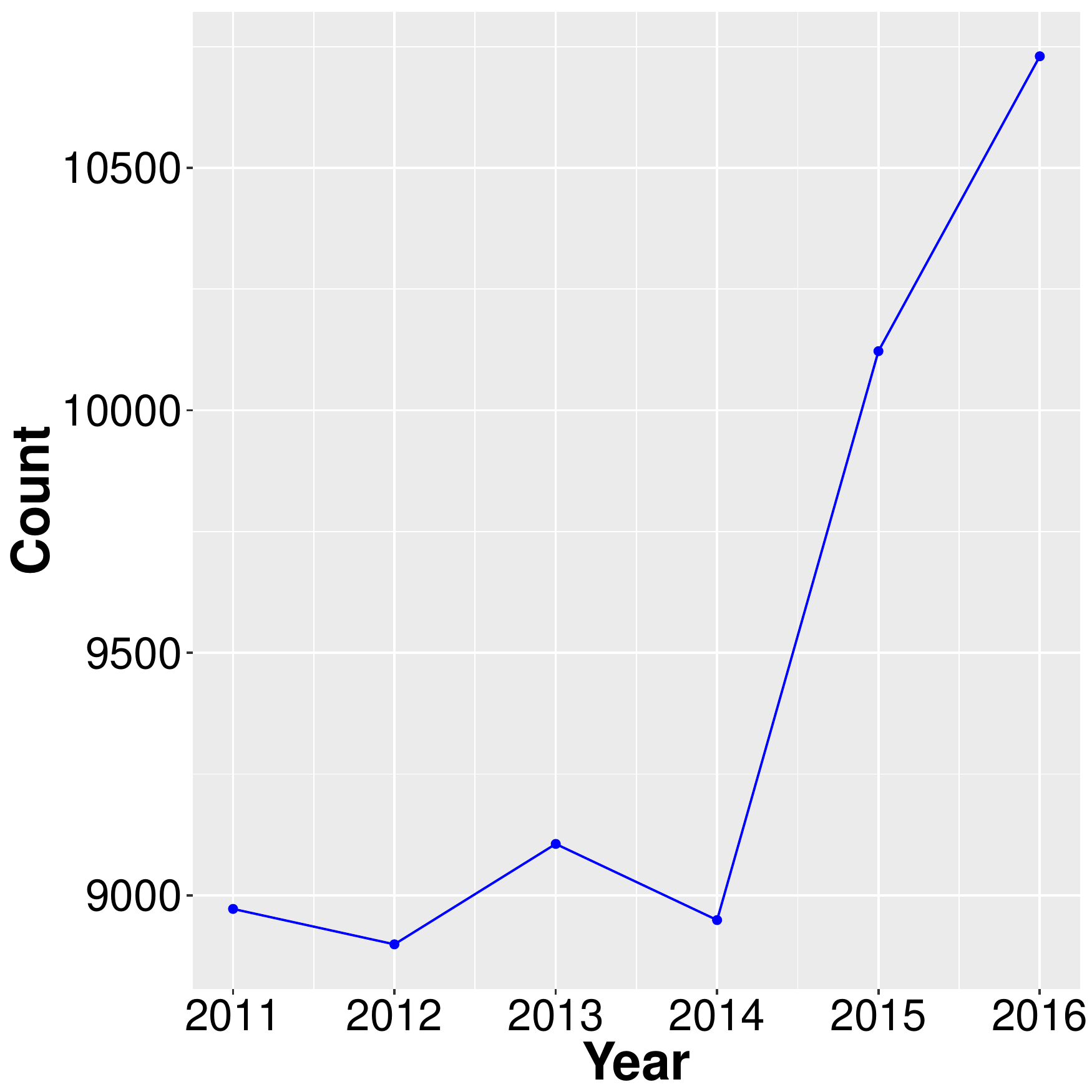}
\caption{Homeless count}
\label{subfig:Homeless}
\end{subfigure}
\begin{subfigure}{.3\textwidth}
  \centering
\includegraphics[width=1\textwidth]{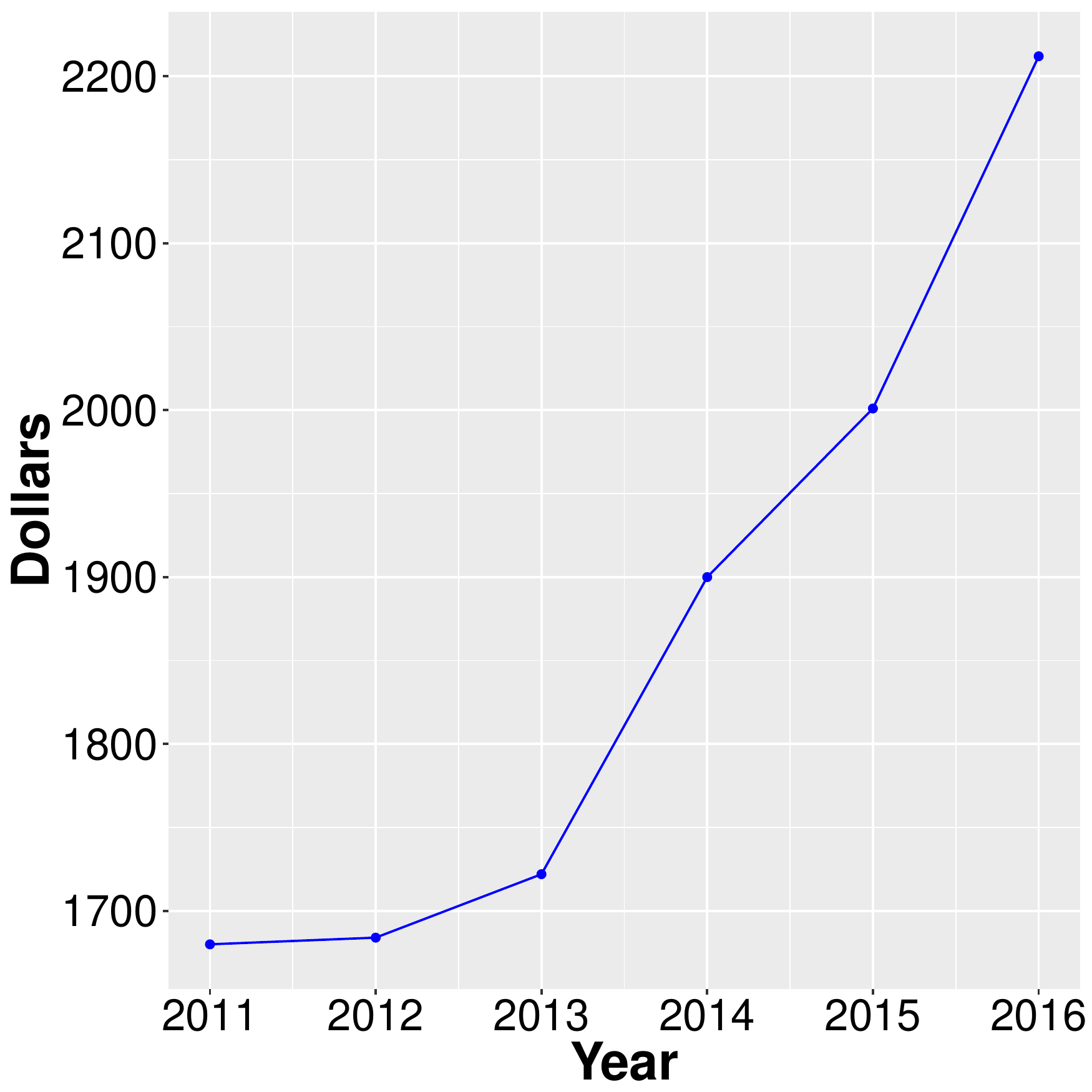}
\caption{ZRI (median rent)}
\label{subfig:ZRI}
\end{subfigure}
\caption{Data from the All Home King County (WA) continuum of care from 2011 - 2016.  Left: the total population in King County has rapidly increased in recent years.  Increased population creates increased demand for rental housing and community services.  Middle: The number of homeless counted in King County has dramatically increased since 2014.  Right: The median rent, as measured by the ZRI, demonstrates the same basic pattern of increases as the count of people experiencing homelessness.}
\label{fig:Seattle_data}
\end{figure}

We also focus on year-over-year changes in the metro-specific ZRI rather than the absolute level of the ZRI itself.  This standardizes the analysis of rental markets across metros.  The result is a data set of synthetic continuums that properly record the counted number of homeless, total population, and changes in rent levels in each metro.    
Because the ZRI is only available after October 2010, the time series for these three quantities are observed from 2011 - 2016 at an annual frequency.  The homeless count data and ZRI are recorded each January, but the intercensal total population estimates from the Census Bureau are dated July 1.  There is a six month temporal mismatch in both the homeless count and ZRI and the total metro population series.  Although we could perform a linear interpolation to align the data, for this analysis we assume the mismatch to be inconsequential.  Figure \ref{fig:Seattle_data} presents these three time series for the All Home King County continuum in Seattle, WA.

The count of homeless in Seattle/King County has dramatically increased since 2014 (Figure \ref{subfig:Homeless}); however, the total population (Figure \ref{subfig:Total_pop}) also significantly increased over that same time period. The King County, WA data demonstrate the need for modeling the homelessness rate to control for increases in the total population.  The ZRI for King County, shown in Figure \ref{subfig:ZRI}, has similarly increased.   

In order to properly calculate the homelessness rate, it is necessary to account for time-variation in the count accuracy.  We define the count accuracy to be the probability that a person who is homeless will be accounted for in the homeless count.  If the count accuracy improves over time, more homeless are likely to be counted.  In a scenario where the homeless count has improved, an increase in the number of homeless counted does not necessarily imply that the total size of the homeless population has increased.  The count may simply represent a higher fraction of the total homeless population.  

\begin{figure}[ht!]
    \centering
    \includegraphics[width=.4\textwidth]{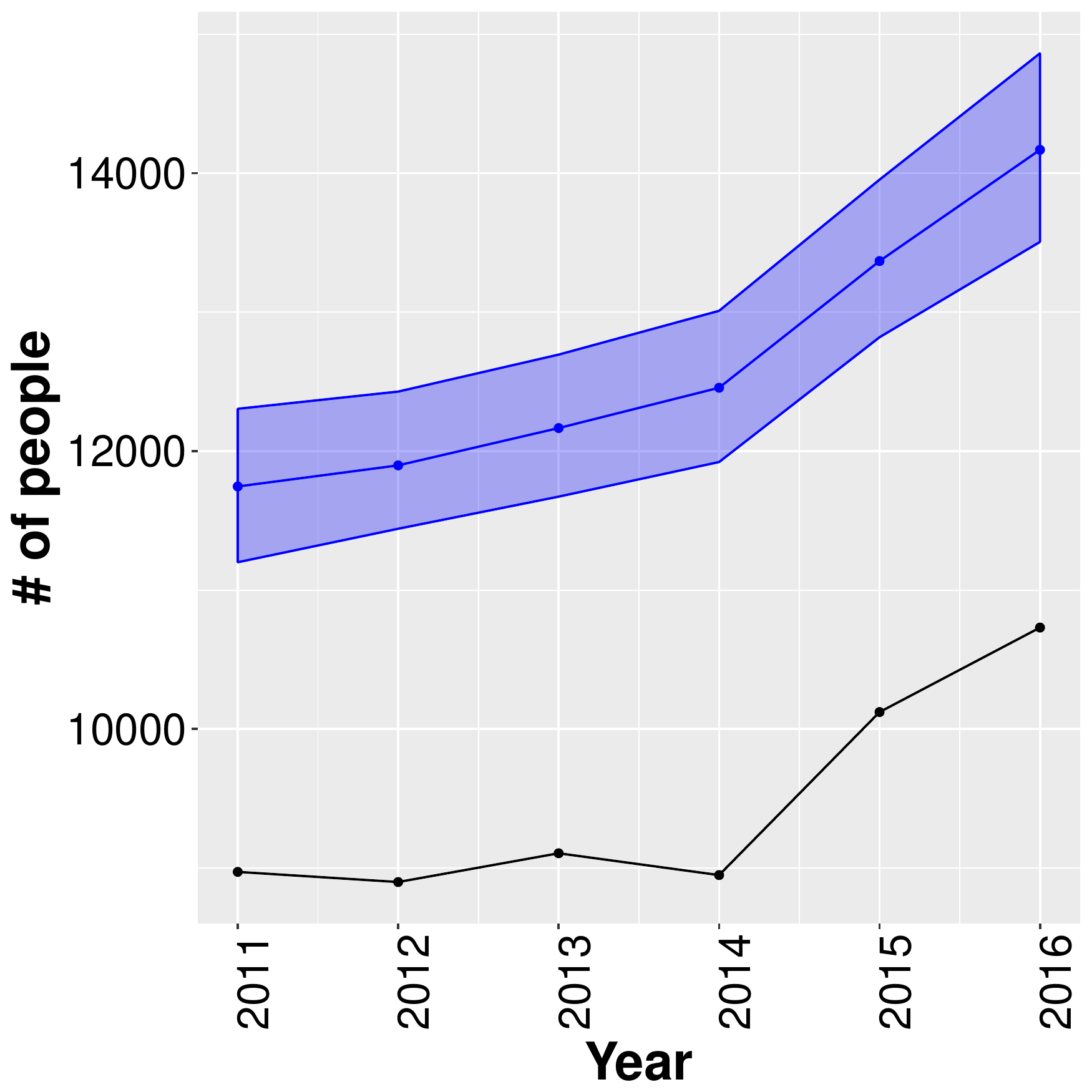}
    \caption{Counted number of homeless (black line) and imputed mean (blue line) and 95\% predicted interval (shaded blue) of the total number of homeless in King County.  In this illustration, we assume the expected count accuracy is constant over time when inferring the distribution of the total number of homeless.}
    \label{fig:Homeless_imputed}
\end{figure}

To account for the homeless not included in the HUD-reported count data, we impute the total homeless population in each metro from 2011-2016 and examine the impact of different trajectories in the count accuracy on the inferred homelessness rate.  Figure \ref{fig:Homeless_imputed} illustrates the distribution of the unobserved total number of homeless over time in King County, WA if we assume that the count accuracy does not improve with time and approximately 75\% of homeless are included in the count.  

It is reasonable to assume time variation in the count accuracy: in many continuums, the count accuracy may incrementally improve each year; in some continuums, the count accuracy could degrade over time due to lack of funding; in others, the accuracy may jump at a single year.  A primary objective of our study is to assess the impact of different trajectories in the count accuracy on the relationship between homelessness rates and changes in ZRI.  The model and prior distribution for different trajectories of the count accuracy will be discussed further in Sections \ref{subsec:Homeless_Count_Model} and \ref{subsec:Prior_CountEff}.   

\section{Model}
\label{sec:Model}
In this section, we develop a joint statistical model for collections of population-level and subpopulation counts.  For each metro, we model (i) the number of homeless counted, (ii) the true number of homeless, and (iii) the total number of people living in the metro.  Of the three quantities, only two are observed: the homeless counted and the total number of people.  The true number of people experiencing homelessness is not observed, and we treat it as missing data.  The total population of a metro (as reported by the Census) is modeled as a noisy observation of the true total population. 

\begin{figure}[ht!]
$$\vcenter{\xymatrix{
& \bar{\nu} \ar[r] & \nu_i \ar[ld] \ar[lld] \ar[rd] \ar[rrd] & & \\
\lambda_{i,1} \ar[d] \ar[r] & \lambda_{i,2} \ar[d] \ar[r] & \cdots \ar[r] & \lambda_{i,t-1} \ar[d] \ar[r] & \ar[d] \lambda_{i,t}\\
N_{i,1} \ar@/^2pc/[ddd] & N_{i,2} \ar@/^2pc/[ddd] & \cdots & N_{i,t-1} \ar@/^2pc/[ddd] & N_{i,t} \ar@/^2pc/[ddd] \\
& \bar{\phi} \ar[r] & \phi_i \ar[ld] \ar[lld] \ar[rd] \ar[rrd] & & \\
\psi_{i,1} \ar[d] \ar[r] & \psi_{i,2} \ar[d] \ar[r] & \cdots \ar[r] & \psi_{i,t-1} \ar[d] \ar[r] & \psi_{i,t} \ar[d] \\
H_{i,1} \ar@/^2pc/[dd]  & H_{i,2}  \ar@/^2pc/[dd] & \cdots & H_{i,t-1} \ar@/^2pc/[dd] & H_{i,t} \ar@/^2pc/[dd] \\
\pi_{i,1} \ar[d] & \pi_{i,2} \ar[d] & \cdots & \pi_{i,t-1} \ar[d] & \pi_{i,t} \ar[d] \\
C_{i,1} & C_{i,2} & \cdots & C_{i,t-1} & C_{i,t}\\
} }$$
\caption{Graphical model of the continuum-level homeless population.  In metro $i$ in year $t$, the total population is modeled by $N_{i,t}$, the homeless population is modeled by $H_{i,t}$, and the number of homeless counted is modeled by $C_{i,t}$.  The dynamical processes and associated parameters are outlined in Sections \ref{subsec:Total_Pop_Model} - \ref{subsec:Homeless_Count_Model}.}
\label{fig:Graph}
\end{figure}
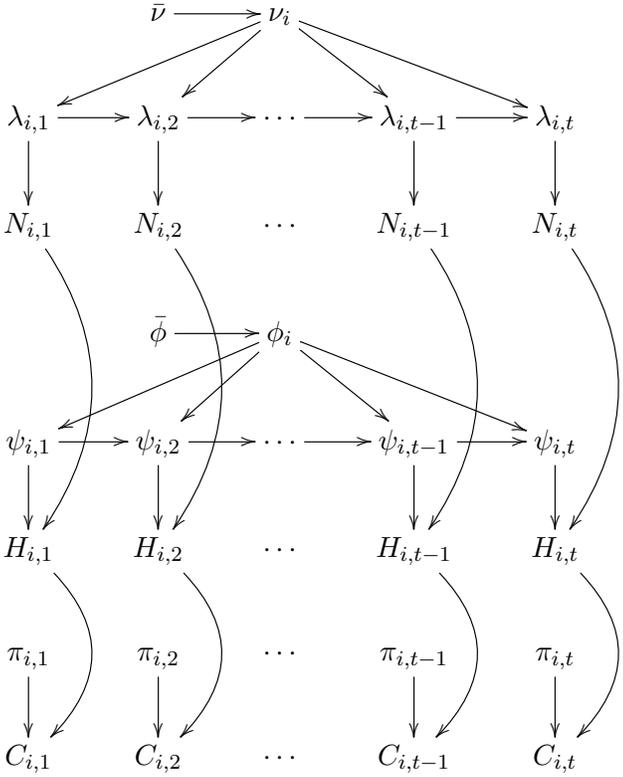

Figure \ref{fig:Graph} is the graphical representation of our dynamic Bayesian hierarchical model.  The random variable $N_{i,t}$ is the total number of people that live in metro $i$ in year $t$.  The $N_{i,1:T}$ variables depend on the dynamic process governing population growth, $\lambda_{i,1:T}$.  The expected increase in population from one year to the next in metro $i$ is modeled by parameter $\nu_i$, and the global population growth is modeled by parameter $\bar{\nu}$.  Section \ref{subsec:Total_Pop_Model} develops the total population model in greater detail.    

The total number of homeless, $H_{i,t}$, depends on $N_{i,t}$ and the probability of being homeless, $p_{i,t}$.  The log odds of homelessness, $\psi_{i,t} = \log \left( \frac{p_{i,t}}{1-p_{i,t}} \right)$, is modeled by a dynamic process that depends on changes in ZRI.  The effect of change in ZRI on log odds of homelessness is modeled hierarchically by parameter $\phi_i$ with global mean $\bar{\phi}$.  The full model for $H_{i,t}$ and the dynamics of $\psi_{i,t}$ are discussed in Section \ref{subsec:Homeless_Pop_Model}.  
The counted number of homeless, $C_{i,t}$, depends on $H_{i,t}$ and the probability that a homeless person is counted, $\pi_{i,t}$.  We call $\pi_{i,t}$ the count accuracy and discuss the count data generating process in Section \ref{subsec:Homeless_Count_Model}.    

\subsection{Total population model}
\label{subsec:Total_Pop_Model}
Significant interest lies in homelessness \textit{rates} that facilitate comparison across different metros.  The total population of the metro, or the denominator in a rate calculation, is uncertain.  The intercensal population estimates reported annually by the U.S. Census Bureau are noisy.  To properly quantify the uncertainty in the estimated homelessness rate of each metro, it is necessary to account for the uncertainty in the total population size.  

A secondary reason for modeling the total population is that it facilitates forecasting.  In order to forecast the size of the homeless population in future years, it is necessary to know the size of the future total population.  A dynamic model for the total population enables a model-based forecast of the homeless population.

The total population size for metro $i$ in year $t$, $N_{i,t}$, is modeled as a time-indexed Poisson random variable that allows for growth and decay in the population of each metro.
\begin{align}
   & N_{i,t} \sim Poisson(\lambda_{i,t}) \label{eq:N_it} \\
   &\lambda_{i,t} = \bar{\lambda}_i \theta_{i,t} \label{eq:lambda_it}
\end{align}

The Poisson rate, $\lambda_{i,t}$, is the product of a static scale factor, $\bar{\lambda}_i$, and a latent time-varying component $\theta_{i,t}$.  The dynamics of $\lambda_{i,t}$ are driven by a dynamic process on the unit interval, $\theta_{i,t} \in (0,1)$.  Modeling $\lambda_{i,t}$ as the product of the scaling factor and the dynamic term $\theta_{i,t}$ provides an intuitive and computationally tractable dynamic model for Poisson counts.  An auxiliary Poisson-Binomial thinning step for efficient computation is discussed in Section \ref{subsec:MCMC_Steps}. The unit-interval-constrained dynamic process $\theta_{i,1:T}$ is constructed with the logistic transformation and a real-valued stochastic process $\eta_{i,1:T}$.  
\begin{align}
    &\theta_{i,t} = \frac{e^{\eta_{i,t}}}{1 + e^{\eta_{i,t}}} \label{eq:theta_it}\\
    &\eta_{i,t} =\eta_{i,t-1} + \nu_{i} + v_{i,t}, \hspace{1cm} v_{i,t} \sim N(0,\sigma^2_{\eta_i}) \label{eq:eta_it}
\end{align}
The nonstationary $\eta_{i,1:T}$ process is a random walk with a metro-specific drift term, $\nu_i$.  The drift component is aimed at modeling population dynamics in cities like Seattle and Detroit.  In Seattle, the population is rapidly growing which would correspond to a positive drift ($\nu_i > 0$).  On the other hand, the population in Detroit has recently decreased, which would correspond to negative drift ($\nu_i < 0)$.  To borrow information across metros, we model the drift components hierarchically.  The parameter $\bar{\nu}$ may be interpreted as the expected drift in population across all metros.  
\begin{align}
    &\nu_{i} = \bar{\nu} + \epsilon_i, \hspace{1cm} \epsilon_i \sim N(0,\sigma^2_{\nu_i}) \\
    &\bar{\nu} \sim N(0,\sigma^2_{\bar{\nu}})
\end{align}

Because $\eta_{i,1:T}$ is nonstationary, the Poisson marginals $p(N_{i,t}|\lambda_{i,t}), \ldots, p(N_{i,T}|\lambda_{i,T})$ are not identically distributed.  Consequently, $N_{i,1:T}|\lambda_{i,1:T}$ is a nonstationary process for the total population counts.  While the PoINAR method of \citet{Aldor-Noiman2016} would be suitable for stationary population modeling, the expected total populations in these metro areas are clearly changing over time.  A second modeling alternative would be to transform the large counts with a natural logarithm and model the transformed response with a Gaussian dynamic model.  Despite the computational simplicity of such a model, we prefer to directly model the count data with discrete, time-varying distributions to more aptly characterize the uncertainty in the observed data.   

\subsection{Homeless population model}
\label{subsec:Homeless_Pop_Model}
In a metro with $N_{i,t}$ total residents, some small fraction of the residents will be homeless.  For this reason, it is natural to model the total number of people experiencing homelessness, $H_{i,t}$, with a time-indexed binomial distribution.  The binomial parameter $p_{i,t}$ is the unobserved probability that a person in metro $i$ is homeless in year $t$ (i.e., the homelessness rate). 
\begin{align}
    H_{i,t} | N_{i,t}, p_{i,t} \sim Binomial(N_{i,t}, p_{i,t}) \label{eq:H_it}
\end{align}
  
One of our primary objectives is to include ZRI as a covariate in the dynamic model for the homeless probability $p_{i,t}$.  We achieve this by modeling the log odds of homelessness.     
\begin{align}
    &p_{i,t} = \frac{e^{\psi_{i,t}}}{1 + e^{\psi_{i,t}}} \label{eq:p_it}\\
    &\psi_{i,t} = \psi_{i,t-1} + \phi_i \Delta ZRI_{i,t} + w_{i,t}, \hspace{1cm} w_{i,t} \sim N(0,\sigma^2_{\psi}) \label{eq:psi_it}
\end{align}
The dynamic process that controls the homelessness rate, $\psi_{i,1:T}$, linearly depends on the year-over-year rate of change in the ZRI, $\Delta ZRI_{i,t}$.
\begin{align}
    \Delta ZRI_{i,t} = \frac{ZRI_{i,t} - ZRI_{i,t-1}}{ZRI_{i,t-1}}
\end{align}
The regression coefficient $\phi_i$ models the relationship between change in rent levels and change in homelessness rates.  As a concrete example, if ZRI increases by $1\%$ in continuum $i$ from one year to the next, the expected log odds of homelessness will increase by $.01 \phi_i$. 

The connection between increased rental costs and homelessness rates is well established in the homelessness literature \citep{hanratty2017local, fargo2013, Culhane2013, stojanovic1999, oflaherty1995, sclar1990homelessness}).  To explicitly model this positive relationship, the regression coefficient $\phi_i$ is truncated at zero.  This guarantees that increasing rent levels result in higher homelessness rates.  The parameter $\phi_i$ is modeled hierarchically across metros to borrow strength and provide a more robust estimation of the impact of rent increases on homelessness. 
\begin{align}
    &\phi_i \sim N(\bar{\phi}, \sigma^2_{\phi_i}) \mathbbm{1}_{\phi_i > 0} \\
    &\bar{\phi} \sim N(m_{\bar{\phi}}, \sigma^2_{\bar{\phi}}) \mathbbm{1}_{\bar{\phi}>0}
\end{align}

As noted at the beginning of Section \ref{sec:Model}, $H_{i,t}$, is not observed.  Only the imperfect homeless count, $C_{i,t}$, is observed.  In our study, we treat $H_{i,t}$ as missing data and impute it to estimate each $\phi_i$.  By modeling the relationship between $\Delta ZRI_{i,t}$ and the imputed $H_{i,t}$, we obtain a more reliable quantification of the uncertainty in the posterior distribution for $\phi_i$ and $\bar{\phi}$.

\subsection{Homeless count model}
\label{subsec:Homeless_Count_Model}
Plant-capture studies and postcount surveys have demonstrated that homeless counts systematically understate the number of people experiencing homelessness \citep{Hopper2008}.  While single night counts are imperfect, it is not clear that there exist feasible alternatives.  Logistics, expenses, and privacy concerns preclude volunteers and continuums from counting every person without a home.    

We model the imperfection in the homeless counts with a binomial thinning step.  Of the true number of homeless, $H_{i,t}$, only $C_{i,t}$ of them are counted.  It is $C_{i,t}$ that we observe.    

\begin{align}
    C_{i,t} &\sim Binomial(H_{i,t}, \pi_{i,t}) \\
    \pi_{i,t} &\sim Beta(a_{i,t}, b_{i,t})
\end{align}

The parameter $\pi_{i,t}$ is the probability that a person experiencing homelessness is counted, and it is modeled with a beta distribution.  In other words, $\pi_{i,t}$ is the accuracy of the count.  If $\pi_{i,t} = 1$, the count in metro $i$ in year $t$ is perfectly accurate and every unsheltered person is counted.  

Observe that we model $\pi_{i,t}$ and $\pi_{i,t-1}$ as independent random variables.  While the count accuracy is surely time-varying and may exhibit trends, we believe there is no clear dependence of $\pi_{i,t}$ on $\pi_{i,t-1}$.  Factors driving count accuracy such as weather and volunteer turnout are unrelated across years.  Even the count methodology utilized by a continuum may change from one year to the next.  As an example, in 2017, the All Home King County continuum of care overhauled its count methodology to enhance the accuracy \citep{Beekman2016}.  Rather than sending volunteers to known areas where homeless congregate, as in previous years, volunteers covered each census tract in the county.  In addition, volunteers were lead by guides who were either currently or recently homeless themselves.  

As a result, the number of homeless counted in January 2017 was significantly higher than the number counted in January 2016.  Due to changes in methodology, it is not necessarily accurate to conclude that the size of the homeless population, $H_{i,t}$, dramatically increased.  By providing a mechanism for changes in count accuracy in each metro from one year to the next, it is possible to more reliably assess the local relationship between increased rental costs and the homeless population.

The count accuracy itself is an unknown quantity, and since we do not observe $H_{i,t}$, it is not possible to learn $\pi_{i,t}$.  Instead of trying to learn $\pi_{i,t}$, we marginalize it out so that $C_{i,t} | H_{i,t} \sim \mbox{Beta-binomial}(H_{i,t},a_{i,t}, b_{i,t})$. Despite the lack of an underlying dynamic model for the count accuracy, we examine the impact of different time trends in $E[\pi_{i,t}]$ on posterior inference for $\phi_i$.  The trends are achieved through specification of the $a_{i,t}$ and $b_{i,t}$ parameters.  Given the sequences of expected values and variances for $\pi_{i,t}$ -- which we assume are provided by the agencies conducting the counts -- the hyperparameters $a_{i,t}$ and $b_{i,t}$ may be computed from \eqref{eq:a_it} and \eqref{eq:b_it}.

\begin{align}
    a_{i,t} &= E[\pi_{i,t}] \left( \frac{(1-E[\pi_{i,t}])E[\pi_{i,t}]}{Var(\pi_{i,t})} - 1\right) \label{eq:a_it}\\
    b_{i,t} &= \frac{Var(\pi_{i,t})}{E[\pi_{i,t}]^2} \left( \frac{a_{i,t}^2}{E[\pi_{i,t}]} + a_{i,t} \right) \label{eq:b_it}
\end{align}
By modeling each $\pi_{i,t}$ with an independent beta distribution, it is possible to easily achieve different types of accuracy trajectories.  We discuss three trajectories of specific interest in Section \ref{subsec:Prior_CountEff}.  
       
\section{Prior Distributions}
\label{sec:Priors}
With limited data for each metro, it is critically important to elicit well-informed prior distributions.  Information from data that predates 2011, existing literature, and the expert opinion of homeless count coordinators have been combined to elicit prior distributions for four components of the model: (i) the count accuracy that is formalized through $\pi_{i,1:T}$ (Section \ref{subsec:Prior_CountEff}); (ii) the relationship between homelessness and rising rental costs, which is modeled with the regression coefficient $\phi_{i}$ (Section \ref{subsec:Prior_Phi}); (iii) the dynamic process $\eta_{i,1:T}$ that governs the total population (Section \ref{subsec:Prior_Eta}); and (iv) the dynamic process $\psi_{i,1:T}$ that governs the total homeless subpopulation dynamics (Section \ref{subsec:Prior_Psi}).  

\subsection{Priors for count accuracy}
\label{subsec:Prior_CountEff}
We base our prior distribution for count accuracy on a study by \citet{Hopper2008}, who report evidence that 60-70\% of unsheltered individuals in New York were visible and included in the city's 2005 count.  They discuss one plant-capture study where only 59\% of participants were counted.  The number of homeless used in our study includes \textit{sheltered} homeless as well.  \citet{Hopper2008} note that counts of sheltered homeless are more reliable than the counts of unsheltered homeless.  To elicit our prior for count accuracy, we compute a weighted average of accuracy for sheltered and unsheltered populations, respectively.  We use homeless counts from 2010, $C_{i,0}$, to compute this weighted average in year $t=0$.

\begin{align}
    E[\pi_{i,0}] = \left(1.0\right) \frac{C^{\mbox{ sheltered}}_{i,0}}{C_{i,0}} + \left(0.6\right) \frac{C^{\mbox{ unsheltered}}_{i,0}}{C_{i,0}}.
\end{align}

Our prior expectation is that the probability that a sheltered homeless person is included in the homeless count is unity.  From the Hopper study, we believe the probability that an unsheltered homeless person is included in the homeless count to be approximately $0.6$.  Because each metro has a different proportion of sheltered and unsheltered homeless, each metro is assigned a unique baseline prior distribution for count accuracy based on the 2010 data.  We develop prior distributions for $\pi_{i,1:T}$ that exhibit different expected trajectories: constant, linear, and step functions in time.  In each of these cases, $Var(\pi_{i,t}) = .0005$ is chosen so that reasonable prior mass covers the $E[\pi_{i,t}] \pm 0.05$ interval.   

\begin{figure}[ht!]
\centering
\begin{subfigure}{.3\textwidth}
  \centering
\includegraphics[width=1\textwidth]{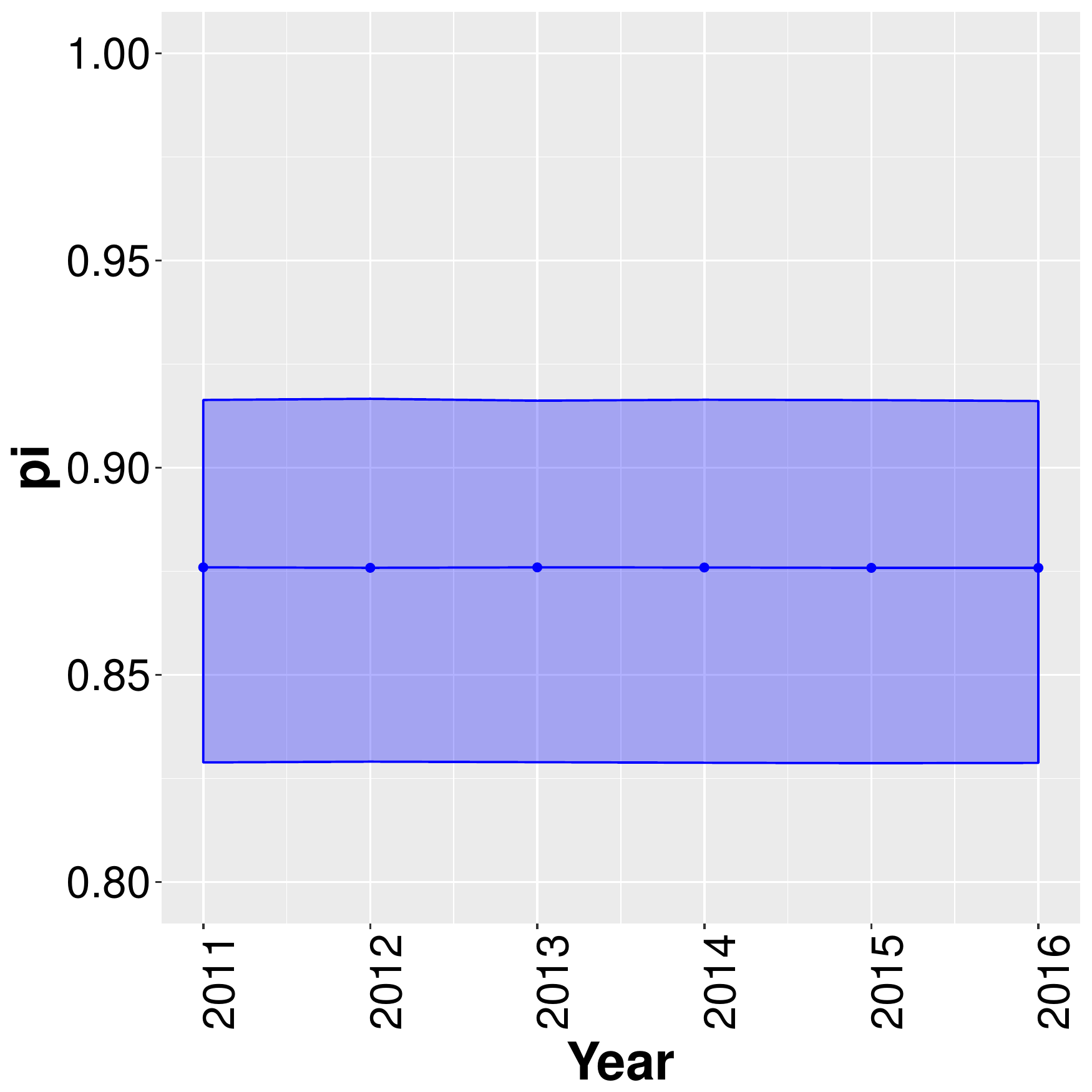}
\caption{Constant accuracy}
\label{subfig:pi_constant}
\end{subfigure}
\begin{subfigure}{.3\textwidth}
  \centering
\includegraphics[width=1\textwidth]{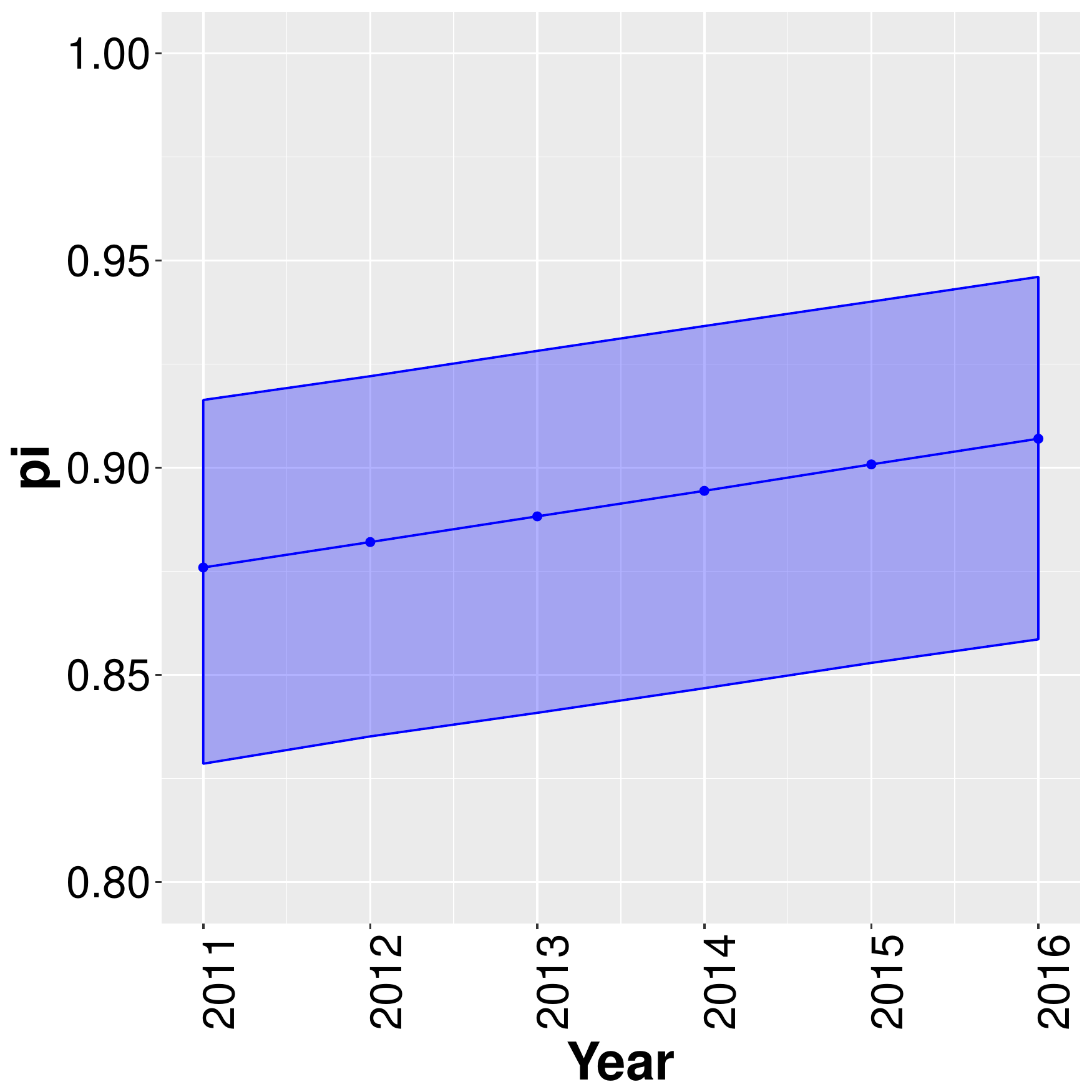}
\caption{Incremental accuracy gains}
\label{subfig:pi_linear}
\end{subfigure}
\begin{subfigure}{.3\textwidth}
  \centering
\includegraphics[width=1\textwidth]{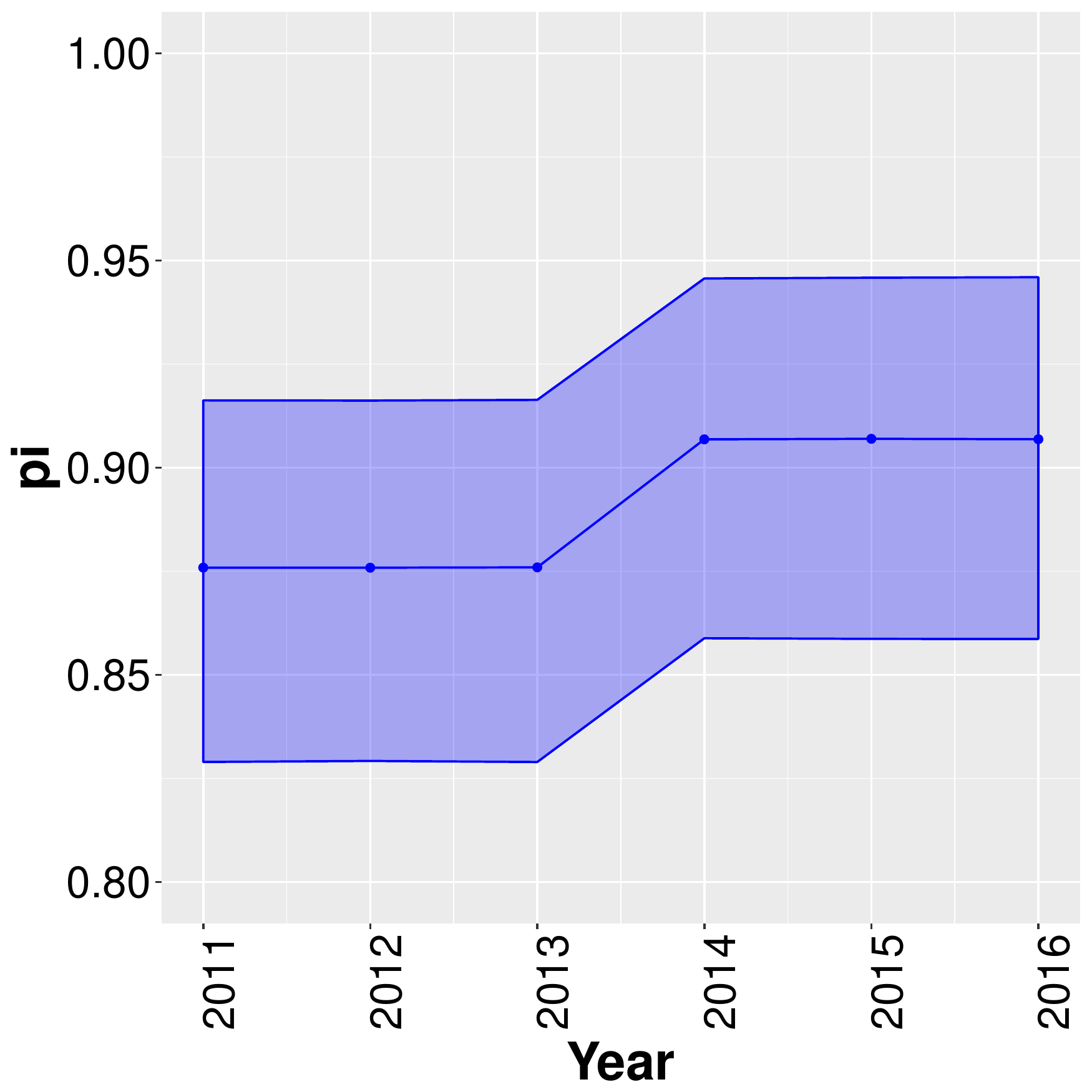}
\caption{Step in accuracy}
\label{subfig:pi_step}
\end{subfigure}
\caption{Different prior beliefs about the trajectory of $\pi_{i,1:T}$ in King County, WA.  The blue lines are the expected count accuracy over time, and the shaded blue interval corresponds to the 95\% prior uncertainty interval.  Left: constant count accuracy.  Middle: incremental (linear) increases in count accuracy.  Right: Step in count accuracy.}
\label{fig:accuracy_Prior}
\end{figure}

The constant case corresponds to a count that utilizes relatively consistent procedures and resources from one year to the next. In this case, the mean and variance of the accuracy are constant over time (for all $t$, $E[\pi_{i,t}] = E[\pi_{i,0}]$).  With $E[\pi_{i,t}]$ and $Var(\pi_{i,t})$, calculation of $a_{i,t}$ and $b_{i,t}$ follows directly from \eqref{eq:a_it} and \eqref{eq:b_it}.  The prior for $\pi_{i,1:T}$ with constant count accuracy in King County, WA is presented in Figure \ref{subfig:pi_constant}.
 
The linear case corresponds to a count where the accuracy incrementally improves by a fixed amount (called $\delta_i$) until it reaches one, as shown in \eqref{eq:pi_it}.  We assume that $\delta_i$ is known and is ideally specified by the agency conducting the count.  Alternatively, $\delta_i$ can be adjusted to examine sensitivity of inference to different accuracy scenarios.  This is the approach adopted here.  As an example, to consider an increase of $\bar{\delta}$ in the accuracy of the unsheltered homeless count, $\delta_i$ is computed as in \eqref{eq:delta_i}.  We assume that sheltered homeless are perfectly counted, and the improvement in accuracy of $\bar{\delta}$ applies only to the unsheltered count.  

\begin{align}
    E[\pi_{i,t+1}] &= \min(E[\pi_{i,t}] + \delta_i,1) \label{eq:pi_it} \\
    \delta_i &= \bar{\delta} \left( \frac{C^{\mbox{ sheltered}}_{i,0}}{C_{i,0}}  \right) \label{eq:delta_i}
\end{align} 

In the step scenario, the accuracy dramatically increases at a specific point in time due to improved count methodology.  This is observed in practice with the All Home King County continuum as discussed in Section \ref{subsec:Homeless_Count_Model}.  In this case, we assume that the year of change for metro $i$, $\tau_i$, is known.  For $t < \tau_i$, $E[\pi_{i,t}] = E[\pi_{i,0}]$.  For $t \geq \tau_i$, $E[\pi_{i,t}] = E[\pi_{i,0}] + \delta_i$.  A step in $E[\pi_{i,t}]$ occurs at time $\tau_i$. Figure \ref{subfig:pi_step} illustrates a hypothetical step in count accuracy for King County in 2014.  It is possible that there could be multiple steps for each metro and that there exists sequences $\tau_i^1, \tau_i^2, \ldots$ and $\delta_i^1, \delta_i^2, \ldots$ where steps of different size occur in different years.  Because we do not know $\tau_i$ for each metro, we do not investigate the step scenario further; however, with consultation of each local coordinator, this may be a very promising area of future work.  

\subsection{Priors for \texorpdfstring{$\phi_{i}$}{} and \texorpdfstring{$\bar{\phi}$}{}}
\label{subsec:Prior_Phi}
We use previous work by \citet{Culhane2013} to form the basis of our prior distribution for $\bar{\phi}$.  \citet{Culhane2013} found that in metropolitan continuums, when median rent increased by \$100, the expected homelessness rate increased by 6.34\%.  The average log odds of homelessness across all continuums in 2010 was $\bar{f}_0:= -5.5$.  The average ZRI across continuums in 2010 was \$1534.  So a \$100 increase in median rent would translate to a percent change in ZRI of $\frac{100}{1534} \approx 6.5\%$.  This leads to calculation of the expectation of $\bar{\phi}$ based on $\bar{f}_0$, the 6.5\% increase in ZRI, and the expected increase in the homelessness rate of 6.34\%:  
\begin{align}
    \frac{1 + \mbox{exp}\left\{ -\bar{f}_0 \right\}}{1 + \mbox{exp}\left\{ -\bar{f}_0 - \frac{\$100}{\$1534} m_{\bar{\phi}} \right\}} = 1.0634.
\end{align}

\begin{figure}[ht!]
\centering
\includegraphics[width=.4\textwidth]{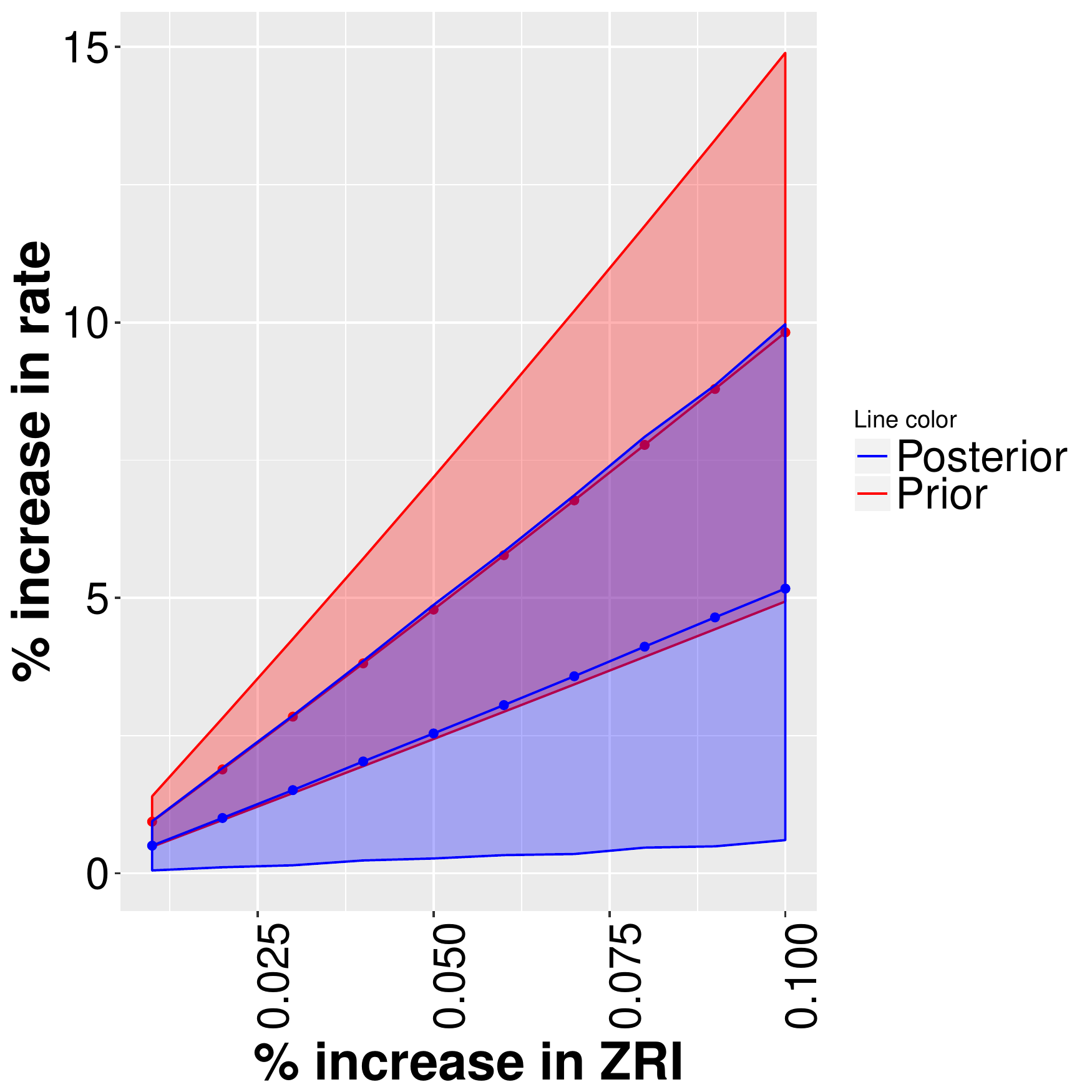}
\caption{Implied prior and posterior distribution of \% change in homelessness rate with increases in ZRI for an arbitrary metro $i$.  The red (blue) line is the prior (posterior) mean, and the shaded red (blue) region is the 95\% prior (posterior) credible interval.} 
\label{fig:HomelessRate_Prior}
\end{figure}

We calculate that $m_{\bar{\phi}} = 0.94$. Because of differences in methodology and data, we use $\sigma^2_{\bar{\phi}} = 0.005$ so that there is reasonable prior uncertainty about $\bar{\phi}$.  We let $\sigma^2_{\phi_i} = 0.05$ so that there is modest shrinkage of each local effect toward the global mean $\bar{\phi}$.  To examine the prior uncertainty in the relationship between increases in ZRI and increases in homelessness implied by our choices of $m_{\bar{\phi}}$, $\sigma^2_{\bar{\phi}}$, and $\sigma^2_{\phi_i}$, we simulate from the marginal prior distribution for percent changes in the homelessness rate (see Figure \ref{fig:HomelessRate_Prior}).  

Although we inform our prior using the results from \citet{Culhane2013}, whose methodology we are trying to advance, notice a few things in Figure \ref{fig:HomelessRate_Prior}.  One is that our prior is diffuse, and it becomes increasingly diffuse with larger percent increases in ZRI.  Second, the inferred posterior concentrates on different values than the prior, indicating that we are indeed learning from data.  The conclusion is that using the Byrne et al. result is a useful way to center our prior.

\subsection{Prior for \texorpdfstring{$\eta_{i,1:T}$}{}}
\label{subsec:Prior_Eta}
The sampling distribution for the total population, $N_{i,t}$, depends on $\eta_{i,t}$ through the Poisson rate, $\lambda_{i,t}$ (refer to \eqref{eq:N_it} - \eqref{eq:theta_it}).  Recall that $\lambda_{i,t}$ is the product of a scaling factor, $\bar{\lambda}_i$, and a dynamic process on the unit interval, $\theta_{i,t}$.  We let the expectation of $\lambda_{i,0}$ be the 2010 population, $N_{i,0}$.  This is achieved by fixing $\bar{\lambda}_i = 2 \times N_{i,0}$ and $E[\theta_{i,0}] = 0.5$ (i.e., $E[\eta_{i,0}] = 0$).  The prior variance of $\eta_{i,0}$ is fixed to be $0.0001$, as we are confident that the Poisson rate of the total population in 2010 is the observed total population.

\begin{figure}[ht!]
\centering
\begin{subfigure}{.3\textwidth}
  \centering
\includegraphics[width=1\textwidth]{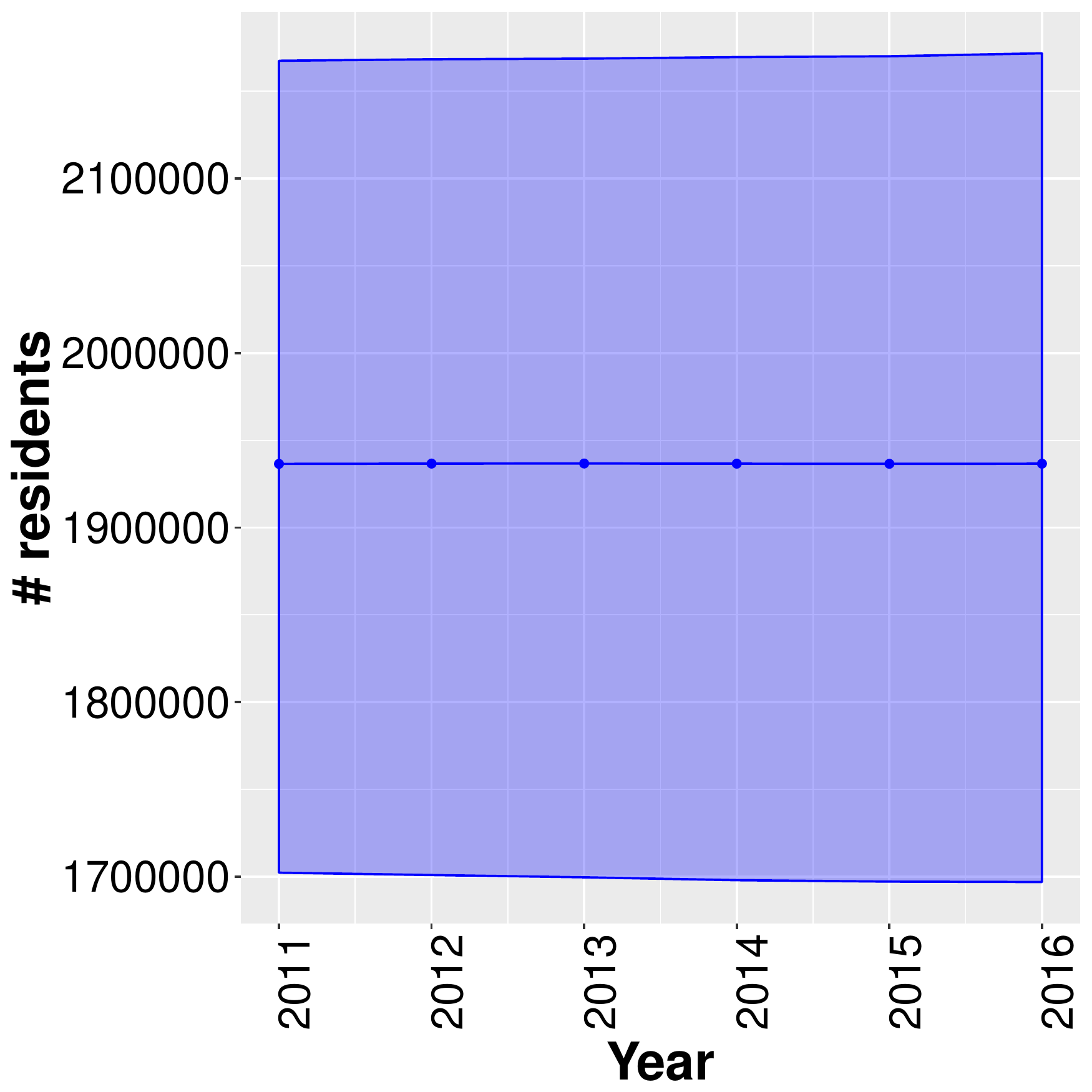}
\caption{Total continuum population}
\label{subfig:Population_Prior}
\end{subfigure}
\begin{subfigure}{.3\textwidth}
  \centering
\includegraphics[width=1\textwidth]{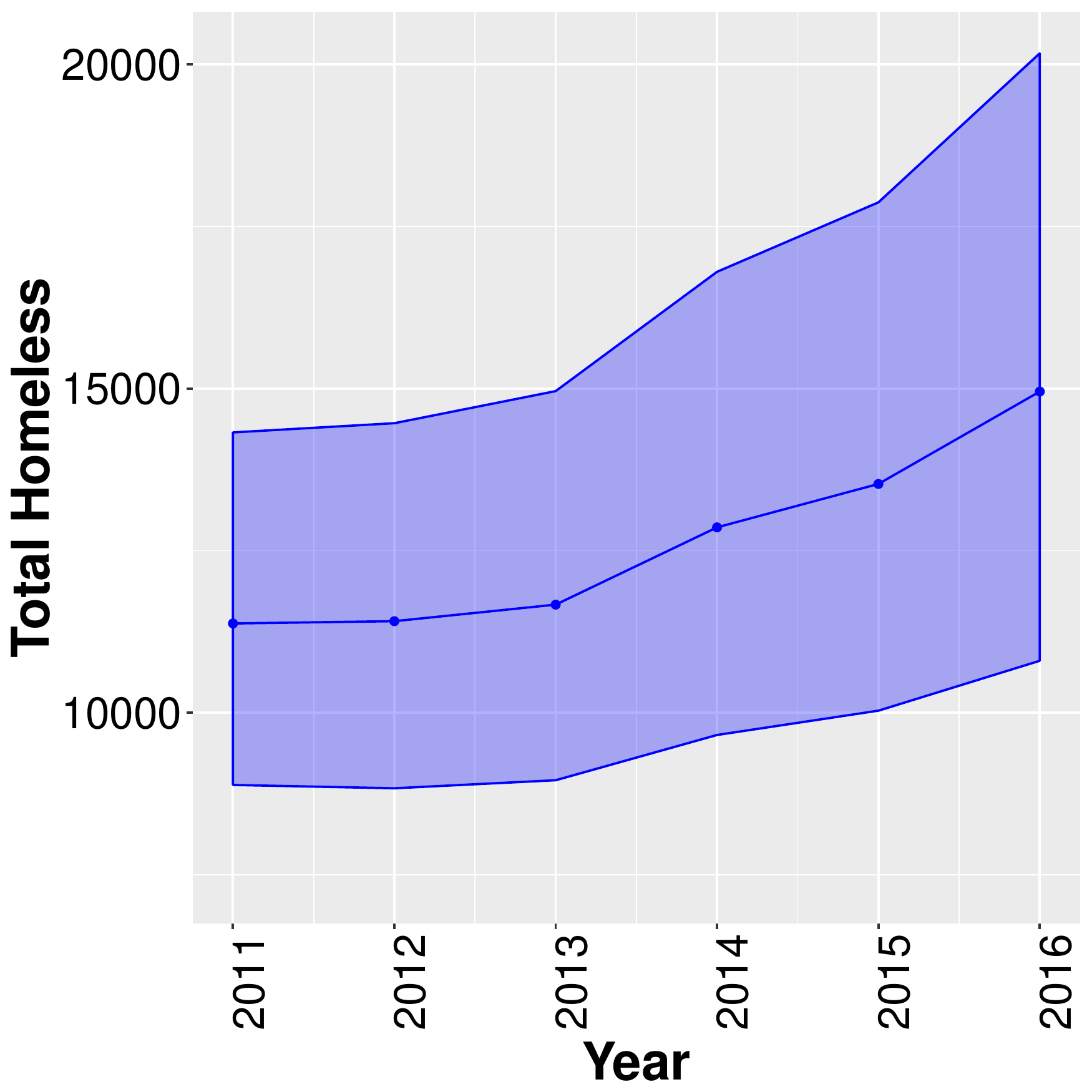}
\caption{Total homeless population}
\label{subfig:Homeless_prior}
\end{subfigure}
\caption{Marginal prior distributions for $N_{i,t}$ and $H_{i,t}|ZRI_{1:T}$ in King County, WA.  The blue lines are the the prior means and the shaded blue regions are the 95\% prior uncertainty intervals.  Left: implied prior distributions for the total population, $N_{i,1:T}$.  Right: Prior for total homeless population, $H_{i,1:T}|ZRI_{1:T}$.  The upward trend in the implied prior for the total homeless population is due to observed increases in ZRI.}
\label{fig:Population_Homeless_Prior}
\end{figure}

We let $\nu_i \sim N(\bar{\nu}, 0.01)$ and $\bar{\nu} \sim N(0,0.005)$.  The innovation variance of the $\eta_{i,1:T}$ process is fixed to be $0.0001$ so that $\nu_i$ primarily drives changes in the Poisson rate.  The implied marginal distribution of $N_{i,1:T}$ in King County, WA is presented in Figure \ref{subfig:Population_Prior}.  Observe that the distribution is centered at the 2010 King County population and allows significant uncertainty over the six year period.  While the prior variance on each $\eta_{i,t}$ is relatively small, the large magnitude of the scaling factor, $\bar{\lambda}_i$, results in a relatively diffuse marginal distribution for $N_{i,t}$.    

\subsection{Prior for \texorpdfstring{$\psi_{i,1:T}$}{}}
\label{subsec:Prior_Psi}
We utilize the counted number of homeless in 2010 to specify the prior expectation $E[\psi_{i,0}]$.  The conditionally binomial sampling distribution for $H_{i,0} | \psi_{i,0}, N_{i,0}$ in \eqref{eq:H_it} yields the expectation $E[H_{i,0}|\psi_{i,0}, N_{i,0}] = \frac{1}{1+e^{-\psi_{i,0}}} N_{i,0}$.  Solving for $\psi_{i,0}$ results in \eqref{eq:psi_i0}.

\begin{align}
    \psi_{i,0} = \log \left( \frac{E[H_{i,0}|\psi_{i,0}]/N_{i,0}}{1 - E[H_{i,0}|\psi_{i,0}]/N_{i,0}} \right) \label{eq:psi_i0}
\end{align}

Because we observe the noisy total population $N_{i,0}$ from the Census estimate in 2010, we can compute a value for $\psi_{i,0}$ given the expectation $E[H_{i,0}|\psi_{i,0}]$.  Though we do not observe $E[H_{i,0}|\psi_{i,0}]$, we use an approximation to center the prior distribution of $\psi_{i,0}$ and compensate for the approximation with moderate prior uncertainty.  We approximate the expected total number of homeless in 2010 as an expected inflation of the observed count, $E[H_{i,0}|\psi_{i,0}] \approx E\left[\frac{1}{\pi_{i,0}}\right] C_{i,0}$.  $C_{i,0}$ is the 2010 homeless count value and $E[1/\pi_{i,0}]$ is the expectation of the reciprocal count accuracy in 2010.  The multiplier $E[1/\pi_{i,0}]$ is evaluated by Monte Carlo simulation.  This leads to a prior expectation as defined in \eqref{eq:E_psi_i0}.

\begin{align}
    E[\psi_{i,0}] := \log \left( \frac{E\left[\frac{1}{\pi_{i,0}}\right] C_{i,0}/N_{i,0}}{1 - E\left[\frac{1}{\pi_{i,0}}\right] C_{i,0}/N_{i,0}} \right) \label{eq:E_psi_i0}
\end{align}

The time zero variance is chosen to be $\sigma^2_{\psi_0} = 0.01$, as this provides a three standard deviation interval of $\pm0.3$ around the prior mean. The result is that $\psi_{i,0} \sim N(E[\psi_{i,0}], \sigma^2_{\psi_0})$.  One reason that the prior specification of $\psi_{i,0}$ is important is that the implied prior distribution for the total homeless population, $H_{i,t}$, depends on $\psi_{i,t}$ (refer to \eqref{eq:H_it} and \eqref{eq:p_it}). The implied prior distribution for $H_{i,1:T}|ZRI_{1:T}$ is presented in Figure \ref{subfig:Homeless_prior}.  The 95\% prior interval spans a very reasonable range for each $H_{i,t}$. 

The innovation variance of the dynamic process is fixed to be $\sigma^2_{\psi} = 0.001$ (see \eqref{eq:psi_it}).  We observe in synthetic data experiments that the innovation variance of $\psi_{i,1:T}$ must be small in order to accurately learn $\phi_i$ and $\bar{\phi}$.  If $\sigma^2_{\psi}$ is large relative to  $Var\left(\phi_{i} \Delta ZRI_{i,t}\right)$, changes in the homelessness rate are modeled as noise in $\psi_{i,t}$ rather than driven by changes in ZRI.  The ratio $\frac{Var(\phi_i \Delta ZRI_{i,t})}{\sigma^2_{\psi}}$ can be thought of as a signal-to-noise ratio for $\psi_{i,t}$.  

We also observe in synthetic data experiments that reliably inferring $\psi_{i,1:T}$ and $\phi_i$ requires that a metro's homelessness rate exceed 0.05\% of the total population.  Because $p_{i,t}$ is the logistic transformation of $\psi_{i,t}$ (see \eqref{eq:p_it}), the derivative $\frac{dp_{i,t}}{d\psi_{i,t}} \rightarrow 0$ as $|\psi_{i,t}|$ increases.  Flat tails of $p_{i,t}$ as a function of $\psi_{i,t}$ mean that in metros with very low homeless rates, practically observed changes in homeless counts are consistent with a wide range of changes in ZRI.  Under such conditions, it is not possible to reliably estimate $\phi_i$.  Inference on $\phi_i$ degrades along the continuum of decreasing $\psi_{i,t}$, but we set a limit based on our empirical studies.  We do not trust inference for metros where the homelessness rate is less than $0.05\%$, or when $\psi_{i,t} < -7.6$.    

\section{Markov chain Monte Carlo}
\label{sec:MCMC}
Our objective is to sample from the posterior distribution
\begin{align}
    p(H_{1:25,1:T}, \eta_{1:25,1:T}, \psi_{1:25,1:T}, \phi_{1:25}, \bar{\phi}, \nu_{1:25}, \bar{\nu} | N_{1:25,1:T}, C_{1:25,1:T}). \label{eq:posterior}
\end{align} 
To sample from the posterior, we develop a custom \Polya-Gamma Gibbs sampler for dynamic Bayesian logistic regression \citep{PSW2013, Windle2013, Windle2014}.  The \Polya-Gamma augmentation strategy allows us to harness a forward filtering and backward sampling (FFBS) algorithm that is commonly used to fit Bayesian dynamic models \citep{fruhwirth-schnatter1994, carter1994}.  We found that a burn-in of 15,000 samples and 25,000 samples collected after burn-in were sufficient for reproducible inferences.  The MCMC simulation took approximately 4 hours to run on a MacBook Pro.

\subsection{Sampling steps}
\label{subsec:MCMC_Steps}
There are ten different sampling steps required in the MCMC algorithm.  The first step is for an auxiliary random variable whose only purpose is to facilitate computation when $N_{i,t} | \tilde{\lambda}_i, \theta_{i,t}$ $\sim Poisson(\tilde{\lambda}_i \theta_{i,t})$ (refer to \eqref{eq:N_it} and \eqref{eq:lambda_it}).  To construct this marginal distribution, we model the auxiliary $Z_{i,t}$ $\sim Poisson(\tilde{\lambda}_i)$ and the observed $N_{i,t}$ conditionally binomial, $N_{i,t} | Z_{i,t}, \theta_{i,t}$ $\sim Binomial(Z_{i,t}, \theta_{i,t})$.  The Binomial-Poisson thinning strategy results in the desired marginal distribution for $N_{i,t}$ and a computationally tractable method for making inference on $\eta_{i,t}$, $\nu_{i}$, and $\bar{\nu}$.  The full conditional for the auxiliary $Z_{i,t}$ is shown in \ref{one}.  

\ref{two} and \ref{six} use \Polya-Gamma data augmentation to allow a forward filtering backward sampling strategy.  \ref{three} and \ref{seven} sample the auxiliary \Polya-Gamma variables $\omega_{i,t}$ and $\zeta_{i,t}$. The collection of auxiliary variables $Z_{1:25,1:T}$, $\omega_{1:25,1:T}$, and $\zeta_{1:25,1:T}$ are numerically integrated out from the posterior by discarding posterior samples.  Each sampling step is outlined below.  

\begin{enumerate}[ref={Step~\arabic*}]
    \item For each $i,t$, sample the auxiliary $Z_{i,t}$ from a shifted Poisson by first sampling\\ $j = Z_{i,t} - N_{i,t} | \tilde{\lambda}_i, \theta_{i,t}$ $\sim Poisson\left( (1-\theta_{i,t}) \tilde{\lambda}_i \right)$ and then fixing $Z_{i,t} = j + N_{i,t}$. \label{one}
    \item \label{two} For each $i$, sample the dynamic process that governs total population growth, $\eta_{i,1:T}|N_{i,1:T},\omega_{i,1:T}$, with an FFBS algorithm. 
    \begin{enumerate}
        \item compute forward filtered distribution $\eta_{i,t} | N_{i,1:t}, Z_{i,1:t}, \nu_i, \omega_{i,1:t} \sim N(m_{i,t}, S_{i,t})$
          \begin{itemize}
            \item $S_{i,t} := \left( \omega_{i,t} + \frac{1}{S_{i,t-1} + \sigma^2_{\eta} } \right)^{-1}$
            \item $m_{i,t} := S_{i,t} \left(N_{i,t} - \frac{1}{2} Z_{i,t} + \frac{m_{i,t-1} + \nu_i}{S_{i,t-1} + \sigma^2_\nu }\right)$
        \end{itemize}
        \item sample recursively $\eta_{i,t} | \eta_{i,t+1}, N_{i,1:t}, \omega_{i,1:t} \sim N(\tilde{m}_{i,t}, \tilde{S}_{i,t})$
        \begin{itemize}
            \item $\tilde{S}_{i,t} := \left(\frac{1}{S_{i,t}} + \frac{1}{\sigma^2_{\eta}} \right)^{-1}$
            \item $\tilde{m}_{i,t} := \tilde{S}_{i,t} \left( \frac{m_{i,t}}{S_{i,t}} + \frac{\eta_{i,t+1} - \nu_i}{\sigma^2_{\eta}} \right)$
        \end{itemize}
    \end{enumerate}
    \item \label{three} For each $i,t$, sample the auxiliary \Polya-Gamma random variates to augment the total population variable, $\omega_{i,t} | Z_{i,t}, \eta_{i,t} \sim PG(Z_{i,t},\eta_{i,t})$.
    \item \label{four} For each $i$, sample the parameter controlling expected population growth in metro $i$, $\nu_i | \bar{\nu}, \eta_{i,1:T} \sim N(\tilde{m}_{\nu_i}, \tilde{\sigma}^2_{\nu_i})$.
    \begin{itemize}
        \item $\tilde{\sigma}^2_{\nu_i} := \left( \frac{1}{C_0 + \sigma^2_{\eta} } + \frac{T-1}{\sigma^2_{\eta}} + \frac{1}{\sigma^2_{\nu_i}} \right)^{-1}$
        \item $\tilde{m}_{\nu_i} := \tilde{\sigma}^2_{\nu_i} \left(\frac{\eta_{i,1}}{C_0 + \sigma^2_{\eta}} + \frac{1}{\sigma^2_{\eta}} \sum_{t=2}^T (\eta_{i,t} - \eta_{i,t-1}) + \frac{\bar{\nu}}{\sigma^2_{\nu_i}} \right)$
    \end{itemize}
  \item \label{five} Sample the expected total population growth globally across metros,\\ $\bar{\nu}|\nu_{1:25} \sim N\left( \left(\frac{N}{\sigma^2_{\nu_i}} + \frac{1}{\sigma^2_{\bar{\nu}}} \right)^{-1}\frac{1}{\sigma^2_{\nu_i}} \sum_{i=1}^{25} \nu_i, \left(\frac{N}{\sigma^2_{\nu_i}} + \frac{1}{\sigma^2_{\bar{\nu}}} \right)^{-1} \right)$.
  \item \label{six} For each $i$, sample the dynamic process for the log odds of homelessness, $\psi_{i,1:T}|N_{i,1:T},H_{i,1:T}, \phi_i,\omega_{i,1:T}$, with an FFBS algorithm.
    \begin{enumerate}
        \item compute forward filtered distribution $\psi_{i,t} | N_{i,1:t}, H_{i,1:t}, \zeta_{i,1:t} \sim N(f_{i,t}, q_{i,t})$
          \begin{itemize}
            \item $q_{i,t} := \left( \zeta_{i,t} + \frac{1}{q_{i,t-1} + \sigma^2_{\psi} } \right)^{-1}$
            \item $f_{i,t} := q_{i,t} \left(H_{i,t} - \frac{1}{2} N_{i,t} + \frac{f_{i,t-1} + \phi_i \Delta ZRI_{i,t}}{q_{i,t-1} + \sigma^2_{\psi} }\right)$
        \end{itemize}
        \item sample recursively $\psi_{i,t} | \psi_{i,t+1}, N_{i,1:t}, H_{i,1:t}, \zeta_{i,1:t} ,\phi_i \sim N(\tilde{f}_{i,t}, \tilde{q}_{i,t})$
        \begin{itemize}
            \item $\tilde{q}_{i,t} := \left(\frac{1}{q_{i,t}} + \frac{1}{\sigma^2_{\psi}} \right)^{-1}$
            \item $\tilde{f}_{i,t} := \tilde{q}_{i,t} \left( \frac{f_{i,t}}{q_{i,t}} + \frac{\psi_{i,t+1} - \phi_i \Delta ZRI_{i,t}}{\sigma^2_{\psi}} \right)$
        \end{itemize}
    \end{enumerate}
    \item \label{seven} For each $i,t$, sample the auxiliary \Polya-Gamma random variates to augment the total homeless variable, $\zeta_{i,t} | N_{i,t}, \psi_{i,t} \sim PG(N_{i,t},\psi_{i,t})$.
    \item \label{eight} For each $i$, sample the parameter governing the relationship between change in ZRI and change in homelessness in metro $i$, $\phi_i | \psi_{i,1:T}, \bar{\phi} \sim N\left( m_{\phi_i}, \Sigma_{\phi_i} \right) \mathbbm{1}_{\phi_i>0}$.
    \begin{itemize}
        \item $\Sigma_{\phi_i} := \left( \frac{(\Delta ZRI_{i,1})^2}{\sigma^2_{\psi_0} + \sigma^2_{\psi}} + \frac{\sum_{t=2}^T(\Delta ZRI_{i,t})^2}{\sigma^2_{\psi}} + \frac{1}{\sigma^2_{\phi}} \right)$
        \item $m_{\phi_i} := \Sigma_{\phi_i} \left( \frac{\bar{\phi}}{\sigma^2_{\phi_i}} + \frac{\Delta ZRI_{i,1} (\psi_{i,1} - f_{i,0})}{\sigma^2_{\psi_0} + \sigma^2_{\psi_i} } + \frac{\sum_{t=2}^T\Delta ZRI_{i,t} (\psi_{i,t} -\psi_{i,t-1})}{\sigma^2_{\psi_i} }   \right)$
    \end{itemize}
    \item \label{nine} Sample the global mean parameter for the change in ZRI and change in homelessness, $\bar{\phi}|\phi_{1:25} \sim N\left( \left(\frac{25}{\sigma^2_{\phi_i}} + \frac{1}{\sigma^2_{\bar{\phi}}} \right)^{-1} \left( \frac{1}{\sigma^2_{\phi_i}} \sum_{i=1}^{25} \phi_i + \frac{m_{\bar{\phi}}}{\sigma^2_{\bar{\phi}}} \right), \left(\frac{25}{\sigma^2_{\phi_i}} + \frac{1}{\sigma^2_{\bar{\phi}}} \right)^{-1} \right)$. 
    \item For each $i,t$, sample the total number of people experiencing homelessness in metro $i$ and year $t$, $H_{i,t}$, from $p(H_{i,t} | N_{i,t}, C_{i,t}, p_{i,t}, \pi_{i,t}) \propto \binom{H_{i,t}}{C_{i,t}}\pi_{i,t}^{C_{i,t}} (1-\pi_{i,t})^{(H_{i,t} - C_{i,t})} \binom{N_{i,t}}{H_{i,t}} p_{i,t}^{H_{i,t}} (1-p_{i,t})^{(N_{i,t} - H_{i,t})} $. \label{ten}
  \end{enumerate}

\ref{ten} requires sampling from a discrete distribution with support $[C_{i,t},N_{i,t}]$.  This large range creates a computational bottleneck as it involves evaluating densities at each value in the support.  In practice, though, posterior probability is concentrated on values much closer to the lower end of the support.  It is possible to speed up computation by setting a threshold after which the support is truncated.  Once posterior probability falls below $1\times 10^{-8}$, we stop evaluating the densities and truncate the support.

\subsection{Posterior Predictive Distributions}
\label{subsec:Posterior_Predictive}
To examine the impact of increased ZRI on total homeless populations and counts, we utilize the posterior predictive distribution for the total homeless population in each metro, $H_{i,t}|C_{1:25,1:T},N_{1:25,1:T}$.  The main quantify of interest is the distribution of the increase in the homeless population when the observed change in ZRI, $\Delta ZRI_{i,t}$, increases by an $x>0$.  The increase is modeled by $\left(H_{i,t}^{x} - H_{i,t}\right)$ $| C_{1:25,1:T},N_{1:25,1:T}$, which is the difference between the predicted homeless total for a change in ZRI of $\Delta ZRI_{i,t} + x$ and the baseline prediction at $\Delta ZRI_{i,t}$.  

We draw samples from this posterior with a three step procedure that approximates the integral:  \begin{align}\small
    &p\left(H_{i,t}^{x} - H_{i,t}| N_{1:25,1:T},C_{1:25,1:T}\right) \nonumber \\
    &=\int p\left(H_{i,t}^{x} - H_{i,t} | \psi_{i,t}, \phi_{i}, N_{1:25,1:T},C_{1:25,1:T}\right) p\left(\psi_{i,t}, \phi_{i} \right) d \psi_{i,t} d \phi_i \label{eq:Homeless_Diff}.
\end{align}
The procedure relies on the $m^{th}$ posterior sample of (i) the relationship between ZRI and homelessness, $\phi_i^{(m)}$, (ii) the log odds of homelessness, $\psi_{i,t}^{(m)}$, and (iii) the Census reported estimate of the total population $N_{i,t}$.  The procedure is detailed below. 

\begin{enumerate}
    \item Construct the $m^{th}$ sample of log odds of homelessness where $\Delta ZRI_{i,t}$ is increased by $x$.
    \begin{equation}
        \psi_{i,t}^{(m),x} = \psi_{i,t}^{(m)} + \phi_i^{(m)} x
    \end{equation}
    \item Generate a prediction for the total homeless population at $\Delta ZRI_{i,t} + x$ by sampling 
    \begin{equation}
        H_{i,t}^{(m),x} \sim Binomial(N_{i,t}, p_{i,t}^{x}) \label{eq:H_Pred}
    \end{equation} 
    where $p_{i,t}^{x}$ is the same logistic transformation of $\psi_{i,t}^{x}$ as in \eqref{eq:p_it}.  
    \item Compute the difference 
    \begin{equation}
        H_{i,t}^{(m),x} - H_{i,t}^{(m)} 
    \end{equation} 
\end{enumerate}

We go one step further and also examine the predicted change in counted homeless under increased ZRI, $\left(C_{i,t}^{*,x} - C_{i,t}^*\right) | C_{1:25,1:T},N_{1:25,1:T}$.  Samples from this distribution are drawn by thinning the $m^{th}$ MCMC samples $H_{i,t}^{(m),x}$ and $H_{i,t}^{(m)}$ with a binomial step 
\begin{align}
C_{i,t}^{(m),*,x} \sim Binomial\left(H_{i,t}^{(m),x}, \pi_{i,t}^{(m)}\right) \label{eq:Count_Diff}
\end{align}
and computing the difference $C_{i,t}^{(m),*,x} - C_{i,t}^{(m),*}$.  The count accuracy is integrated out by sampling $\pi_{i,t}^{(m)} \sim Beta(a_{i,t}, b_{i,t})$ from the prior distribution.

\subsection{Reproducible MCMC inference}
\label{subsec:MCMC_Reproducibility}
\begin{figure}[!ht]
    \centering
    \includegraphics[width = .4\textwidth]{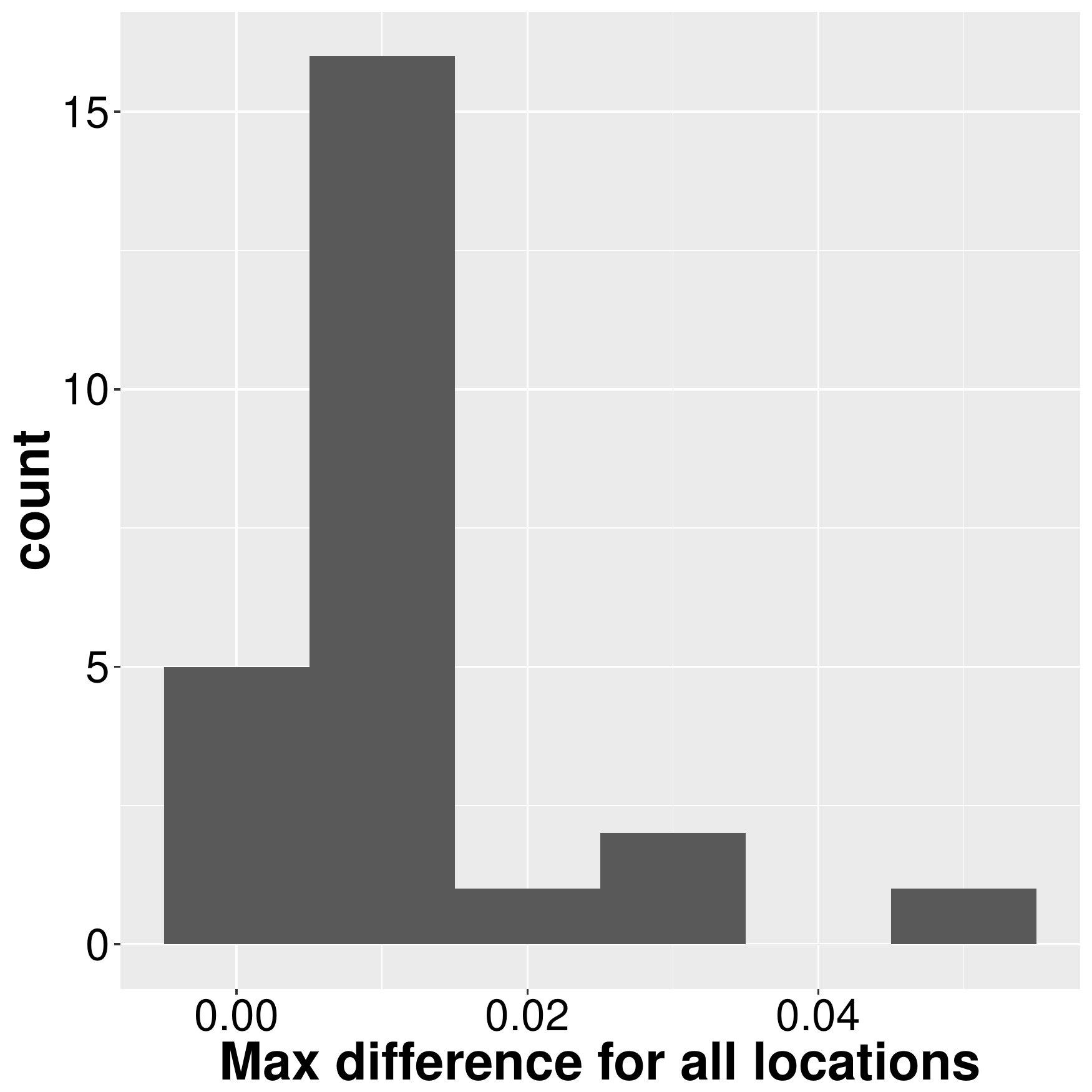}
    \caption{The maximum difference in posterior means of $\phi_1, \ldots, \phi_{25}$ for 10 different MCMC simulations.  Each of the 25 points in the histogram corresponds to $\max_{j} |\phi_i^{(j)} - \phi_i^{(1)}|, $ for each of the 25 metros denoted by $i$.  The superscript index $j$ denotes the MCMC simulation.  The very small differences indicate that our MCMC simulations generate reproducible inference in $\phi_i$.}
    \label{fig:Reproducible_Phi}
\end{figure}

We verify that our MCMC simulation generates reproducible inference about the relationship between increases in homelessness and increases in ZRI by examining the posterior distribution for $\phi_i$.  Ten different MCMC simulations are run, and inferences from two simulations $j$ and $j'$ are compared by computing $|E[\phi_{i}^{(j)}] - E[\phi_i^{(j')}]|$.  Figure \ref{fig:Reproducible_Phi} illustrates the largest deviation across simulations by computing $\max_{j}|E[\phi_i^{(j)}] - E[\phi_i^{(1)}]|$ for each metro.  Each point in the histogram corresponds to the largest difference in posterior mean in reference to the first simulation for each of the 25 metros.  The small values of these maximum differences in Figure \ref{fig:Reproducible_Phi} give us confidence that our MCMC simulation generates reproducible inferences.  

\section{Results}
\label{sec:Results}
We seek to answer five questions, \ref{Q1} - \ref{Q5}.  Each of these questions is answered (in order) in Sections \ref{subsec:Results_Q1} - \ref{subsec:Results_Q5}.  In Section \ref{subsec:Results_Q1}, we examine changes in homelessness rates from 2011-2016 across all metropolitan areas.  In Section \ref{subsec:Results_Q2}, the inferred relationship between increased ZRI and increases in homelessness is presented.  Posterior predictive distributions for additional homeless counts are presented in Section \ref{subsec:Results_Q3}, and the imputed distributions for the total number of homeless in each metro are presented in Section \ref{subsec:Results_Q4}.  Though the one-night-counts for January 2017 have already occurred, the results have not been fully released.  Section \ref{subsec:Results_Q5} discusses our forecasts for the total homeless populations in 2017.  

\subsection{Percent changes in the homelessness rate}
\label{subsec:Results_Q1}
\begin{figure}[!ht]
    \centering
        \begin{subfigure}{.4\textwidth}
        \centering
         \includegraphics[width = 1\textwidth]{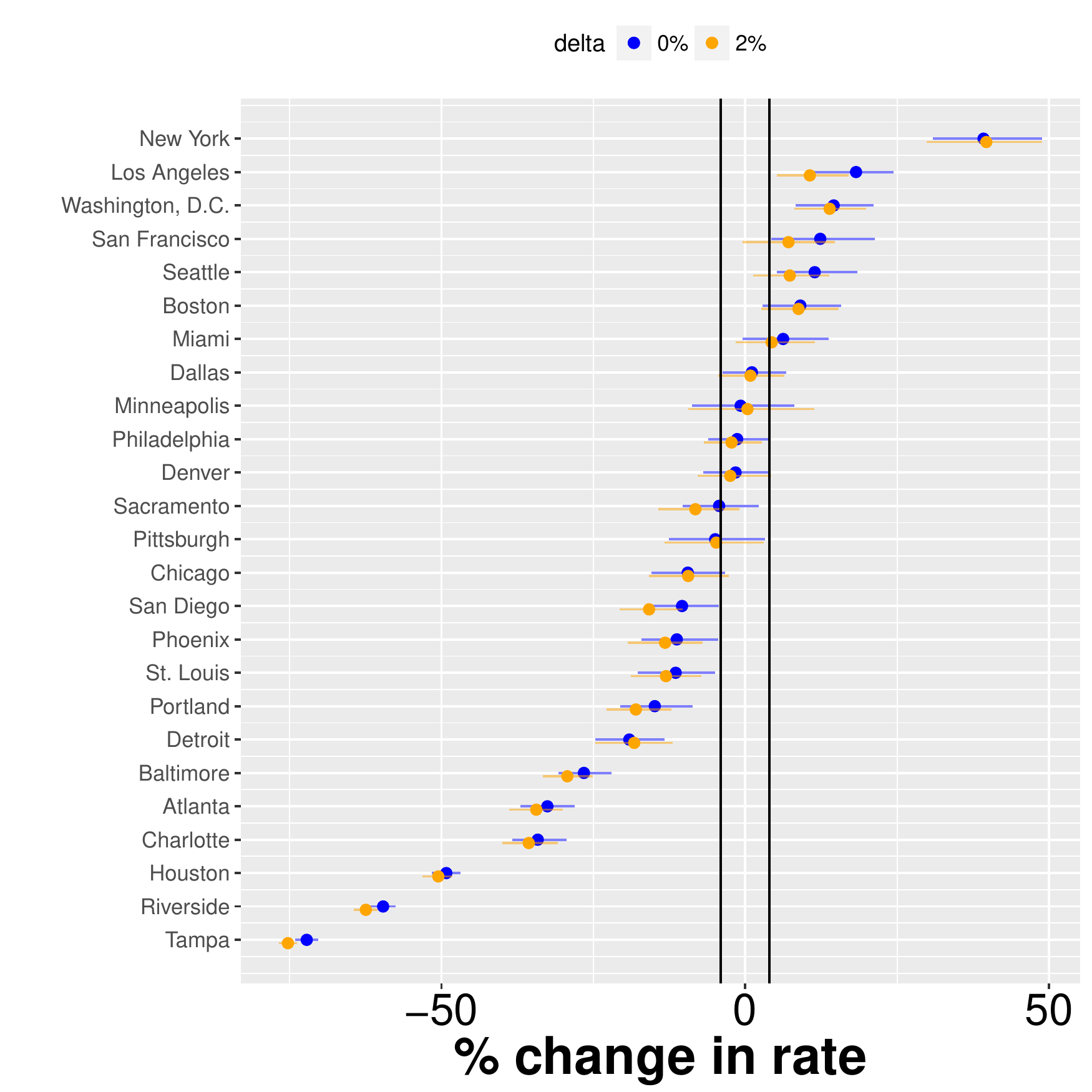}
        \caption{Rate Change 2011-2016}
        \label{subfig:Rate_Change_11_16}
        \end{subfigure}
        \begin{subfigure}{.4\textwidth}
        \centering
        \includegraphics[width = 1\textwidth]{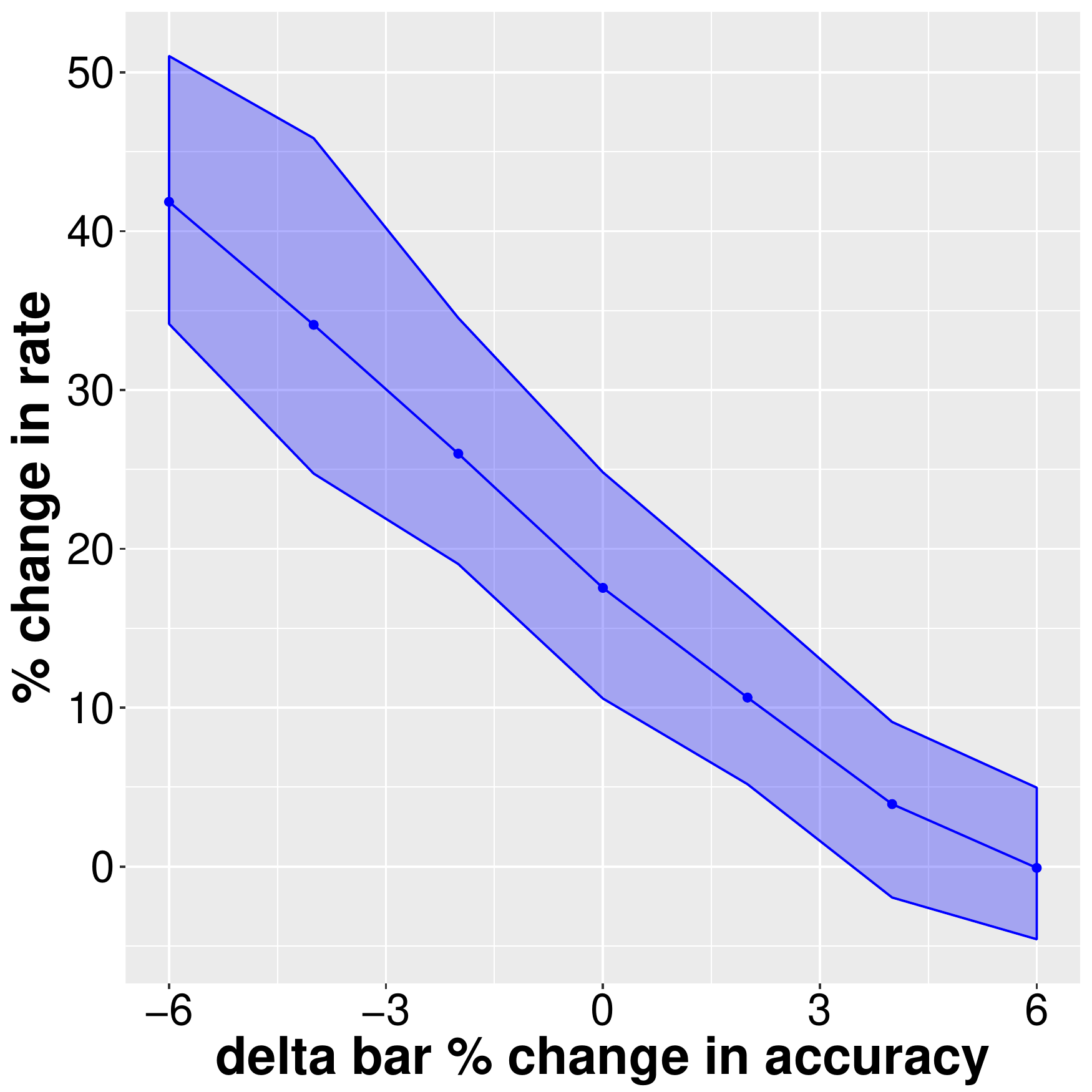}
        \caption{LA Rate Sensitivity}
        \label{subfig:LA_Rate_Sensitivity}
        \end{subfigure}
    \caption{Left: Posterior distribution in percent change in homelessness rate from 2011 to 2016.  The middle point in each segment is the posterior mean and the line segment encompasses the 95\% posterior credible interval.  For each metro, there is a posterior presented when the count accuracy is modeled with a constant mean over time ($\bar{\delta}=0$ in blue) and a posterior when the count accuracy of unsheltered homeless improves by 2\% each year ($\bar{\delta} = .02$ in orange).  The solid vertical lines mark the $\pm 4\%$ boundaries for distinguishing significant increases and decreases in the homelessness rate. Right: Sensitivity of the percent increase in homelessness rate from 2011 - 2016 to different choices of $\bar{\delta}$ in Los Angeles.}
\end{figure}

The inferred increases in homelessness rates from 2011 - 2016 are illustrated in Figure \ref{subfig:Rate_Change_11_16}.  We present results under two scenarios for the trajectory of the count accuracy: (i) the mean of the count accuracy is constant over time (i.e. $\bar{\delta}$ in \eqref{eq:delta_i} is zero); and (ii) the mean of the unsheltered count accuracy increases by $2\%$ annually until it reaches 100\% (i.e. $\bar{\delta}=.02$).

Metros where the rate of homelessness increased by at least 4\% include New York, Los Angeles, Washington, D.C., San Francisco, and Seattle when $\bar{\delta}=0$.  For these cities, the 95\% posterior credible interval for the percent change in the homelessness rate is bounded below by 4\%.  In response to its growing homeless population, the City of Seattle has declared an official state of emergency \citep{Beekman2015}.  We adopt this moniker and characterize these metros as in similar states of emergency.

Metros where the homelessness rate has decreased by at least $4\%$ include San Diego, Phoenix, St. Louis, Portland, Detroit, Baltimore, Atlanta, Charlotte, Houston, Riverside, and Tampa when $\bar{\delta}=0$.  For these cities, the 95\% posterior credible interval for the percent change in the homelessness rate is bounded above by $-4\%$.  It seems as though real progress has been made in reducing homelessness in this group, with the caveat that our method does not account for homeless relocation.  It is possible that people experiencing homelessness relocate to another nearby metro or continuum of care.  Though homelessness in one continuum may decrease, it may increase in another. We view relocation and network effects as outside the scope of our present study. 

A third group of cities exists where the percent change in the homelessness rate has neither significantly increased nor decreased in either scenario.  For these cities, the 95\% posterior credible interval is not bounded away from the $\pm4\%$ interval.  These cities are Boston, Miami, Dallas, Minneapolis, Philadelphia, Denver, Sacramento, Pittsburgh, and Chicago.  The situation remains largely unchanged in this group, and the current homelessness rate is the status quo.        

Observe in Figure \ref{subfig:Rate_Change_11_16} that New York and Los Angeles exhibit different sensitivities to change in the count accuracy over time.  In New York, a city with a predominantly sheltered population, the inferred percent increase is essentially unchanged between the two scenarios.  In Los Angeles, a warm-weather city with a large unsheltered population, the difference between the $\bar{\delta}=0$ and $\bar{\delta}=0.02$ cases is large, as demonstrated by separation of the posterior means. Equation \eqref{eq:delta_i} demonstrates that, in metros with large unsheltered populations, a $\bar{\delta}$ increase in the accuracy of an unsheltered homeless count leads to large changes in the overall count accuracy, $\pi_{i,t}$.

\begin{figure}[!ht]
    \centering
        \begin{subfigure}{.4\textwidth}
        \centering
        \includegraphics[width = 1\textwidth]{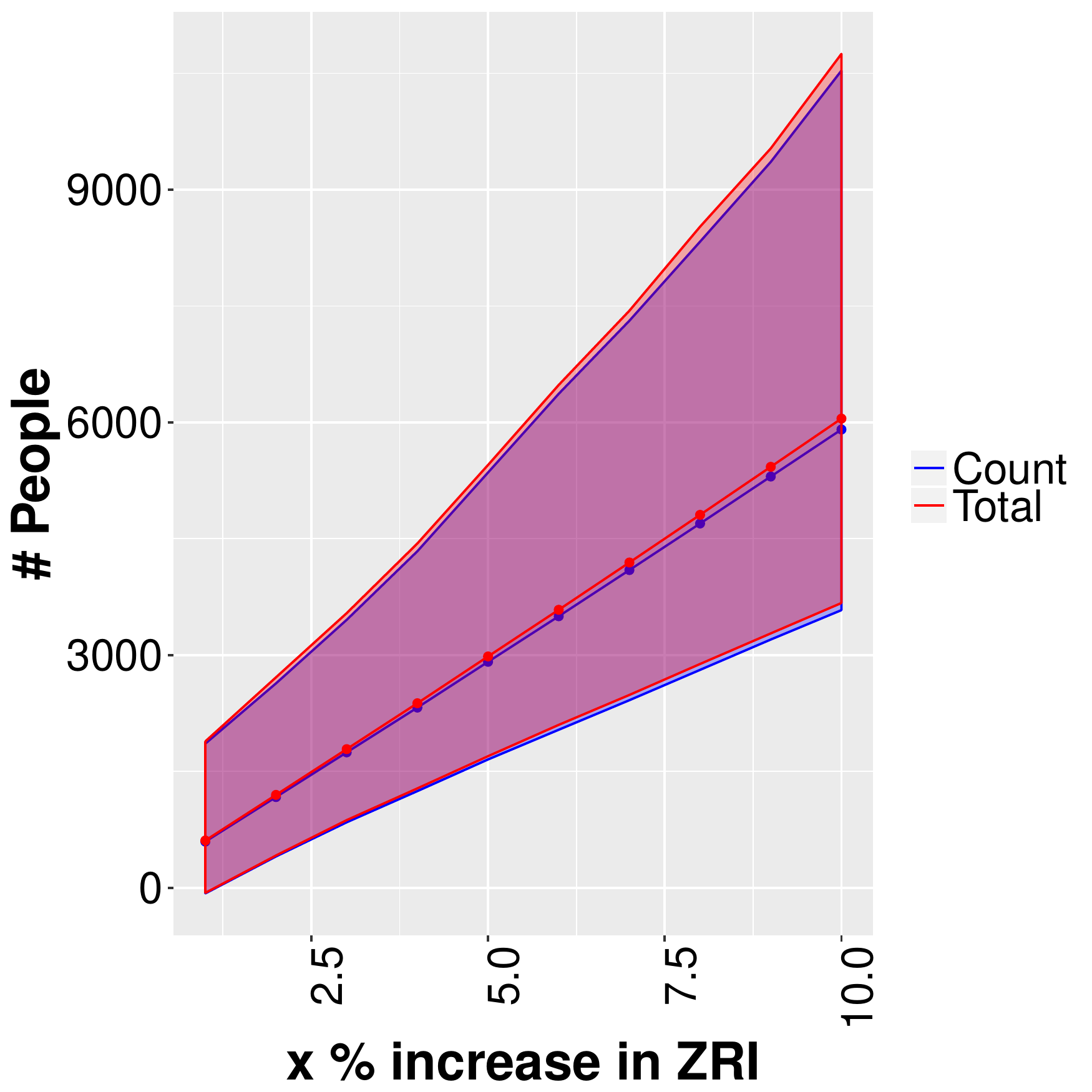}
        \caption{New York}
        \label{subfig:NY_Homeless_Increase}
        \end{subfigure}
        \begin{subfigure}{.4\textwidth}
        \centering
         \includegraphics[width = 1\textwidth]{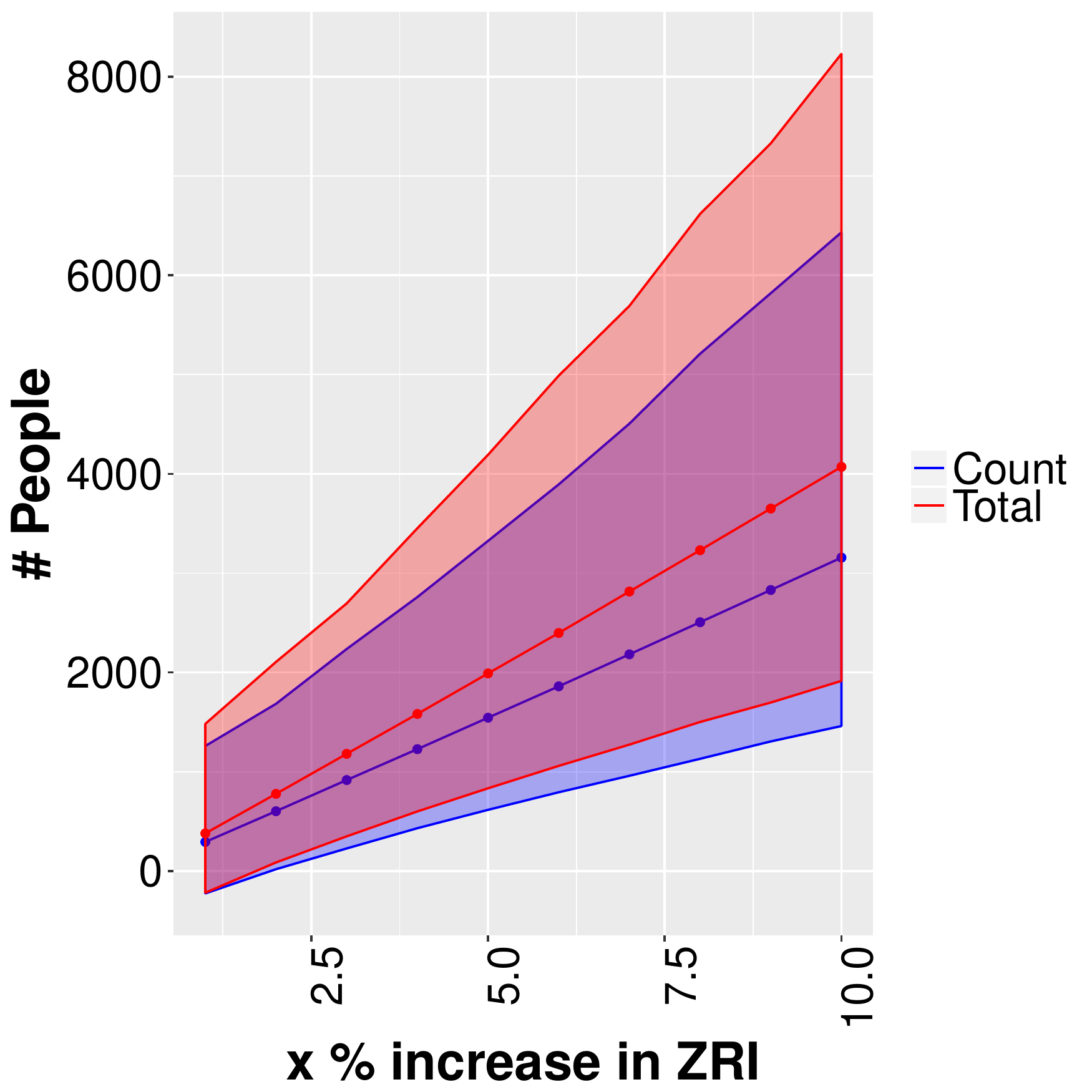}
        \caption{Los Angeles}
        \label{subfig:LA_Homeless_Increase}
        \end{subfigure}
    \caption{Posterior predictive distributions of increased homeless counts ($C_{i,t}^{*,x} - C_{i,T}^* | C_{1:25,1:T},N_{1:25,1:T}$) and total homeless populations ($H_{i,t}^{x} - H_{i,T} | C_{1:25,1:T},N_{1:25,1:T}$) associated with increases in ZRI for both New York and Los Angeles when $\bar{\delta} = 0$.  The blue lines and shaded intervals correspond to the posterior mean and 95\% predictive interval of increases in the homeless count.  The red lines and shaded intervals correspond to the posterior mean and 95\% predictive interval of increases in the total homeless population.}
    \label{fig:Homeless_Increase}
\end{figure}

In Figure \ref{subfig:LA_Rate_Sensitivity}, we examine how inference on the change in homelessness rates from 2011 - 2016 can change with different values of $\bar{\delta}$.  We focus in on Los Angeles, above considered in a 
\lq\lq state of emergency\rq\rq and see that if $\bar{\delta} = 4\%$ instead of 2\%, the posterior credible interval contains 0; in this case, we are unable to confidently say that the homelessness rate increased over this time period.  The conclusion again is that any inferences drawn about changes in homeless populations are highly sensitive to assumptions about the count accuracy.  Sensitivity analyses similar to the one presented in Figure \ref{subfig:LA_Rate_Sensitivity} are presented for each metro in Appendix \ref{app:A}.

\subsection{Rental costs and homelessness}
\label{subsec:Results_Q2}
To examine the predicted increase in homelessness and homeless counts as ZRI increases, we focus on the posterior predictive distributions $\left(H_{i,t}^{x} - H_{i,t}\right)$ $| C_{1:25,1:T},N_{1:25,1:T}$ and $\left(C_{i,t}^{*,x} - C_{i,t}^*\right) $ $|C_{1:25,1:T},N_{1:25,1:T}$. Section \ref{subsec:Posterior_Predictive} provides complete details for sampling from these distributions.  

\begin{figure}[ht!]
\centering
\begin{subfigure}{.4\textwidth}
  \centering
\includegraphics[width=1\textwidth]{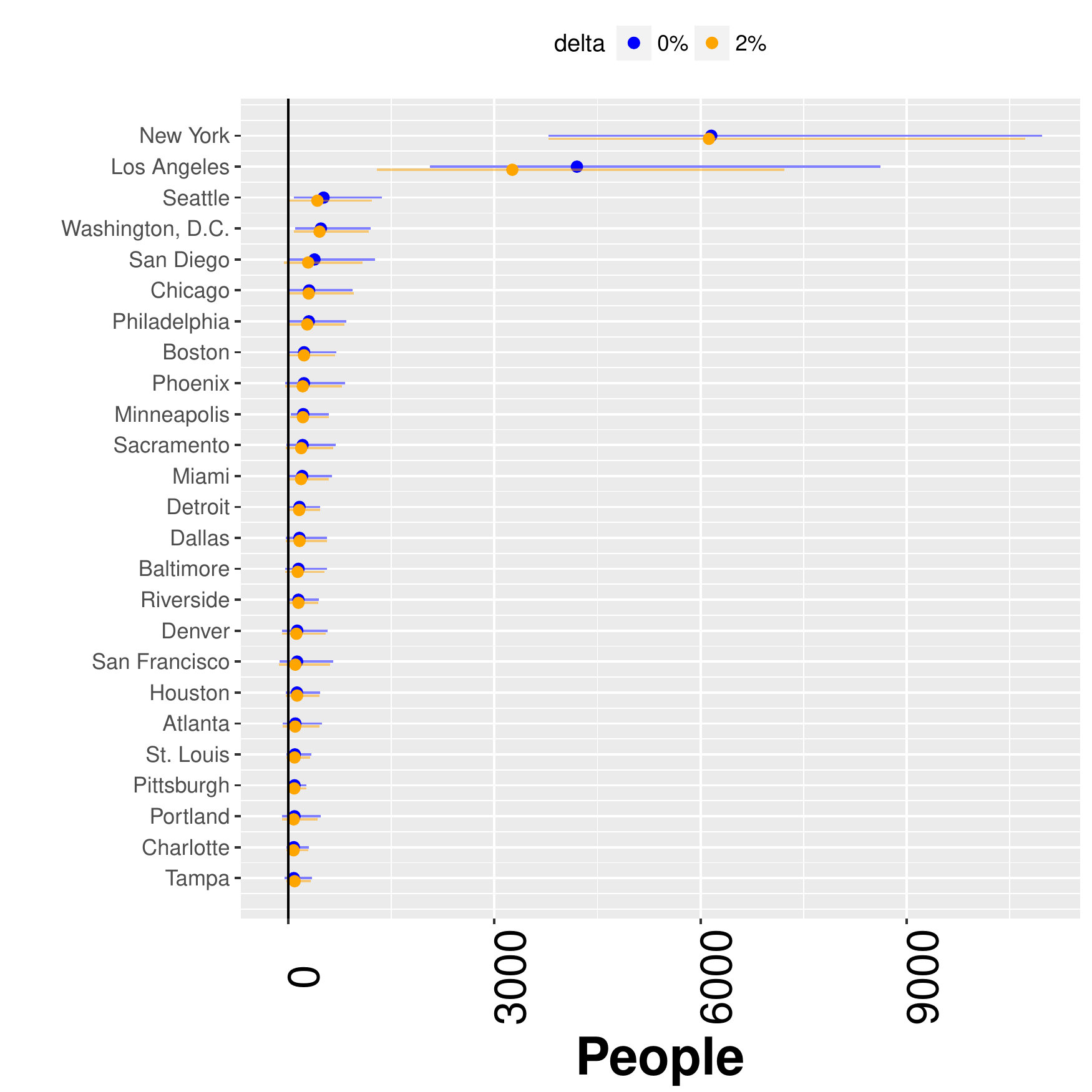}
\caption{25 largest metros}
\label{subfig:Summary_ZRI_increase}
\end{subfigure}
\begin{subfigure}{.4\textwidth}
  \centering
\includegraphics[width=1\textwidth]{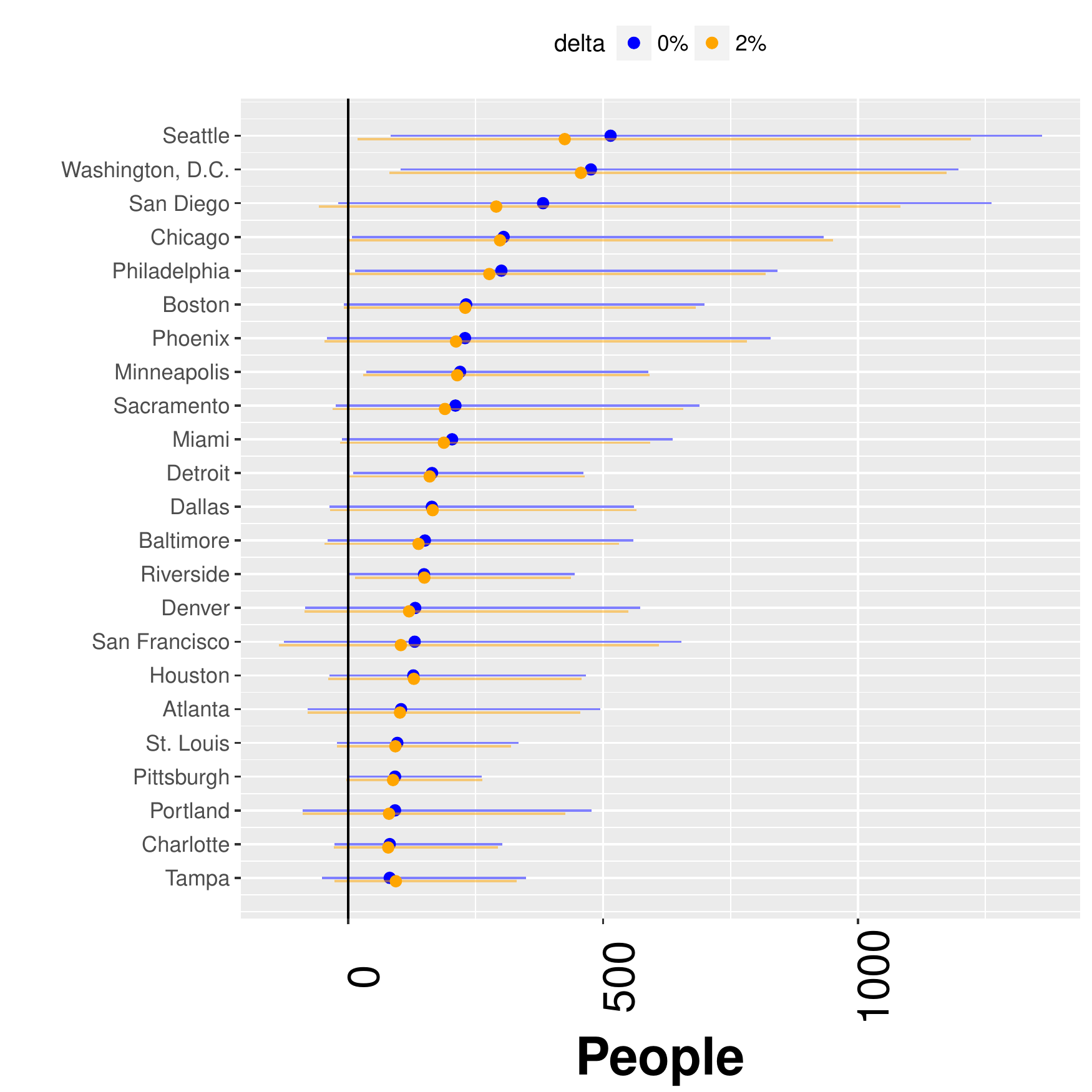}
\caption{Excluding NY and LA}
\label{subfig:Summary_ZRI_increase_exNYLA}
\end{subfigure}
\caption{Effect of increasing ZRI by $x=10\%$ on the homeless population in 2016.  The points are the posterior mean of $\left(H_{i,T}^{x} - H_{i,T}\right)$ $| C_{1:25,1:T},N_{1:25,1:T}$.  The line segment spans the one-sided (right-tail) 95\% posterior credible interval.  In the left panel, results are presented for all metros.  In the right panel, New York and Los Angeles are excluded for more careful inspection of the remaining 23 metros.}
\label{fig:Summary_ZRI}
\end{figure}

We find that, for a fixed percent increase in ZRI of $x=10\%$, the predicted increase in homelessness is largest in New York and Los Angeles (see Figure \ref{fig:Summary_ZRI}).  Predicted increases in homeless counts are robust to whether we set $\bar{\delta} = 0$ or $\bar{\delta}=.02$, as one would hope; however, the predicted count increases map to different increases in total homelessness under different prior beliefs about $\pi_{i,t}$.  In Figure \ref{fig:Homeless_Increase}, the posterior predictive distributions for the increase in total and counted homeless are illustrated for different increases in ZRI in New York and Los Angeles. 

In New York, the large sheltered population and high count accuracy imply that the distributions of increased counts and total homeless populations are nearly identitical (Figure \ref{subfig:NY_Homeless_Increase}).  If the ZRI in New York increases by $x=10\%$, given 2016 levels of homelessness, we expect that the homeless population will increase by 6,048 people, with 95\% posterior probability of the homelessness increase in New York being more than 3,680 people and less than 10,712 people.  In Los Angeles, the lower overall count accuracy implies more separation between the distributions of increased counted and total homeless (Figure \ref{subfig:LA_Homeless_Increase}).  Under the same $x=10\%$ increase in ZRI in Los Angeles, we expect that 4,072 people will become homeless, with 95\% posterior probability of more than 1,930 people and less than 8,268 people.  

Figure \ref{subfig:Summary_ZRI_increase} summarizes the predicted increase in the total homeless population when ZRI increases by $x=10\%$ across all metros.  The distributions of increases presented in Figure \ref{fig:Summary_ZRI} account for the different sizes of metros with binomial sampling as shown in \eqref{eq:H_Pred} and \eqref{eq:Count_Diff} (i.e. the values $N_{i,t}$ are larger for larger metros).  We expect the largest increases to occur in the largest metros (New York and Los Angeles), and this is confirmed by our analysis of the data.  For the increase in the homeless population associated with increases in ZRI, we report the one-sided 95\% posterior credible interval to shed light on the far right tail of the distribution.      

In Seattle (523 people), Washington, D.C. (474 people), Chicago (331 people), Philadelphia (303 people), Minneapolis (218 people), Detroit (166 people), and Pittsburgh (92 people), we also see statistically meaningful increases in the predicted homeless populations when the ZRI increases by 10\% and $\bar{\delta}=0$ (see Figure \ref{subfig:Summary_ZRI_increase_exNYLA}).  For these metros, the 95\% posterior credible intervals on the increase in the homeless population are bounded below by zero. 

In the remaining metros, the data do not support concluding that the homeless populations will meaningfully increase when a 10\% increase in ZRI occurs.  While the expected increases in these metros may be large (e.g., San Diego), the variance in the posterior predictive distribution precludes us from confidently concluding that the predicted increases are strictly greater than zero.  Predicted increases in homelessness as a function of increases in ZRI, as shown in Figure \ref{fig:Homeless_Increase} for New York and Los Angeles, are available in Appendix \ref{app:A} for each metro.

\subsection{Additional homeless counts}
\label{subsec:Results_Q3}
\begin{figure}[!ht]
    \centering
        \begin{subfigure}{.4\textwidth}
        \centering
         \includegraphics[width = 1\textwidth]{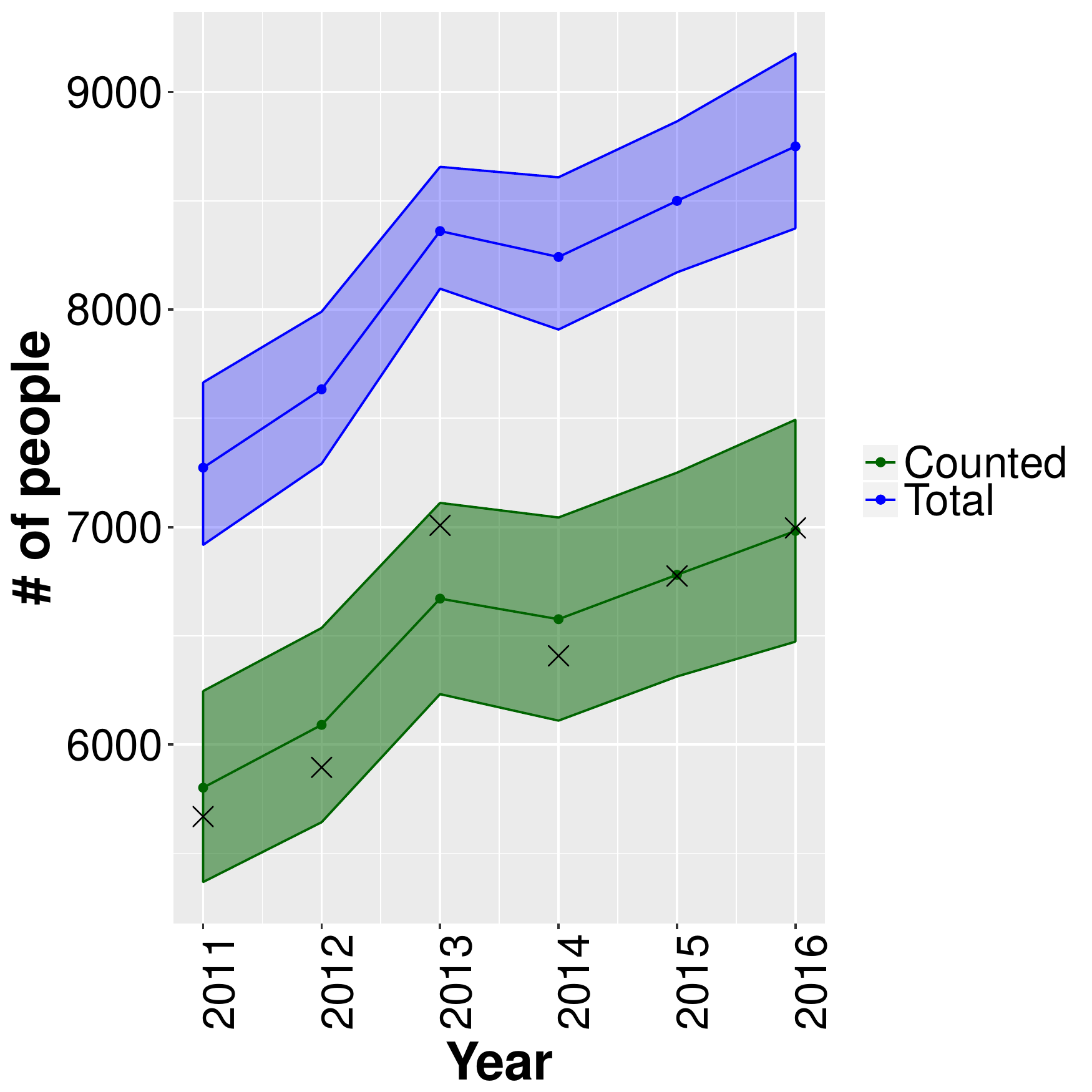}
        \caption{Predicted Homeless}
        \label{subfig:SanFrancisco_Homeless}
        \end{subfigure}
        \begin{subfigure}{.4\textwidth}
        \centering
       \includegraphics[width = 1\textwidth]{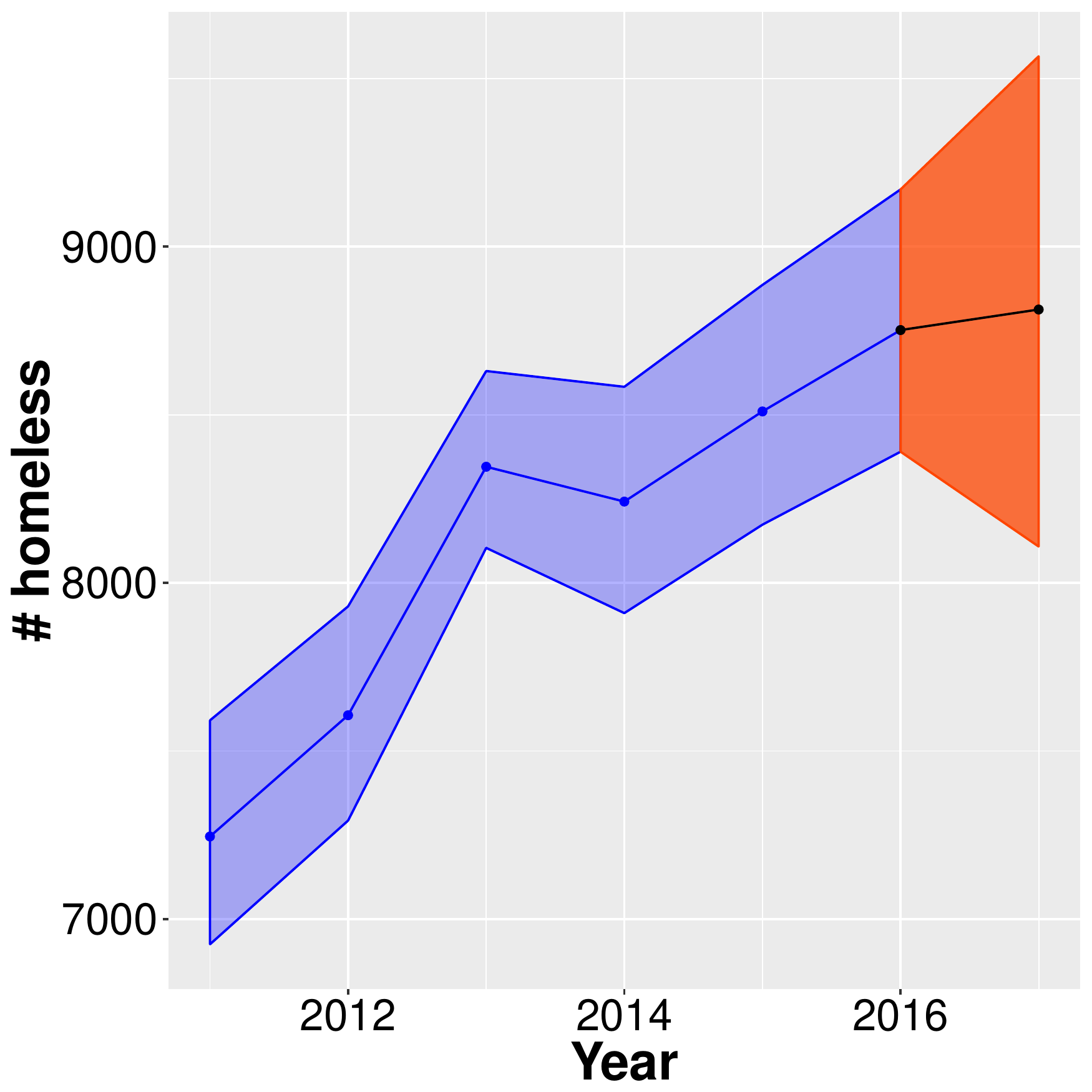}
        \caption{2017 Forecast}
        \label{subfig:SanFran_Pred}
        \end{subfigure}
    \caption{Predicted homeless totals for San Francisco, CA.  Left: Predicted number of counted homeless in additional (hypothetical) counts and predicted number of total homeless in San Francisco.  The black \lq x\rq marks are the actual HUD reported counts, and the green line is the filtered and retrospectively smoothed mean of the posterior predictive distribution $C_{i,1:T}^* | C_{1:25,1:T}, N_{1:25,1:T}$.  The blue line is the mean of the posterior predictive distribution $H_{i,1:T} | C_{1:25,1:T}, N_{1:25,1:T}$, and the shaded blue region is the 95\% predictive interval.  Right: Predicted number of total homeless in San Francisco, CA in 2017, $H_{i,T+1}|C_{1:25,1:T},N_{1:25,1:T}, ZRI_{i,T+1}$ The black line is mean of the out-of-sample prediction for 2017, and the orange shaded region is the out-of-sample predictive interval.}
\end{figure}

The number of homeless that HUD reports in each continuum is from an annual point-in-time count conducted in January.  Expense and logistical challenges preclude more frequent counts in many metros.  In this section, we predict the outcome of a second hypothetical count in each metro each year.  We report the posterior predictive distribution for this hypothetical second homeless count.  The prediction, denoted by $C_{i,t}^*|C_{1:25,1:T},N_{1:25,1:T}$, conditions on both the observed counts and the census reported total populations in all metros.  Figure \ref{subfig:SanFrancisco_Homeless} presents the predicted outcome from additional homeless counts for San Francisco, a metro with one of the larger increases in the homelessness rate from 2011-2016.  Observe that the posterior mean of $C_{i,t}^*|C_{1:25,1:T},N_{1:25,1:T}$ is a filtered and retrospectively smoothed quantity.  The smoothing is apparent in 2013, when the HUD reported count appears to be an outlier relative to prior and subsequent HUD reported counts.  Though the 2013 posterior mean is pulled slightly upward toward the reported count, the model does not overfit the data.  In the remaining years, the posterior mean closely tracks the reported HUD counts.  In 2016, the HUD reported count of homeless in San Francisco was 6,996.  If a second count were conducted in 2016, we expect the counted number of homeless would have been 6,984, with 95\% posterior probability of being more than 6,499 and less than 7,492.

 The predictive distribution $C_{i,t}^*$ $|C_{1:25,1:T},N_{1:25,1:T}$ provides policymakers and resource constrained counting agencies with a principled and data-driven way of conducting synthetic \lq\lq additional\rq\rq homeless counts.  The posterior predictive distributions for additional 2016 counts in all metros are summarized in Table \ref{tab:Results_Table}.  Each metro has its own version of Figure \ref{subfig:SanFrancisco_Homeless} in Appendix \ref{app:A}.

\begin{table}[!ht]
\centering
\small
\begin{tabular}{rlllll}
  \hline
 Metro & HUD & Synthetic count & Total homeless & Forecast (2017) \\ 
  \hline
New York & 73,523 & 74,633  (69,423, 79,426) & 76,411  (73,683, 80,735) & 76,341  (70,375, 83,079) \\ 
  Los Angeles & 46,874 & 46,149  (42,950, 49,279) & 59,508  (57,024, 61,842) & 61,398 (56,400, 66,508) \\ 
  Chicago &  6,841 & 7,158  (6,657, 7,679) & 7,614  (7,271, 8,085) & 7,641  (6,997, 8,362) \\ 
  Dallas &  3,810 & 3,782  (3,537, 3,967) & 3,866  (3,811, 4,017) & 4,019  (3,671, 4,406) \\ 
  Philadelphia &  6,112 & 6,082  (5,691, 6,428) & 6,281  (6,132, 6,567) & 6,345  (5,840, 6,898) \\ 
  Houston &  4,031 & 4,364  (4,070, 4,617) & 5,032  (4,818, 5,152) & 5,224  (4,788, 5,677) \\ 
  Washington, D.C. &  8,350 & 8,273  (7,740, 8,661) & 8,498  (8,358, 8,794) & 8,703  (8,047, 9,408) \\ 
  Miami &  4,235 & 4,263  (3,979, 4,546) & 4,624  (4,437, 4,852) & 4,701  (4,296, 5,139) \\ 
  Atlanta &  4,546 & 4,775  (4,427, 5,116) & 5,447  (5,164, 5,722) & 5,605  (5,109, 6,129) \\ 
  Boston &  6,240 & 6,291  (5,849, 6,696) & 6,418  (6,242, 6,770) & 6,557  (6,029, 7,150) \\ 
  San Francisco &  6,996 & 6,984  (6,499, 7,492) & 8,752  (8,390, 9,170) & 8,815  (8,104, 9,581) \\ 
  Detroit &  2,612 & 2,778  (2,538, 3,022) & 2,872  (2,678, 3,097) & 2,898  (2,602, 3,217) \\ 
  Riverside &  2,165 & 2,368  (2,174, 2,545) & 3,207  (3,034, 3,316) & 3,352  (3,036, 3,669) \\ 
  Phoenix &  5,702 & 5,840  (5,430, 6,259) & 6,918  (6,604, 7,270) & 7,162  (6,548, 7,838) \\ 
  Seattle & 10,730 & 10,720  (9,991, 11,458) & 12,240  (11,734, 12,848) & 12,763  (11,677, 13,940) \\ 
  Minneapolis &  3,056 & 3,250  (2,977, 3,532) & 3,359  (3,141, 3,622) & 3,531  (3,178, 3,923) \\ 
  San Diego &  8,669 & 8,775  (8,096, 9,446) & 11,149  (10,572, 11,720) & 11,455  (10,416, 12,524) \\ 
  St. Louis &  1,713 & 1,730  (1,611, 1,848) & 1,879  (1,802, 1,976) & 1,926  (1,740, 2,125) \\ 
  Tampa &  1,817 & 1,974  (1,794, 2,150) & 3,090  (2,919, 3,226) & 3,204  (2,909, 3,514) \\ 
  Baltimore &  3,488 & 3,508  (3,267, 3,753) & 4,088  (3,914, 4,285) & 4,121  (3,762, 4,504) \\ 
  Denver &  5,728 & 5,830  (5,431, 6,243) & 6,320  (6,047, 6,670) & 6,457  (5,917, 7,056) \\ 
  Pittsburgh &  1,156 & 1,268  (1,153, 1,367) & 1,318  (1,217, 1,401) & 1,375  (1,225, 1,529) \\ 
  Portland &  3,914 & 3,972  (3,688, 4,255) & 4,674  (4,458, 4,906) & 4,807  (4,374, 5,264) \\ 
  Charlotte &  1,818 & 1,913  (1,768, 2,050) & 2,139  (2,027, 2,251) & 2,249  (2,035, 2,475) \\ 
  Sacramento &  4,145 & 4,182  (3,884, 4,481) & 5,107  (4,881, 5,350) & 5,288  (4,820, 5,796) \\ 
  \hline
\end{tabular}
\caption{Summary of posterior distributions across metros for 2016.  The number of counted homeless reported by HUD is presented the first column. The Synthetic counts column corresponds to the posterior predictive distribution for a second hypothetical count in metro $i$ in 2016, $C_{i,T}^* | C_{1:25,1:T}, N_{1:25,1:T}$.  The Total homeless column corresponds to the posterior predictive distribution for the total number of homeless in metro $i$ in 2016, $H_{i,T}|C_{1:25,1:T}, N_{1:25, 1:T}$.  The Forecasted homeless (2017) column is the posterior predictive distribution for the total homeless population in 2017, $H_{i,T+1} | C_{1:25, 1:T}, N_{1:25,1:T}$.  In all cases, the first number reported is the posterior mean, with the 95\% posterior predictive interval in parenthesis.}
\label{tab:Results_Table}
\end{table}

\subsection{Imputed total number of homeless}        
\label{subsec:Results_Q4}
Imperfect count accuracy leads to count totals that are less than the size of the total homeless population.  By modeling the mechanism of count accuracy, we are able to include the uncounted number of homeless in our estimate of the size of the total homeless population.  In this section, we predict the total number of homeless in each metro and year.  We report the posterior distribution $H_{i,t} | C_{1:25,1:T}, N_{1:25,1:T}$.  Observe that the posterior distribution does not condition on the count accuracy parameter, $\pi_{i,t}$.  The count accuracy has been integrated out; however, the variance of $H_{i,t} | C_{1:25,1:T}, N_{1:25,1:T}$ is inextricably linked to the prior variance of the count accuracy, $\pi_{i,t}$.  In this analysis, we fixed the prior variance to be 0.0005 so that prior mass would span the $\pm 0.05$ interval.  Though it is appealing to specify a diffuse prior for count accuracy, we found in practice that such a prior does not provide sufficient regularization.  In settings with overly diffuse priors for count accuracy, inference for $\phi_i$ was not reproducible across MCMC simulations.  This highlights the importance of reliable prior information about count accuracy as it pertains to estimating the relationship between trends in ZRI and homelessness.

In Figure \ref{subfig:SanFrancisco_Homeless}, observe that, because the counting process is imperfect, the expected total number of homeless is more than the counted number of homeless.  In San Francisco, we expect that, in 2016, there were 8,752 people experiencing homelessness, with 95\% posterior probability that there were more than 8,390 and fewer than 9,170.  Table \ref{tab:Results_Table} presents the posterior mean and 95\% credible interval for the total number of homeless in 2016 for each metro. 

\subsection{Forecasts for 2017}
\label{subsec:Results_Q5}
Resources to address the needs of a homeless population are budgeted well in advance of the January point-in-time count.  In order to allocate resources in communities with growing (shrinking) homeless populations, a forecast of the next year's total homeless population is needed.  In this section, we forecast the total homeless population in each metro in January 2017.  

Our forecasts of the homeless population for 2017 take into account both predicted increases in the 2017 total population and the January 2017 ZRI value.  We report the one-year-ahead forecast $H_{i,T+1} | C_{1:25,1:T}, N_{1:25,1:T}, ZRI_{1:25,T+1}$.  For true out-of-sample forecasting when $ZRI_{i,T+1}$ is not yet available, we could utilize Zillow's forecasted ZRI for metro $i$,  $\hat{ZRI}_{i,T+1}$.

Figure \ref{subfig:SanFran_Pred} illustrates the year-ahead forecast in San Francisco.  Forecasting the total homeless population one year in the future requires forecasting both the year-ahead total metro population and log odds of homelessness.  The uncertainty in these component year-ahead forecasts accumulates, and the result is an uncertainty interval for the 2017 year-ahead forecast that increases relative to the intervals presented from 2011-2016.  This is observed in Figure \ref{subfig:SanFran_Pred} as the orange interval fans out relative to the blue interval.

We predict that 8,815 people experienced homelessness in San Francisco on any given January night in 2017, with 95\% posterior probability of more than 8,104 and less than 9,581 people. Although the January 2017 ZRI decreased 3.6\% relative to its January 2016 value, we still expect a slight increase in San Francisco's total homeless population.  The increase is largely driven by the model-based forecasted increase in San Francisco's total population.  Each metro has a figure corresponding to Figure \ref{subfig:SanFran_Pred} presented in Appendix \ref{app:A}.  The forecasted mean and 95\% posterior predictive intervals for each metro are shown in Table \ref{tab:Results_Table}.

\section{Discussion}
\label{sec:Discussion}
We presented statistical evidence that the relationship between rental costs and homelessness depends on one's beliefs about the time-varying accuracy of homeless counts.  We highlight this fact to encourage public policy researchers, policymakers, and continuum leaders to carefully quantify their beliefs and uncertainty about count accuracy or the inferences drawn from studies relying on these counts.  While the prior beliefs about count accuracy that we elicit in this paper are informed by existing literature and our discussions with count coordinators and homelessness experts from around the country, we believe that collecting expert opinions from every continuum can lead to a more robust and informed study.  We encourage other researchers in this area to explicitly model variation in count accuracy when conducting their own analyses.  

We found in synthetic data experiments that making accurate inference on the relationship between ZRI and homelessness with the model outlined in this paper requires homelessness rates that exceed 0.05\% of the total population (Section \ref{subsec:Prior_Psi}).  To work with counts as large as possible, this implies that sheltered and unsheltered homeless totals should be combined in a single analysis of a metro's total homeless population.  Furthermore, reliable estimates of the entire homeless population are significantly aided by a metro having either a large sheltered population or high count accuracy of the unsheletered population; utilizing data on unsheltered homeless populations alone does not yield reliable results.  This observation fits with our broader theme of data quality.

Modeling count accuracy is another place where we have directly addressed data quality challenges.  Acknowledging that a continuum's homeless counts are imperfect and that the accuracy varies from one year to the next should not be viewed in any way as a failure of the count coordinators or volunteers.  We view quantifying the count accuracy as an important step in accounting for the uncertainty inherent in such a difficult undertaking.

In our analysis, we have used the time-varying count accuracy to impute the size of the total homeless population.  We believe it is natural in this application to think of the total homeless population as missing data.  Assuming that the total population size is the observed count has two flaws.  First, it understates the size of the homeless population.  Second, it leads to overly confident estimates of regression coefficients $\phi_1,\ldots,\phi_{25}$.  Imputing the missing homeless population size naturally resolves both of these problems.  

In this application, proper uncertainty quantification is critical.  Counties, city governments, shelters, and health care providers are likely to benefit from an expected range of the homeless population size when they budget resources.  By reporting the 95\% posterior credible intervals on the total homeless population predictions and the 2017 forecasts, we emphasize the uncertain size of homeless populations now and in the future.   

We provide evidence that the homelessness rate increased by at least 4\% from 2011 - 2016 in five metros: New York, Los Angeles, Washington, D.C., San Francisco, and Seattle.  We also found that large increases in ZRI lead to significant increases in the homeless population in four of those five metros: New York, Los Angeles, Washington, D.C., and Seattle.  Homelessness in these four metros has significantly increased in recent years and is severely affected by increases in rental costs. 

Metro-specific estimates of the relationship between rental costs and homelessness allow for each metro to make more informed policy decisions about affordable housing initiatives.  Forecasts of year-ahead homeless populations can guide resource allocation.  While we are not public policy experts, we believe that our results provide context for policy discussions that are happening in cities across the United States.  

There are a few limitations of our current approach.  One is that it does not account for relocation in homeless populations.  It is possible that people experiencing homelessness move to cities with more services and away from cities with fewer services.  In this scenario, increases in the population of one metro are driven by decreases in another.  Our present approach does not take into account network effects, and we assume that homeless relocation patterns are not a significant driver of trends across metros.  At present, we do not know of data that would allow us to further investigate network effects.  A second limitation of our work is that it relies on January count data.  It is likely that seasonal patterns in homelessness exist, though our current annual data set does not provide insight into such seasonal fluctuations.  Modeling network effects and seasonal fluctuations in homeless populations are important areas of future work.  
  
\bibliographystyle{apalike}
\bibliography{main}

\newpage
\appendix
\section{Results for remaining metros}
\label{app:A}
The figures for each metropolitan area (ordered by descending population) are presented below. 

\begin{figure}[ht!]
\centering
\begin{subfigure}{.4\textwidth}
  \centering
\includegraphics[width=1\textwidth]{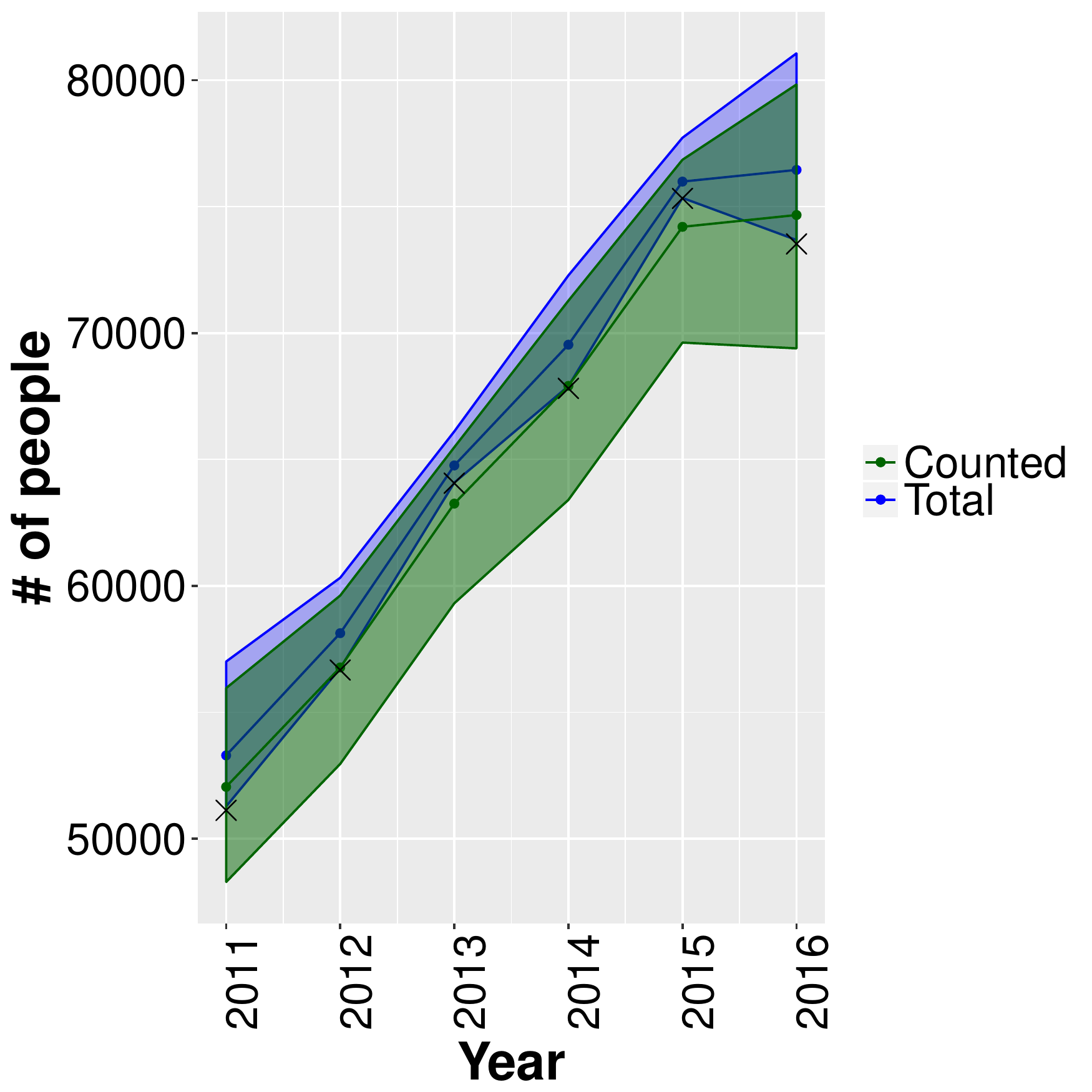}
\caption{\# of homeless}
\end{subfigure}
\begin{subfigure}{.4\textwidth}
  \centering
\includegraphics[width=1\textwidth]{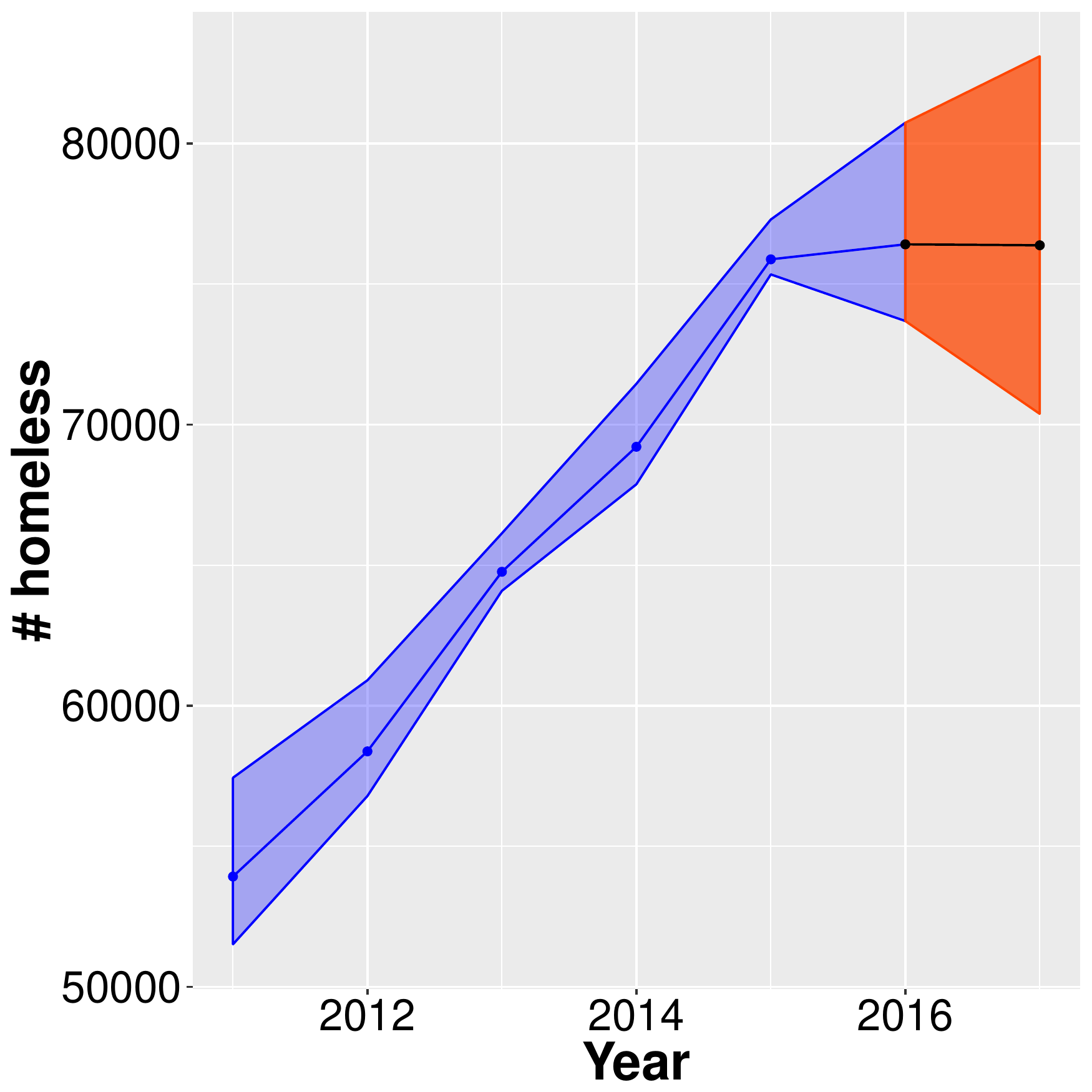}
\caption{2017 forecast}
\end{subfigure}
\begin{subfigure}{.4\textwidth}
  \centering
\includegraphics[width=1\textwidth]{NY_Homeless_Increase.pdf}
\caption{ZRI effect}
\end{subfigure}
\begin{subfigure}{.4\textwidth}
  \centering
\includegraphics[width=1\textwidth]{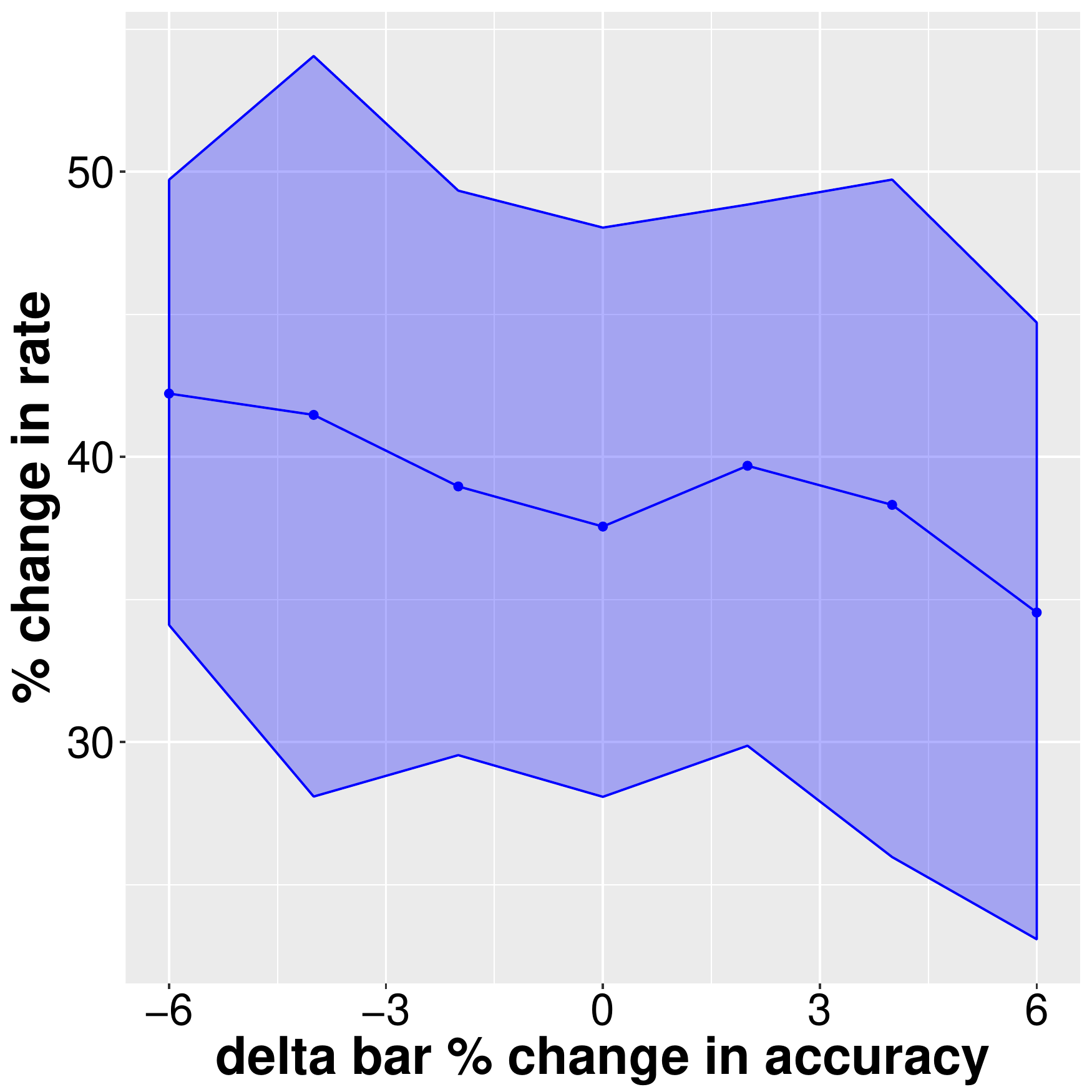}
\caption{Rate}
\end{subfigure}
\caption{Results for New York, NY.  Top left (a): Posterior predictive distribution for homeless counts, $C_{i,1:T}^* | C_{1:25,1:T}, N_{1:25,1:T}$, in green, and the imputed total homeless population size, $H_{i,1:T}|C_{1:25,1:T}, H_{1:25,1:T}$, in blue.  The black 'x' marks correspond to the observed (raw) homeless count by year.  The count accuracy is modeled with a constant expectation.  Top right (b): Predictive distribution for total homeless population in 2017, $H_{i,2017} | C_{1:25,1:T}, N_{1:25,1:T}$.  Bottom left (c):  Posterior distribution of increase in total homeless population with increases in ZRI.  Bottom right (d): Sensitivity of the inferred increase in the homelessness rate from 2011 - 2016 to different annual changes in count accuracy.}
\label{fig:NY_Results}
\end{figure}

\begin{figure}[ht!]
\centering
\begin{subfigure}{.4\textwidth}
  \centering
\includegraphics[width=1\textwidth]{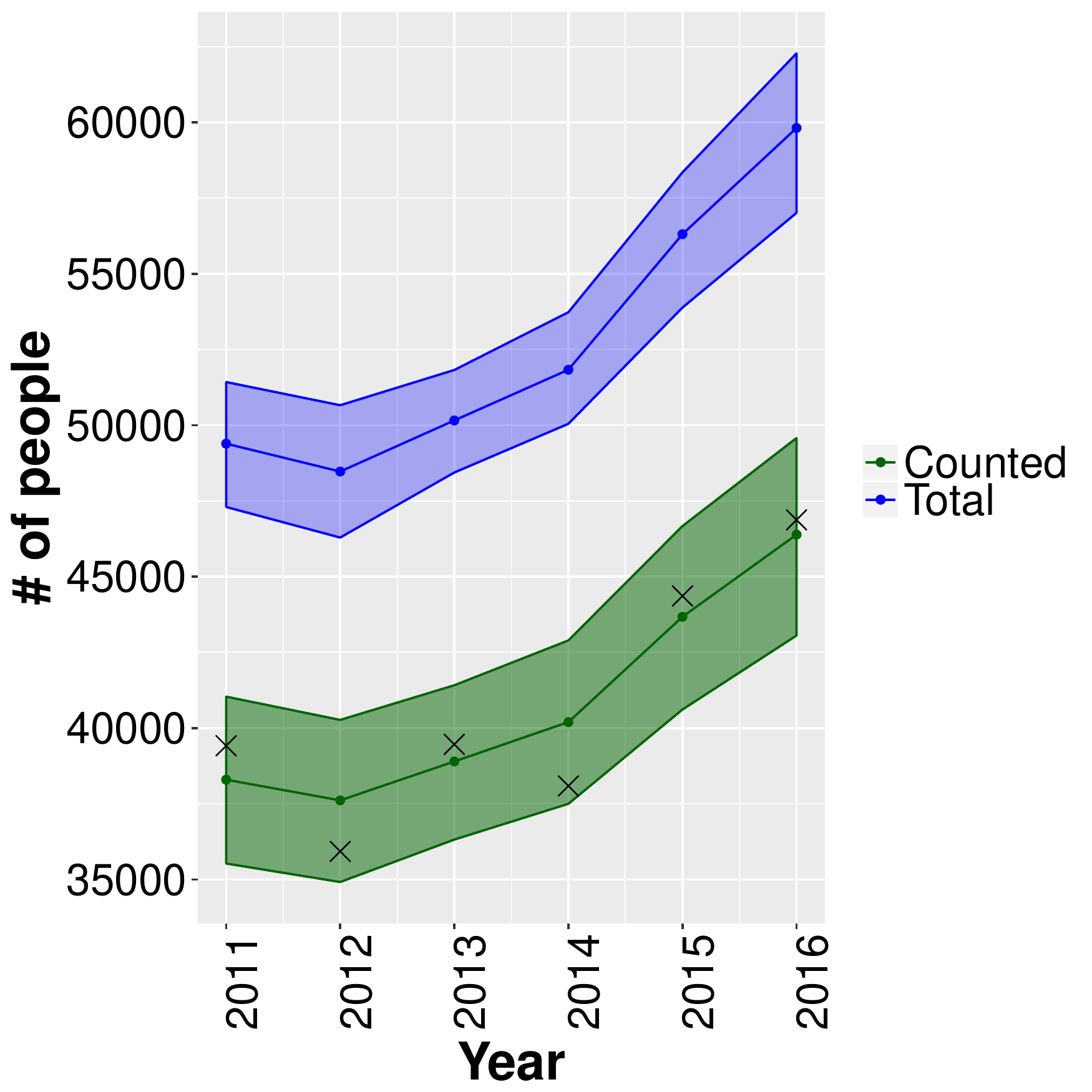}
\caption{\# of homeless}
\end{subfigure}
\begin{subfigure}{.4\textwidth}
  \centering
\includegraphics[width=1\textwidth]{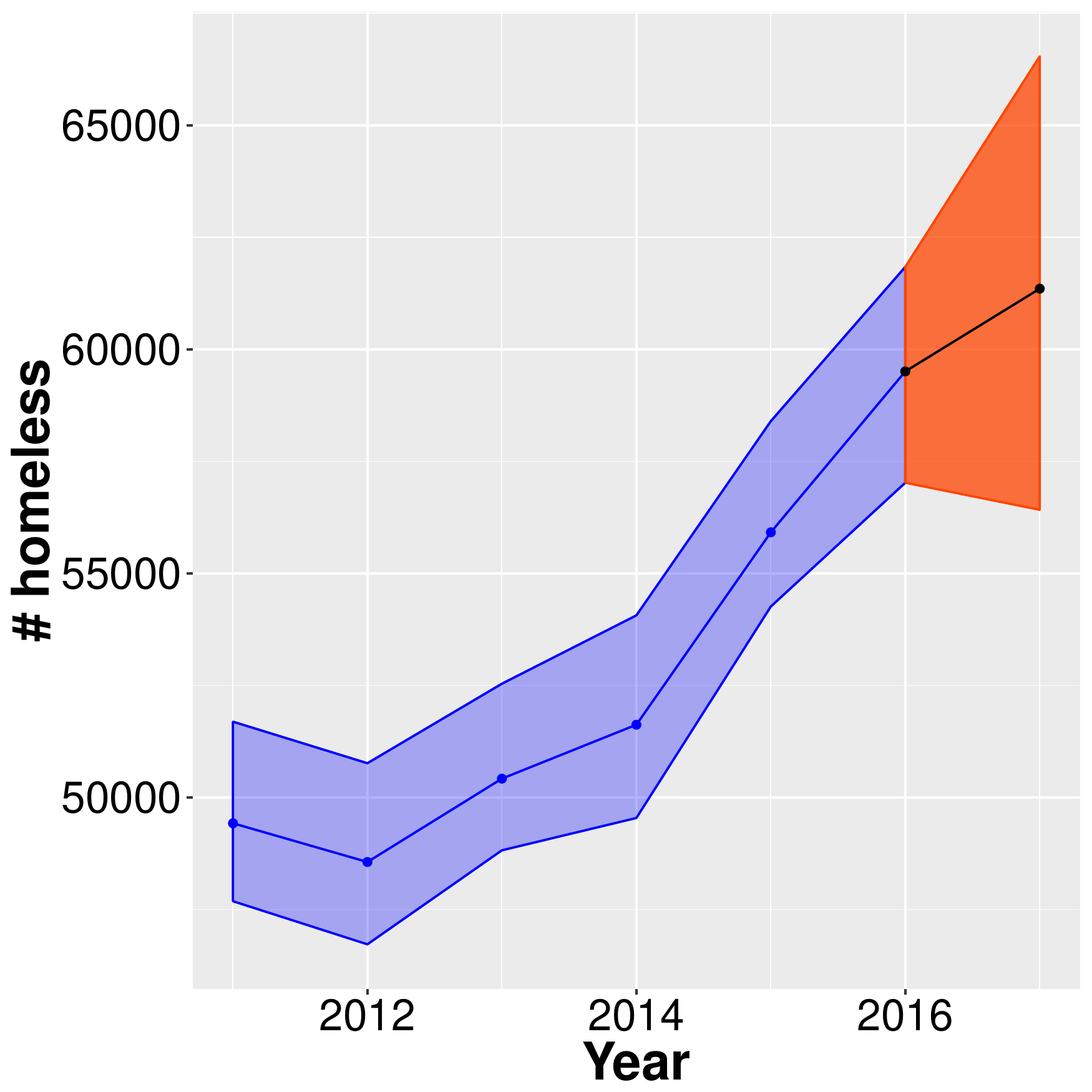}
\caption{2017 forecast}
\end{subfigure}
\begin{subfigure}{.4\textwidth}
  \centering
\includegraphics[width=1\textwidth]{LA_Homeless_Increase.pdf}
\caption{ZRI effect}
\end{subfigure}
\begin{subfigure}{.4\textwidth}
  \centering
\includegraphics[width=1\textwidth]{LA_psi_sensitivity.pdf}
\caption{Rate}
\end{subfigure}
\caption{Results for Los Angeles, CA.  Top left (a): Posterior predictive distribution for homeless counts, $C_{i,1:T}^* | C_{1:25,1:T}, N_{1:25,1:T}$, in green, and the imputed total homeless population size, $H_{i,1:T}|C_{1:25,1:T}, H_{1:25,1:T}$, in blue.  The black 'x' marks correspond to the observed (raw) homeless count by year.  The count accuracy is modeled with a constant expectation.  Top right (b): Predictive distribution for total homeless population in 2017, $H_{i,2017} | C_{1:25,1:T}, N_{1:25,1:T}$.  Bottom left (c):  Posterior distribution of increase in total homeless population with increases in ZRI.  Bottom right (d): Sensitivity of the inferred increase in the homelessness rate from 2011 - 2016 to different annual changes in count accuracy.}
\label{fig:LA_Results}
\end{figure}

\begin{figure}[ht!]
\centering
\begin{subfigure}{.4\textwidth}
  \centering
\includegraphics[width=1\textwidth]{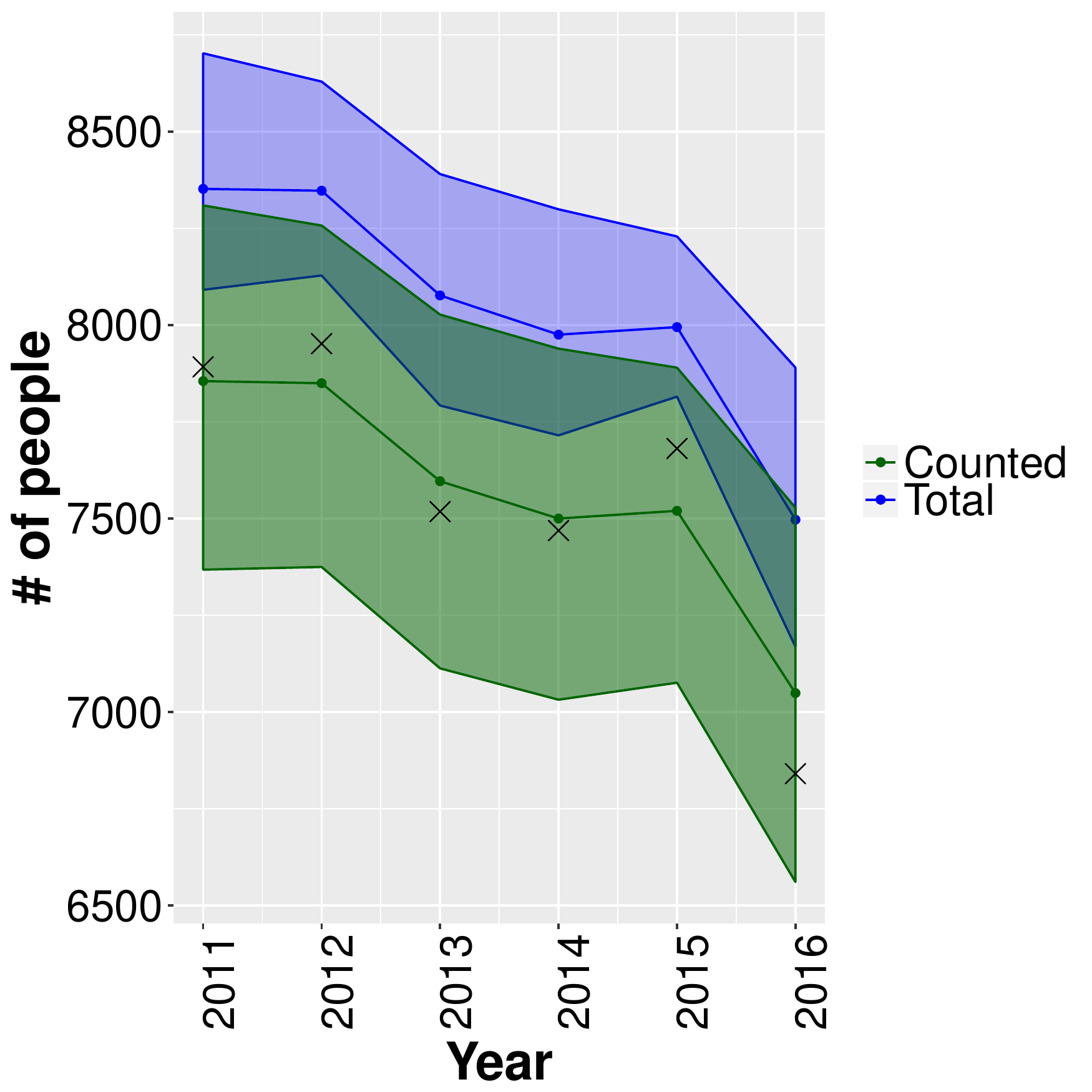}
\caption{\# of homeless}
\end{subfigure}
\begin{subfigure}{.4\textwidth}
  \centering
\includegraphics[width=1\textwidth]{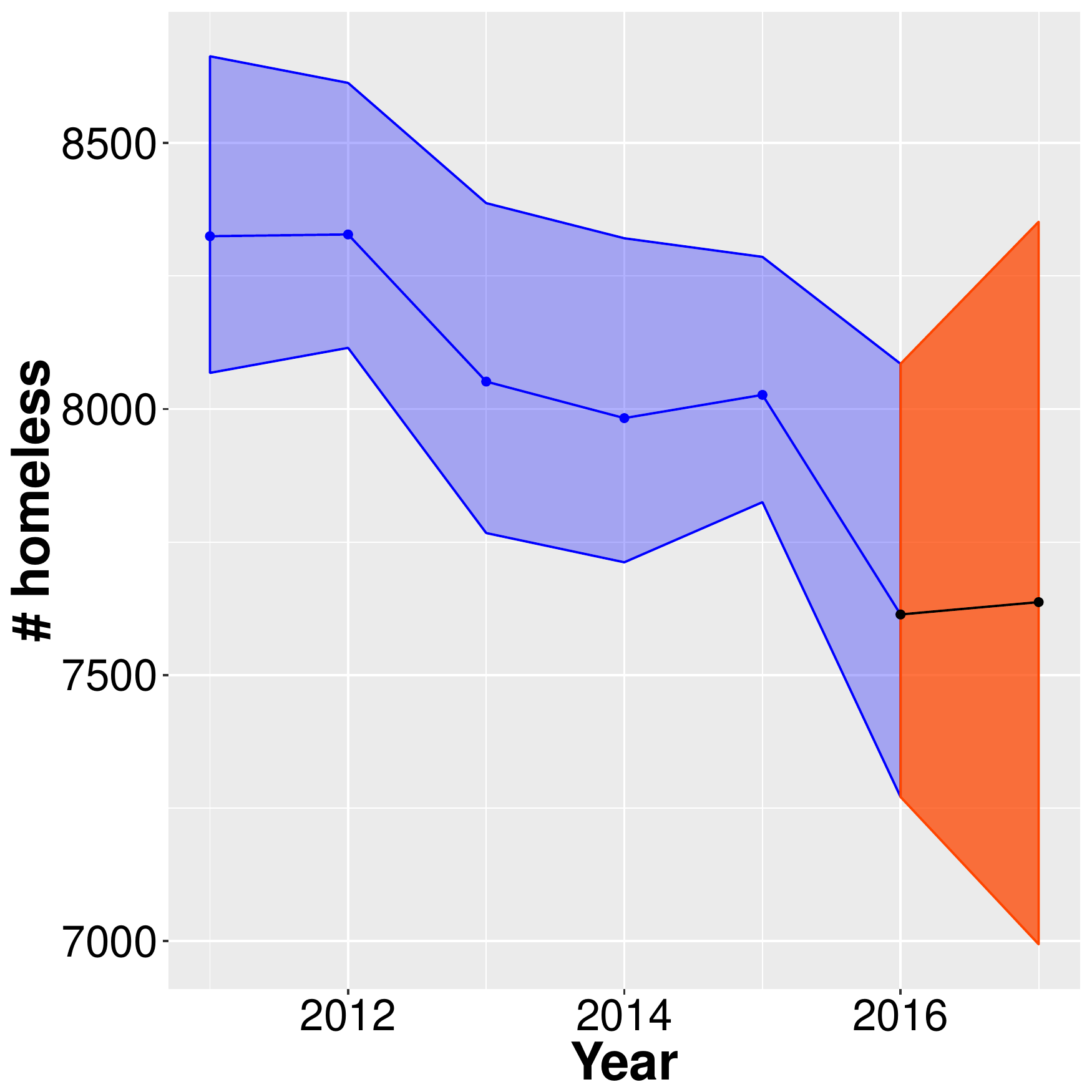}
\caption{2017 forecast}
\end{subfigure}
\begin{subfigure}{.4\textwidth}
  \centering
\includegraphics[width=1\textwidth]{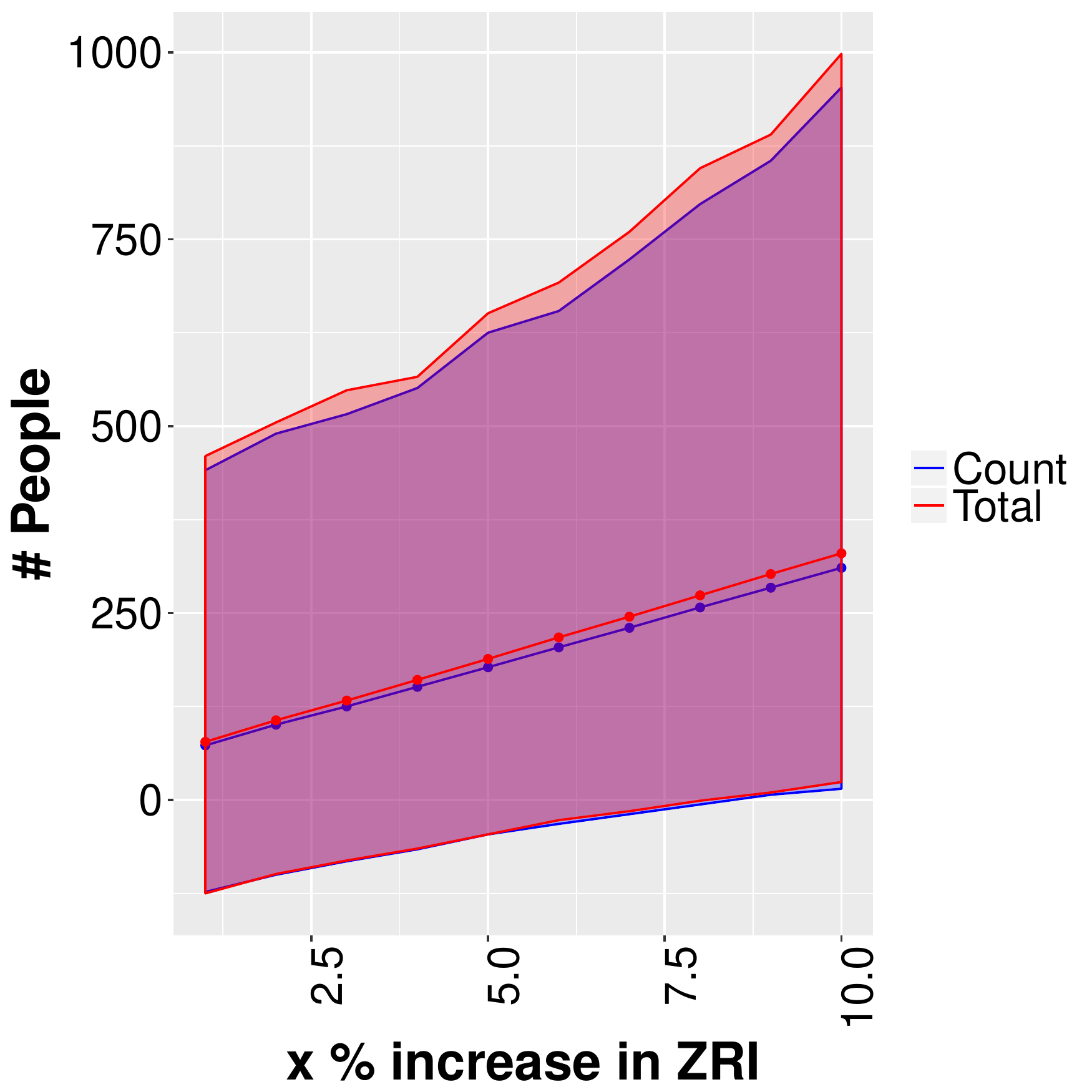}
\caption{ZRI effect}
\end{subfigure}
\begin{subfigure}{.4\textwidth}
  \centering
\includegraphics[width=1\textwidth]{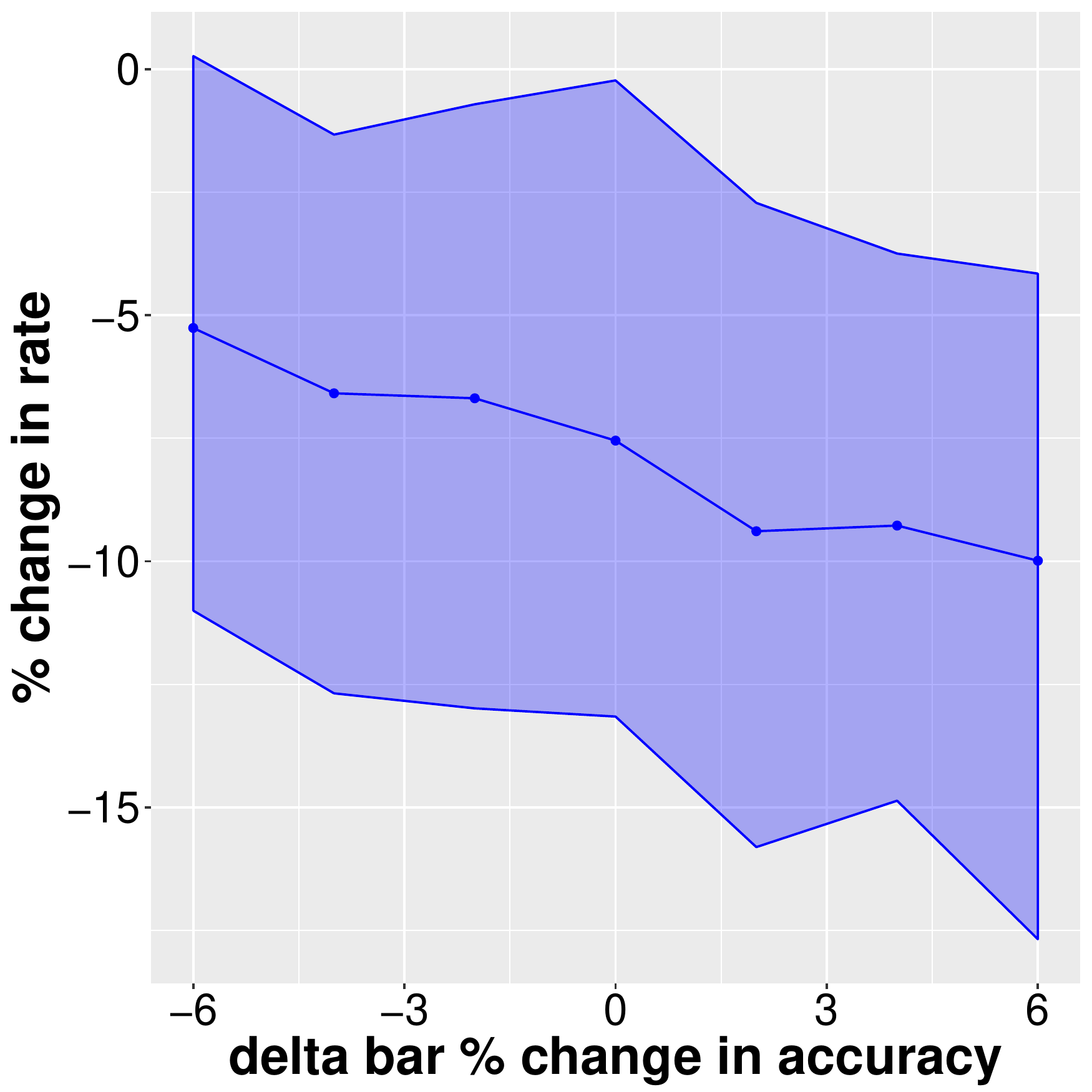}
\caption{Rate}
\end{subfigure}
\caption{Results for Chicago, IL.  Top left (a): Posterior predictive distribution for homeless counts, $C_{i,1:T}^* | C_{1:25,1:T}, N_{1:25,1:T}$, in green, and the imputed total homeless population size, $H_{i,1:T}|C_{1:25,1:T}, H_{1:25,1:T}$, in blue.  The black 'x' marks correspond to the observed (raw) homeless count by year.  The count accuracy is modeled with a constant expectation.  Top right (b): Predictive distribution for total homeless population in 2017, $H_{i,2017} | C_{1:25,1:T}, N_{1:25,1:T}$.  Bottom left (c):  Posterior distribution of increase in total homeless population with increases in ZRI.  Bottom right (d): Sensitivity of the inferred increase in the homelessness rate from 2011 - 2016 to different annual changes in count accuracy.}
\label{fig:Chicago_Results}
\end{figure}

\begin{figure}[ht!]
\centering
\begin{subfigure}{.4\textwidth}
  \centering
\includegraphics[width=1\textwidth]{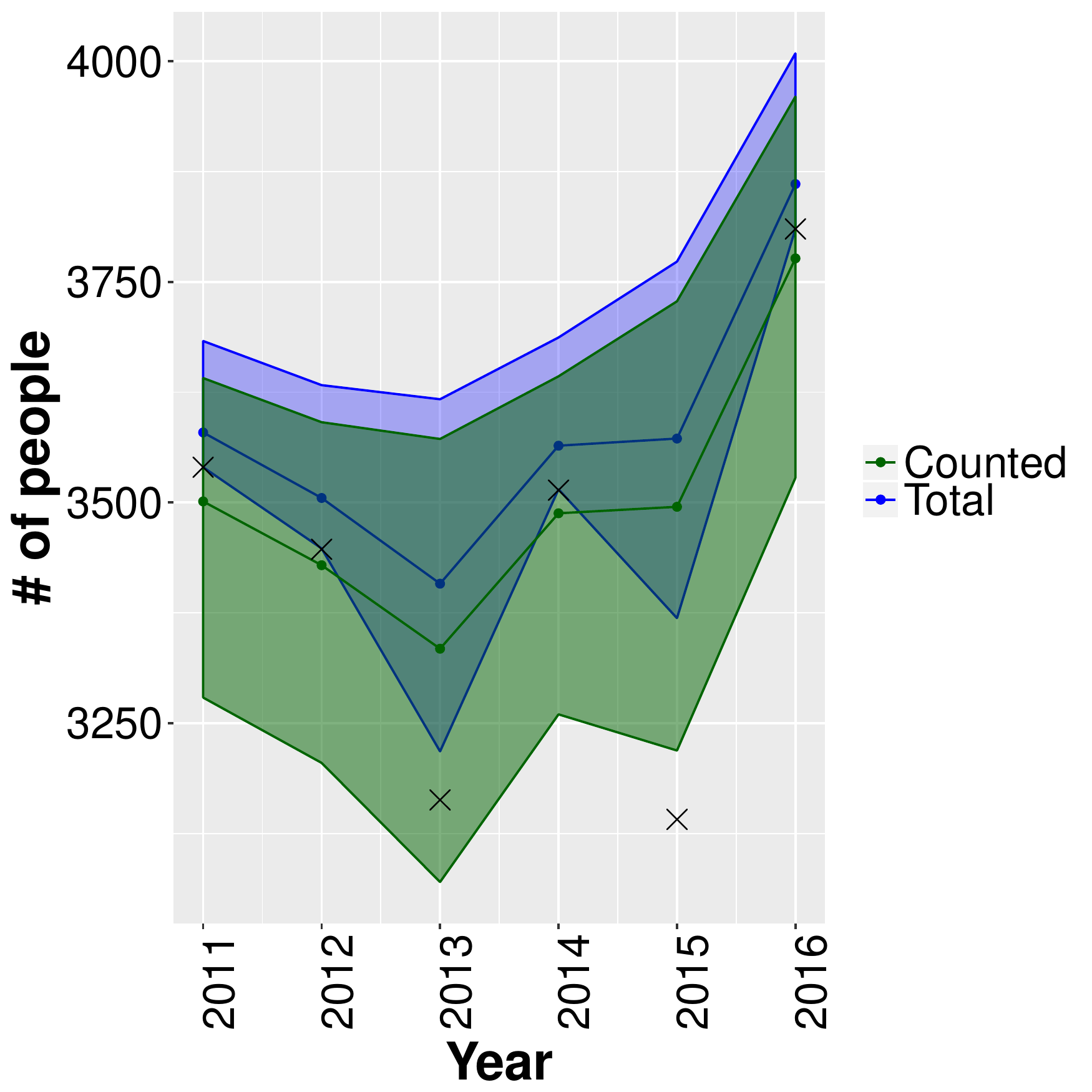}
\caption{\# of homeless}
\end{subfigure}
\begin{subfigure}{.4\textwidth}
  \centering
\includegraphics[width=1\textwidth]{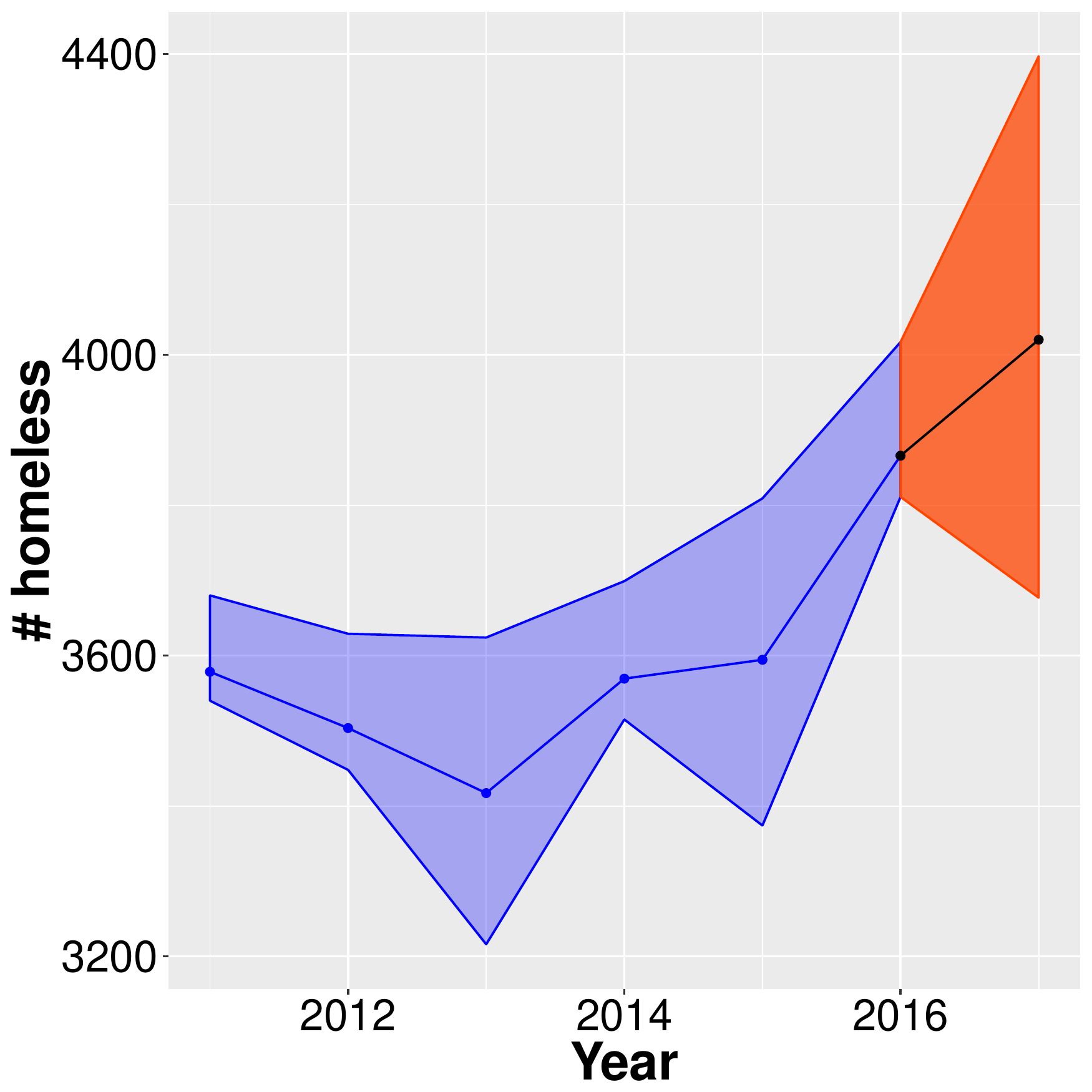}
\caption{2017 forecast}
\end{subfigure}
\begin{subfigure}{.4\textwidth}
  \centering
\includegraphics[width=1\textwidth]{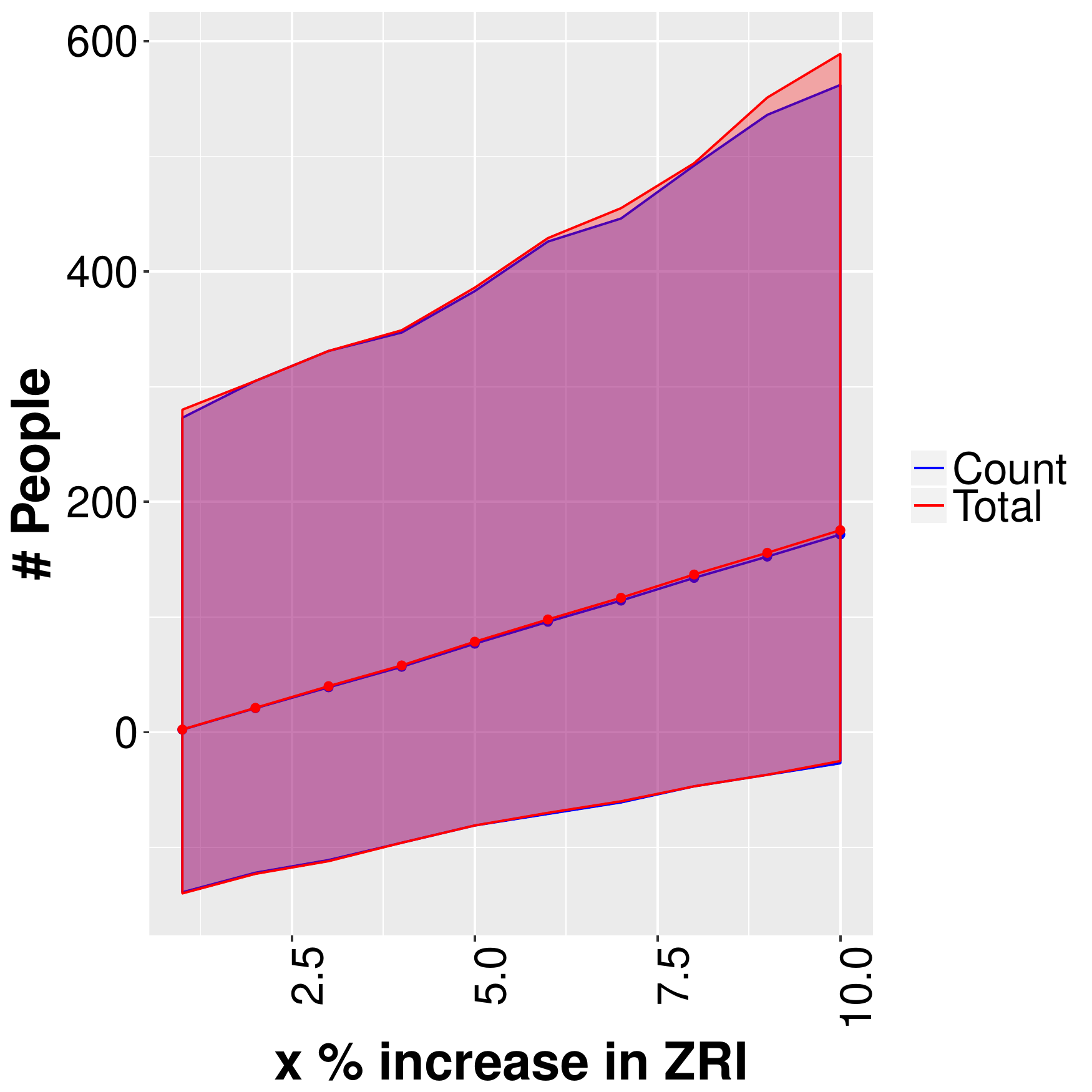}
\caption{ZRI effect}
\end{subfigure}
\begin{subfigure}{.4\textwidth}
  \centering
\includegraphics[width=1\textwidth]{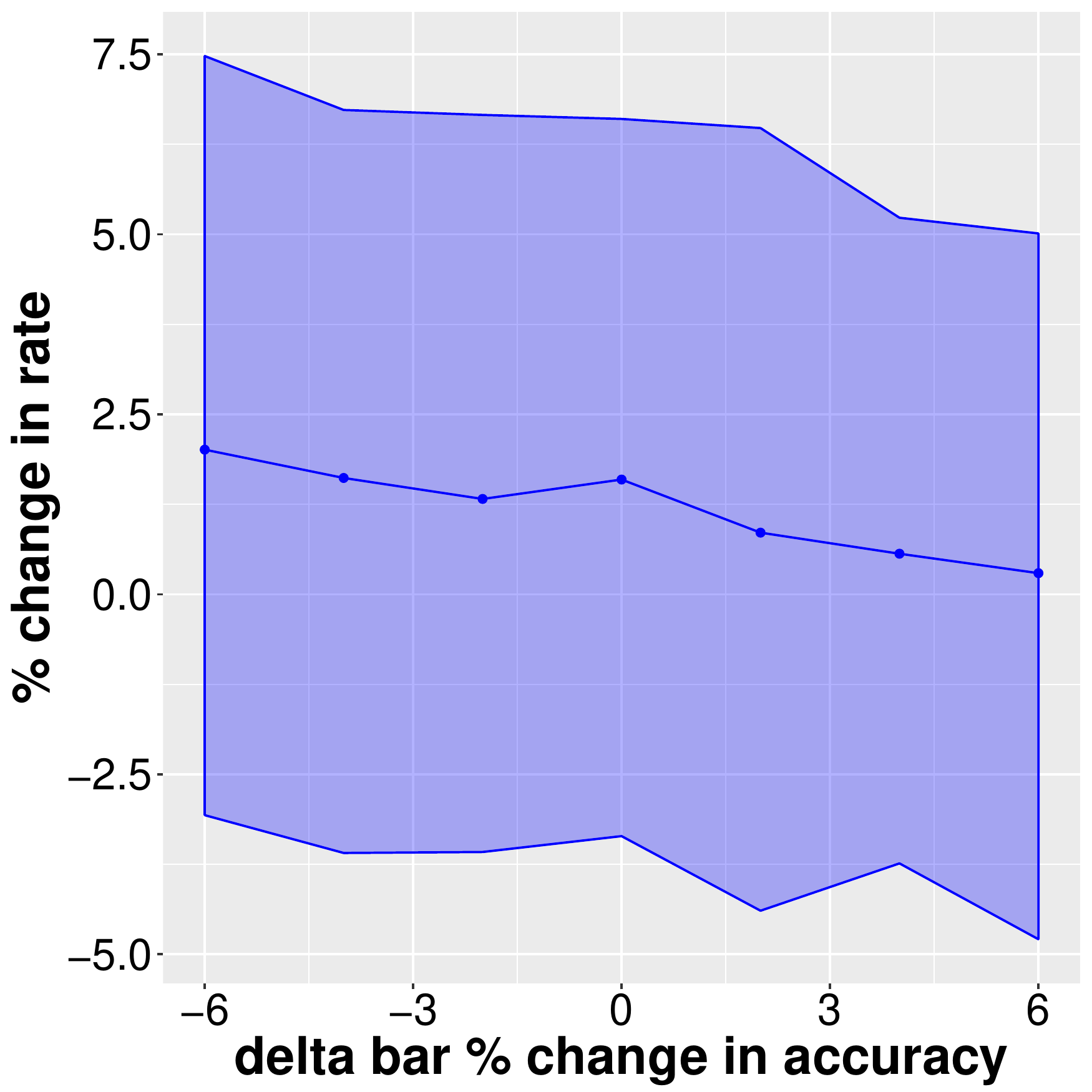}
\caption{Rate}
\end{subfigure}
\caption{Results for Dallas, TX.  Top left (a): Posterior predictive distribution for homeless counts, $C_{i,1:T}^* | C_{1:25,1:T}, N_{1:25,1:T}$, in green, and the imputed total homeless population size, $H_{i,1:T}|C_{1:25,1:T}, H_{1:25,1:T}$, in blue.  The black 'x' marks correspond to the observed (raw) homeless count by year.  The count accuracy is modeled with a constant expectation.  Top right (b): Predictive distribution for total homeless population in 2017, $H_{i,2017} | C_{1:25,1:T}, N_{1:25,1:T}$.  Bottom left (c):  Posterior distribution of increase in total homeless population with increases in ZRI.  Bottom right (d): Sensitivity of the inferred increase in the homelessness rate from 2011 - 2016 to different annual changes in count accuracy.}
\label{fig:Dallas_Results}
\end{figure}

\begin{figure}[ht!]
\centering
\begin{subfigure}{.4\textwidth}
  \centering
\includegraphics[width=1\textwidth]{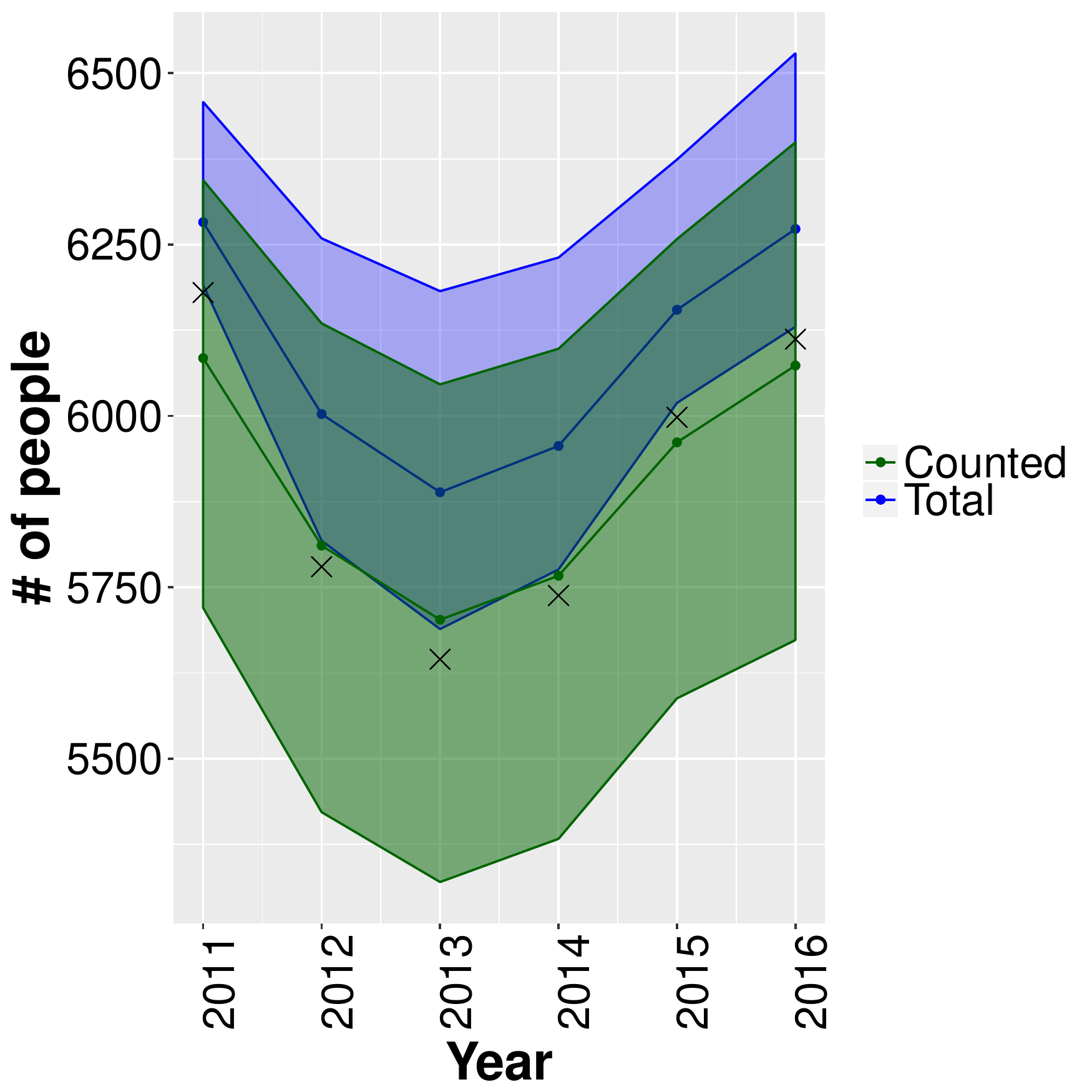}
\caption{\# of homeless}
\end{subfigure}
\begin{subfigure}{.4\textwidth}
  \centering
\includegraphics[width=1\textwidth]{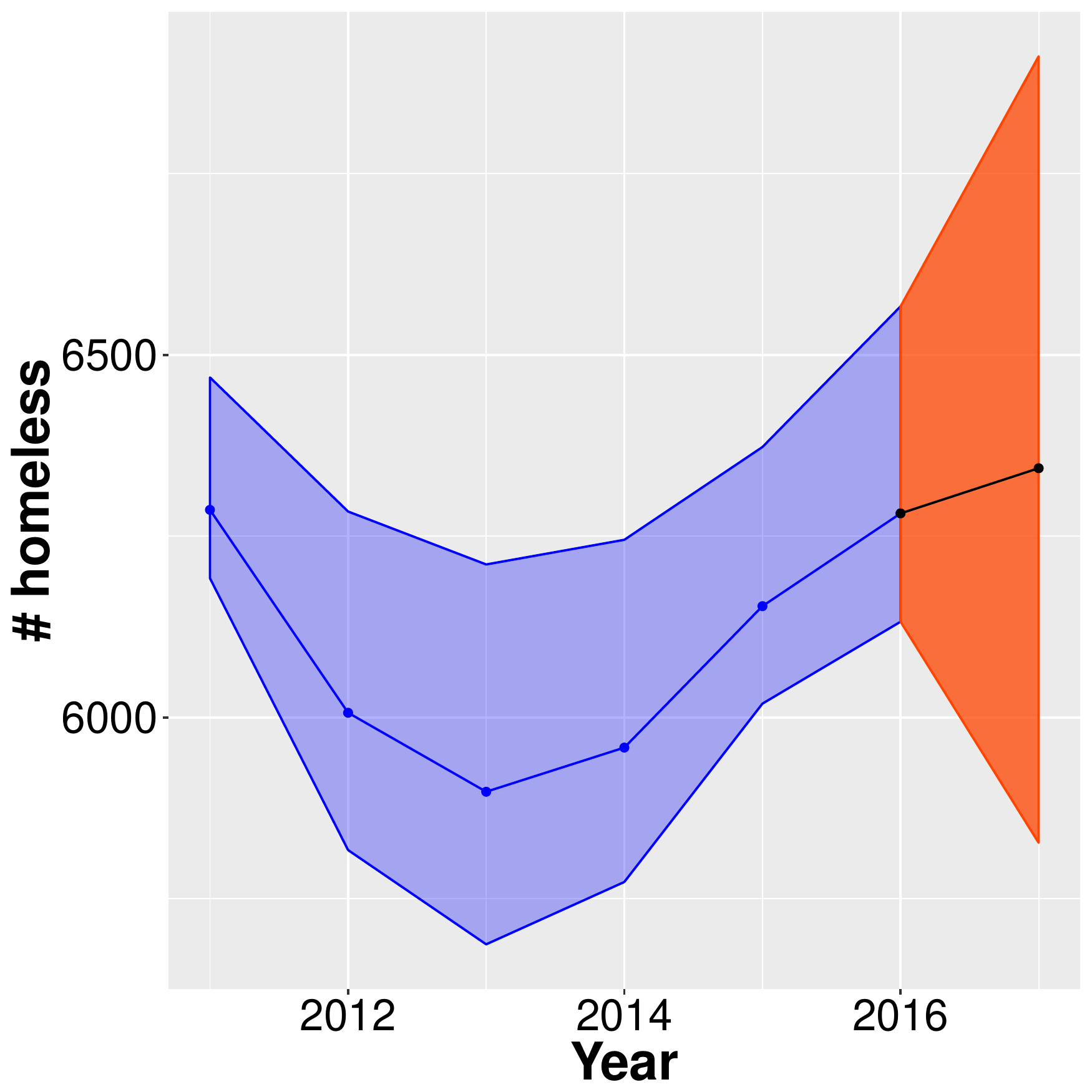}
\caption{2017 forecast}
\end{subfigure}
\begin{subfigure}{.4\textwidth}
  \centering
\includegraphics[width=1\textwidth]{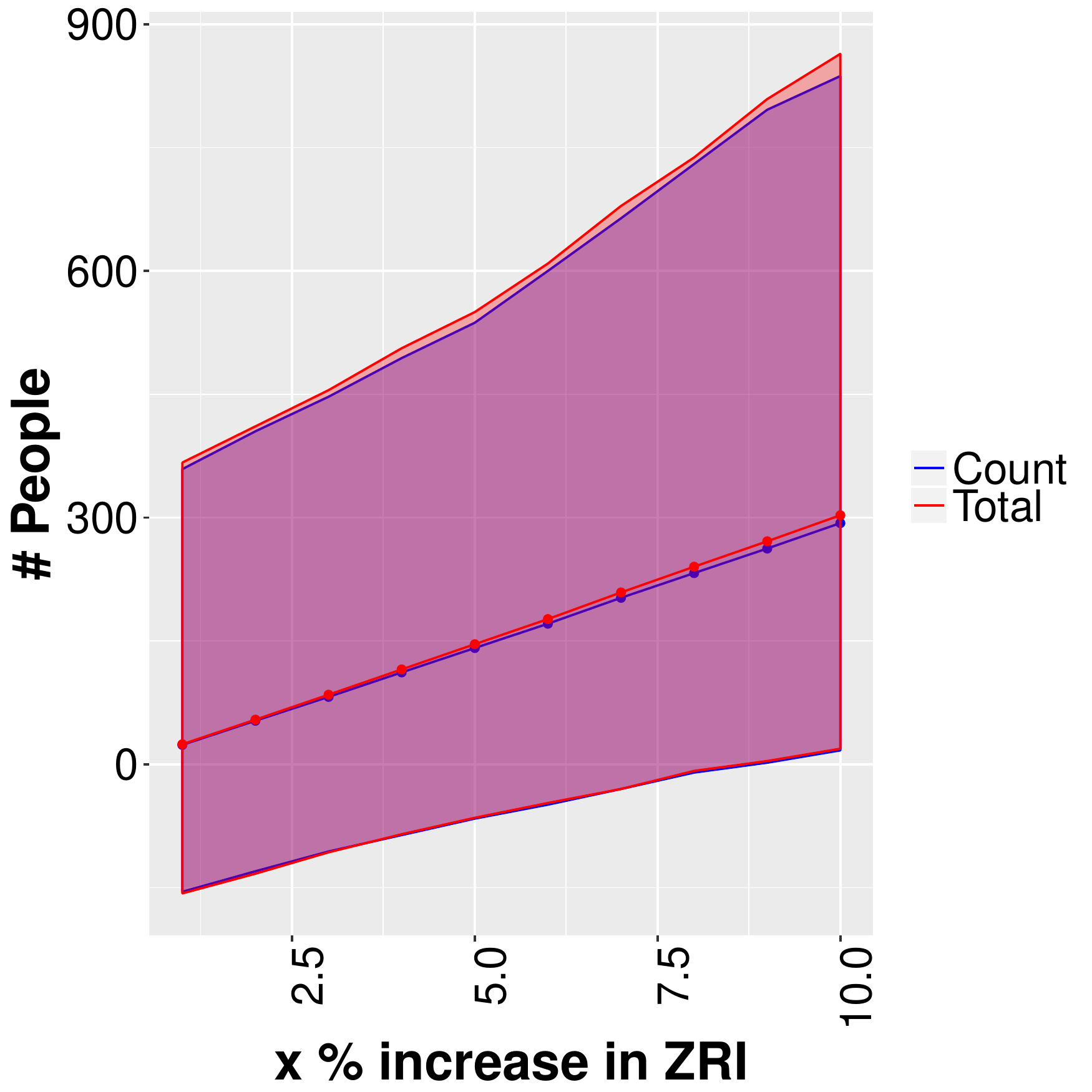}
\caption{ZRI effect}
\end{subfigure}
\begin{subfigure}{.4\textwidth}
  \centering
\includegraphics[width=1\textwidth]{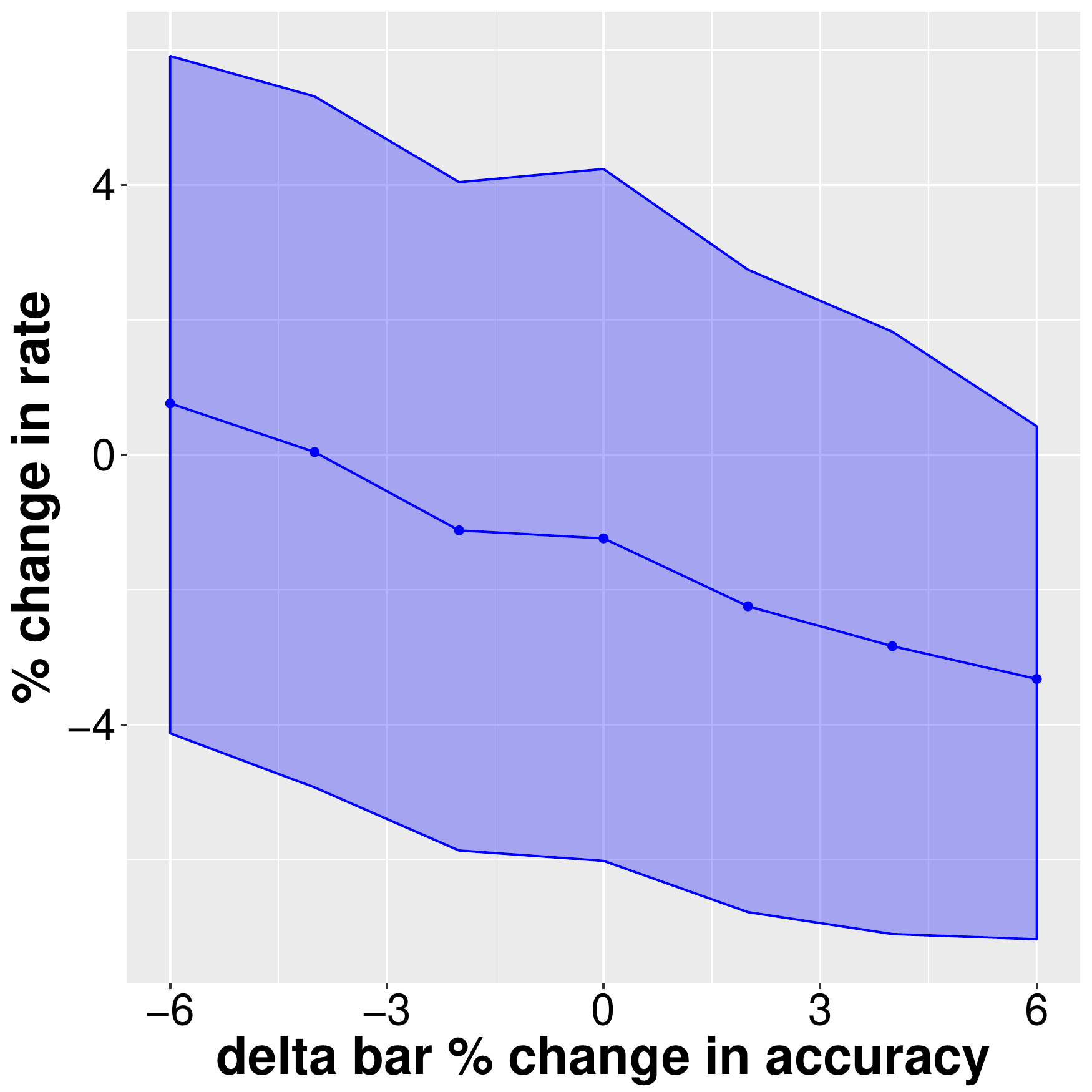}
\caption{Rate}
\end{subfigure}
\caption{Results for the Philadelphia, PA.  Top left (a): Posterior predictive distribution for homeless counts, $C_{i,1:T}^* | C_{1:25,1:T}, N_{1:25,1:T}$, in green, and the imputed total homeless population size, $H_{i,1:T}|C_{1:25,1:T}, H_{1:25,1:T}$, in blue.  The black 'x' marks correspond to the observed (raw) homeless count by year.  The count accuracy is modeled with a constant expectation.  Top right (b): Predictive distribution for total homeless population in 2017, $H_{i,2017} | C_{1:25,1:T}, N_{1:25,1:T}$.  Bottom left (c):  Posterior distribution of increase in total homeless population with increases in ZRI.  Bottom right (d): Sensitivity of the inferred increase in the homelessness rate from 2011 - 2016 to different annual changes in count accuracy.}
\label{fig:Philadelphia_Results}
\end{figure}

\begin{figure}[ht!]
\centering
\begin{subfigure}{.4\textwidth}
  \centering
\includegraphics[width=1\textwidth]{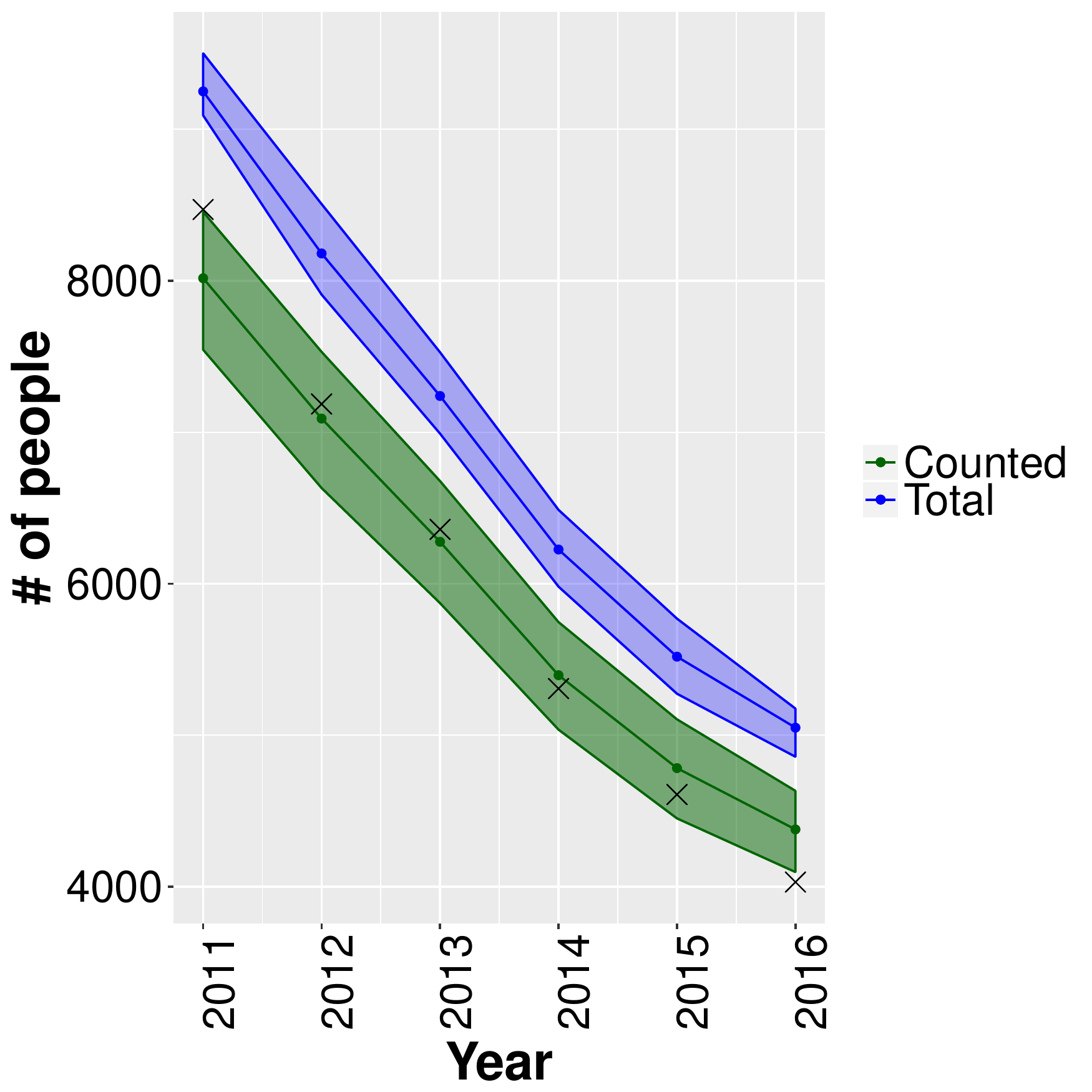}
\caption{\# of homeless}
\end{subfigure}
\begin{subfigure}{.4\textwidth}
  \centering
\includegraphics[width=1\textwidth]{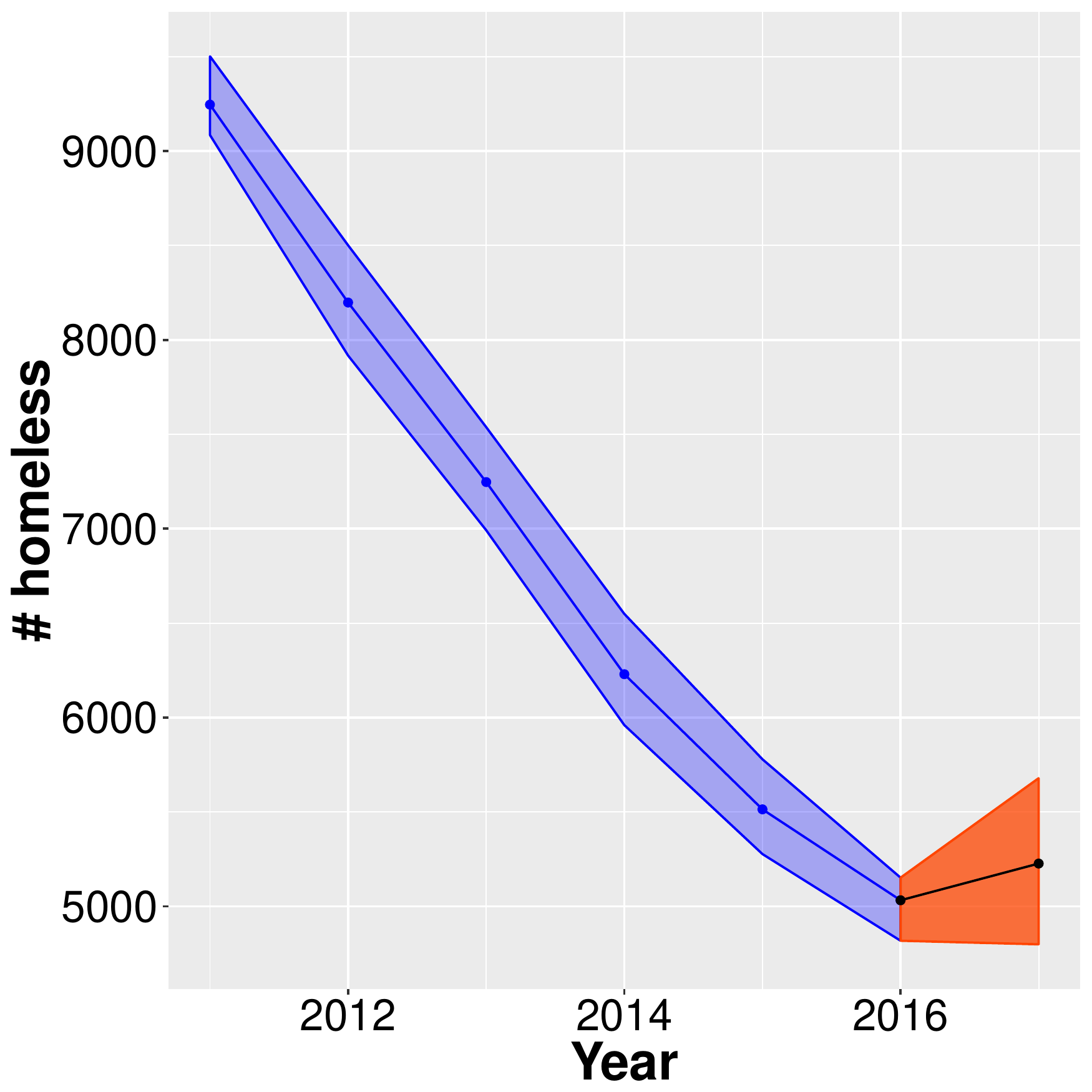}
\caption{2017 forecast}
\end{subfigure}
\begin{subfigure}{.4\textwidth}
  \centering
\includegraphics[width=1\textwidth]{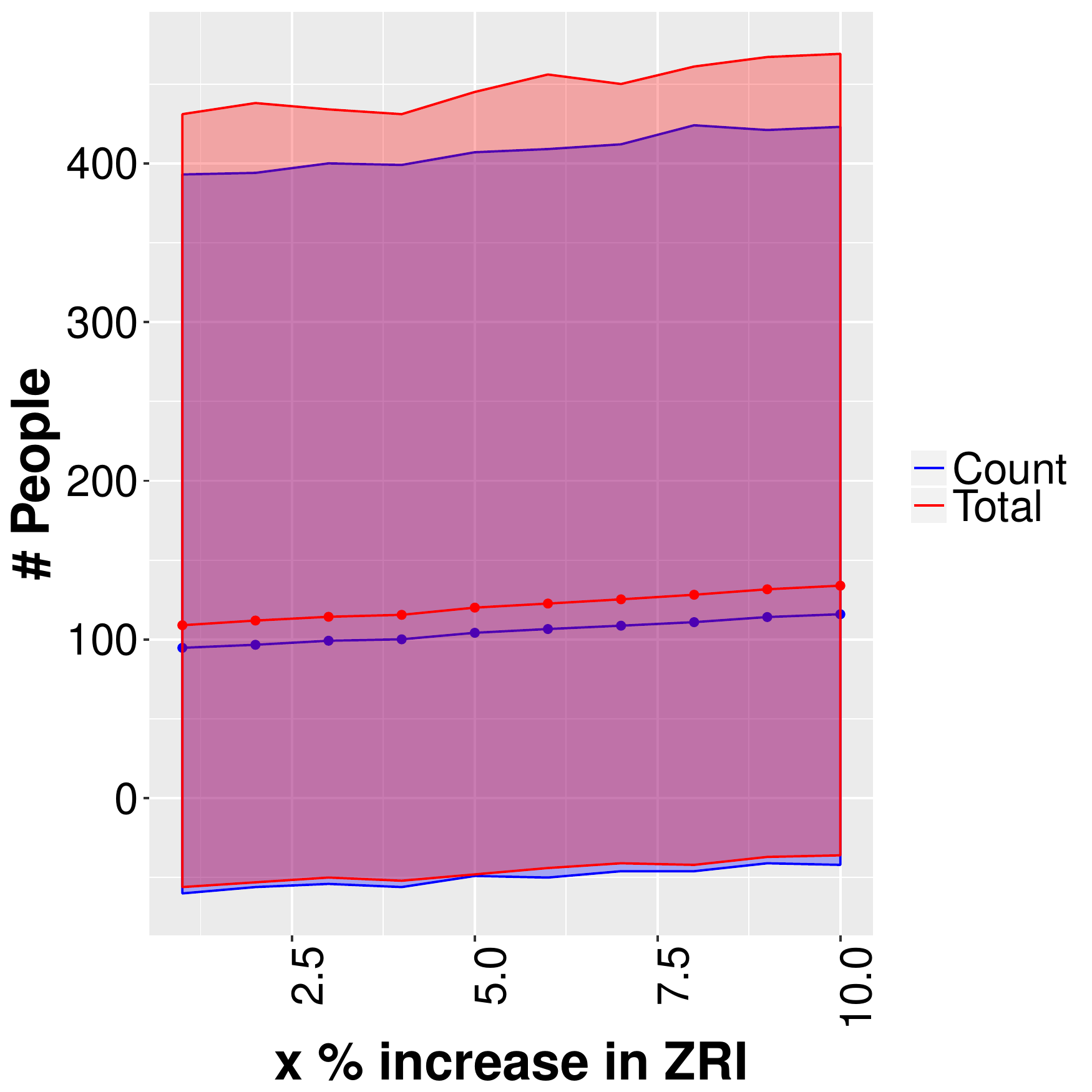}
\caption{ZRI effect}
\end{subfigure}
\begin{subfigure}{.4\textwidth}
  \centering
\includegraphics[width=1\textwidth]{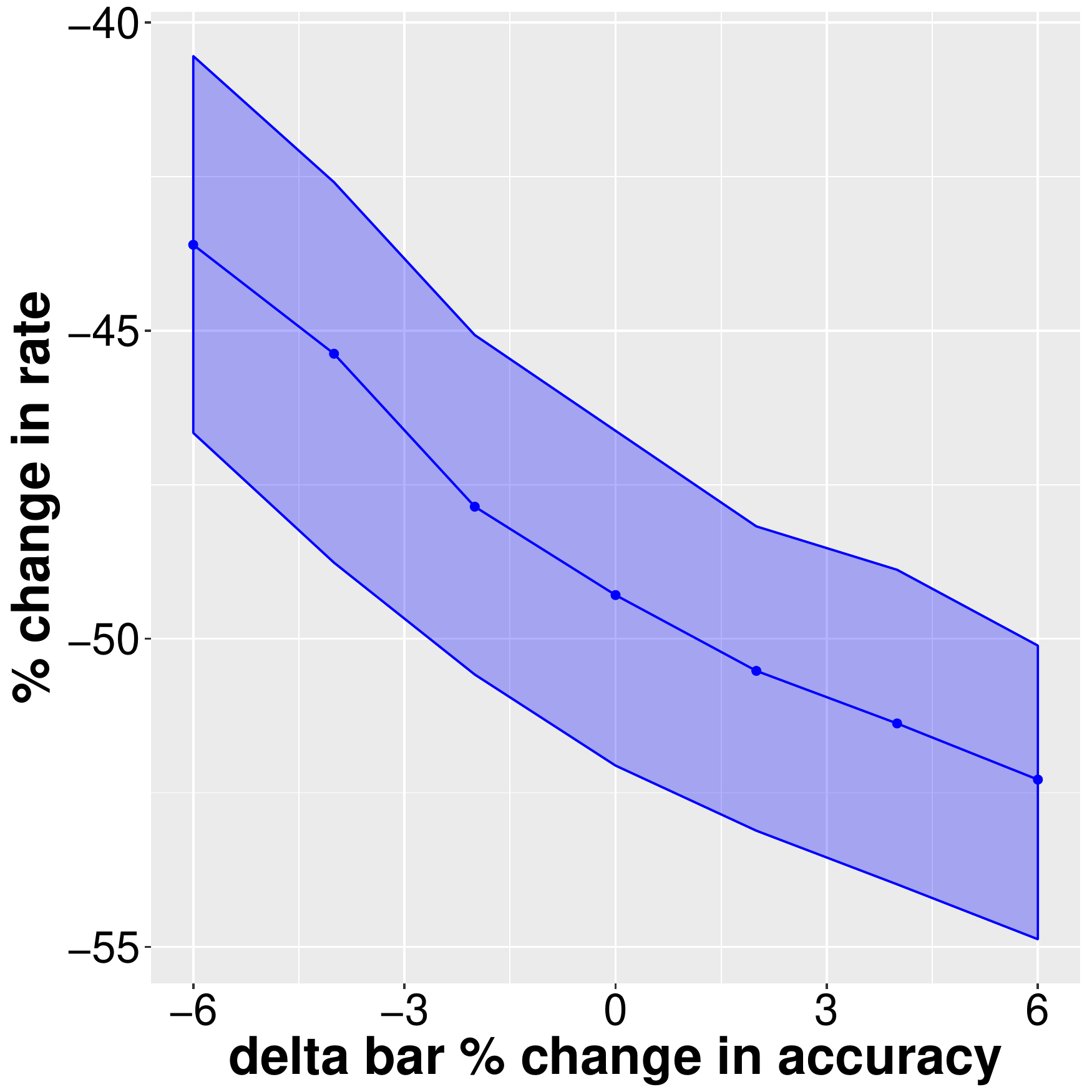}
\caption{Rate}
\end{subfigure}
\caption{Results for Houston, TX.  Top left (a): Posterior predictive distribution for homeless counts, $C_{i,1:T}^* | C_{1:25,1:T}, N_{1:25,1:T}$, in green, and the imputed total homeless population size, $H_{i,1:T}|C_{1:25,1:T}, H_{1:25,1:T}$, in blue.  The black 'x' marks correspond to the observed (raw) homeless count by year.  The count accuracy is modeled with a constant expectation.  Top right (b): Predictive distribution for total homeless population in 2017, $H_{i,2017} | C_{1:25,1:T}, N_{1:25,1:T}$.  Bottom left (c):  Posterior distribution of increase in total homeless population with increases in ZRI.  Bottom right (d): Sensitivity of the inferred increase in the homelessness rate from 2011 - 2016 to different annual changes in count accuracy.}
\label{fig:Houston_Results}
\end{figure}

\begin{figure}[ht!]
\centering
\begin{subfigure}{.4\textwidth}
  \centering
\includegraphics[width=1\textwidth]{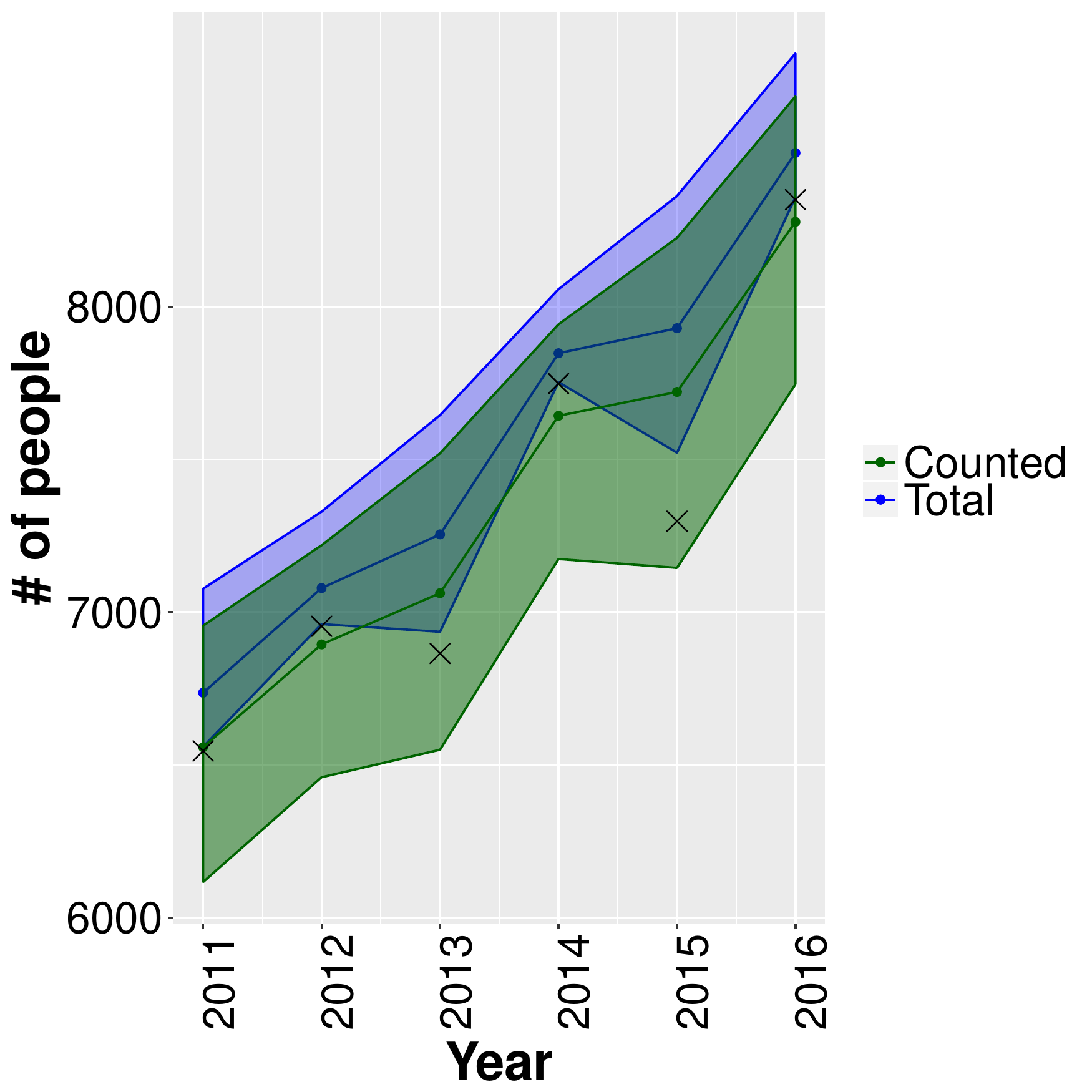}
\caption{\# of homeless}
\end{subfigure}
\begin{subfigure}{.4\textwidth}
  \centering
\includegraphics[width=1\textwidth]{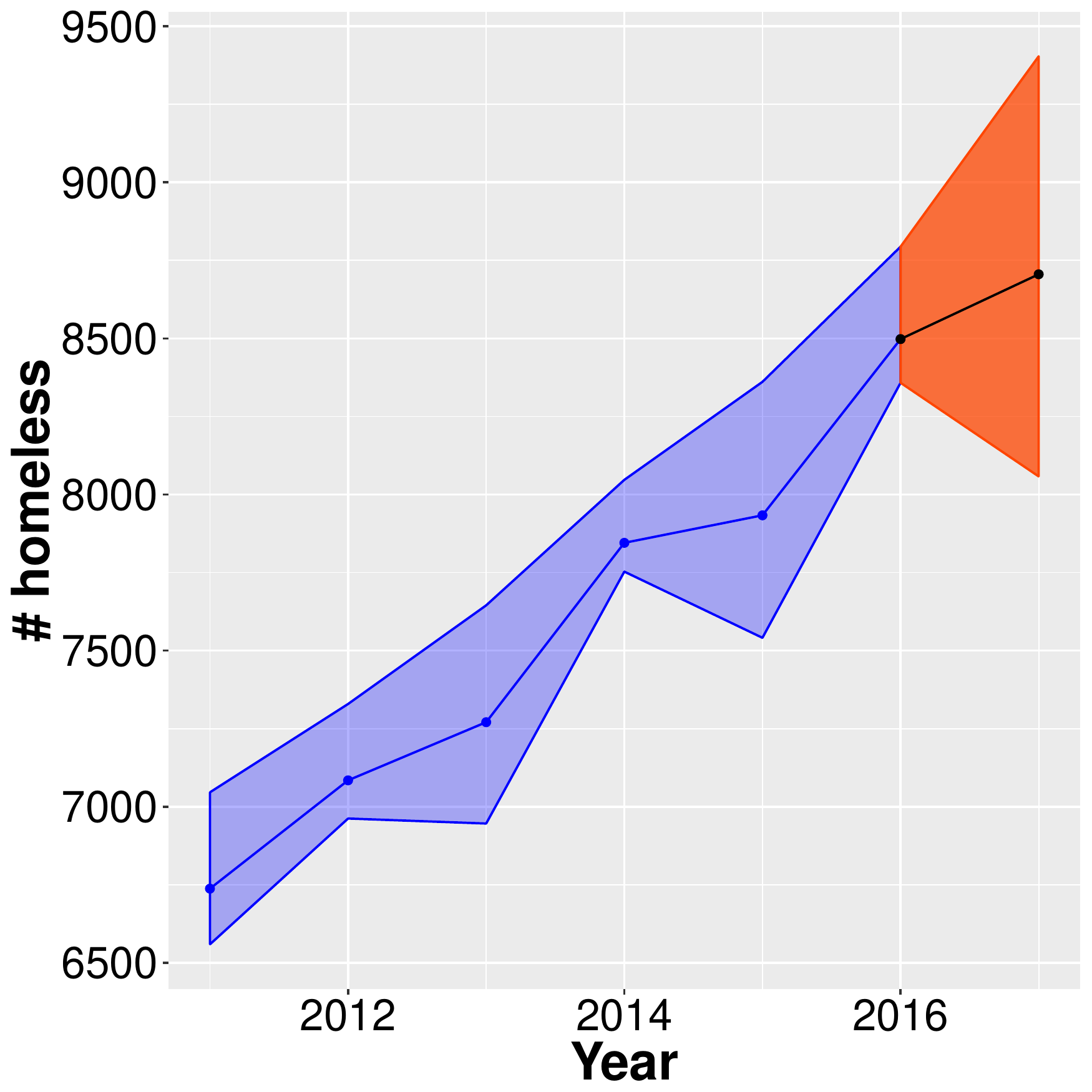}
\caption{2017 forecast}
\end{subfigure}
\begin{subfigure}{.4\textwidth}
  \centering
\includegraphics[width=1\textwidth]{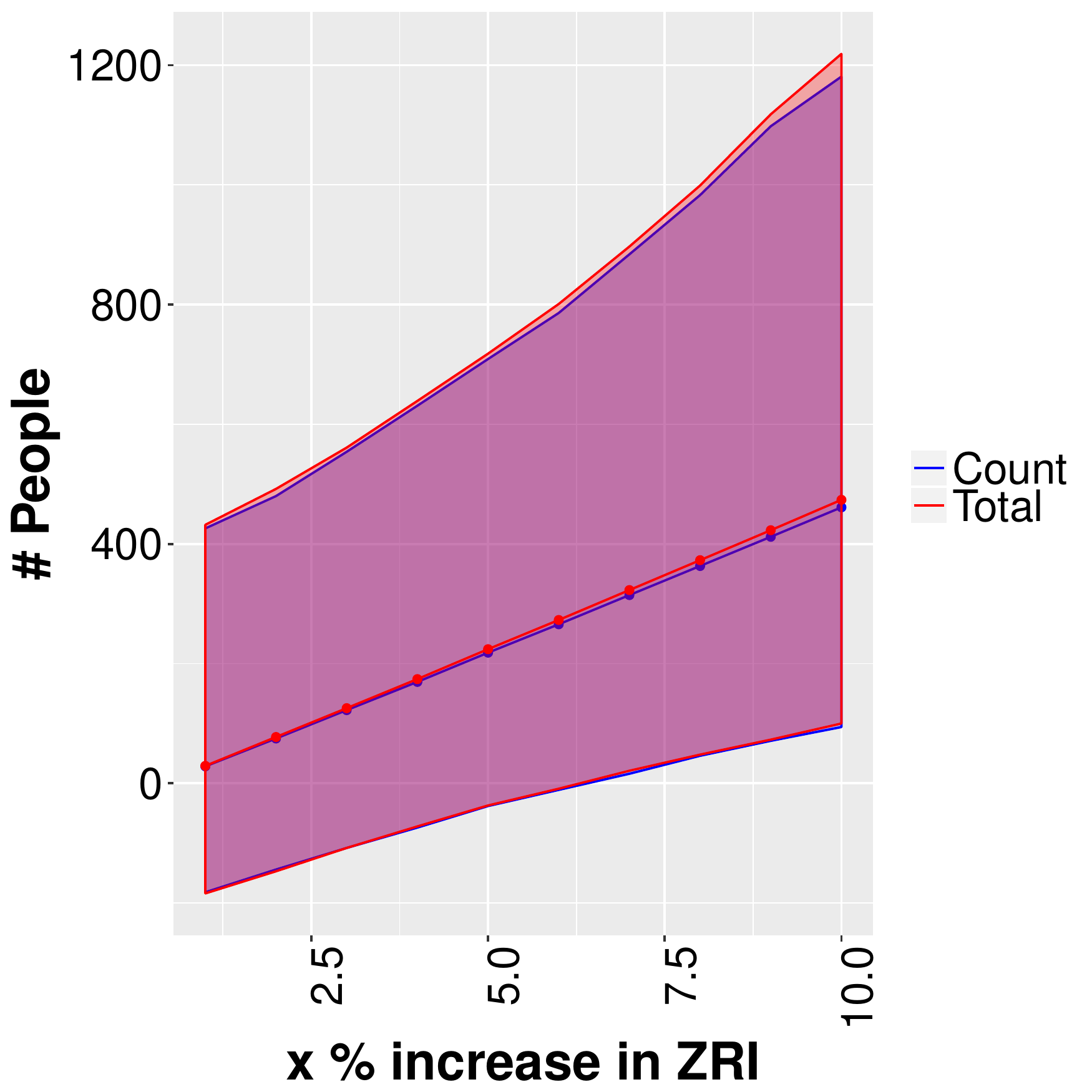}
\caption{ZRI effect}
\end{subfigure}
\begin{subfigure}{.4\textwidth}
  \centering
\includegraphics[width=1\textwidth]{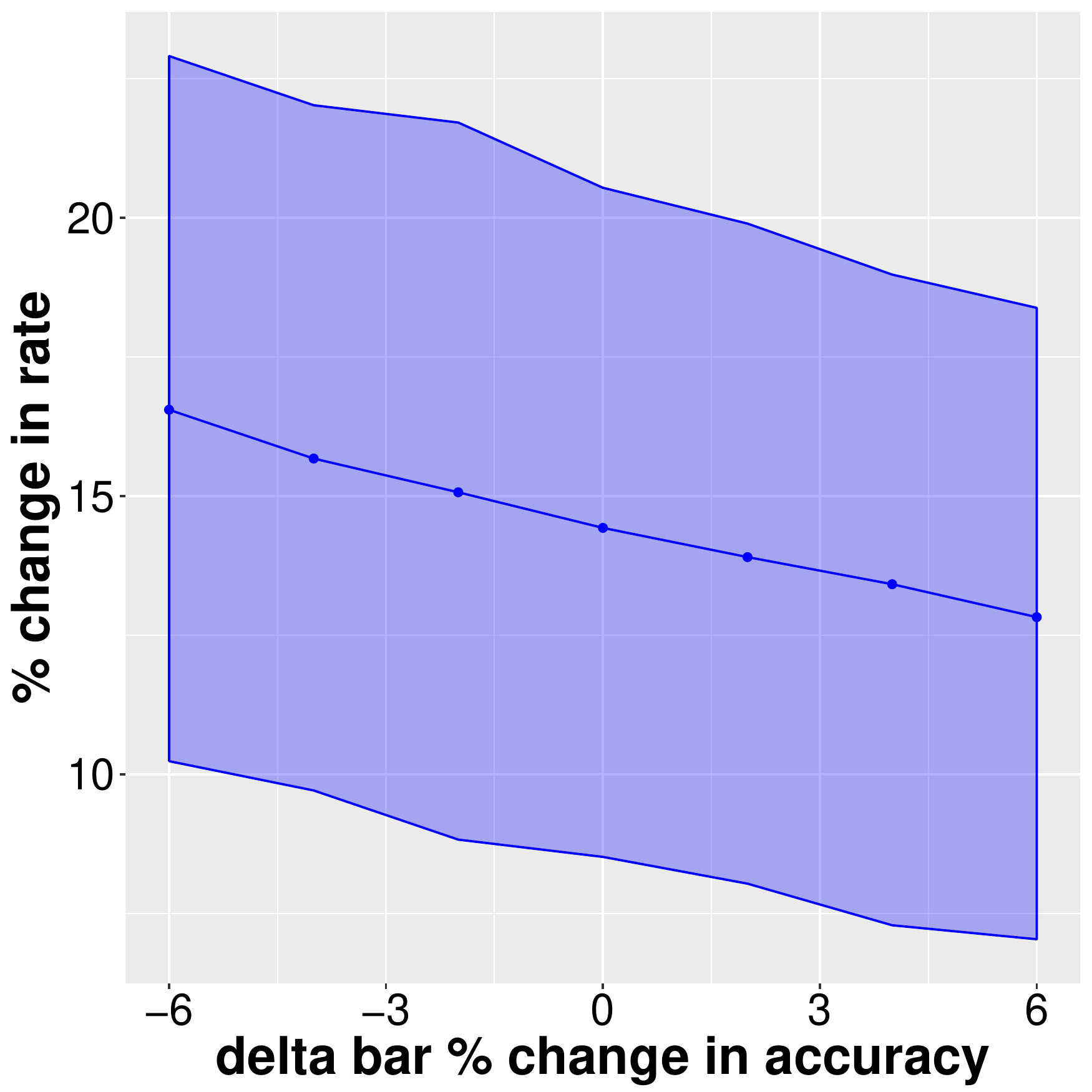}
\caption{Rate}
\end{subfigure}
\caption{Results for Washington, D.C..  Top left (a): Posterior predictive distribution for homeless counts, $C_{i,1:T}^* | C_{1:25,1:T}, N_{1:25,1:T}$, in green, and the imputed total homeless population size, $H_{i,1:T}|C_{1:25,1:T}, H_{1:25,1:T}$, in blue.  The black 'x' marks correspond to the observed (raw) homeless count by year.  The count accuracy is modeled with a constant expectation.  Top right (b): Predictive distribution for total homeless population in 2017, $H_{i,2017} | C_{1:25,1:T}, N_{1:25,1:T}$.  Bottom left (c):  Posterior distribution of increase in total homeless population with increases in ZRI.  Bottom right (d): Sensitivity of the inferred increase in the homelessness rate from 2011 - 2016 to different annual changes in count accuracy.}
\label{fig:DC_Results}
\end{figure}

\begin{figure}[ht!]
\centering
\begin{subfigure}{.4\textwidth}
  \centering
\includegraphics[width=1\textwidth]{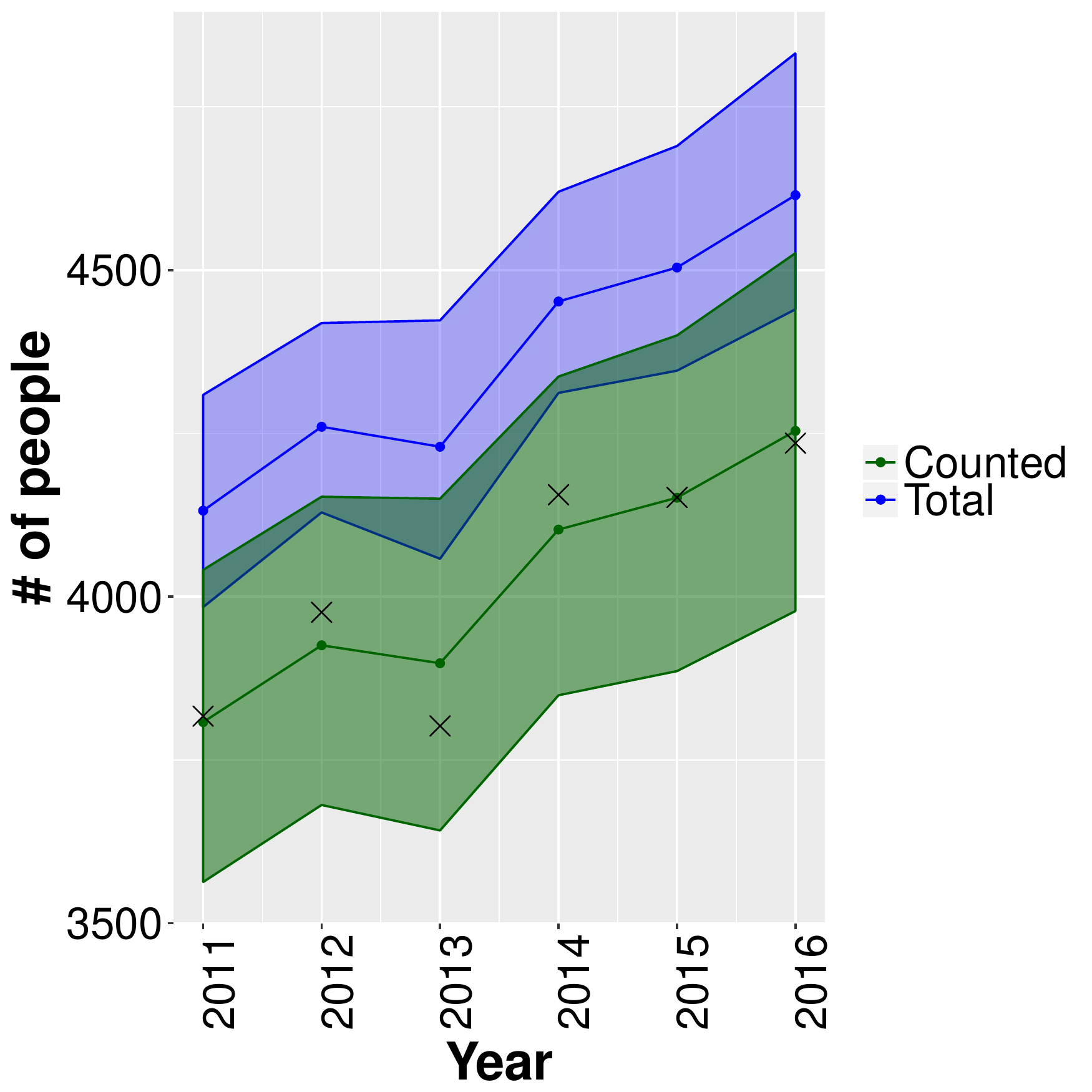}
\caption{\# of homeless}
\end{subfigure}
\begin{subfigure}{.4\textwidth}
  \centering
\includegraphics[width=1\textwidth]{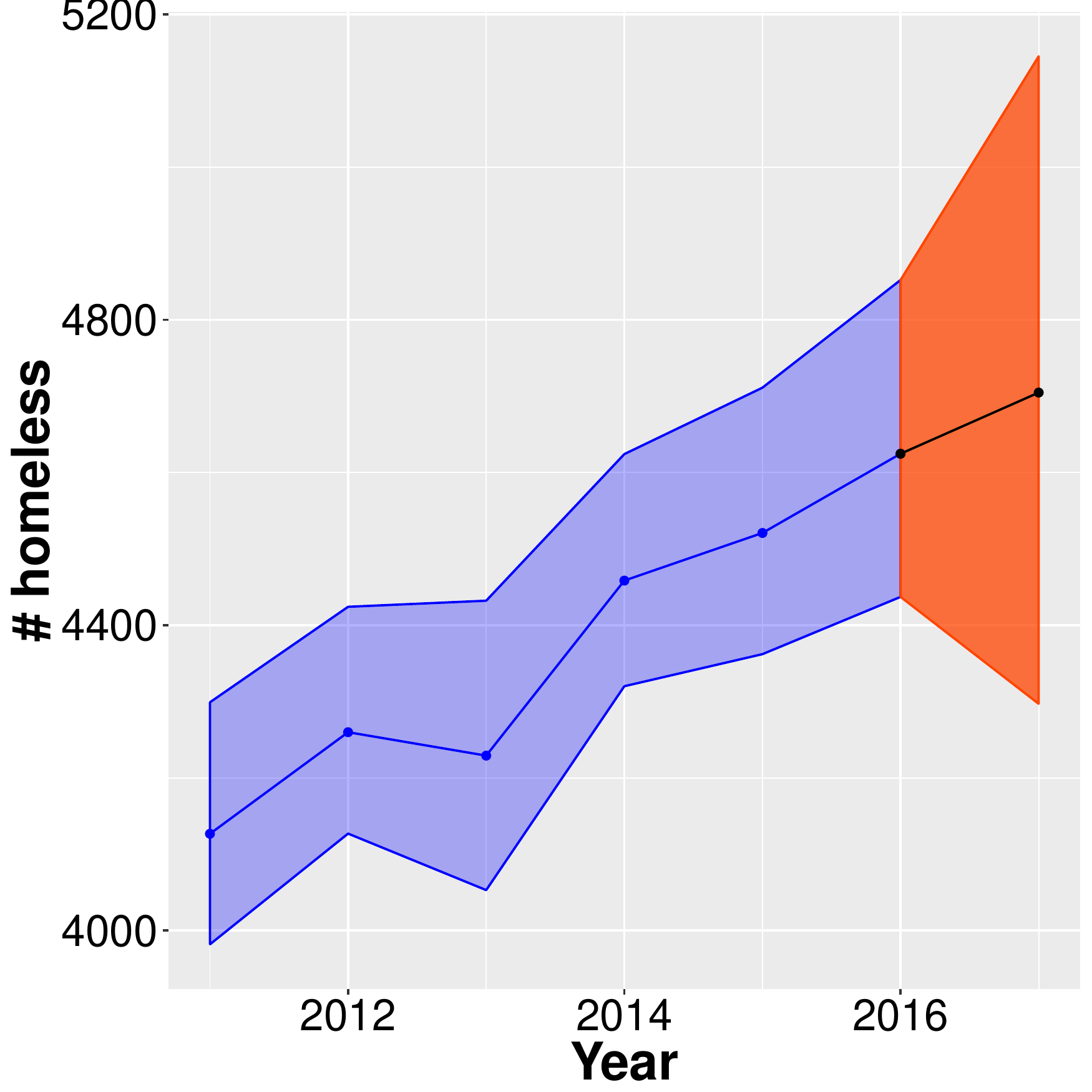}
\caption{2017 forecast}
\end{subfigure}
\begin{subfigure}{.4\textwidth}
  \centering
\includegraphics[width=1\textwidth]{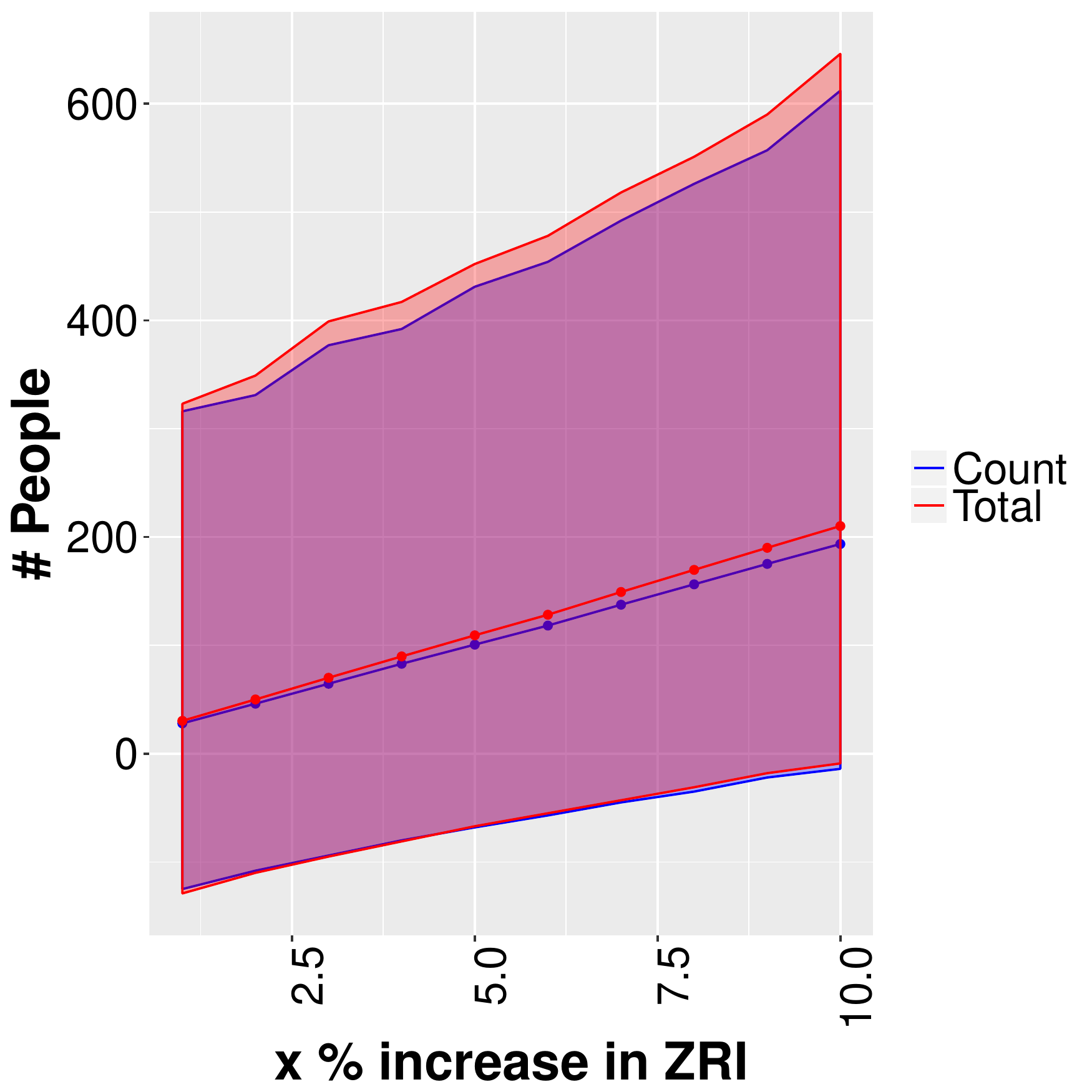}
\caption{ZRI effect}
\end{subfigure}
\begin{subfigure}{.4\textwidth}
  \centering
\includegraphics[width=1\textwidth]{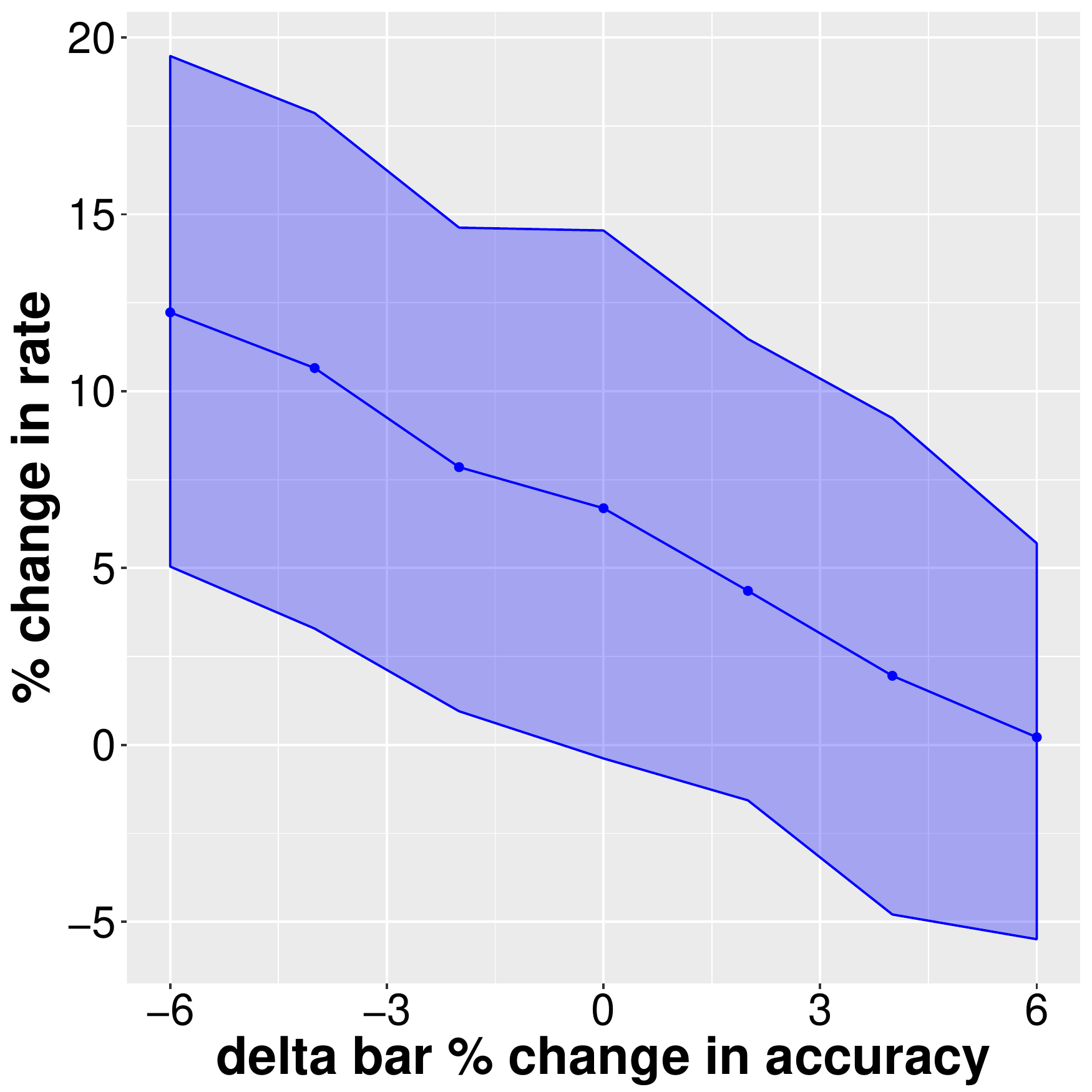}
\caption{Rate}
\end{subfigure}
\caption{Results for Miami, FL.  Top left (a): Posterior predictive distribution for homeless counts, $C_{i,1:T}^* | C_{1:25,1:T}, N_{1:25,1:T}$, in green, and the imputed total homeless population size, $H_{i,1:T}|C_{1:25,1:T}, H_{1:25,1:T}$, in blue.  The black 'x' marks correspond to the observed (raw) homeless count by year.  The count accuracy is modeled with a constant expectation.  Top right (b): Predictive distribution for total homeless population in 2017, $H_{i,2017} | C_{1:25,1:T}, N_{1:25,1:T}$.  Bottom left (c):  Posterior distribution of increase in total homeless population with increases in ZRI.  Bottom right (d): Sensitivity of the inferred increase in the homelessness rate from 2011 - 2016 to different annual changes in count accuracy.}
\label{fig:Miami_Results}
\end{figure}

\begin{figure}[ht!]
\centering
\begin{subfigure}{.4\textwidth}
  \centering
\includegraphics[width=1\textwidth]{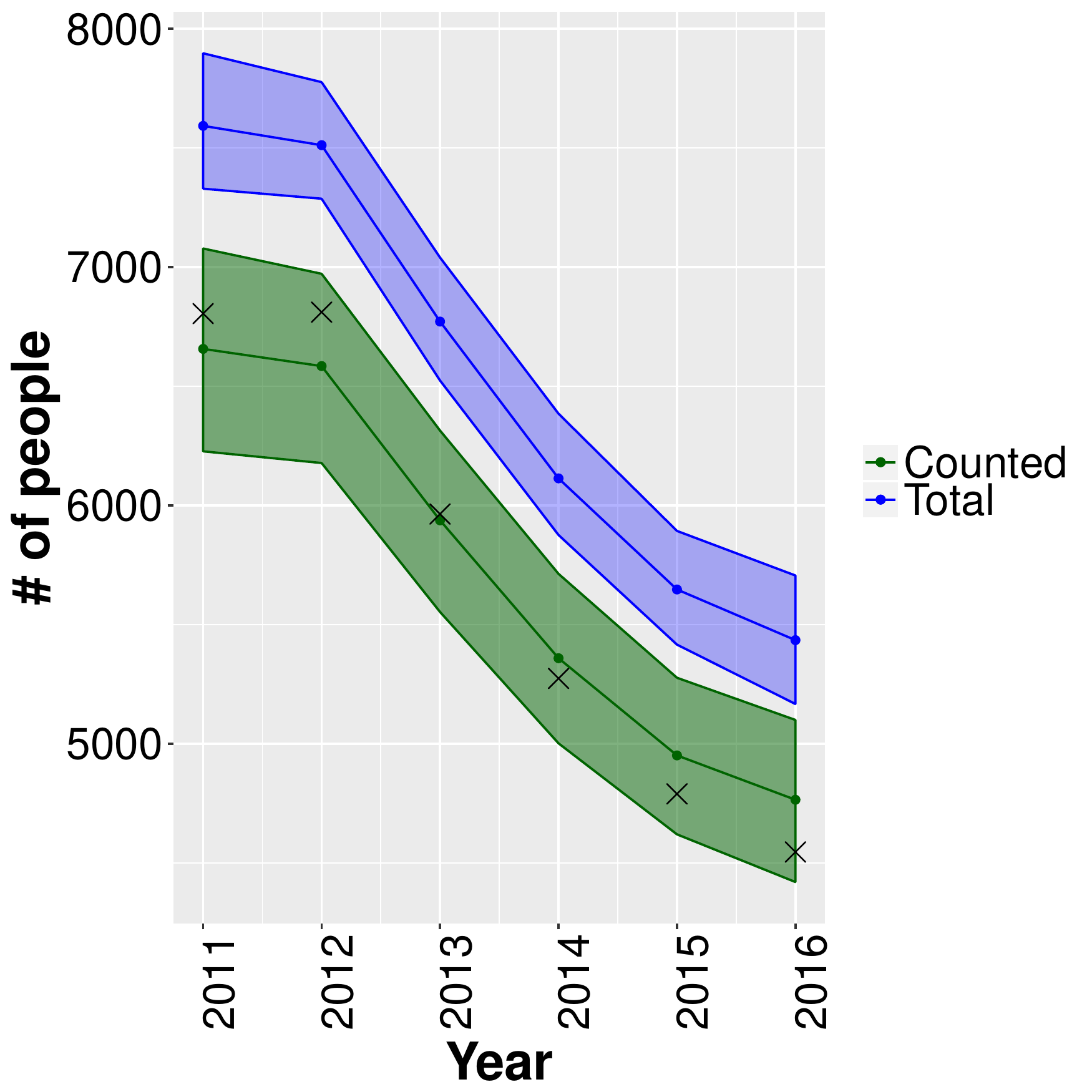}
\caption{\# of homeless}
\end{subfigure}
\begin{subfigure}{.4\textwidth}
  \centering
\includegraphics[width=1\textwidth]{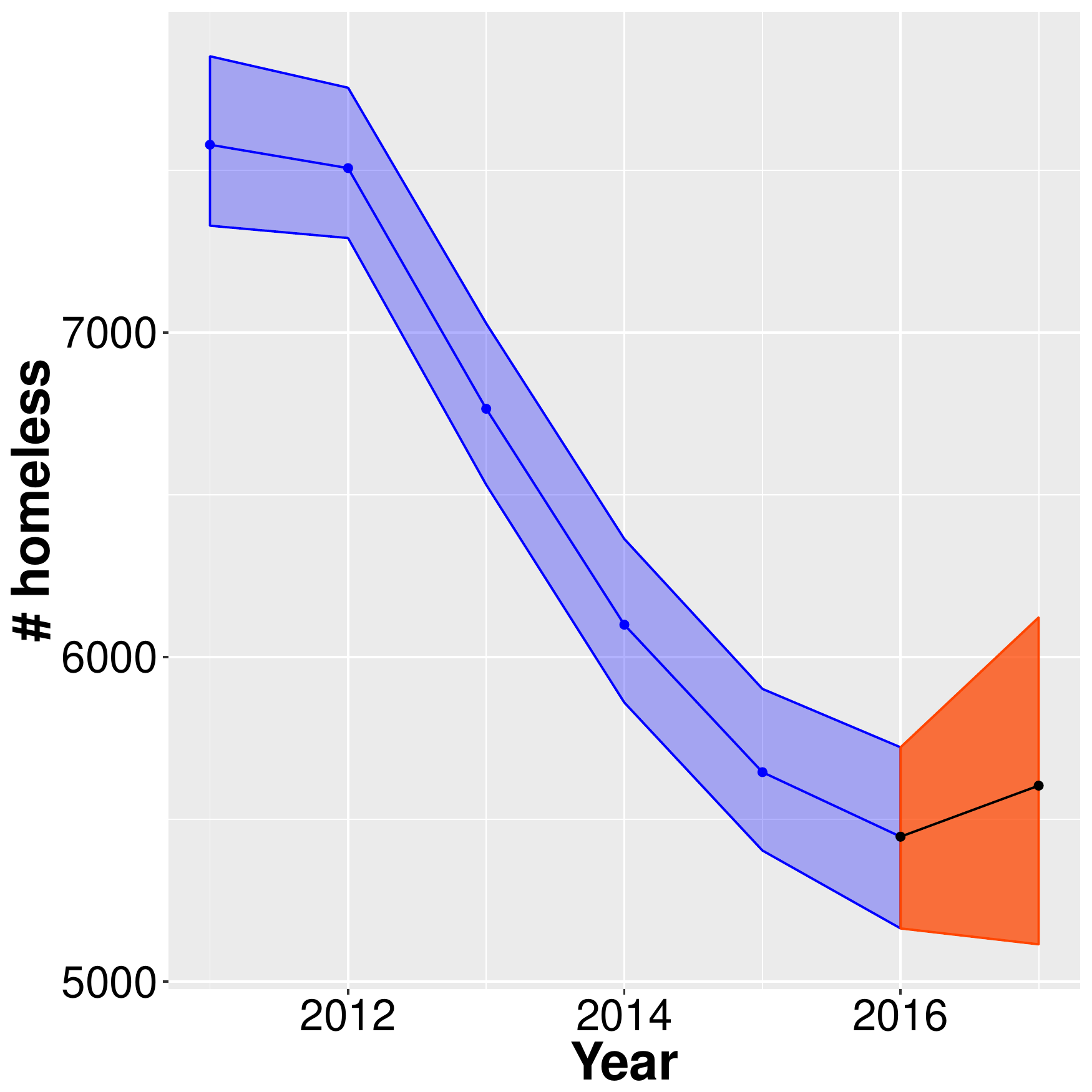}
\caption{2017 forecast}
\end{subfigure}
\begin{subfigure}{.4\textwidth}
  \centering
\includegraphics[width=1\textwidth]{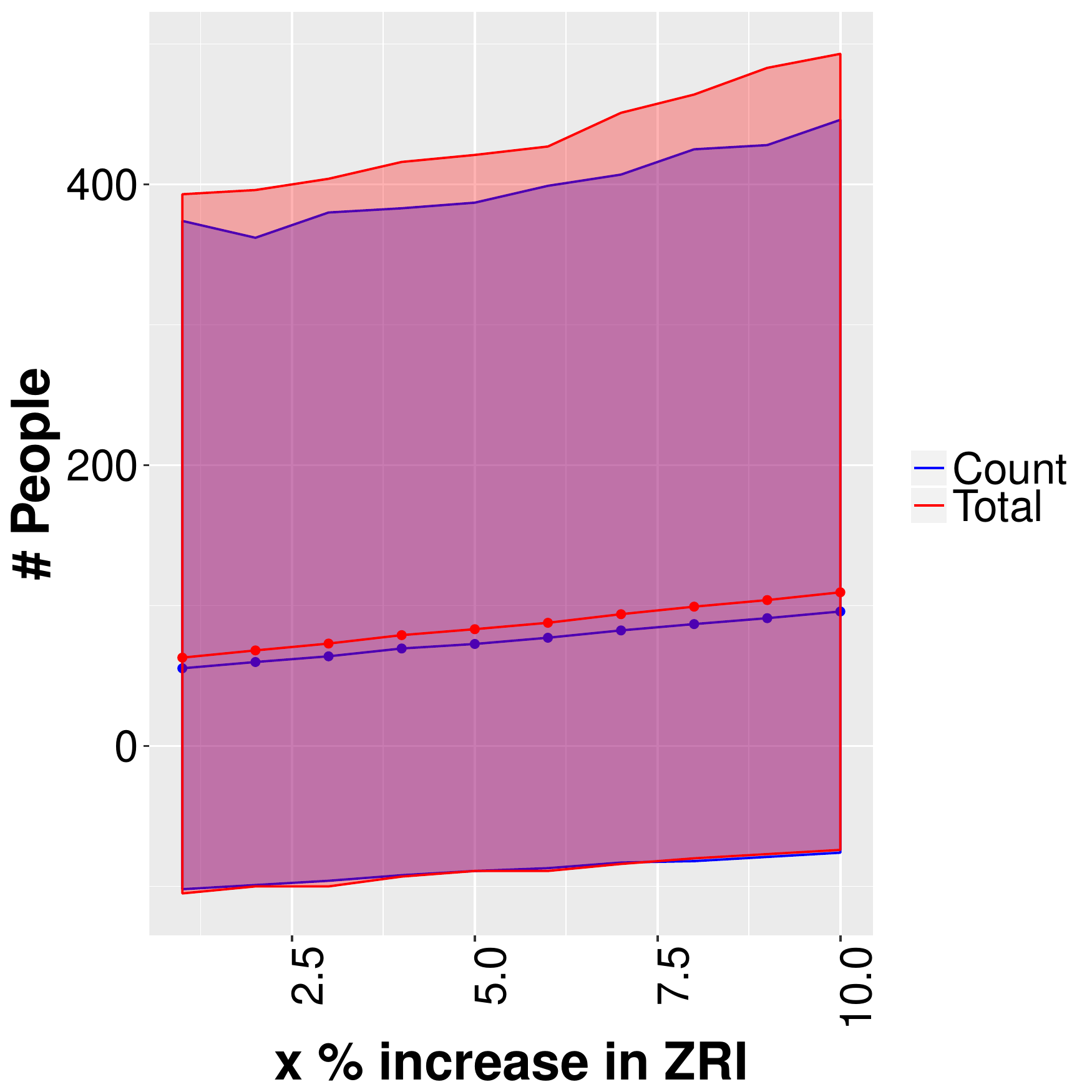}
\caption{ZRI effect}
\end{subfigure}
\begin{subfigure}{.4\textwidth}
  \centering
\includegraphics[width=1\textwidth]{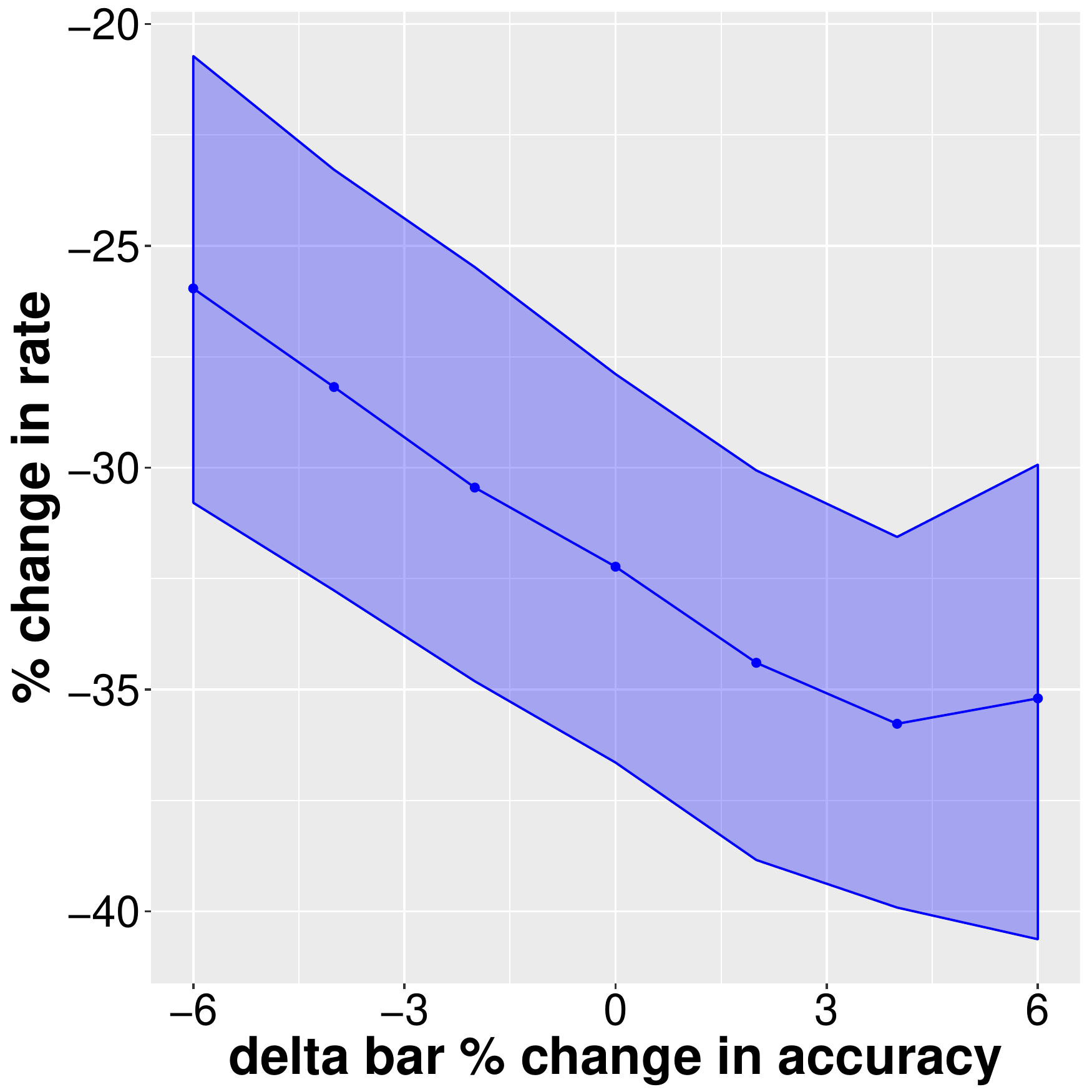}
\caption{Rate}
\end{subfigure}
\caption{Results for Atlanta, GA.  Top left (a): Posterior predictive distribution for homeless counts, $C_{i,1:T}^* | C_{1:25,1:T}, N_{1:25,1:T}$, in green, and the imputed total homeless population size, $H_{i,1:T}|C_{1:25,1:T}, H_{1:25,1:T}$, in blue.  The black 'x' marks correspond to the observed (raw) homeless count by year.  The count accuracy is modeled with a constant expectation.  Top right (b): Predictive distribution for total homeless population in 2017, $H_{i,2017} | C_{1:25,1:T}, N_{1:25,1:T}$.  Bottom left (c):  Posterior distribution of increase in total homeless population with increases in ZRI.  Bottom right (d): Sensitivity of the inferred increase in the homelessness rate from 2011 - 2016 to different annual changes in count accuracy.}
\label{fig:Atlanta_Results}
\end{figure}

\begin{figure}[ht!]
\centering
\begin{subfigure}{.4\textwidth}
  \centering
\includegraphics[width=1\textwidth]{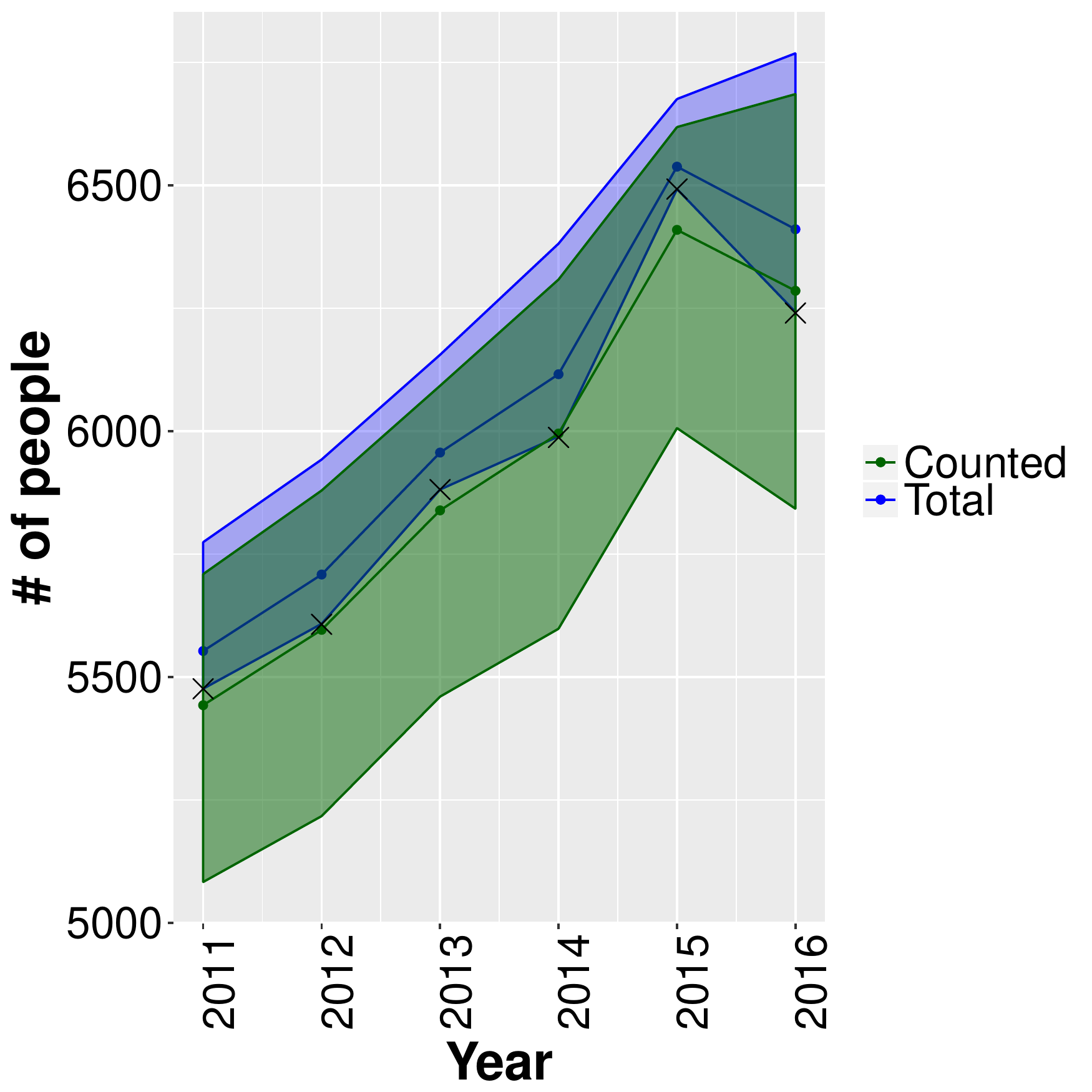}
\caption{\# of homeless}
\end{subfigure}
\begin{subfigure}{.4\textwidth}
  \centering
\includegraphics[width=1\textwidth]{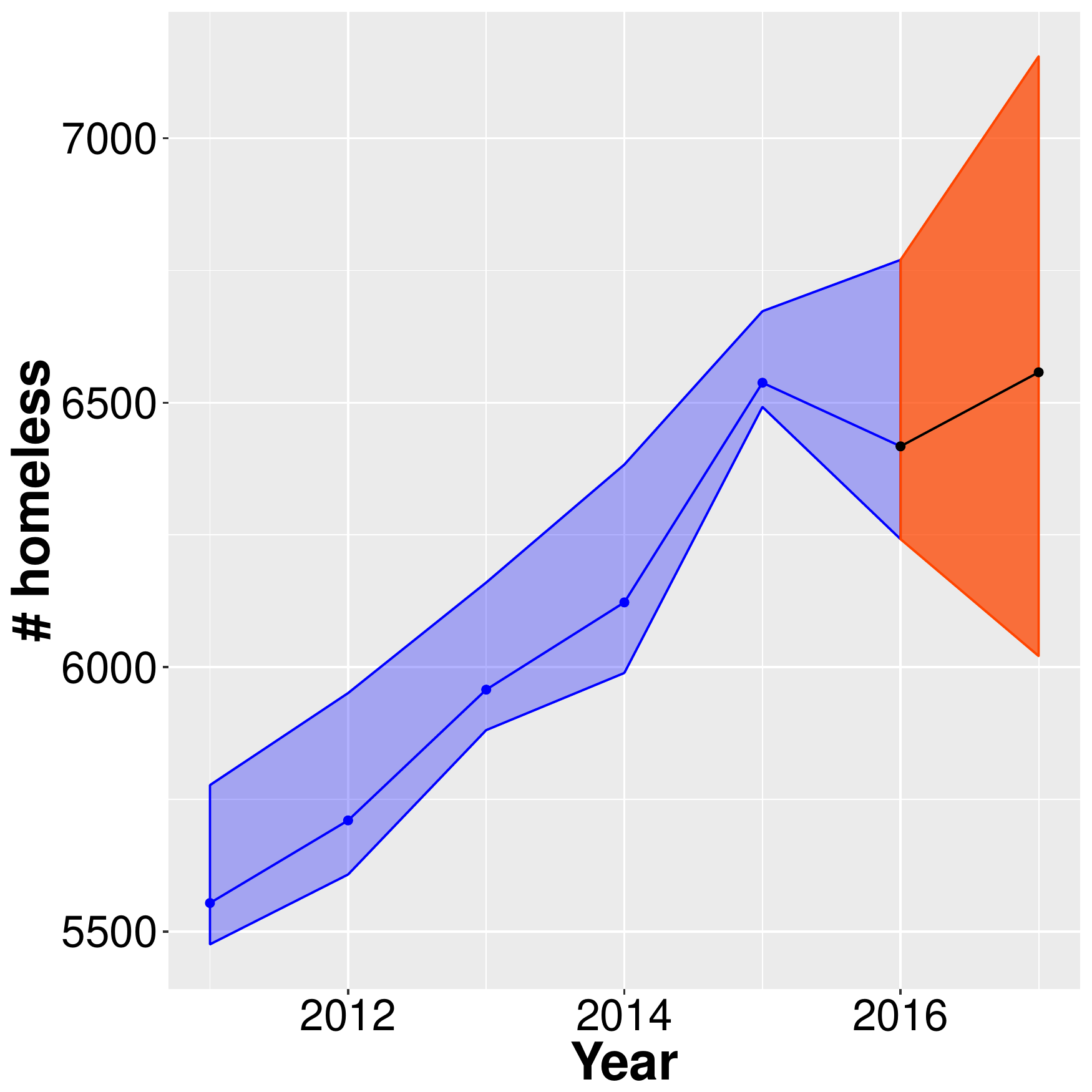}
\caption{2017 forecast}
\end{subfigure}
\begin{subfigure}{.4\textwidth}
  \centering
\includegraphics[width=1\textwidth]{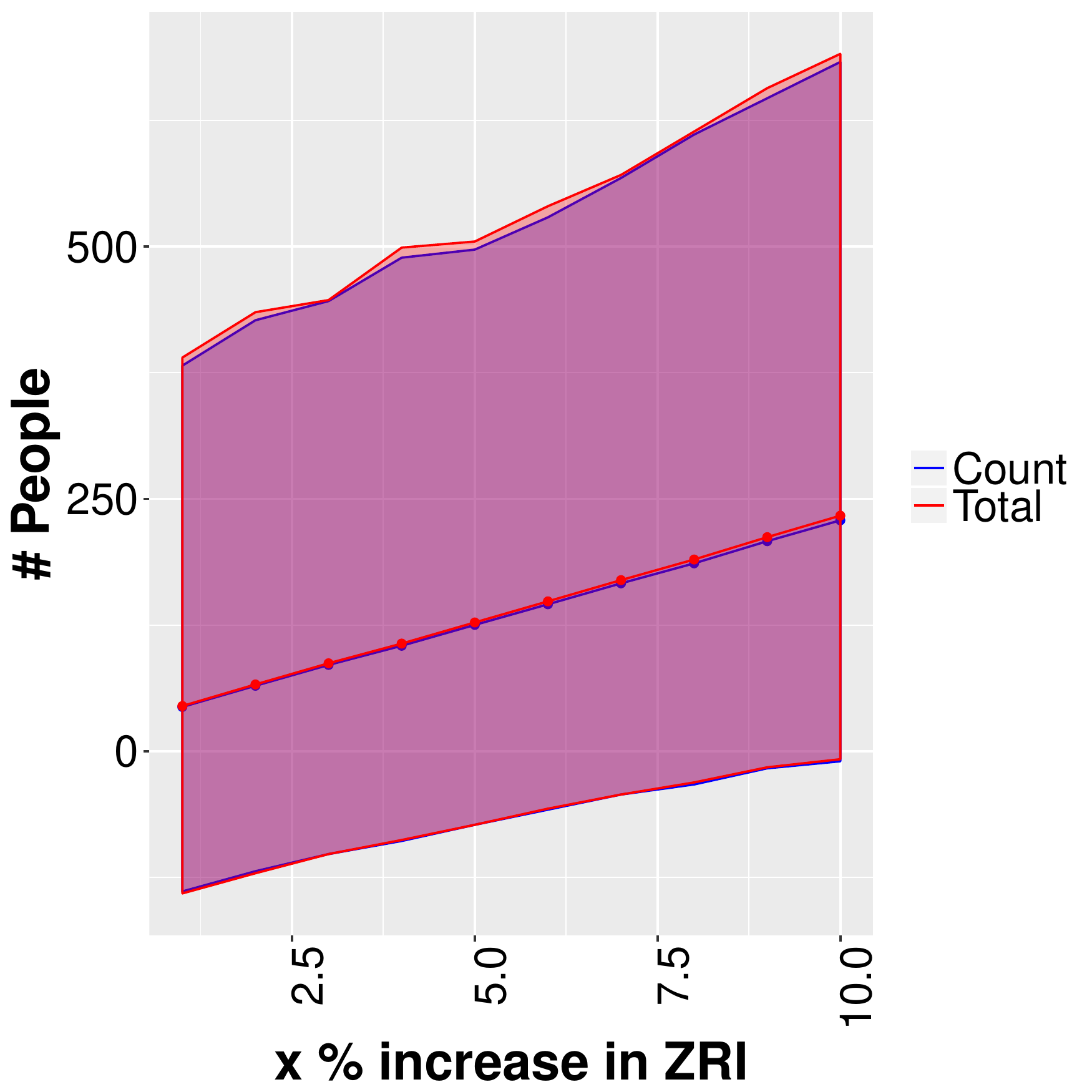}
\caption{ZRI effect}
\end{subfigure}
\begin{subfigure}{.4\textwidth}
  \centering
\includegraphics[width=1\textwidth]{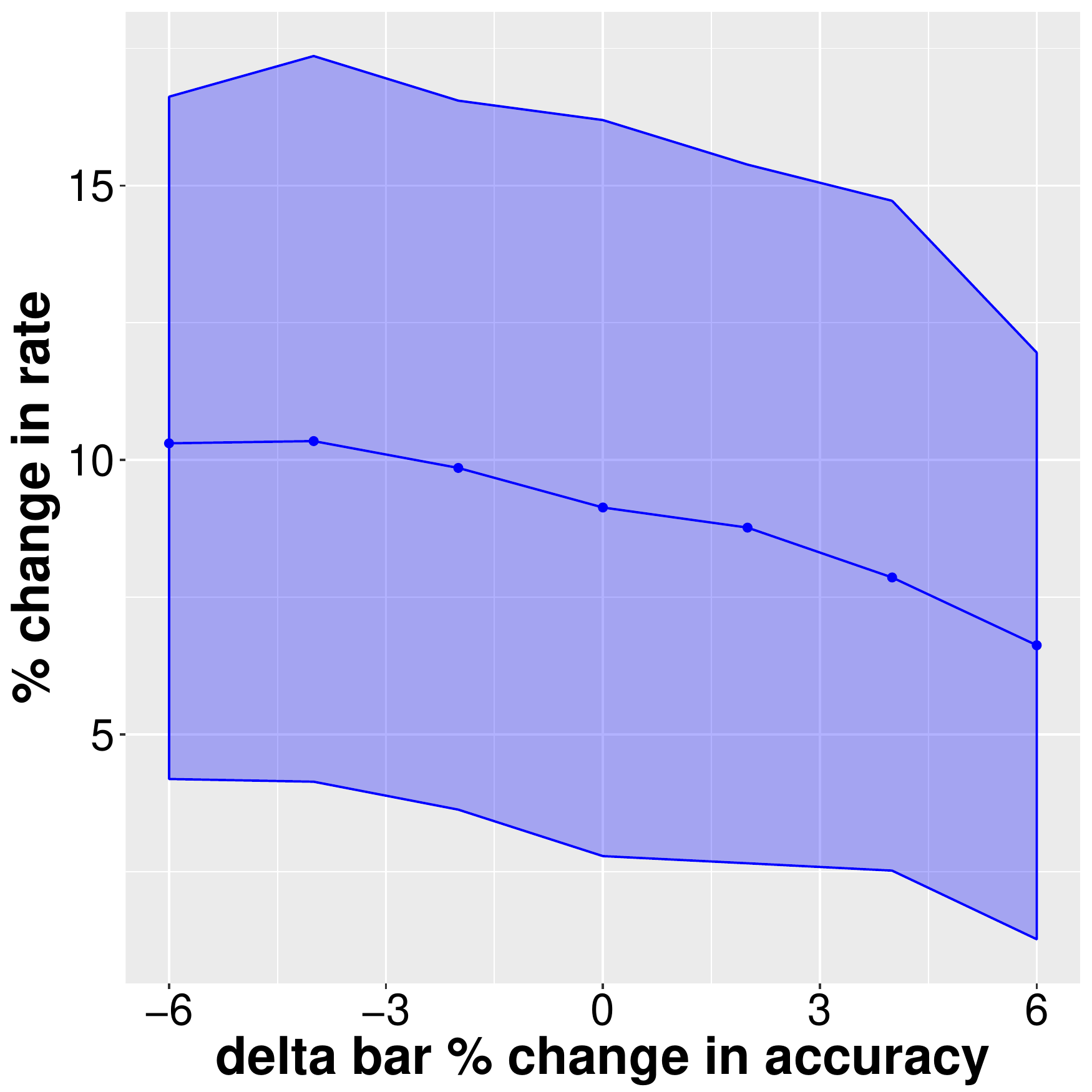}
\caption{Rate}
\end{subfigure}
\caption{Results for Boston, MA.  Top left (a): Posterior predictive distribution for homeless counts, $C_{i,1:T}^* | C_{1:25,1:T}, N_{1:25,1:T}$, in green, and the imputed total homeless population size, $H_{i,1:T}|C_{1:25,1:T}, H_{1:25,1:T}$, in blue.  The black 'x' marks correspond to the observed (raw) homeless count by year.  The count accuracy is modeled with a constant expectation.  Top right (b): Predictive distribution for total homeless population in 2017, $H_{i,2017} | C_{1:25,1:T}, N_{1:25,1:T}$.  Bottom left (c):  Posterior distribution of increase in total homeless population with increases in ZRI.  Bottom right (d): Sensitivity of the inferred increase in the homelessness rate from 2011 - 2016 to different annual changes in count accuracy.}
\label{fig:Boston_Results}
\end{figure}

\begin{figure}[ht!]
\centering
\begin{subfigure}{.4\textwidth}
  \centering
\includegraphics[width=1\textwidth]{SF_Homeless.pdf}
\caption{\# of homeless}
\end{subfigure}
\begin{subfigure}{.4\textwidth}
  \centering
\includegraphics[width=1\textwidth]{SF_Homeless_Forecast.pdf}
\caption{2017 forecast}
\end{subfigure}
\begin{subfigure}{.4\textwidth}
  \centering
\includegraphics[width=1\textwidth]{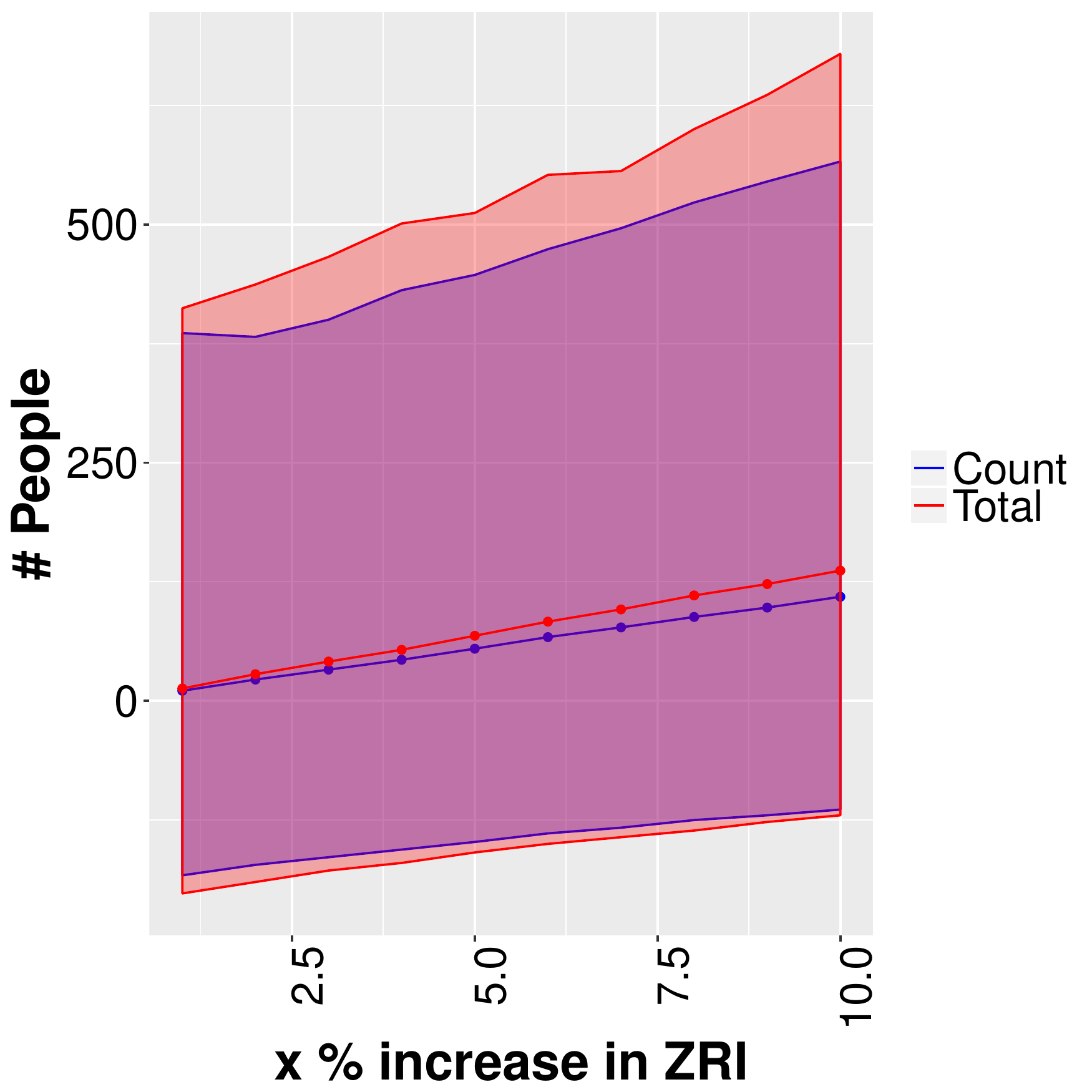}
\caption{ZRI effect}
\end{subfigure}
\begin{subfigure}{.4\textwidth}
  \centering
\includegraphics[width=1\textwidth]{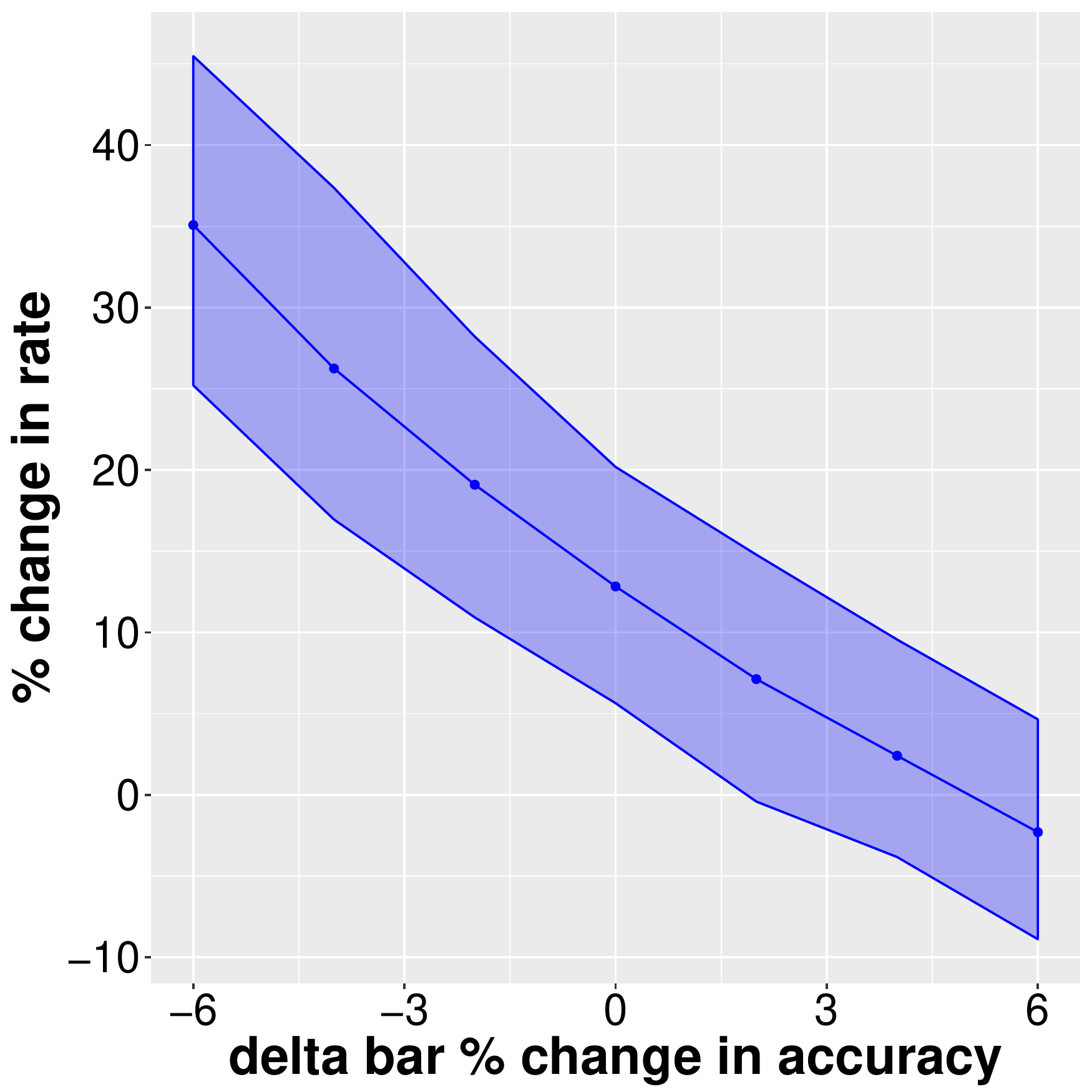}
\caption{Rate}
\end{subfigure}
\caption{Results for San Francisco, CA.  Top left (a): Posterior predictive distribution for homeless counts, $C_{i,1:T}^* | C_{1:25,1:T}, N_{1:25,1:T}$, in green, and the imputed total homeless population size, $H_{i,1:T}|C_{1:25,1:T}, H_{1:25,1:T}$, in blue.  The black 'x' marks correspond to the observed (raw) homeless count by year.  The count accuracy is modeled with a constant expectation.  Top right (b): Predictive distribution for total homeless population in 2017, $H_{i,2017} | C_{1:25,1:T}, N_{1:25,1:T}$.  Bottom left (c):  Posterior distribution of increase in total homeless population with increases in ZRI.  Bottom right (d): Sensitivity of the inferred increase in the homelessness rate from 2011 - 2016 to different annual changes in count accuracy.}
\label{fig:SF_Results}
\end{figure}

\begin{figure}[ht!]
\centering
\begin{subfigure}{.4\textwidth}
  \centering
\includegraphics[width=1\textwidth]{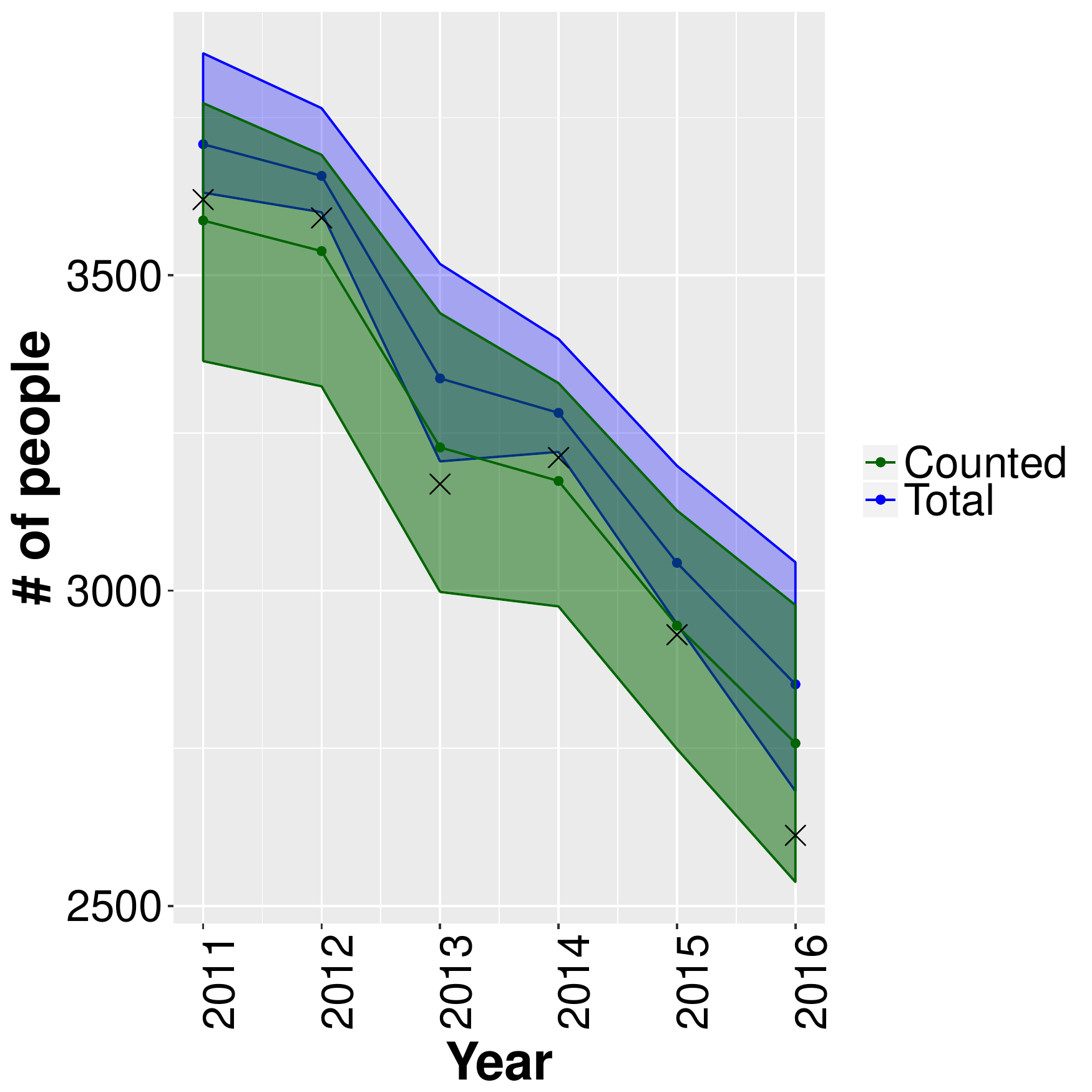}
\caption{\# of homeless}
\end{subfigure}
\begin{subfigure}{.4\textwidth}
  \centering
\includegraphics[width=1\textwidth]{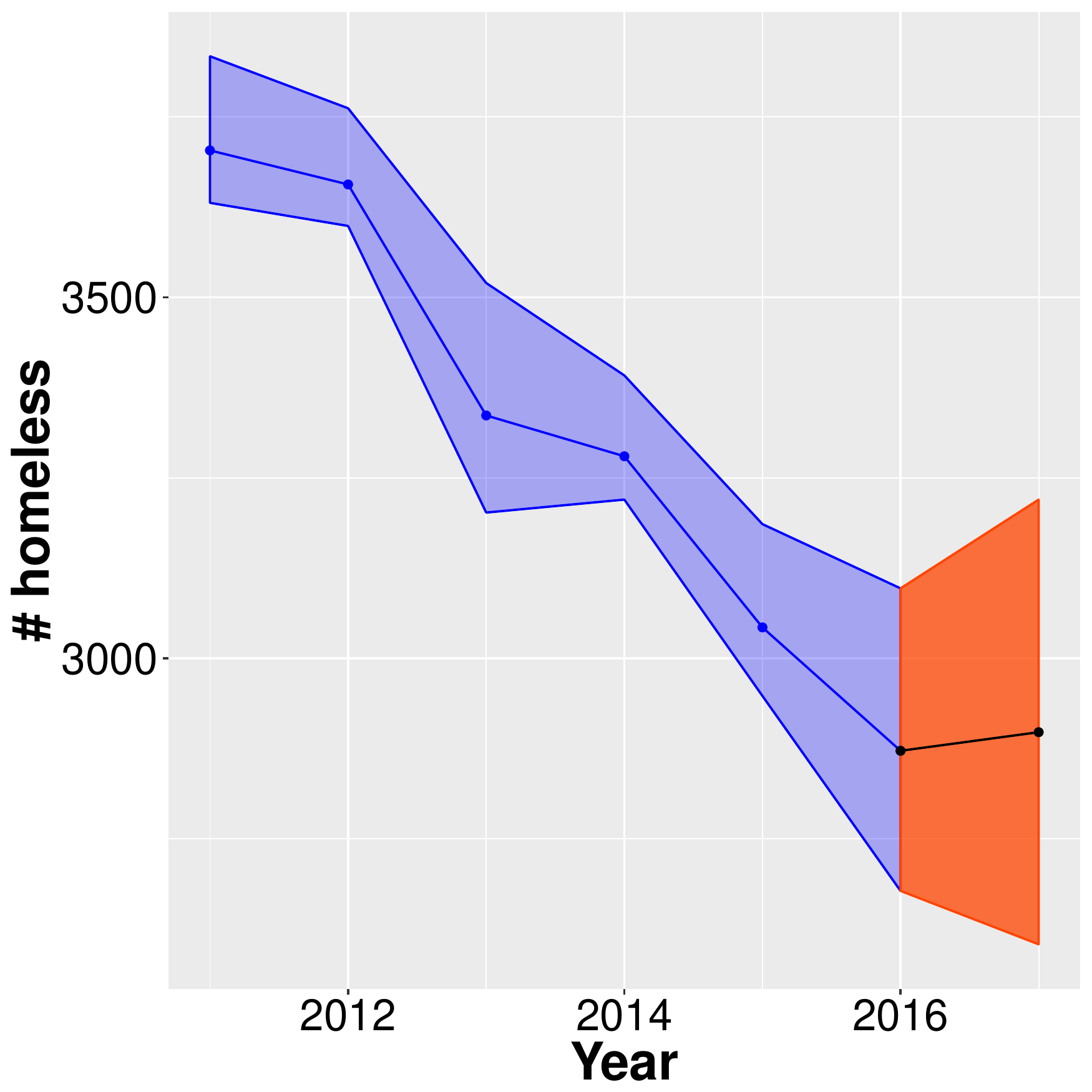}
\caption{2017 forecast}
\end{subfigure}
\begin{subfigure}{.4\textwidth}
  \centering
\includegraphics[width=1\textwidth]{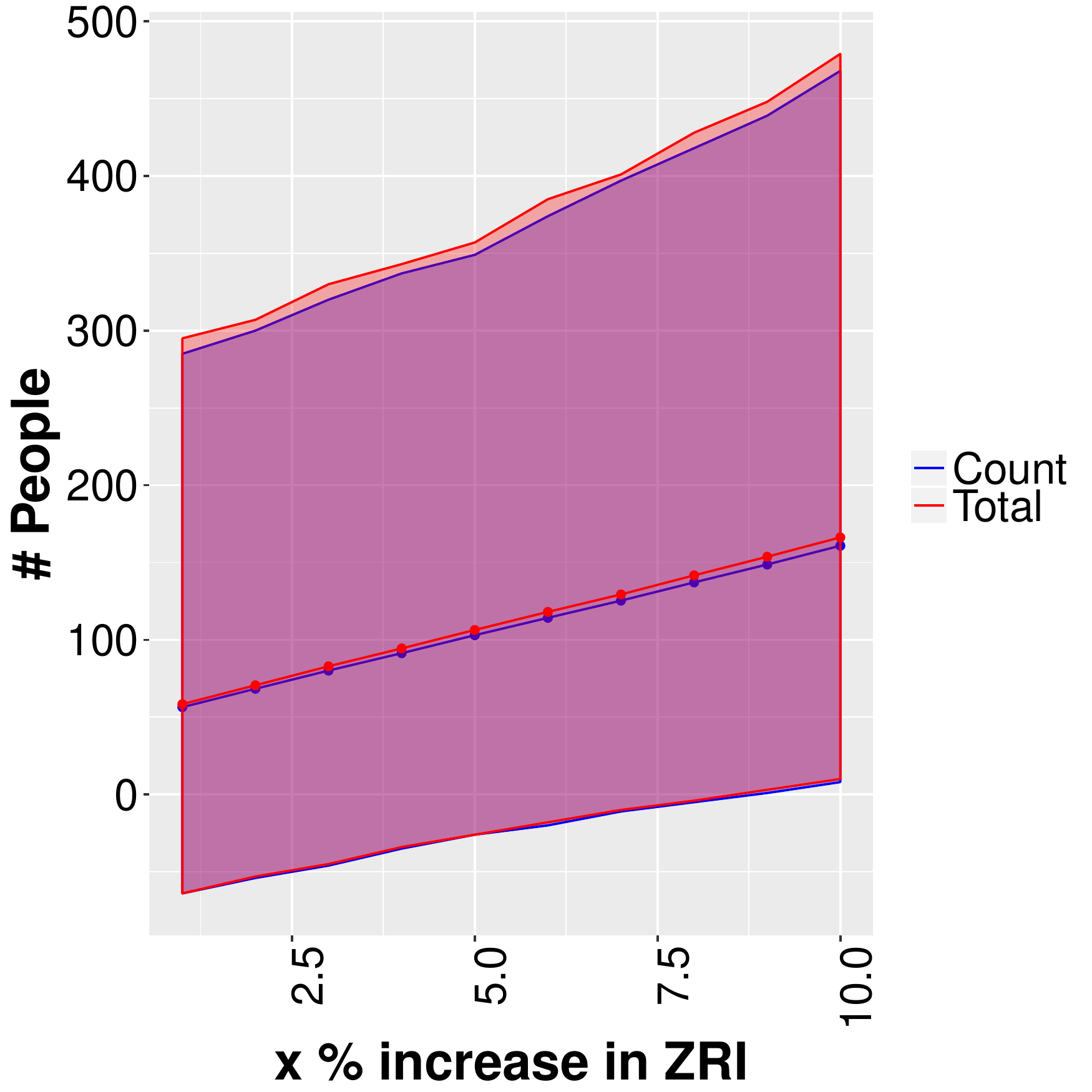}
\caption{ZRI effect}
\end{subfigure}
\begin{subfigure}{.4\textwidth}
  \centering
\includegraphics[width=1\textwidth]{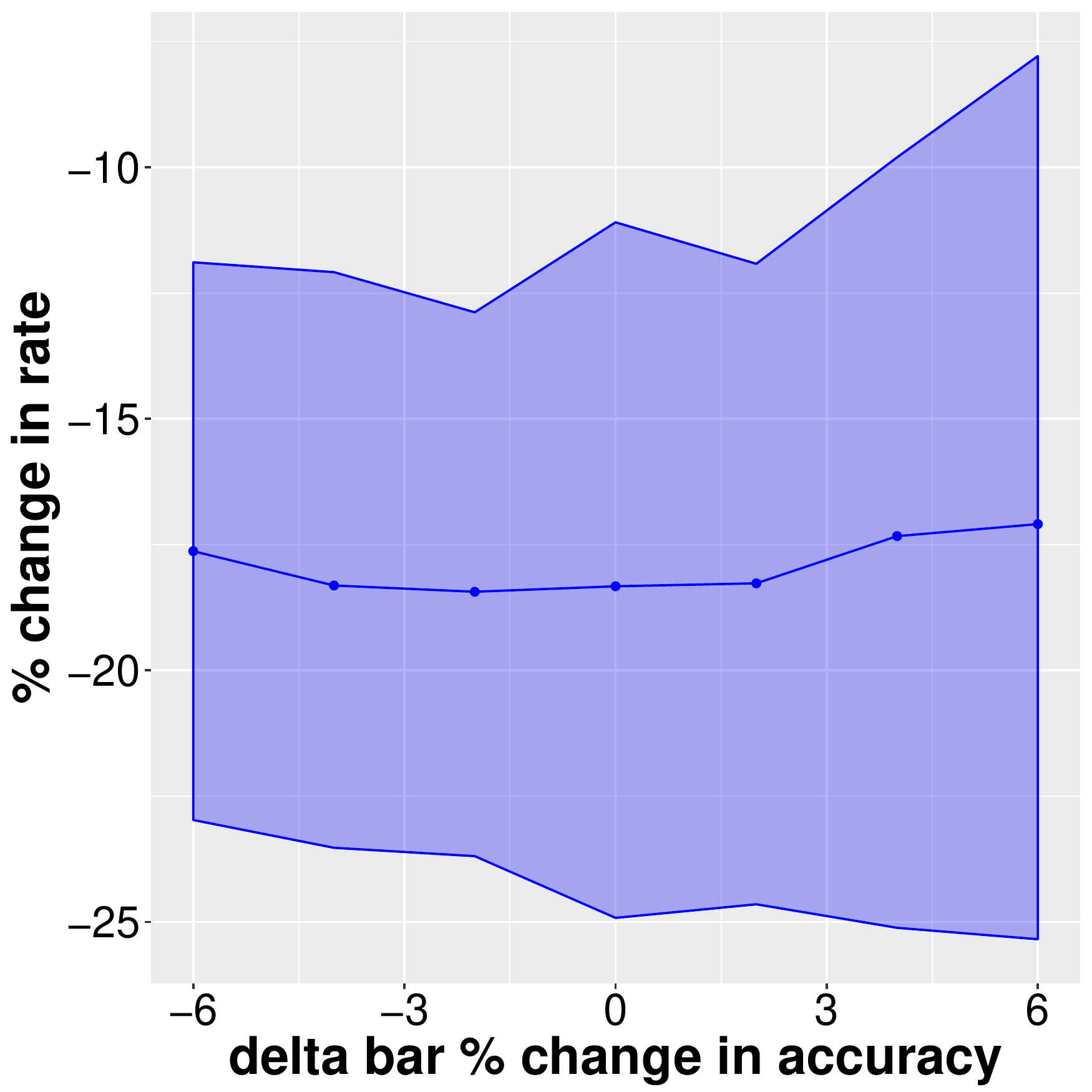}
\caption{Rate}
\end{subfigure}
\caption{Results for Detroit, MI.  Top left (a): Posterior predictive distribution for homeless counts, $C_{i,1:T}^* | C_{1:25,1:T}, N_{1:25,1:T}$, in green, and the imputed total homeless population size, $H_{i,1:T}|C_{1:25,1:T}, H_{1:25,1:T}$, in blue.  The black 'x' marks correspond to the observed (raw) homeless count by year.  The count accuracy is modeled with a constant expectation.  Top right (b): Predictive distribution for total homeless population in 2017, $H_{i,2017} | C_{1:25,1:T}, N_{1:25,1:T}$.  Bottom left (c):  Posterior distribution of increase in total homeless population with increases in ZRI.  Bottom right (d): Sensitivity of the inferred increase in the homelessness rate from 2011 - 2016 to different annual changes in count accuracy.}
\label{fig:Detroit_Results}
\end{figure}

\begin{figure}[ht!]
\centering
\begin{subfigure}{.4\textwidth}
  \centering
\includegraphics[width=1\textwidth]{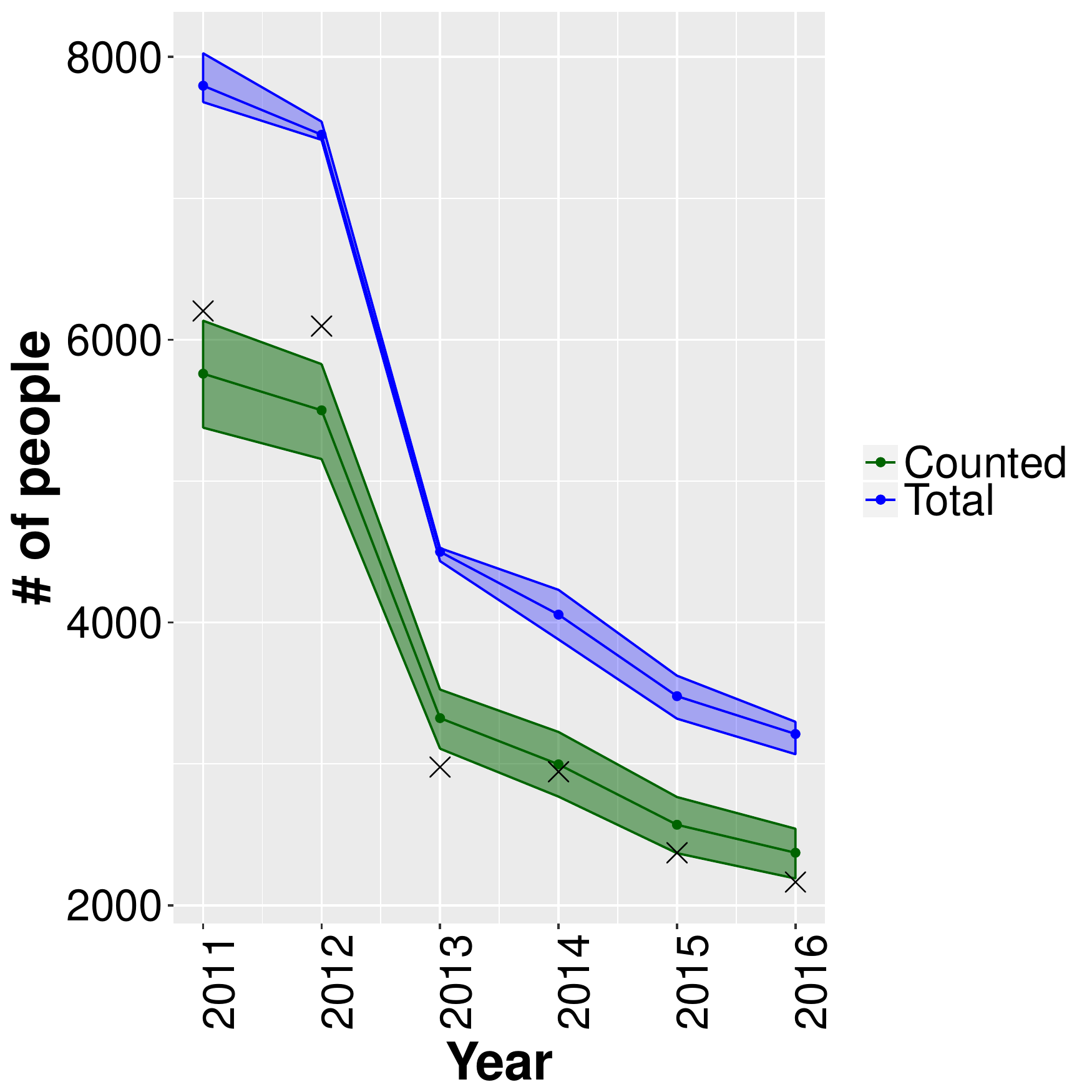}
\caption{\# of homeless}
\end{subfigure}
\begin{subfigure}{.4\textwidth}
  \centering
\includegraphics[width=1\textwidth]{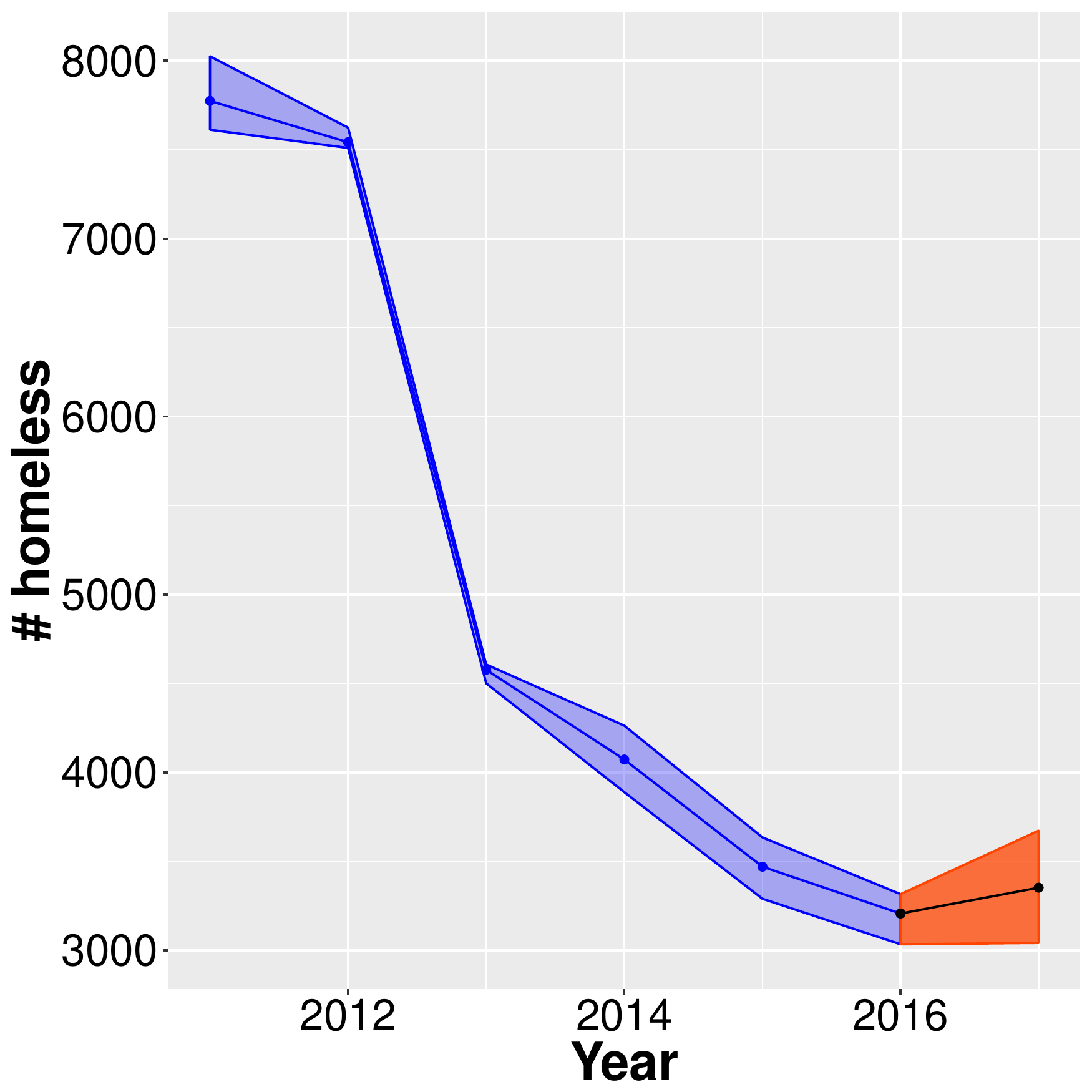}
\caption{2017 forecast}
\end{subfigure}
\begin{subfigure}{.4\textwidth}
  \centering
\includegraphics[width=1\textwidth]{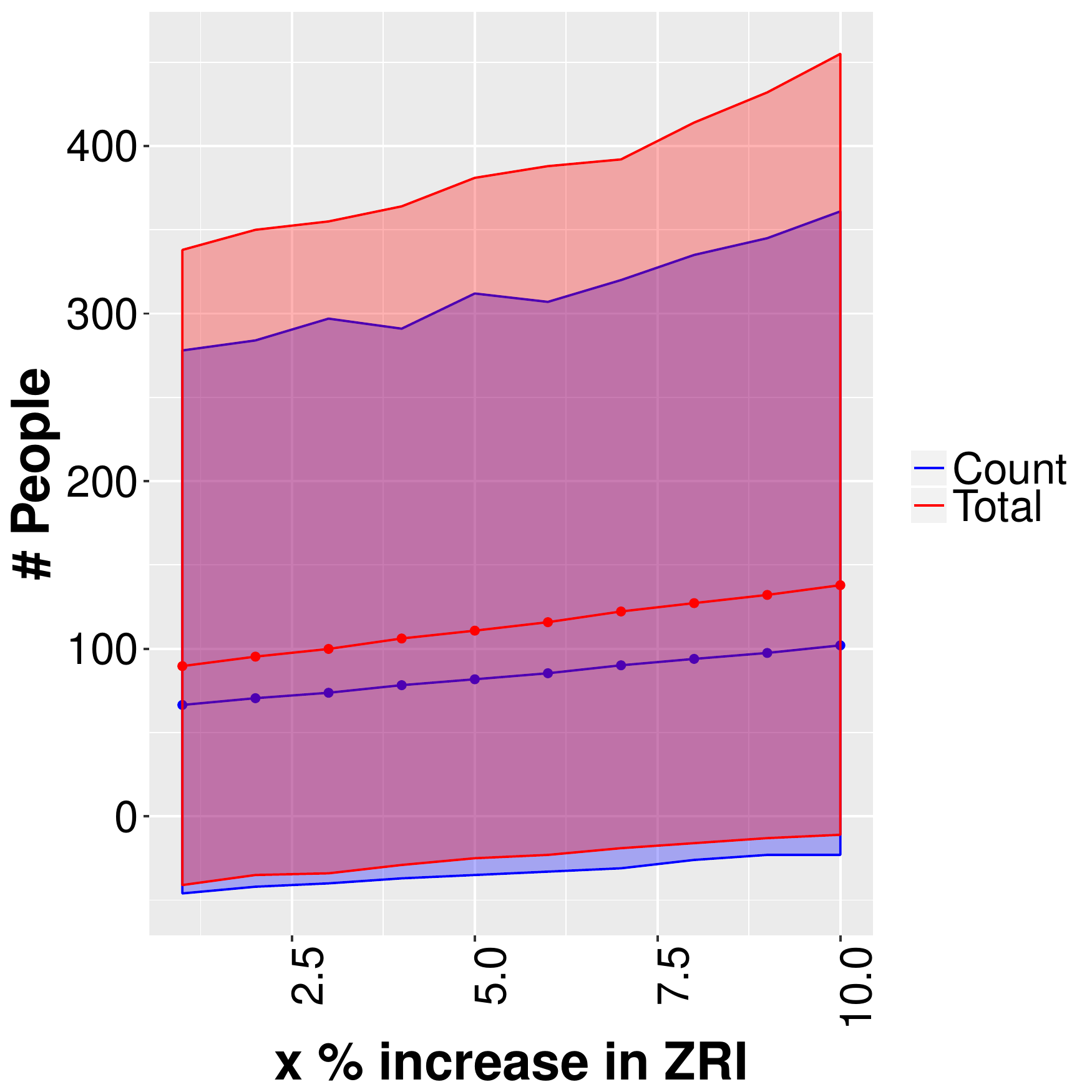}
\caption{ZRI effect}
\end{subfigure}
\begin{subfigure}{.4\textwidth}
  \centering
\includegraphics[width=1\textwidth]{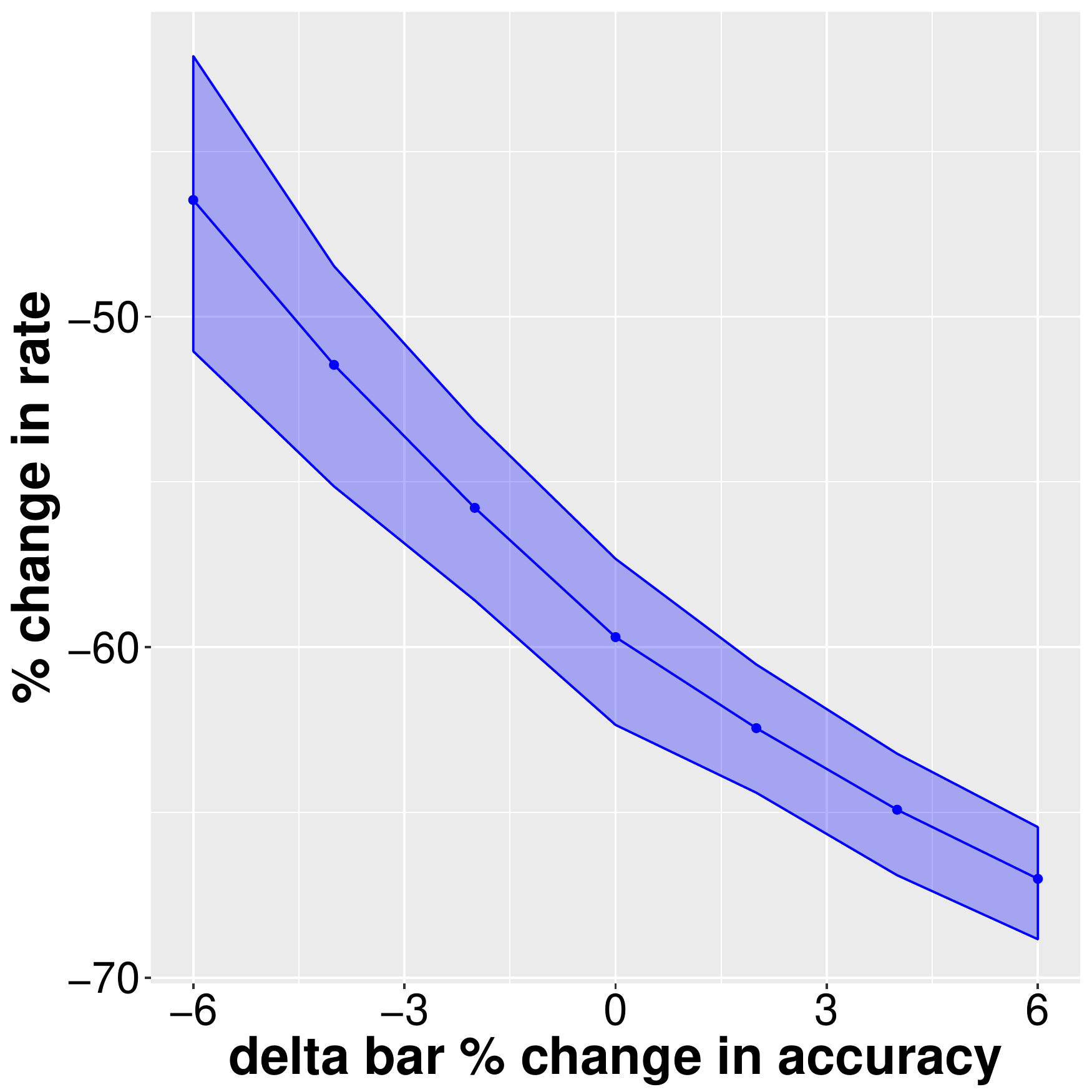}
\caption{Rate}
\end{subfigure}
\caption{Results for Riverside, CA.  Top left (a): Posterior predictive distribution for homeless counts, $C_{i,1:T}^* | C_{1:25,1:T}, N_{1:25,1:T}$, in green, and the imputed total homeless population size, $H_{i,1:T}|C_{1:25,1:T}, H_{1:25,1:T}$, in blue.  The black 'x' marks correspond to the observed (raw) homeless count by year.  The count accuracy is modeled with a constant expectation.  Top right (b): Predictive distribution for total homeless population in 2017, $H_{i,2017} | C_{1:25,1:T}, N_{1:25,1:T}$.  Bottom left (c):  Posterior distribution of increase in total homeless population with increases in ZRI.  Bottom right (d): Sensitivity of the inferred increase in the homelessness rate from 2011 - 2016 to different annual changes in count accuracy.}
\label{fig:Riverside_Results}
\end{figure}

\begin{figure}[ht!]
\centering
\begin{subfigure}{.4\textwidth}
  \centering
\includegraphics[width=1\textwidth]{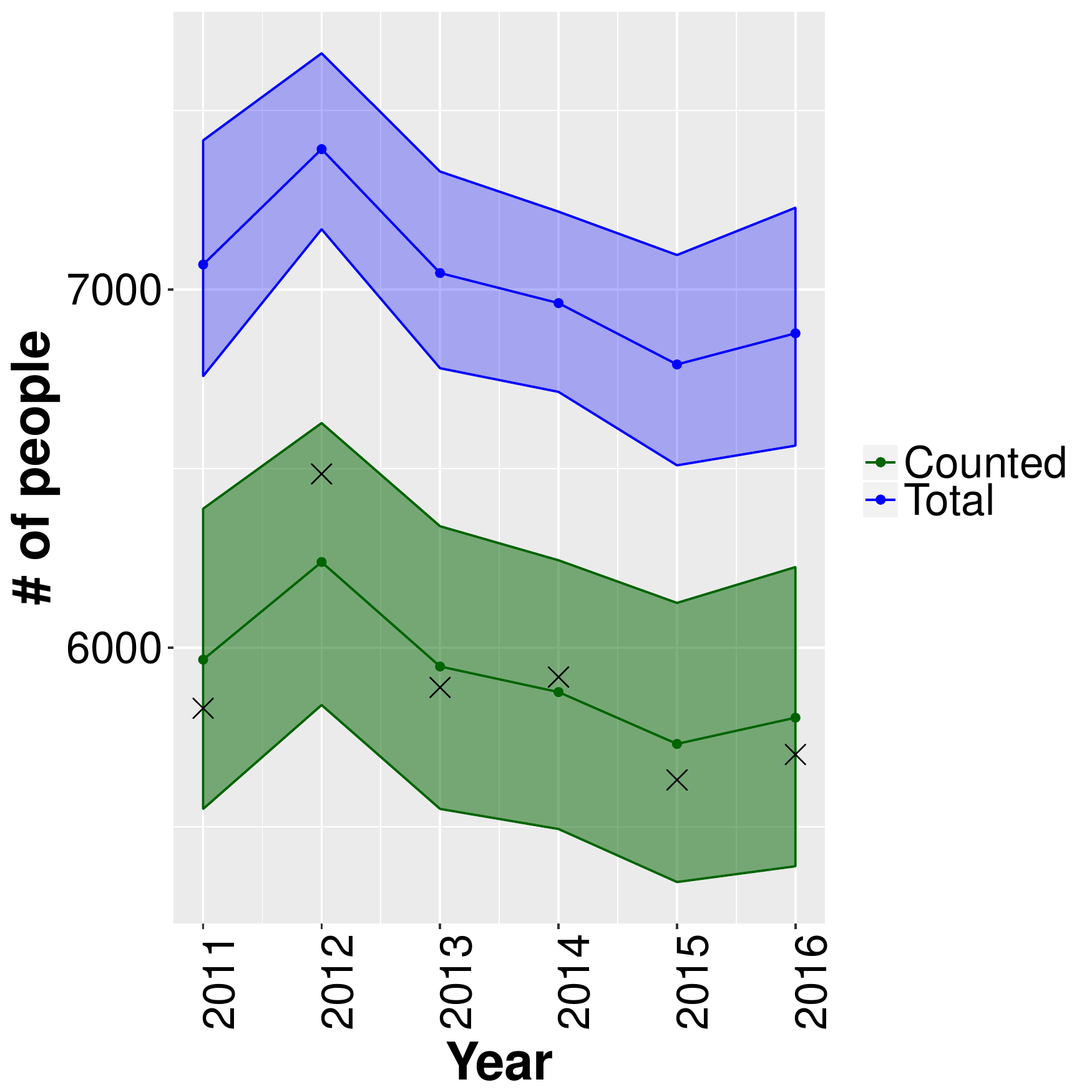}
\caption{\# of homeless}
\end{subfigure}
\begin{subfigure}{.4\textwidth}
  \centering
\includegraphics[width=1\textwidth]{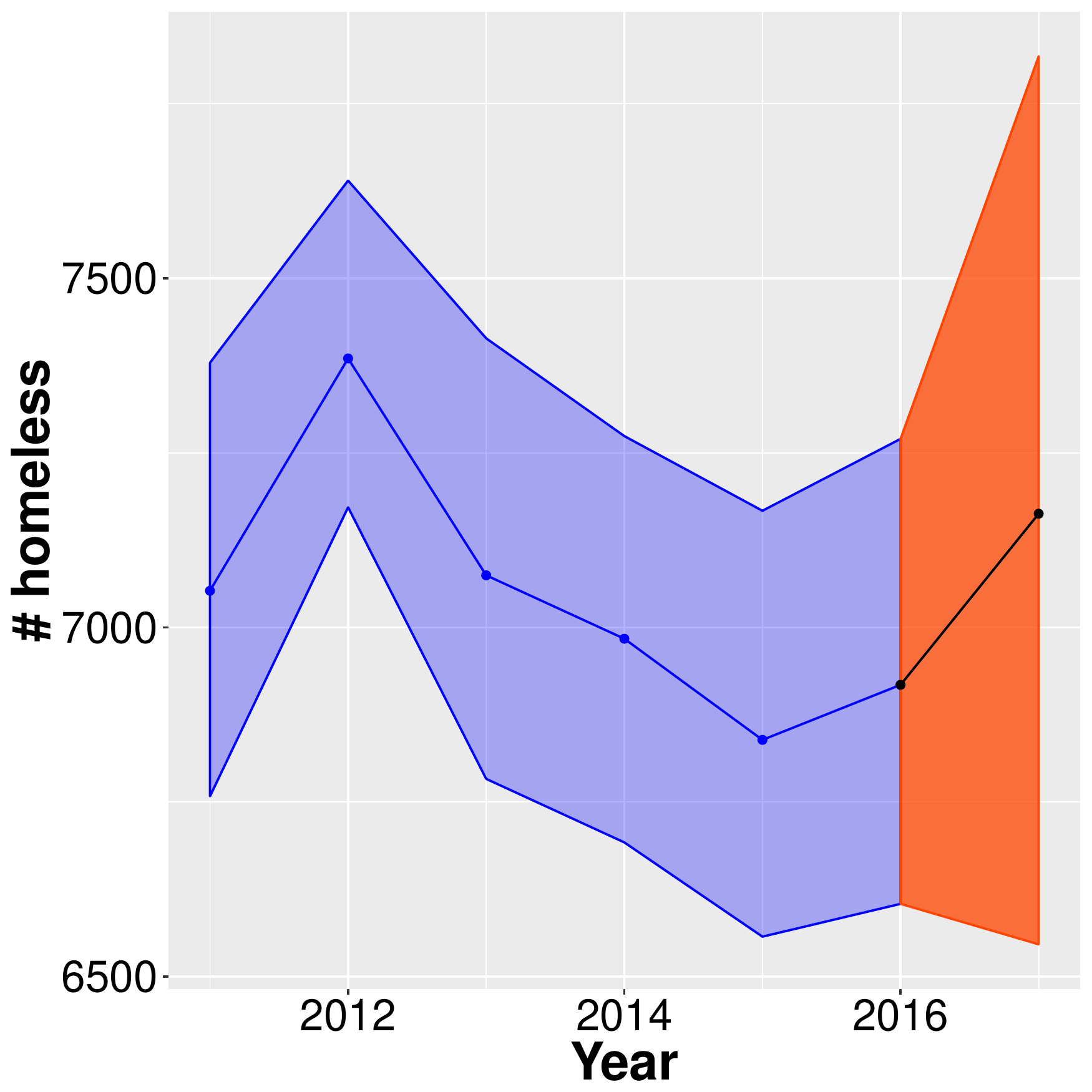}
\caption{2017 forecast}
\end{subfigure}
\begin{subfigure}{.4\textwidth}
  \centering
\includegraphics[width=1\textwidth]{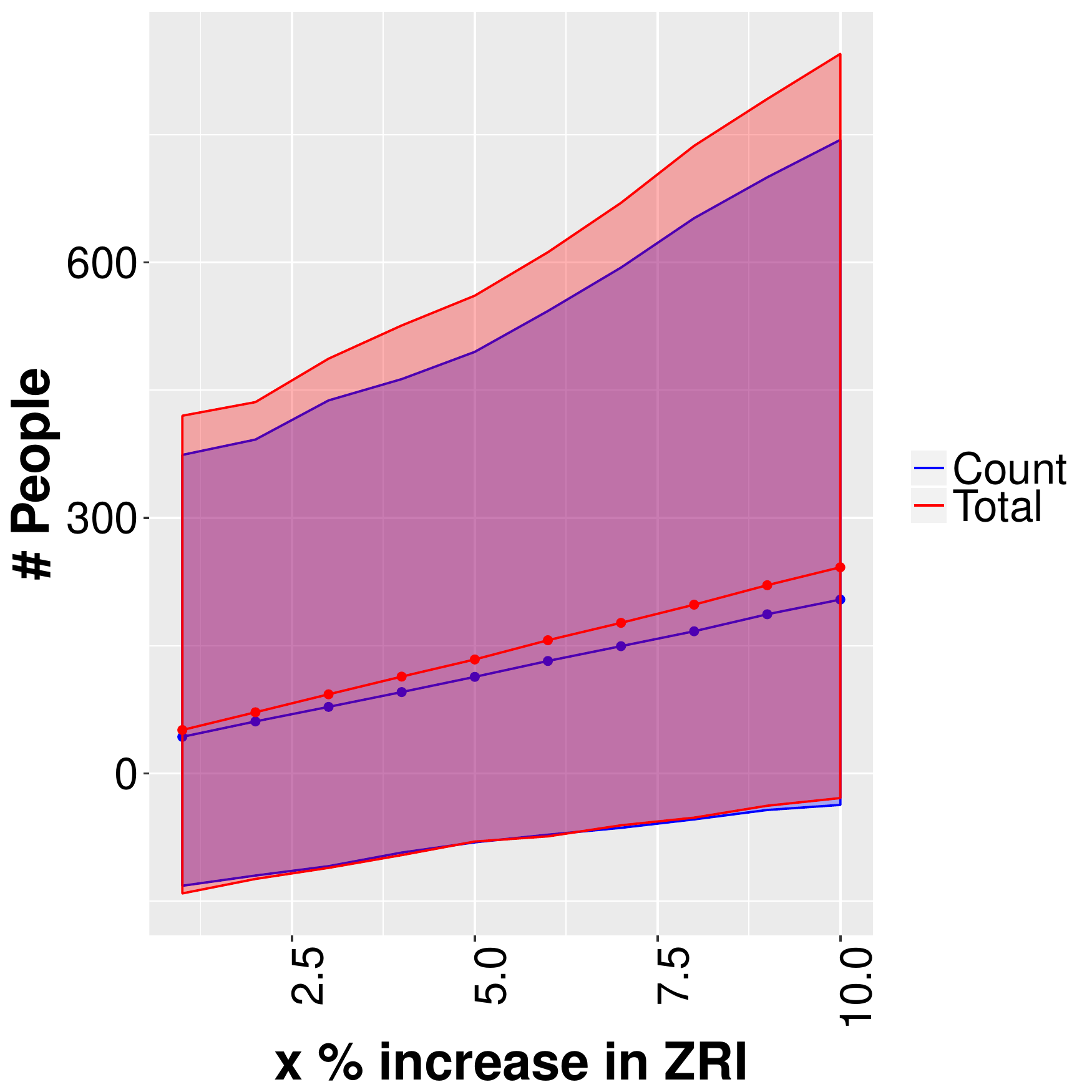}
\caption{ZRI effect}
\end{subfigure}
\begin{subfigure}{.4\textwidth}
  \centering
\includegraphics[width=1\textwidth]{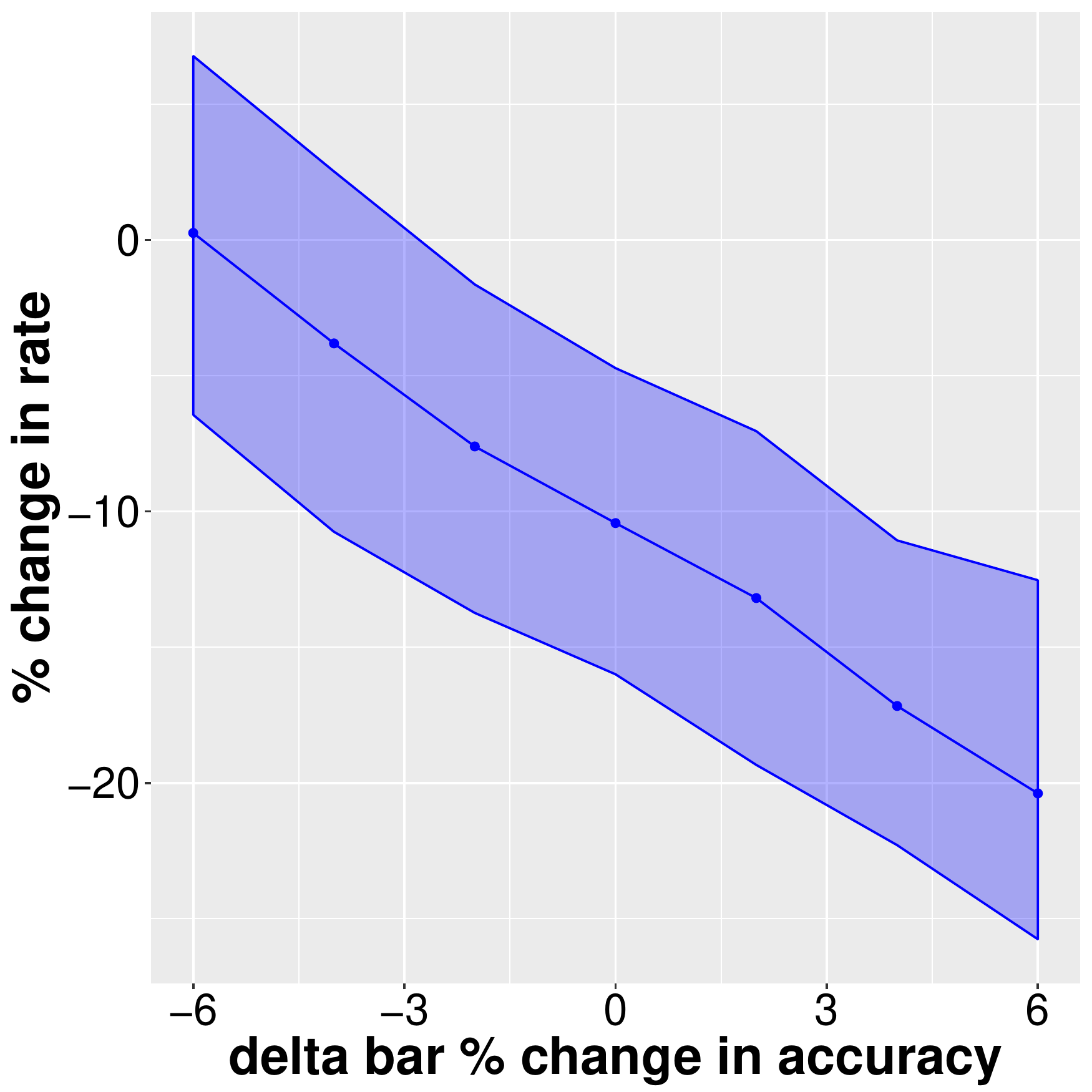}
\caption{Rate}
\end{subfigure}
\caption{Results for Phoenix, AZ.  Top left (a): Posterior predictive distribution for homeless counts, $C_{i,1:T}^* | C_{1:25,1:T}, N_{1:25,1:T}$, in green, and the imputed total homeless population size, $H_{i,1:T}|C_{1:25,1:T}, H_{1:25,1:T}$, in blue.  The black 'x' marks correspond to the observed (raw) homeless count by year.  The count accuracy is modeled with a constant expectation.  Top right (b): Predictive distribution for total homeless population in 2017, $H_{i,2017} | C_{1:25,1:T}, N_{1:25,1:T}$.  Bottom left (c):  Posterior distribution of increase in total homeless population with increases in ZRI.  Bottom right (d): Sensitivity of the inferred increase in the homelessness rate from 2011 - 2016 to different annual changes in count accuracy.}
\label{fig:Phoenix_Results}
\end{figure}

\begin{figure}[ht!]
\centering
\begin{subfigure}{.4\textwidth}
  \centering
\includegraphics[width=1\textwidth]{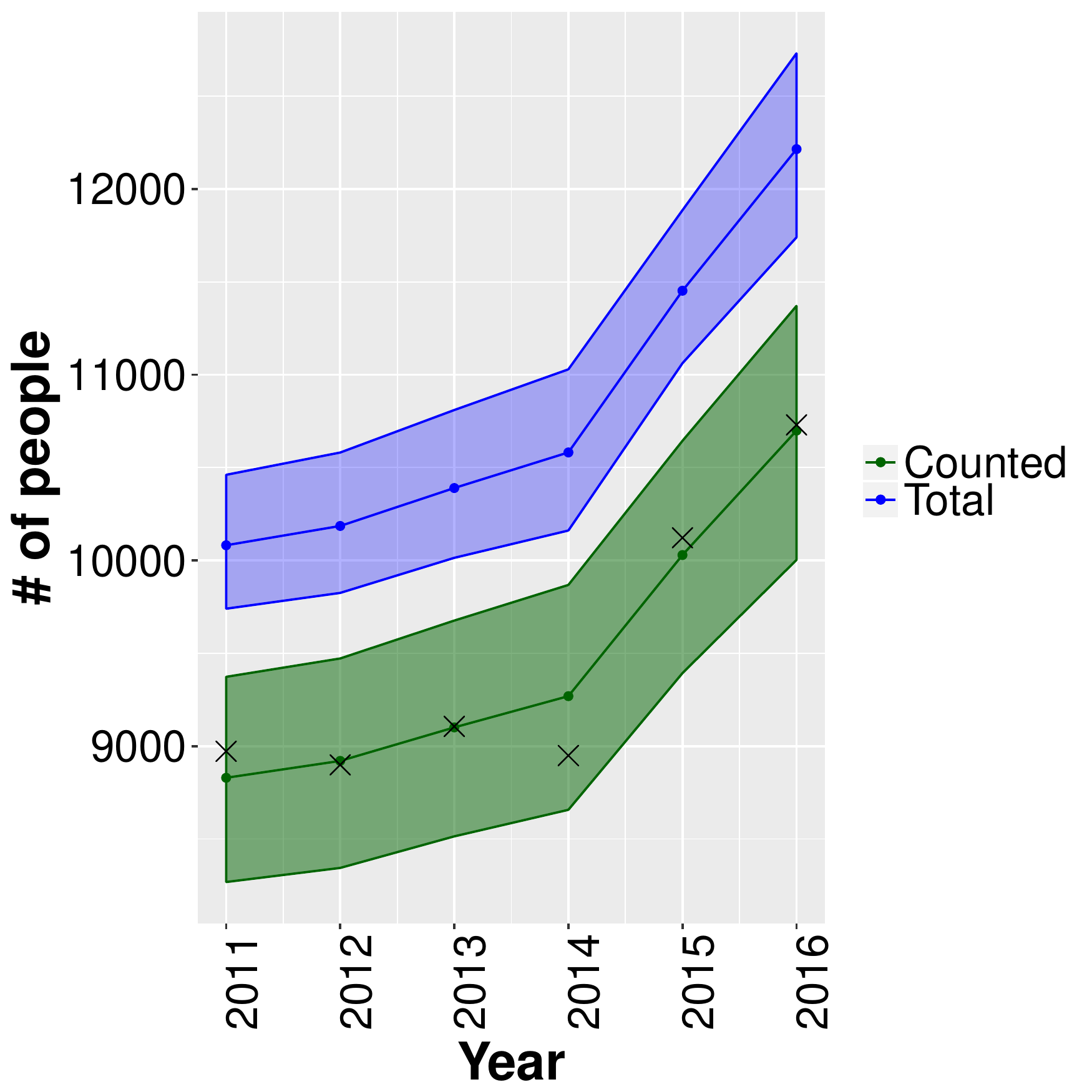}
\caption{\# of homeless}
\end{subfigure}
\begin{subfigure}{.4\textwidth}
  \centering
\includegraphics[width=1\textwidth]{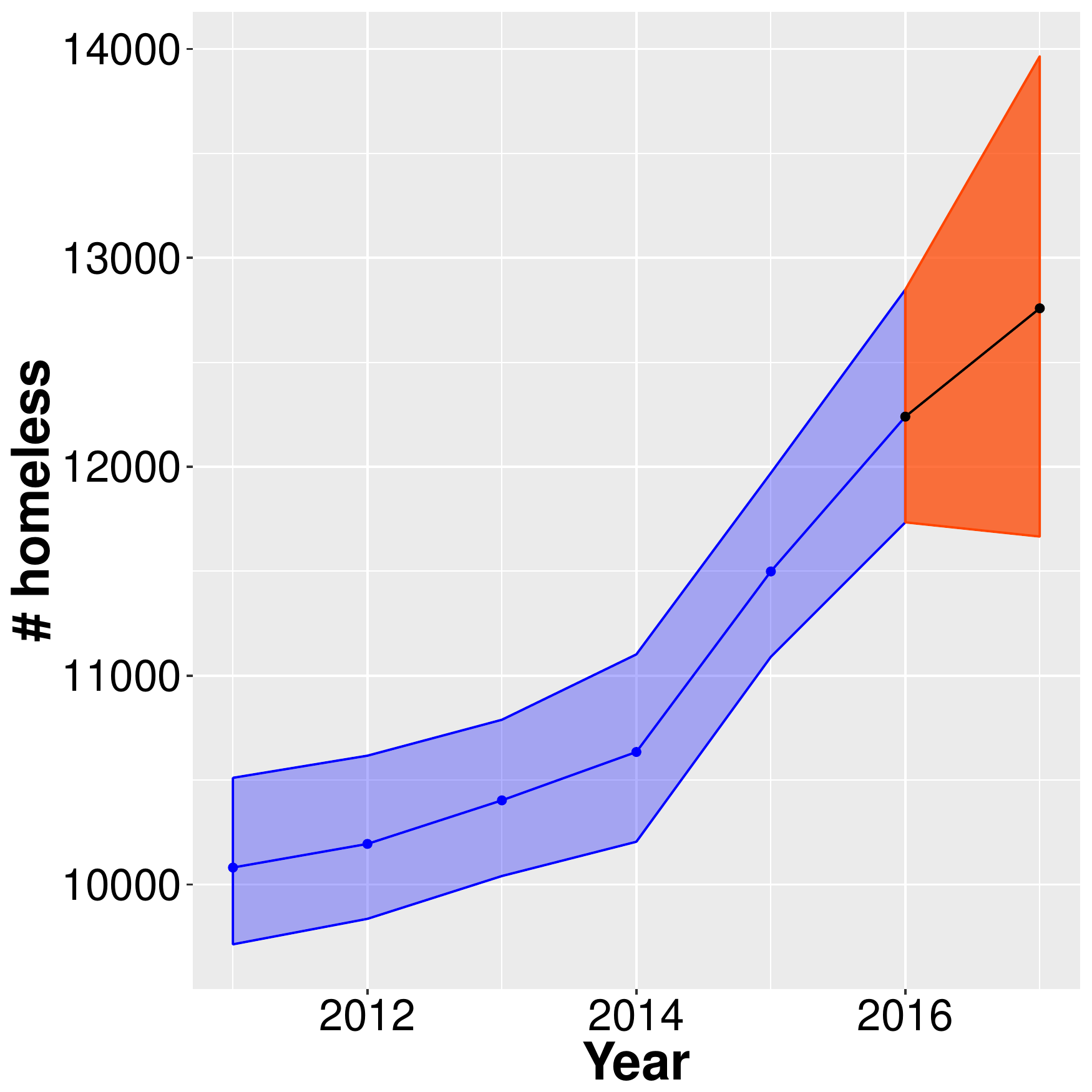}
\caption{2017 forecast}
\end{subfigure}
\begin{subfigure}{.4\textwidth}
  \centering
\includegraphics[width=1\textwidth]{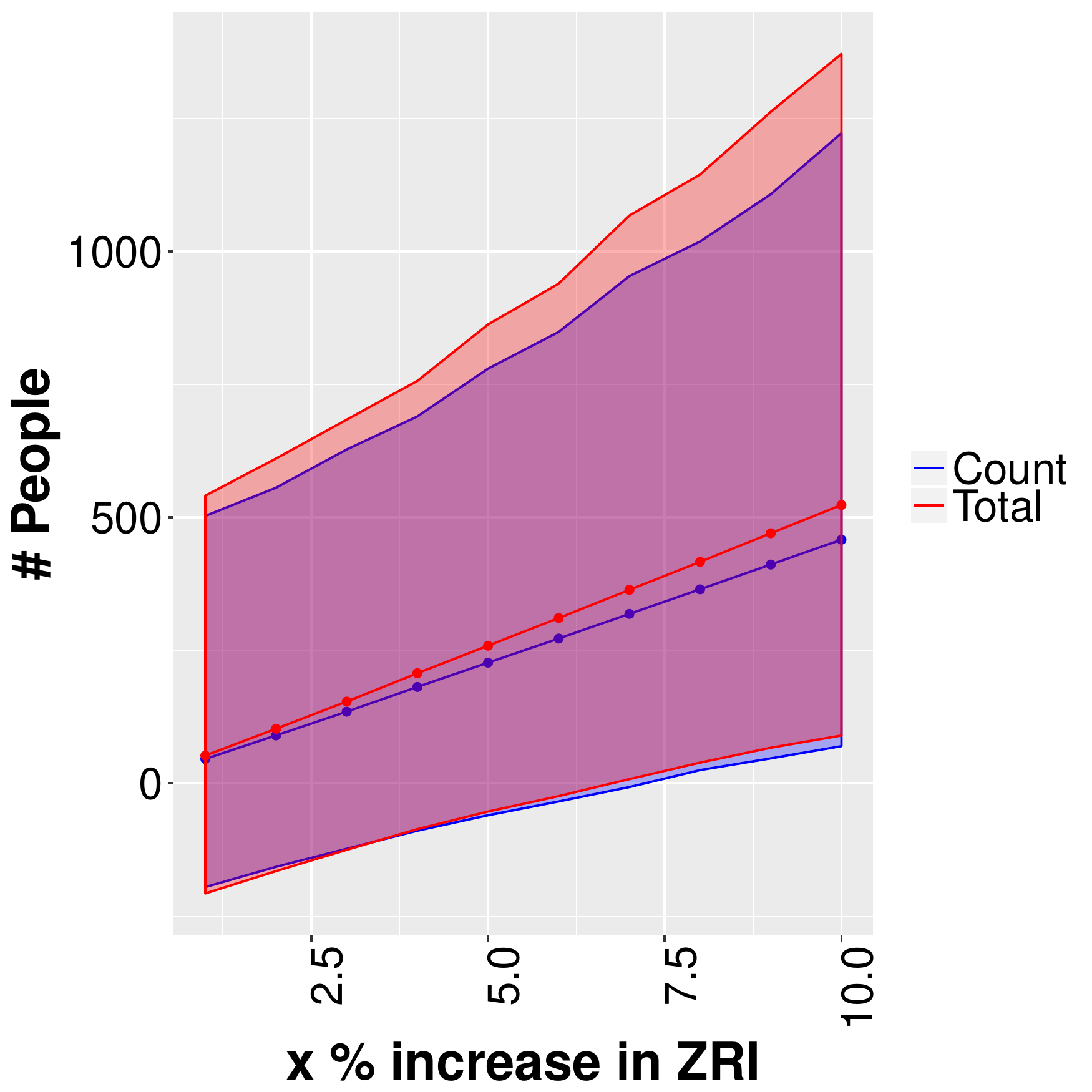}
\caption{ZRI effect}
\end{subfigure}
\begin{subfigure}{.4\textwidth}
  \centering
\includegraphics[width=1\textwidth]{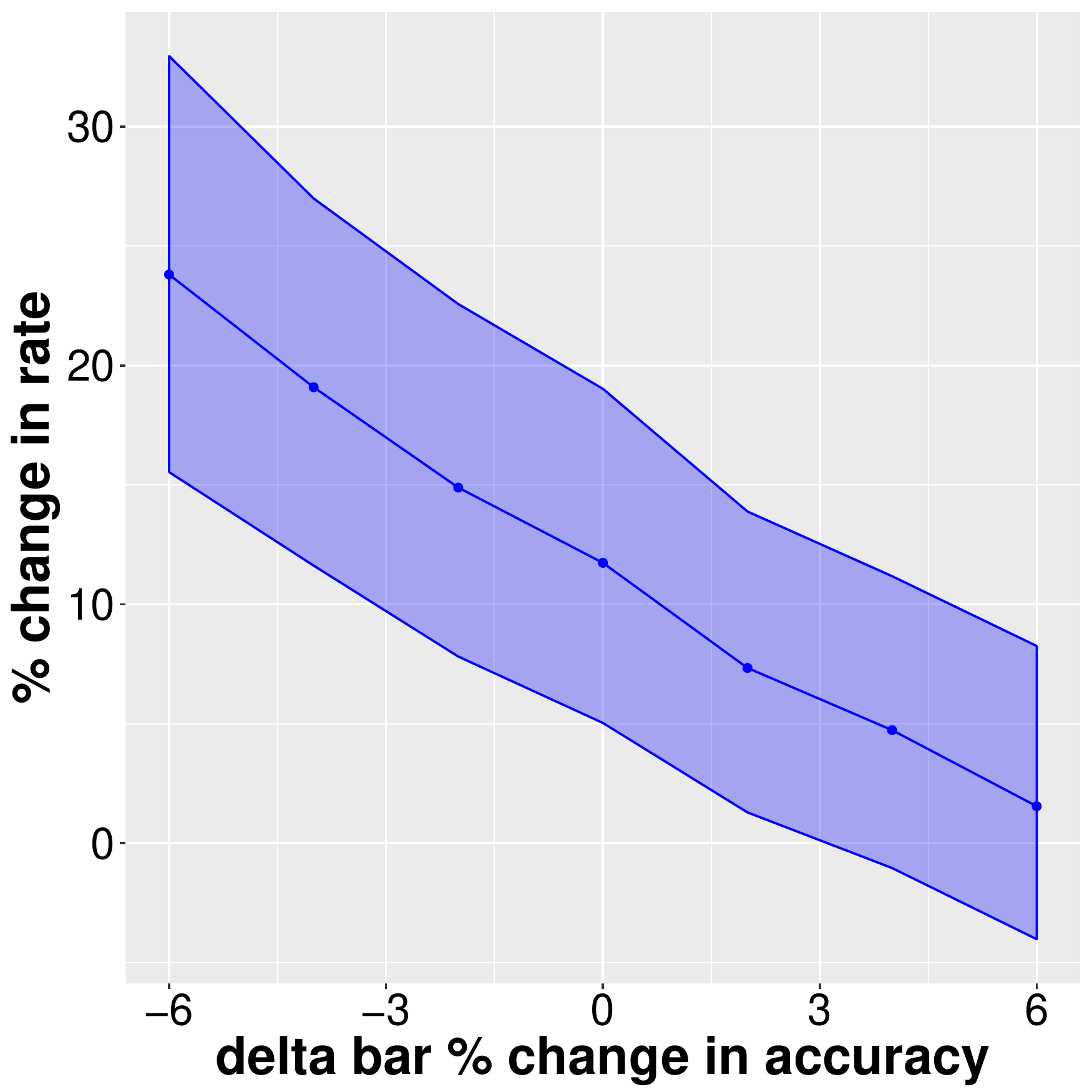}
\caption{Rate}
\end{subfigure}
\caption{Results for Seattle / King County, WA.  Top left (a): Posterior predictive distribution for homeless counts, $C_{i,1:T}^* | C_{1:25,1:T}, N_{1:25,1:T}$, in green, and the imputed total homeless population size, $H_{i,1:T}|C_{1:25,1:T}, H_{1:25,1:T}$, in blue.  The black 'x' marks correspond to the observed (raw) homeless count by year.  The count accuracy is modeled with a constant expectation.  Top right (b): Predictive distribution for total homeless population in 2017, $H_{i,2017} | C_{1:25,1:T}, N_{1:25,1:T}$.  Bottom left (c):  Posterior distribution of increase in total homeless population with increases in ZRI.  Bottom right (d): Sensitivity of the inferred increase in the homelessness rate from 2011 - 2016 to different annual changes in count accuracy.}
\label{fig:Seattle_Results}
\end{figure}

\begin{figure}[ht!]
\centering
\begin{subfigure}{.4\textwidth}
  \centering
\includegraphics[width=1\textwidth]{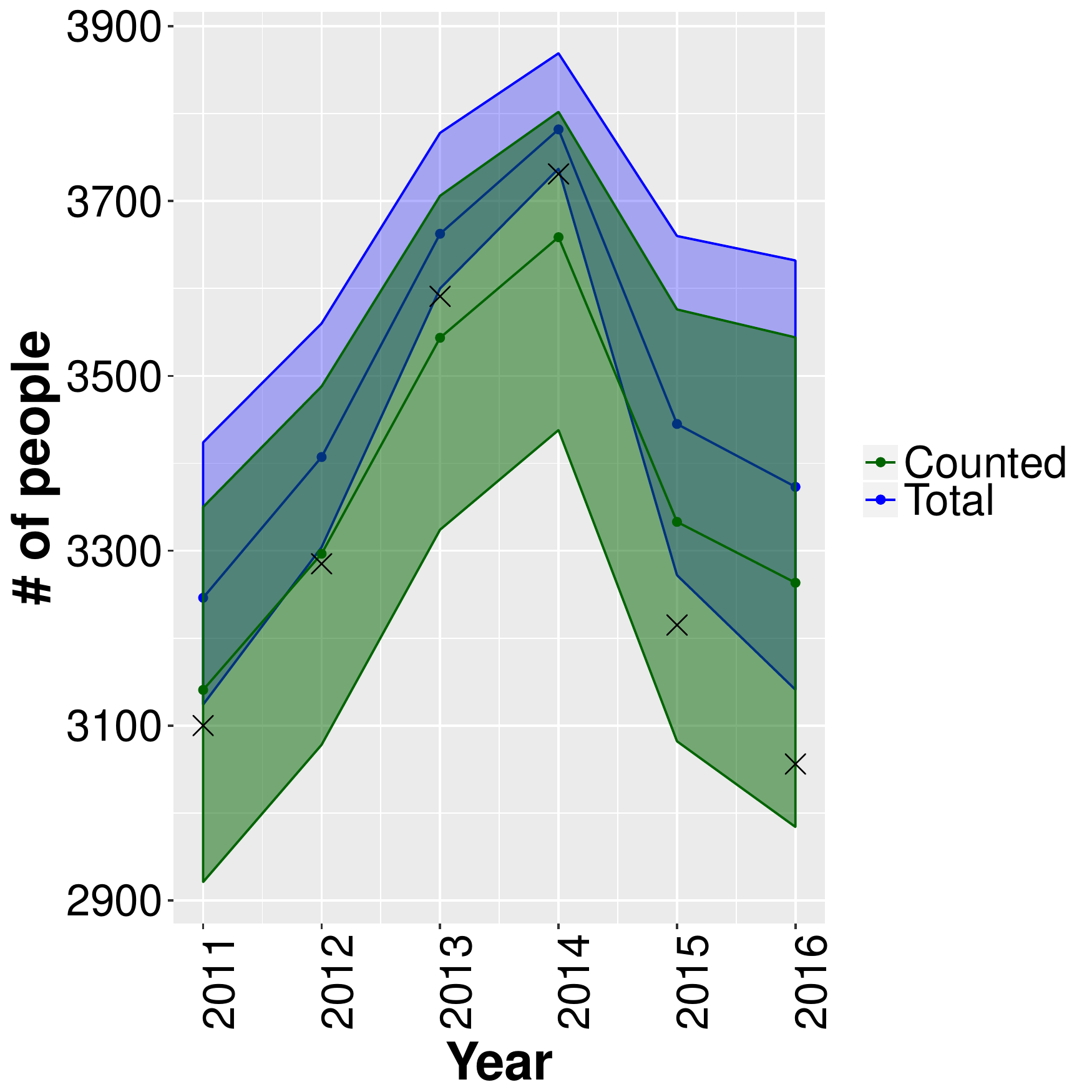}
\caption{\# of homeless}
\end{subfigure}
\begin{subfigure}{.4\textwidth}
  \centering
\includegraphics[width=1\textwidth]{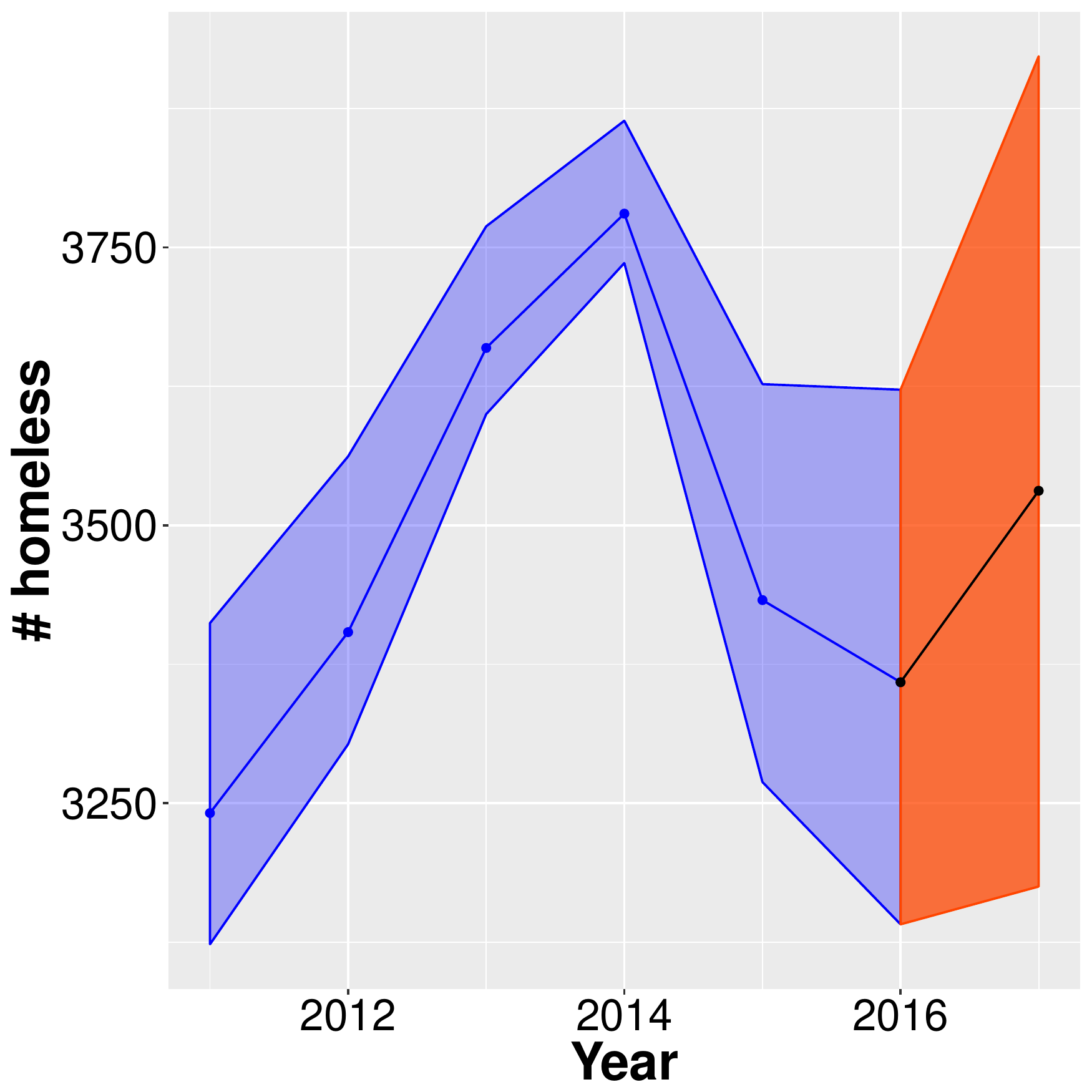}
\caption{2017 forecast}
\end{subfigure}
\begin{subfigure}{.4\textwidth}
  \centering
\includegraphics[width=1\textwidth]{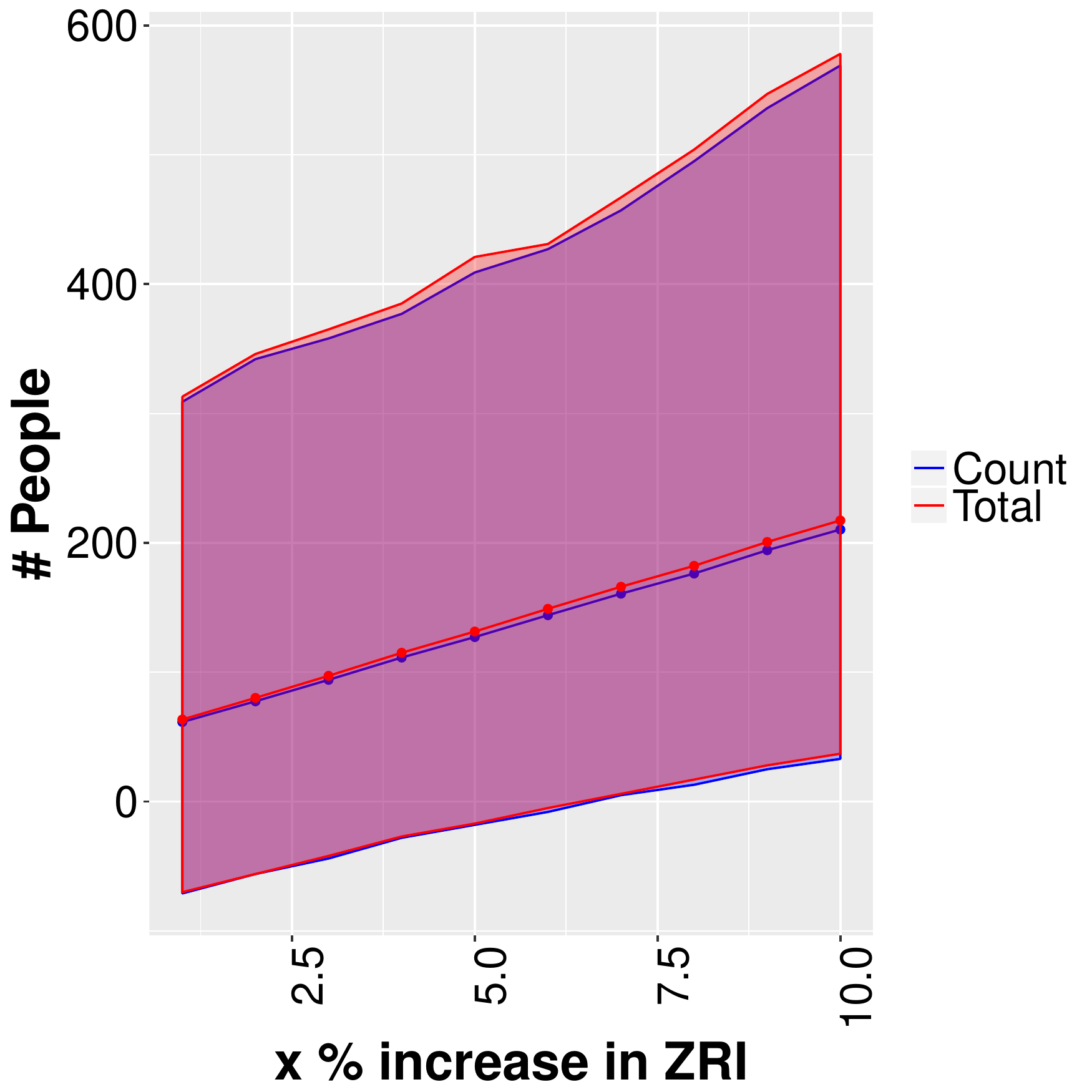}
\caption{ZRI effect}
\end{subfigure}
\begin{subfigure}{.4\textwidth}
  \centering
\includegraphics[width=1\textwidth]{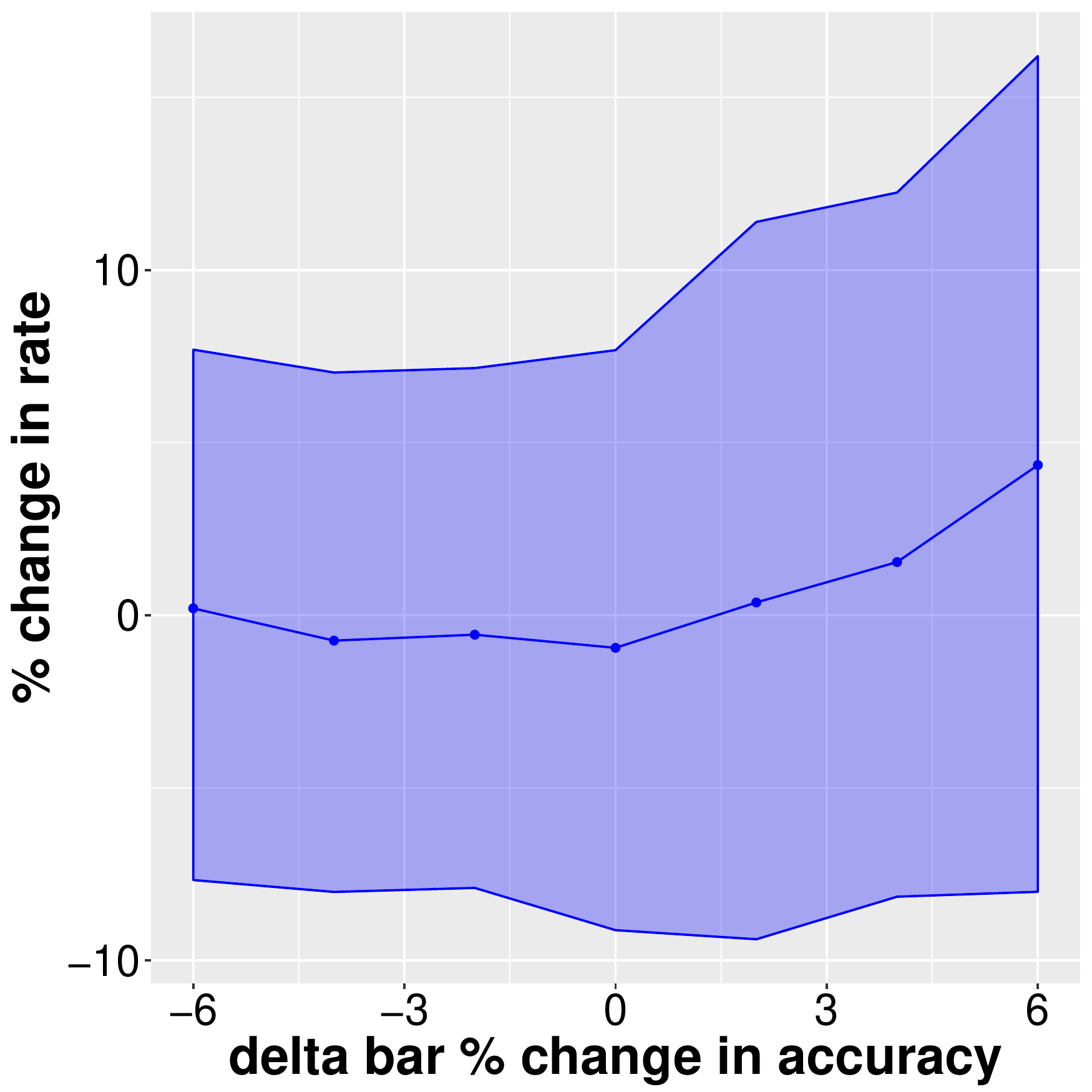}
\caption{Rate}
\end{subfigure}
\caption{Results for Minneapolis, MN.  Top left (a): Posterior predictive distribution for homeless counts, $C_{i,1:T}^* | C_{1:25,1:T}, N_{1:25,1:T}$, in green, and the imputed total homeless population size, $H_{i,1:T}|C_{1:25,1:T}, H_{1:25,1:T}$, in blue.  The black 'x' marks correspond to the observed (raw) homeless count by year.  The count accuracy is modeled with a constant expectation.  Top right (b): Predictive distribution for total homeless population in 2017, $H_{i,2017} | C_{1:25,1:T}, N_{1:25,1:T}$.  Bottom left (c):  Posterior distribution of increase in total homeless population with increases in ZRI.  Bottom right (d): Sensitivity of the inferred increase in the homelessness rate from 2011 - 2016 to different annual changes in count accuracy.}
\label{fig:Minneapolis_Results}
\end{figure}

\begin{figure}[ht!]
\centering
\begin{subfigure}{.4\textwidth}
  \centering
\includegraphics[width=1\textwidth]{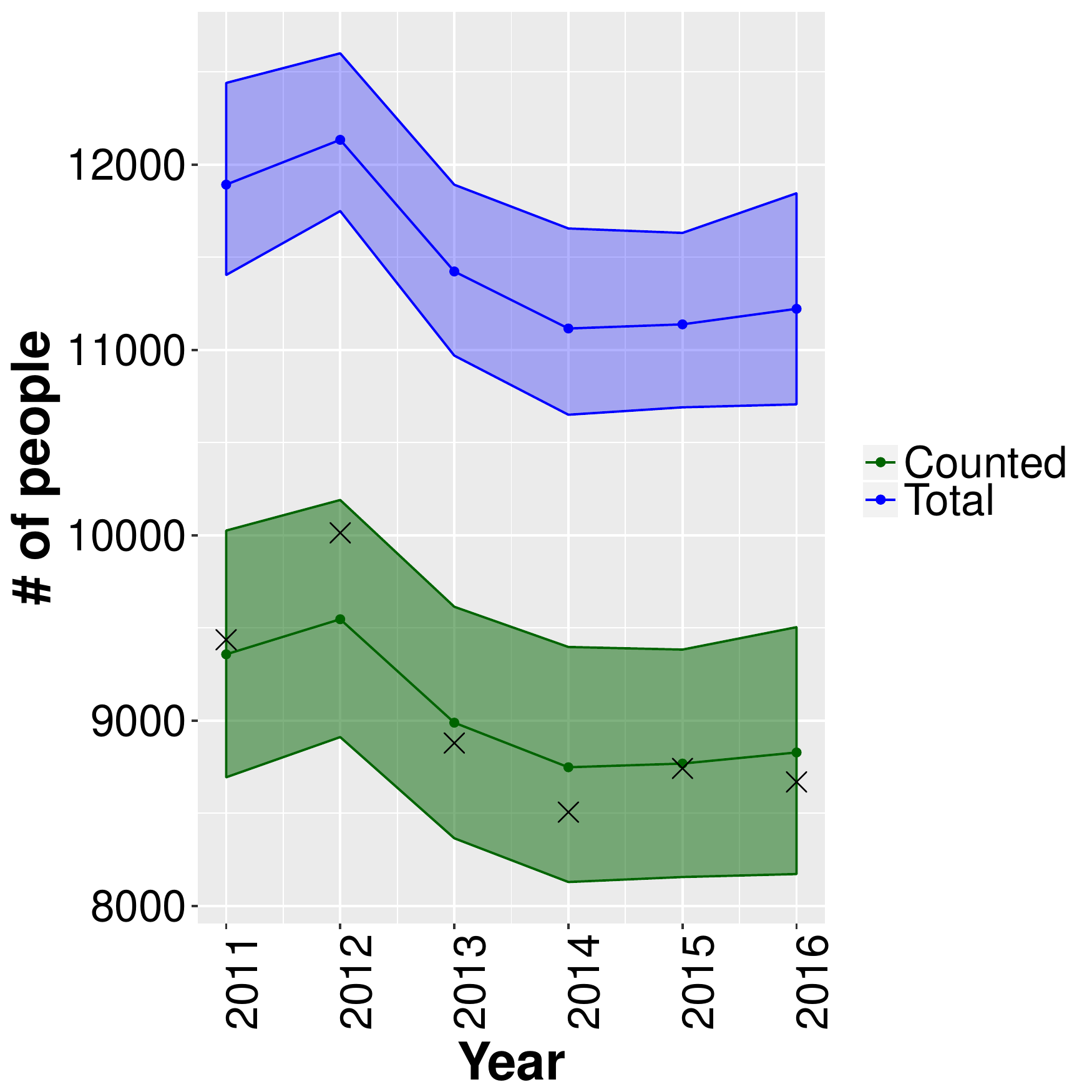}
\caption{\# of homeless}
\end{subfigure}
\begin{subfigure}{.4\textwidth}
  \centering
\includegraphics[width=1\textwidth]{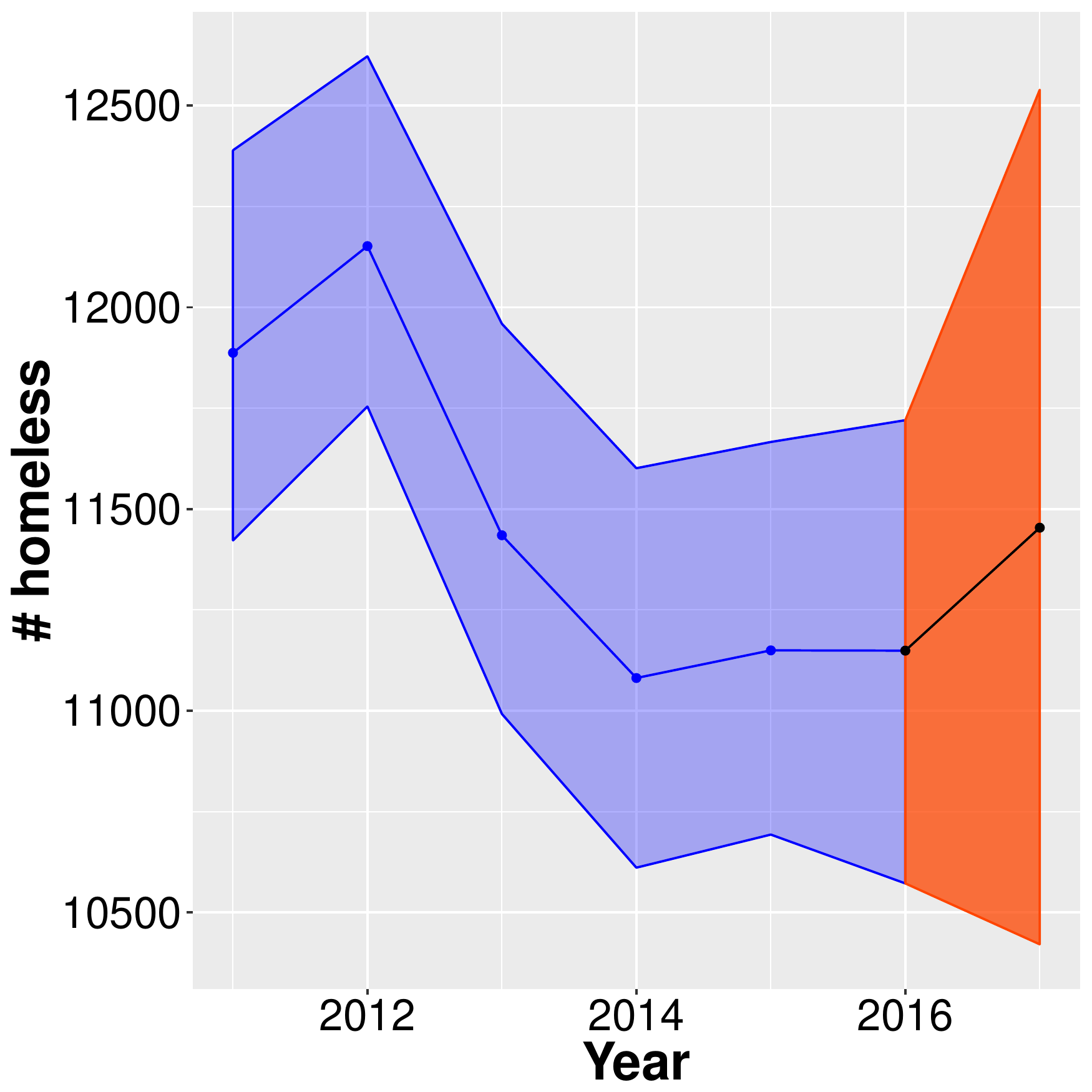}
\caption{2017 forecast}
\end{subfigure}
\begin{subfigure}{.4\textwidth}
  \centering
\includegraphics[width=1\textwidth]{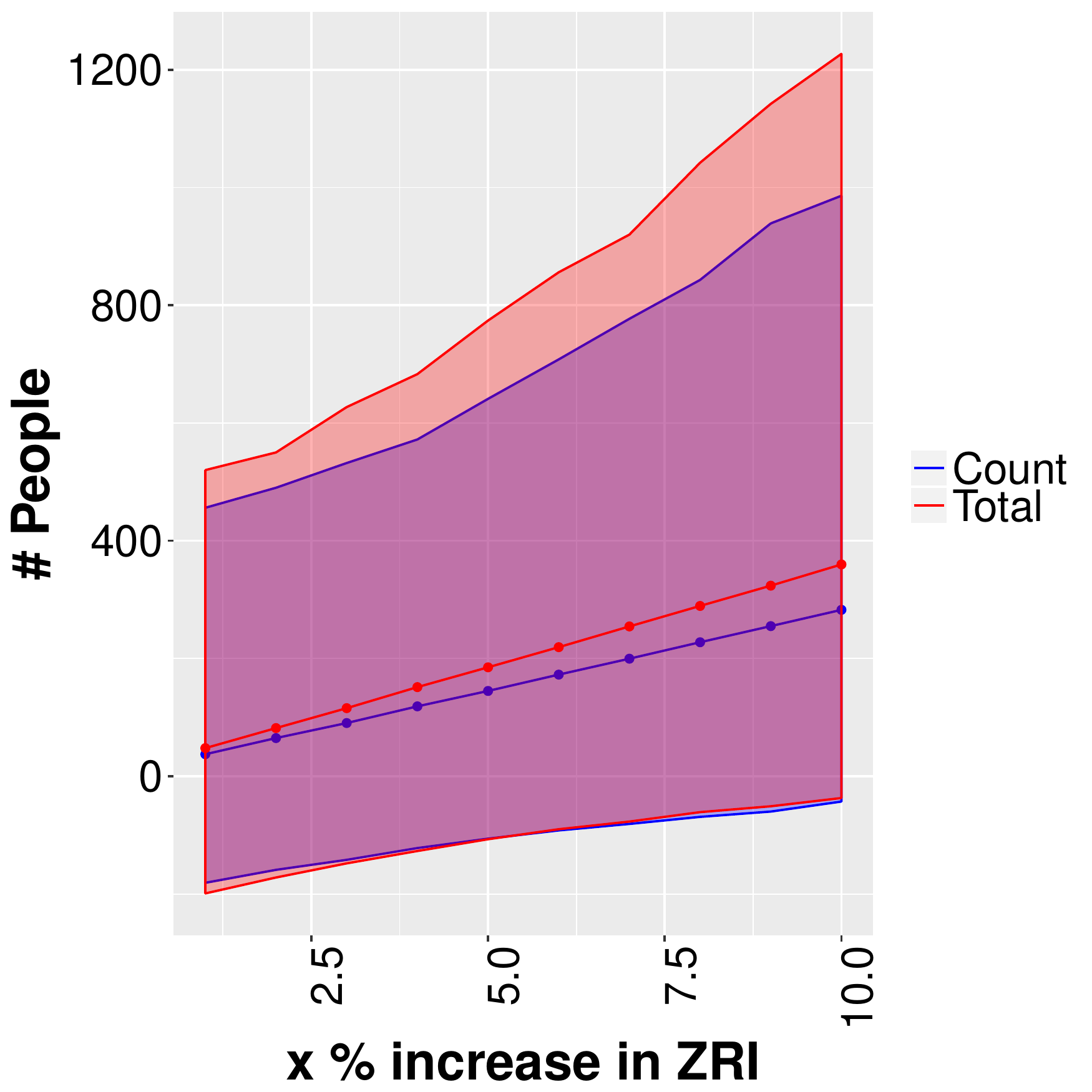}
\caption{ZRI effect}
\end{subfigure}
\begin{subfigure}{.4\textwidth}
  \centering
\includegraphics[width=1\textwidth]{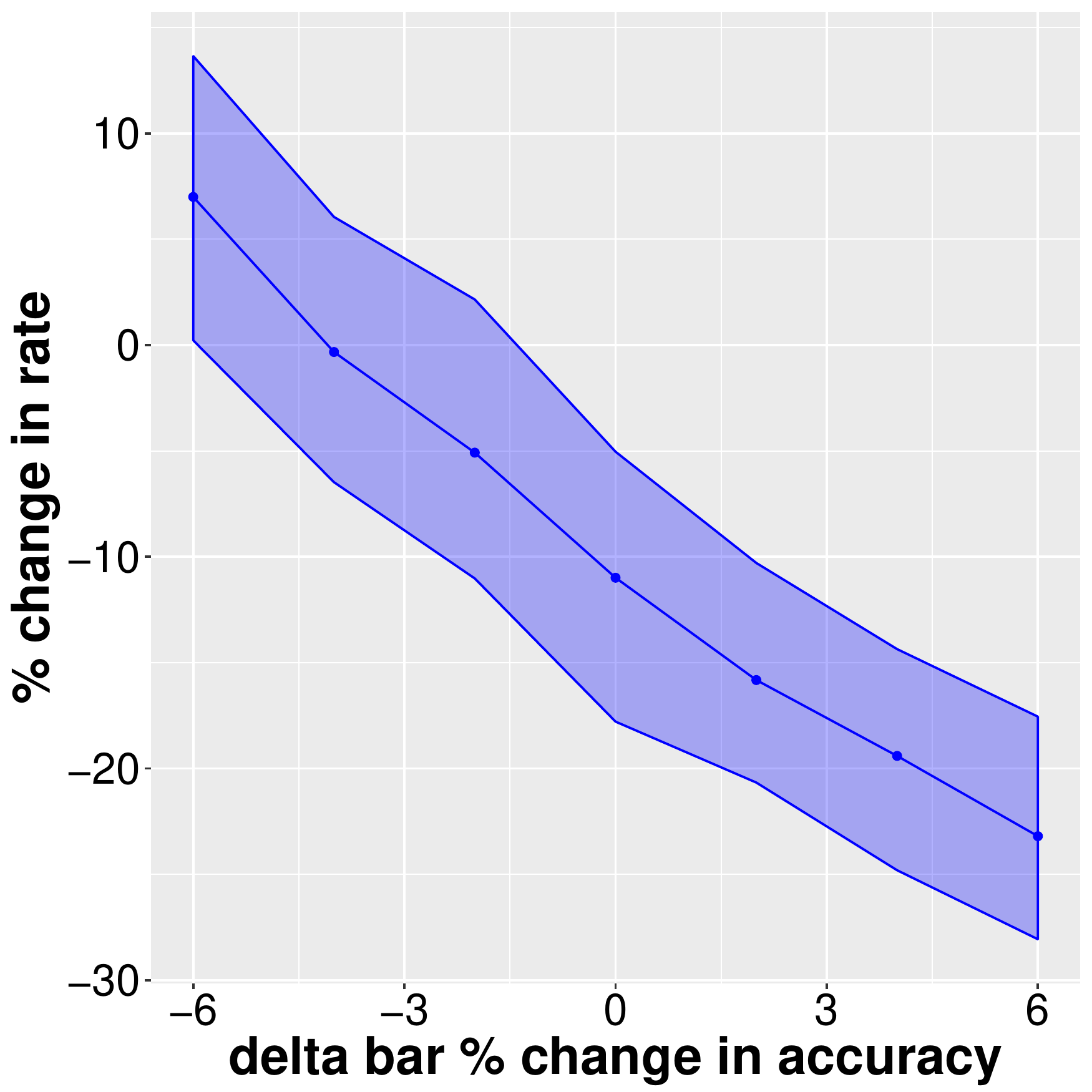}
\caption{Rate}
\end{subfigure}
\caption{Results for San Diego, CA.  Top left (a): Posterior predictive distribution for homeless counts, $C_{i,1:T}^* | C_{1:25,1:T}, N_{1:25,1:T}$, in green, and the imputed total homeless population size, $H_{i,1:T}|C_{1:25,1:T}, H_{1:25,1:T}$, in blue.  The black 'x' marks correspond to the observed (raw) homeless count by year.  The count accuracy is modeled with a constant expectation.  Top right (b): Predictive distribution for total homeless population in 2017, $H_{i,2017} | C_{1:25,1:T}, N_{1:25,1:T}$.  Bottom left (c):  Posterior distribution of increase in total homeless population with increases in ZRI.  Bottom right (d): Sensitivity of the inferred increase in the homelessness rate from 2011 - 2016 to different annual changes in count accuracy.}
\label{fig:SD_Results}
\end{figure}

\begin{figure}[ht!]
\centering
\begin{subfigure}{.4\textwidth}
  \centering
\includegraphics[width=1\textwidth]{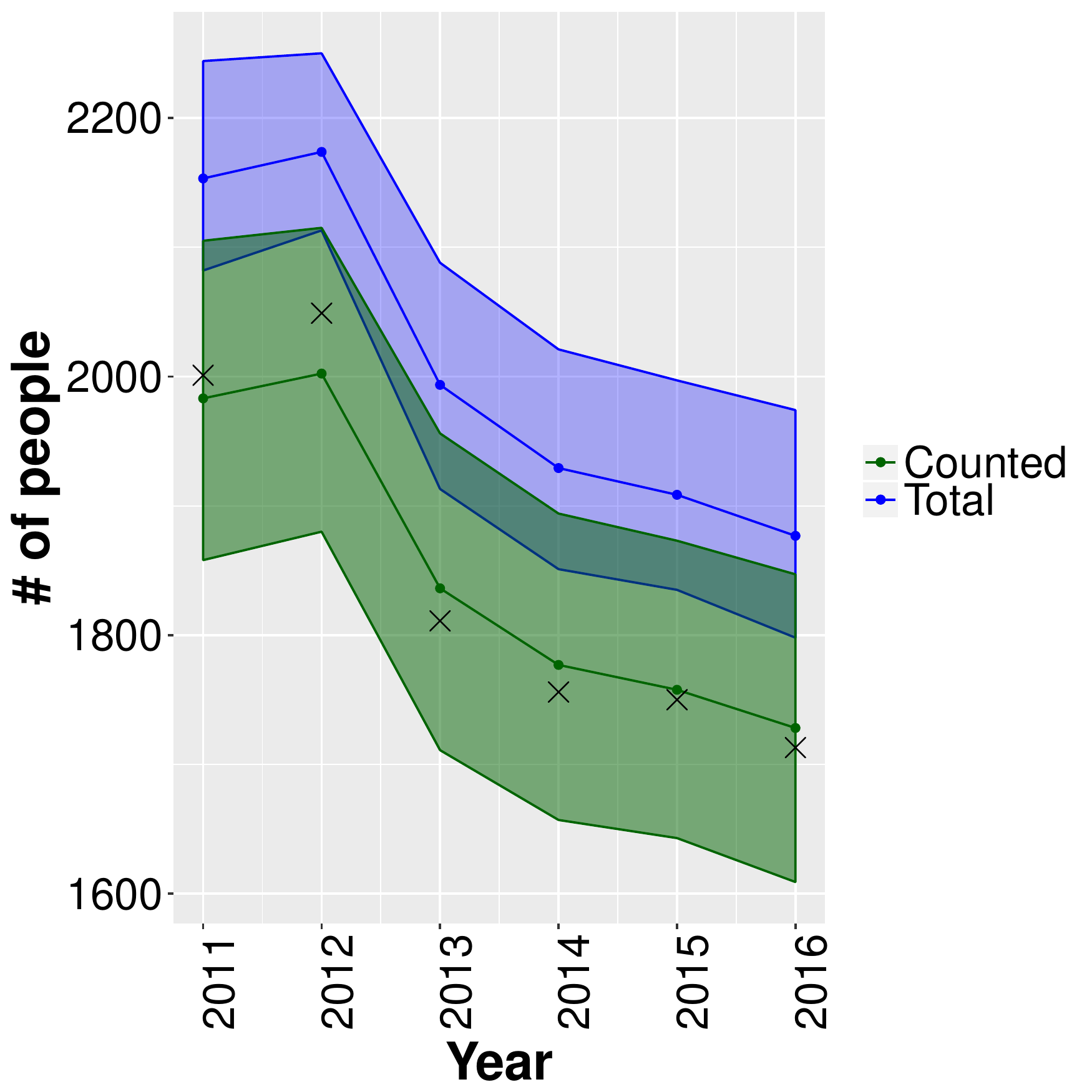}
\caption{\# of homeless}
\end{subfigure}
\begin{subfigure}{.4\textwidth}
  \centering
\includegraphics[width=1\textwidth]{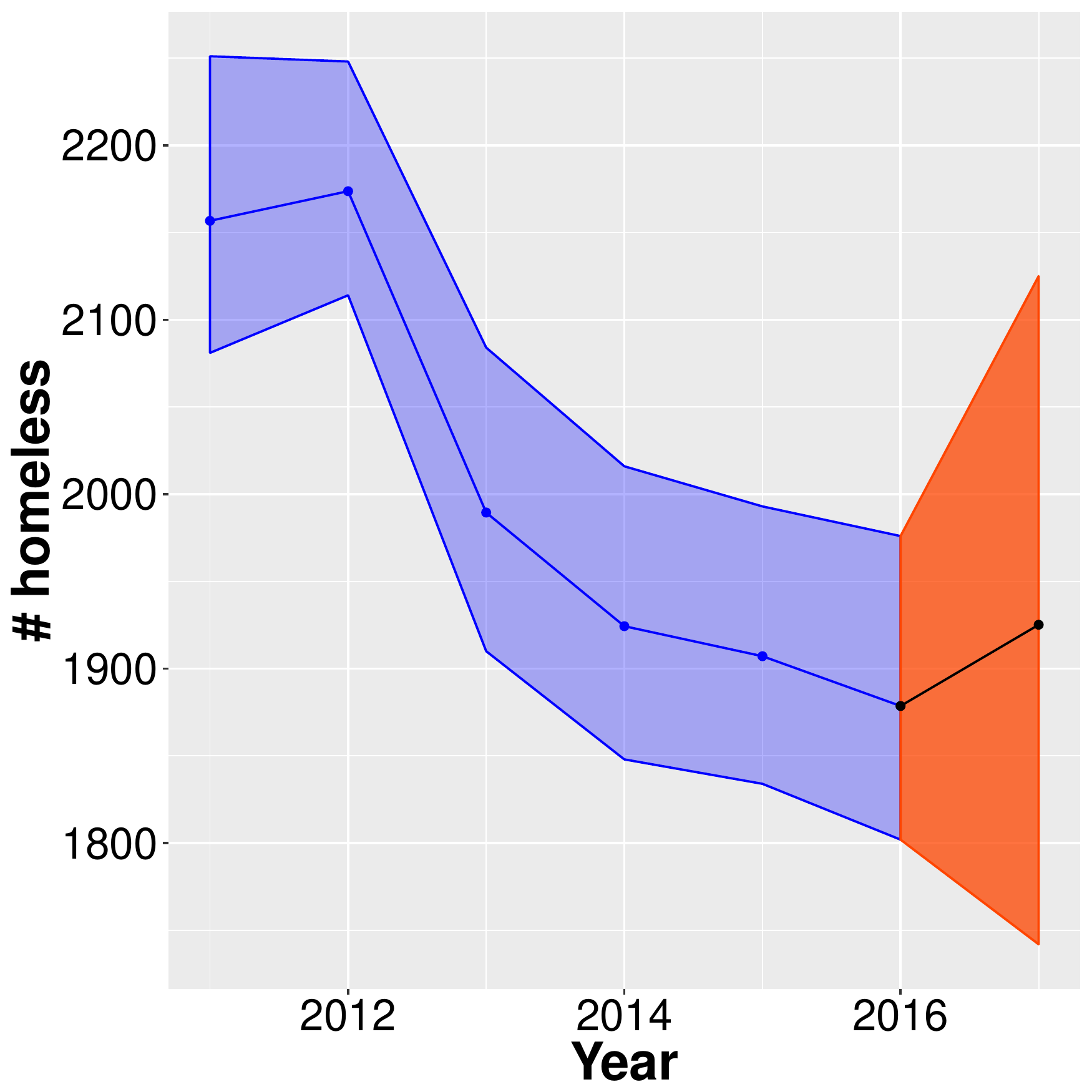}
\caption{2017 forecast}
\end{subfigure}
\begin{subfigure}{.4\textwidth}
  \centering
\includegraphics[width=1\textwidth]{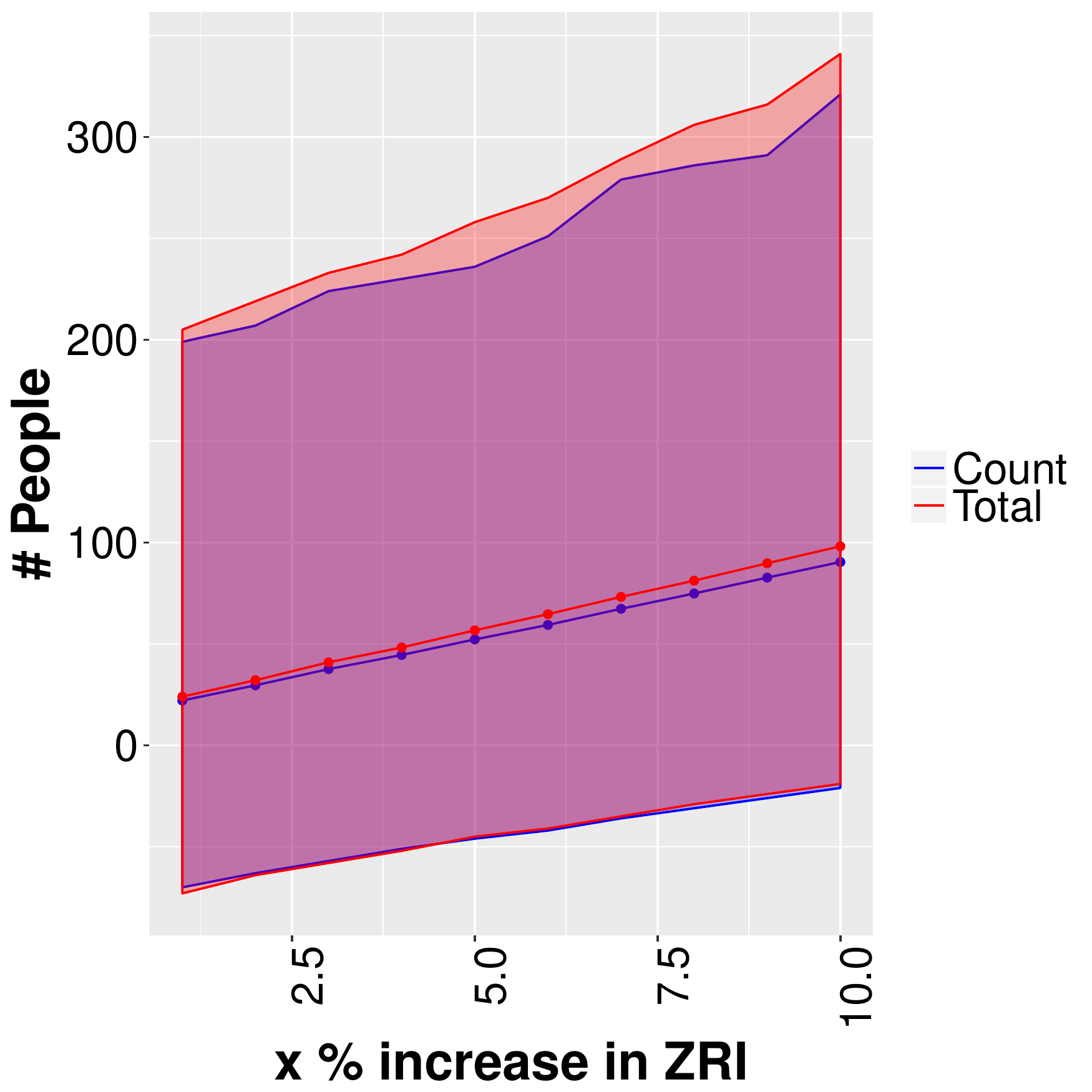}
\caption{ZRI effect}
\end{subfigure}
\begin{subfigure}{.4\textwidth}
  \centering
\includegraphics[width=1\textwidth]{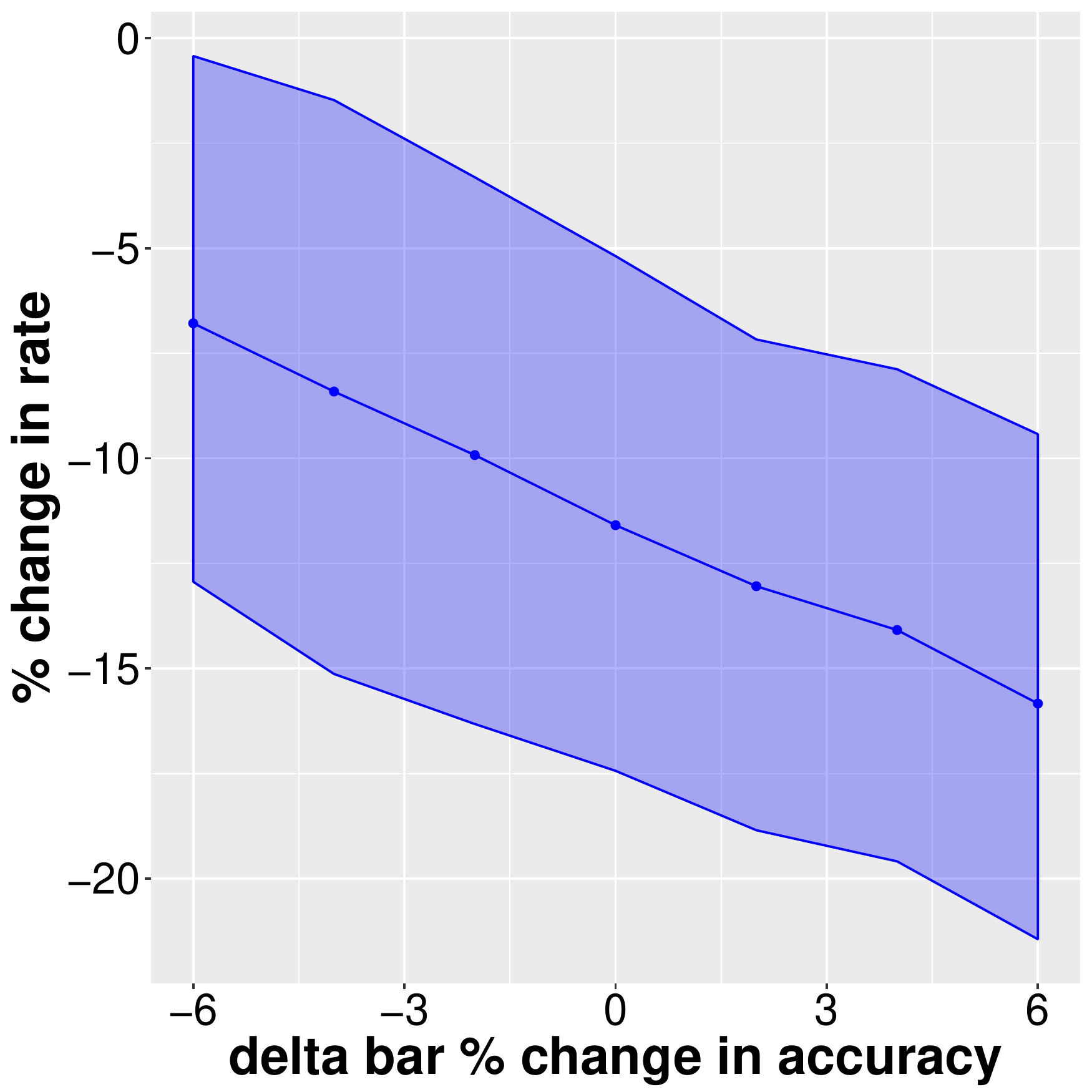}
\caption{Rate}
\end{subfigure}
\caption{Results for St. Louis, MO.  Top left (a): Posterior predictive distribution for homeless counts, $C_{i,1:T}^* | C_{1:25,1:T}, N_{1:25,1:T}$, in green, and the imputed total homeless population size, $H_{i,1:T}|C_{1:25,1:T}, H_{1:25,1:T}$, in blue.  The black 'x' marks correspond to the observed (raw) homeless count by year.  The count accuracy is modeled with a constant expectation.  Top right (b): Predictive distribution for total homeless population in 2017, $H_{i,2017} | C_{1:25,1:T}, N_{1:25,1:T}$.  Bottom left (c):  Posterior distribution of increase in total homeless population with increases in ZRI.  Bottom right (d): Sensitivity of the inferred increase in the homelessness rate from 2011 - 2016 to different annual changes in count accuracy.}
\label{fig:StLouis_Results}
\end{figure}

\begin{figure}[ht!]
\centering
\begin{subfigure}{.4\textwidth}
  \centering
\includegraphics[width=1\textwidth]{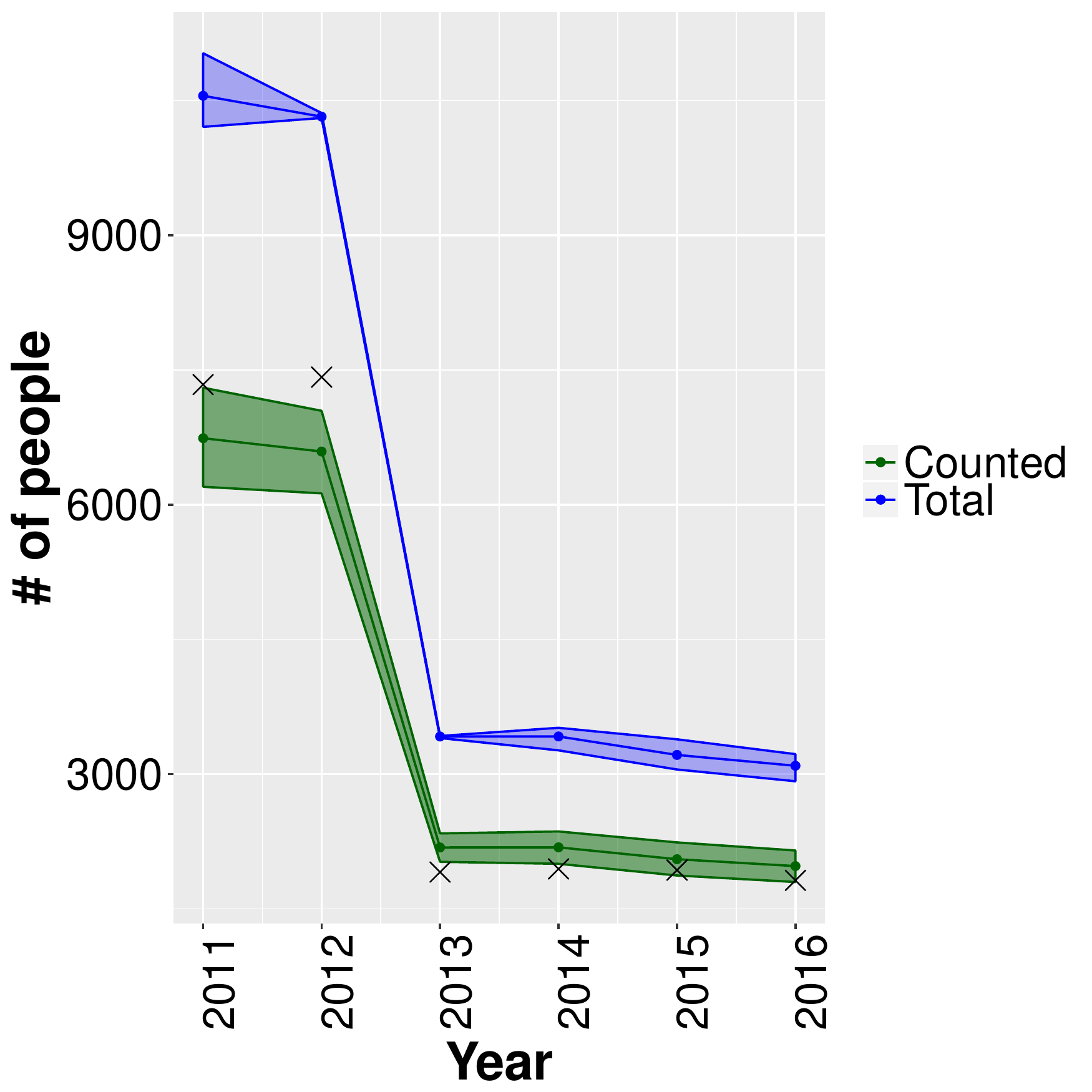}
\caption{\# of homeless}
\end{subfigure}
\begin{subfigure}{.4\textwidth}
  \centering
\includegraphics[width=1\textwidth]{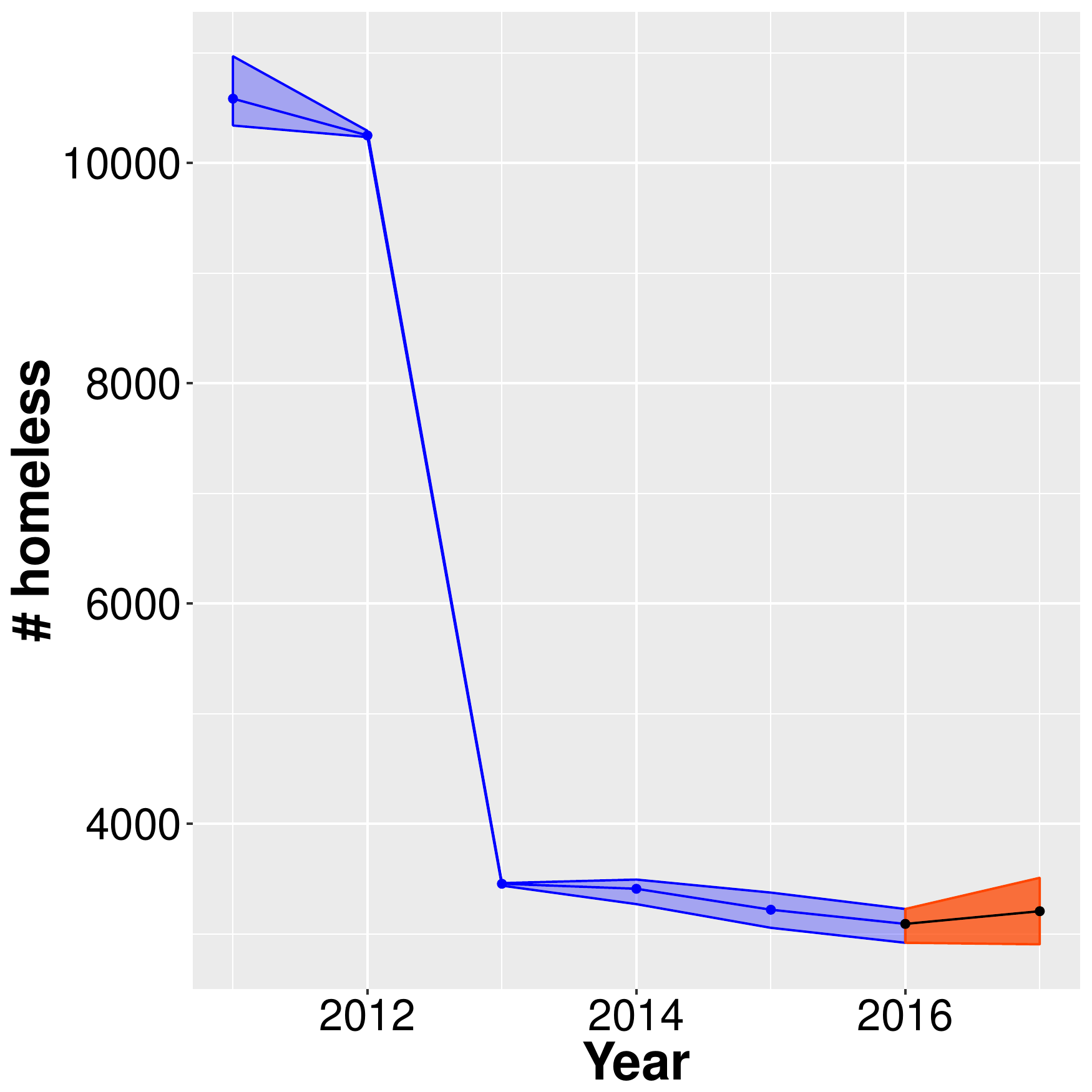}
\caption{2017 forecast}
\end{subfigure}
\begin{subfigure}{.4\textwidth}
  \centering
\includegraphics[width=1\textwidth]{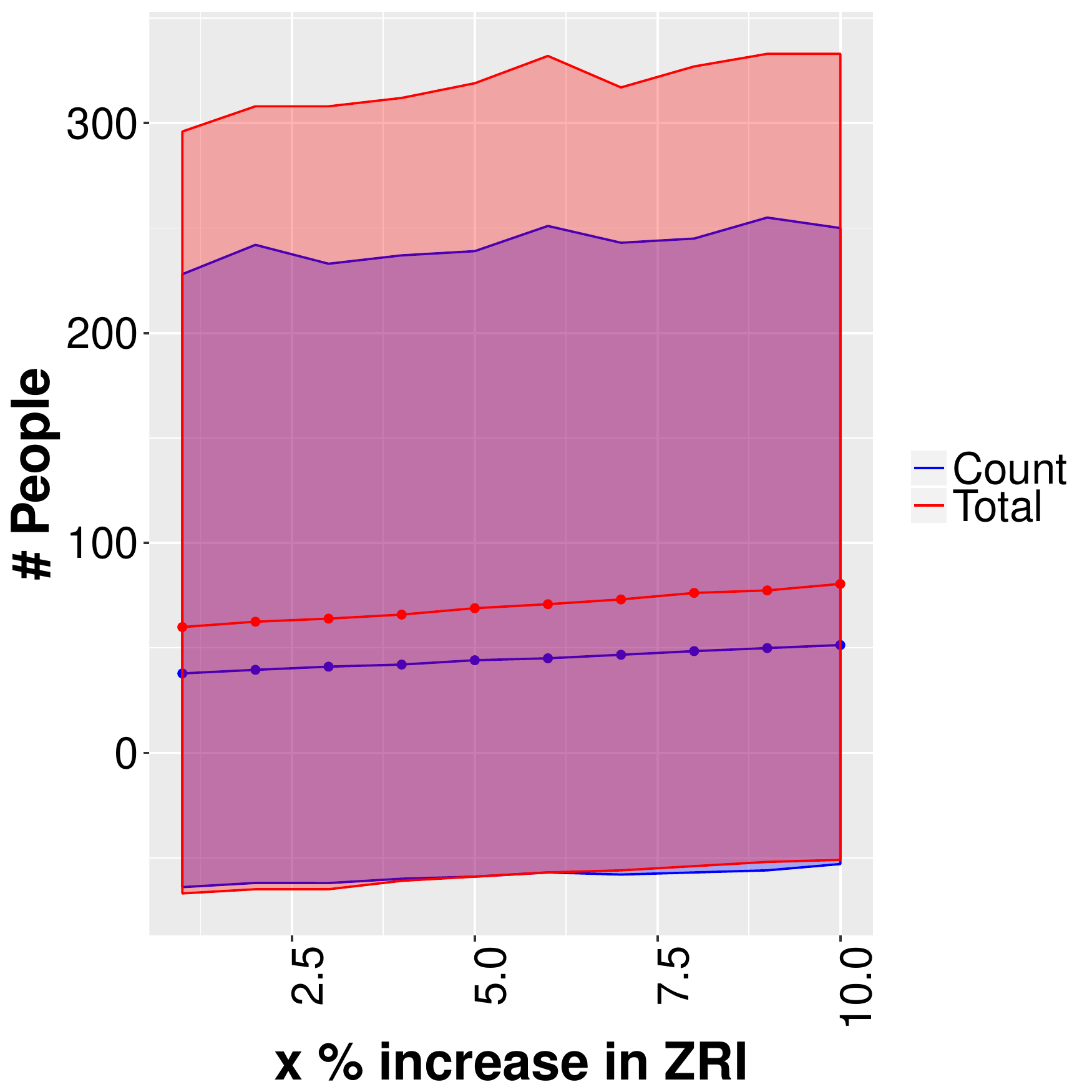}
\caption{ZRI effect}
\end{subfigure}
\begin{subfigure}{.4\textwidth}
  \centering
\includegraphics[width=1\textwidth]{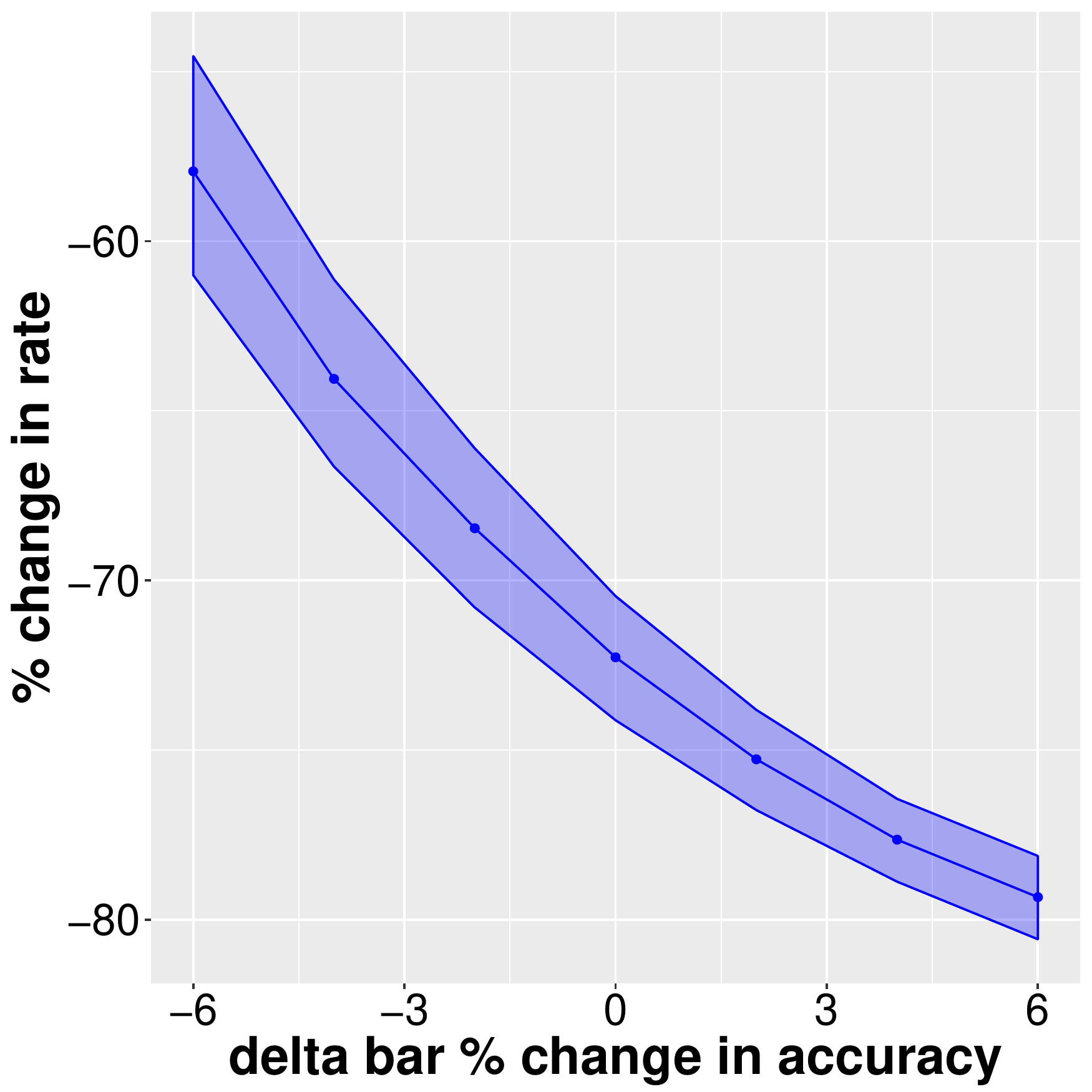}
\caption{Rate}
\end{subfigure}
\caption{Results for Tampa, FL.  Top left (a): Posterior predictive distribution for homeless counts, $C_{i,1:T}^* | C_{1:25,1:T}, N_{1:25,1:T}$, in green, and the imputed total homeless population size, $H_{i,1:T}|C_{1:25,1:T}, H_{1:25,1:T}$, in blue.  The black 'x' marks correspond to the observed (raw) homeless count by year.  The count accuracy is modeled with a constant expectation.  Top right (b): Predictive distribution for total homeless population in 2017, $H_{i,2017} | C_{1:25,1:T}, N_{1:25,1:T}$.  Bottom left (c):  Posterior distribution of increase in total homeless population with increases in ZRI.  Bottom right (d): Sensitivity of the inferred increase in the homelessness rate from 2011 - 2016 to different annual changes in count accuracy.}
\label{fig:Tampa_Results}
\end{figure}

\begin{figure}[ht!]
\centering
\begin{subfigure}{.4\textwidth}
  \centering
\includegraphics[width=1\textwidth]{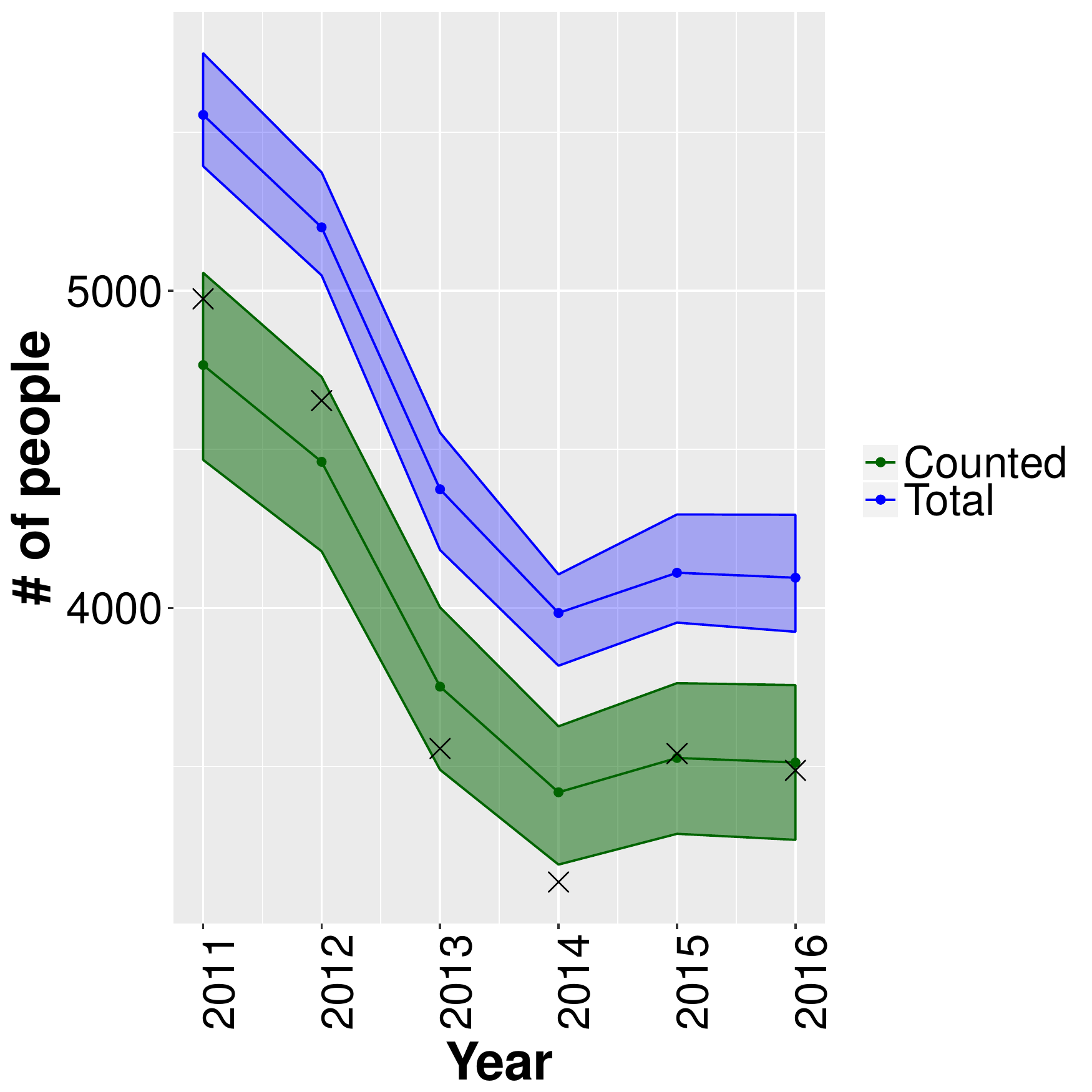}
\caption{\# of homeless}
\end{subfigure}
\begin{subfigure}{.4\textwidth}
  \centering
\includegraphics[width=1\textwidth]{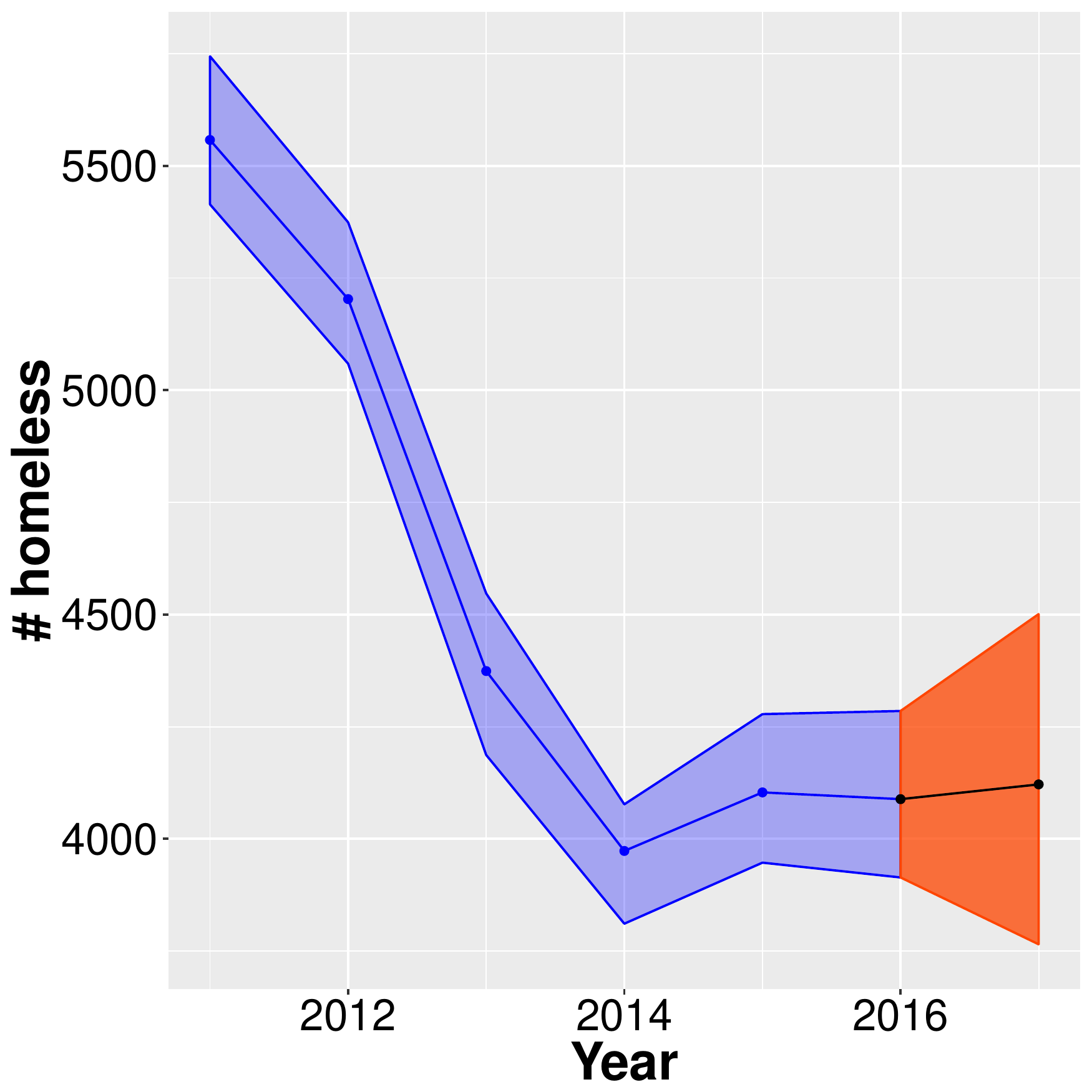}
\caption{2017 forecast}
\end{subfigure}
\begin{subfigure}{.4\textwidth}
  \centering
\includegraphics[width=1\textwidth]{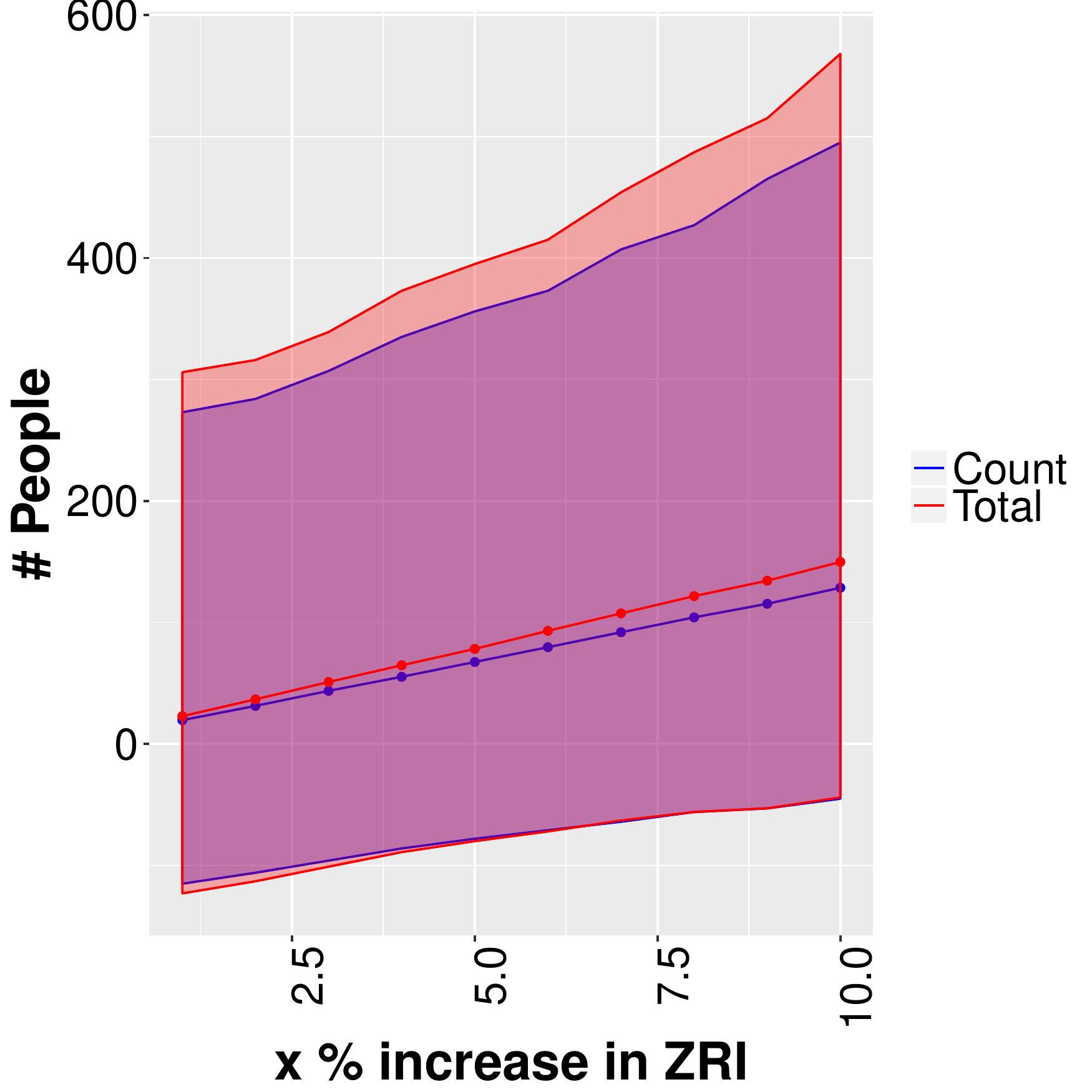}
\caption{ZRI effect}
\end{subfigure}
\begin{subfigure}{.4\textwidth}
  \centering
\includegraphics[width=1\textwidth]{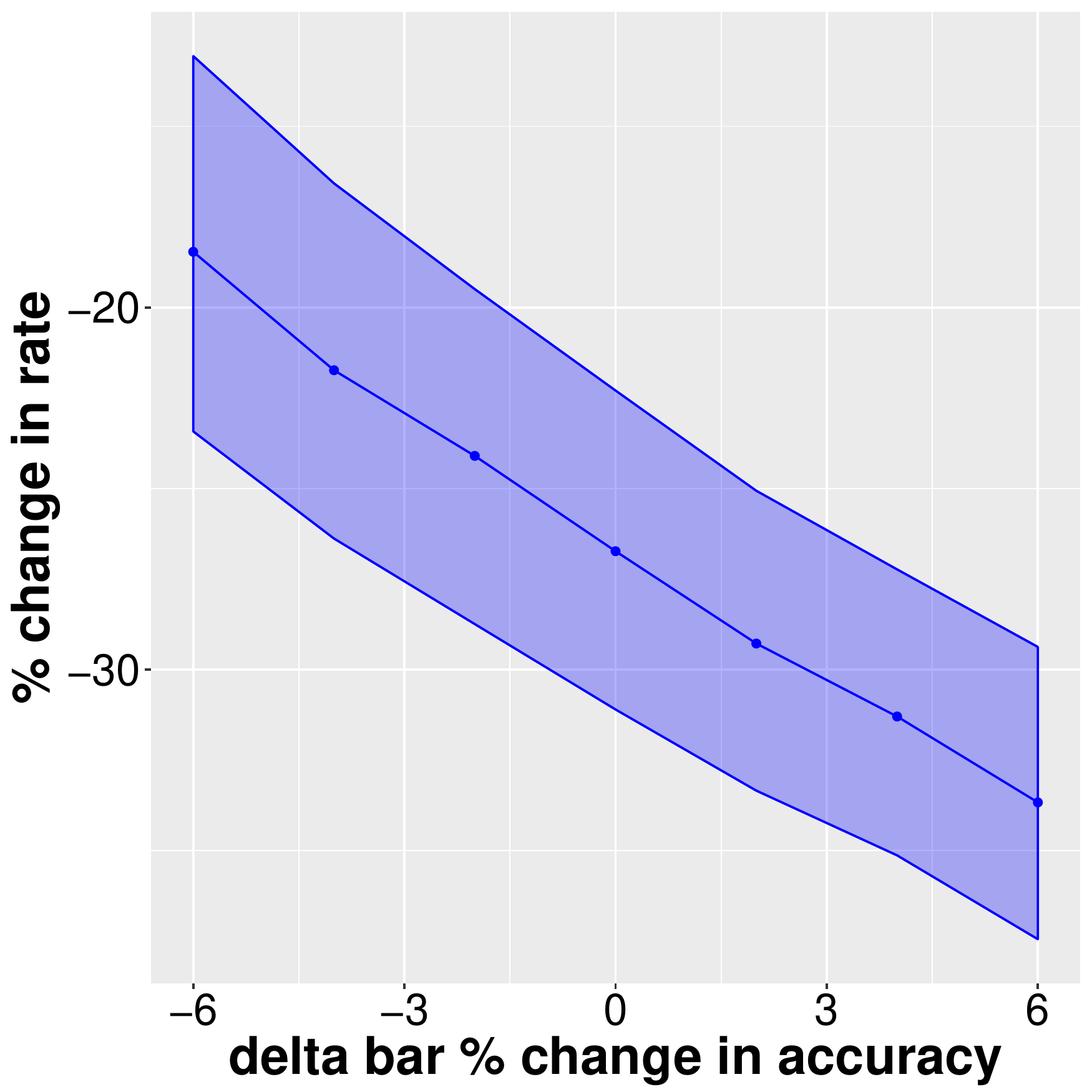}
\caption{Rate}
\end{subfigure}
\caption{Results for Baltimore, MD.  Top left (a): Posterior predictive distribution for homeless counts, $C_{i,1:T}^* | C_{1:25,1:T}, N_{1:25,1:T}$, in green, and the imputed total homeless population size, $H_{i,1:T}|C_{1:25,1:T}, H_{1:25,1:T}$, in blue.  The black 'x' marks correspond to the observed (raw) homeless count by year.  The count accuracy is modeled with a constant expectation.  Top right (b): Predictive distribution for total homeless population in 2017, $H_{i,2017} | C_{1:25,1:T}, N_{1:25,1:T}$.  Bottom left (c):  Posterior distribution of increase in total homeless population with increases in ZRI.  Bottom right (d): Sensitivity of the inferred increase in the homelessness rate from 2011 - 2016 to different annual changes in count accuracy.}
\label{fig:Baltimore_Results}
\end{figure}

\begin{figure}[ht!]
\centering
\begin{subfigure}{.4\textwidth}
  \centering
\includegraphics[width=1\textwidth]{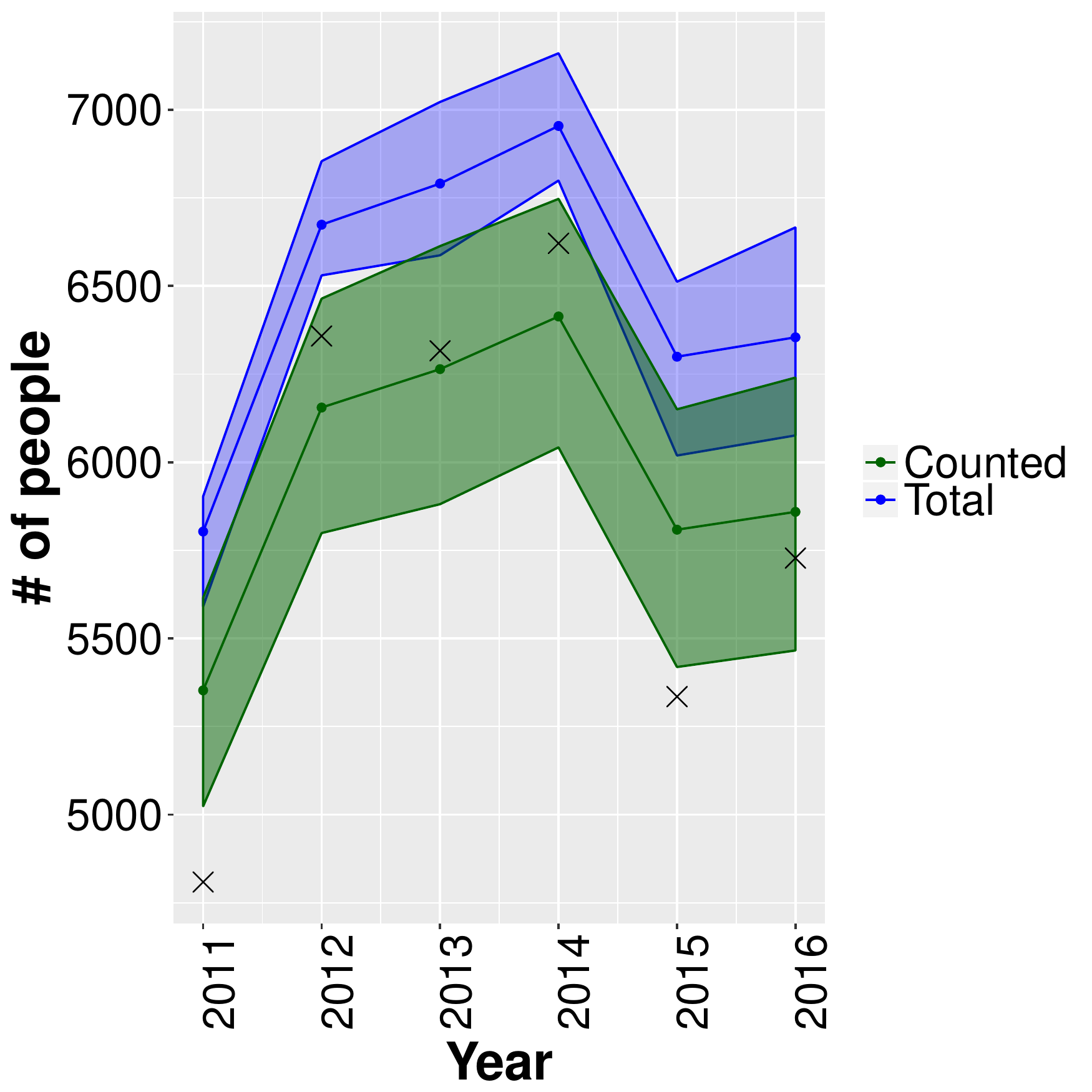}
\caption{\# of homeless}
\end{subfigure}
\begin{subfigure}{.4\textwidth}
  \centering
\includegraphics[width=1\textwidth]{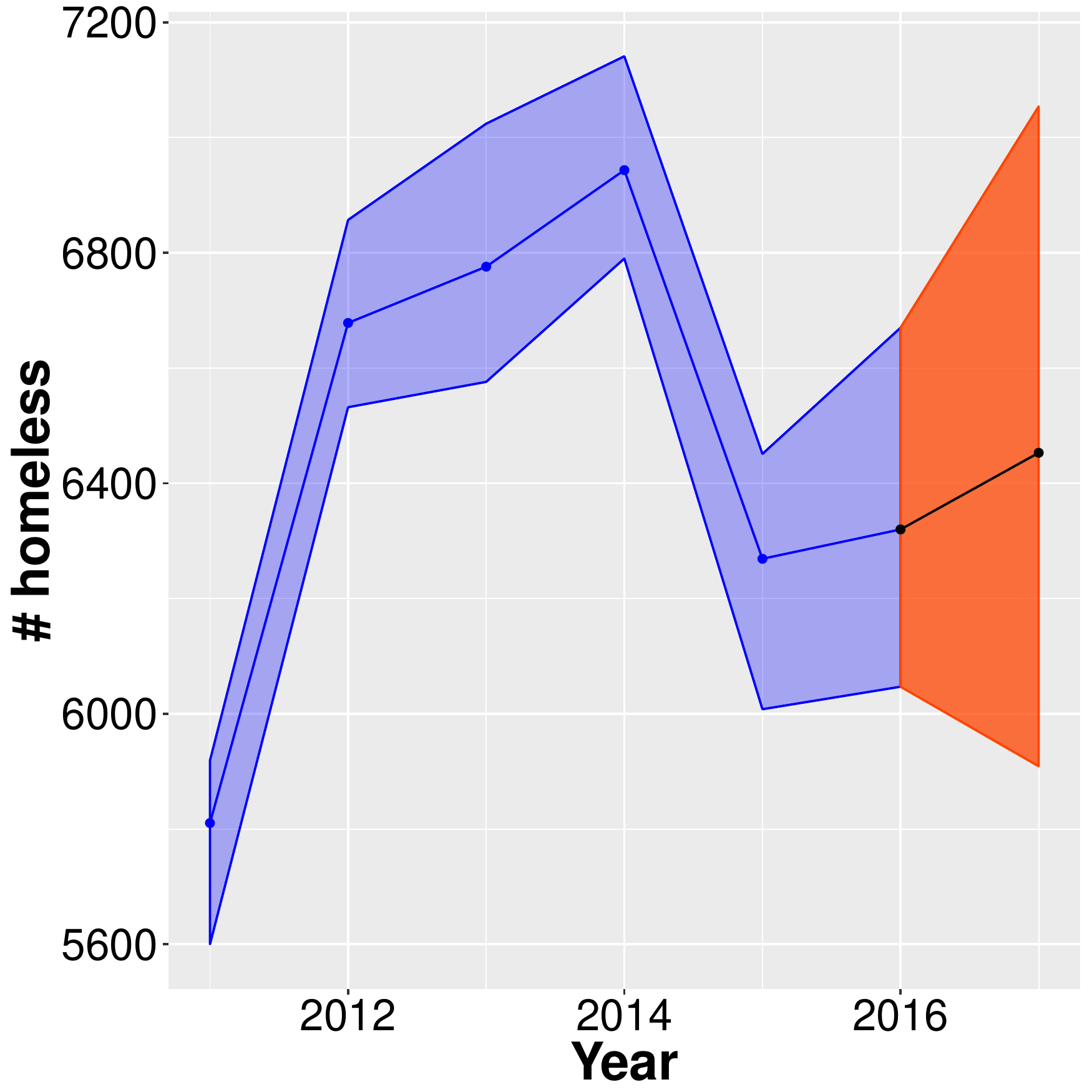}
\caption{2017 forecast}
\end{subfigure}
\begin{subfigure}{.4\textwidth}
  \centering
\includegraphics[width=1\textwidth]{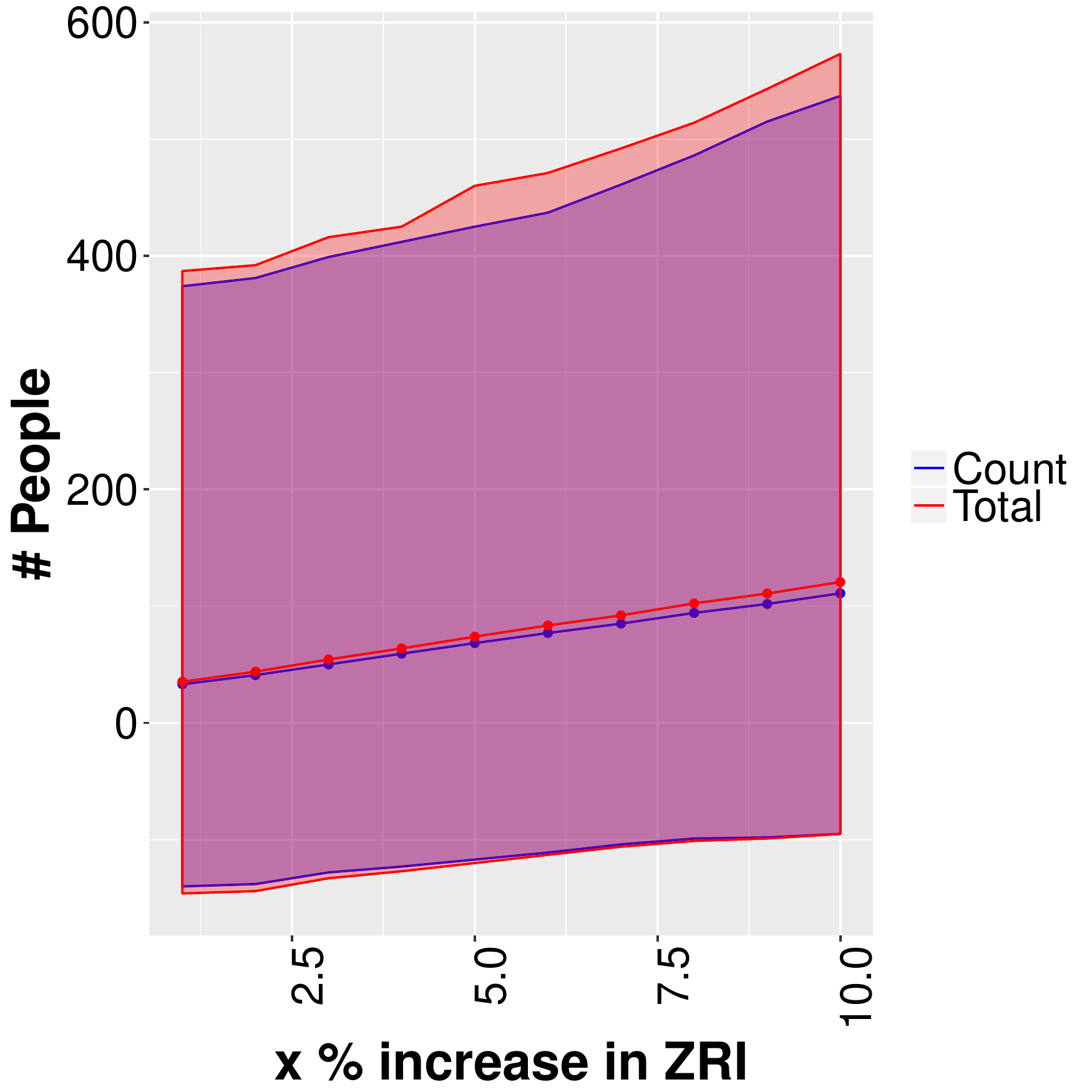}
\caption{ZRI effect}
\end{subfigure}
\begin{subfigure}{.4\textwidth}
  \centering
\includegraphics[width=1\textwidth]{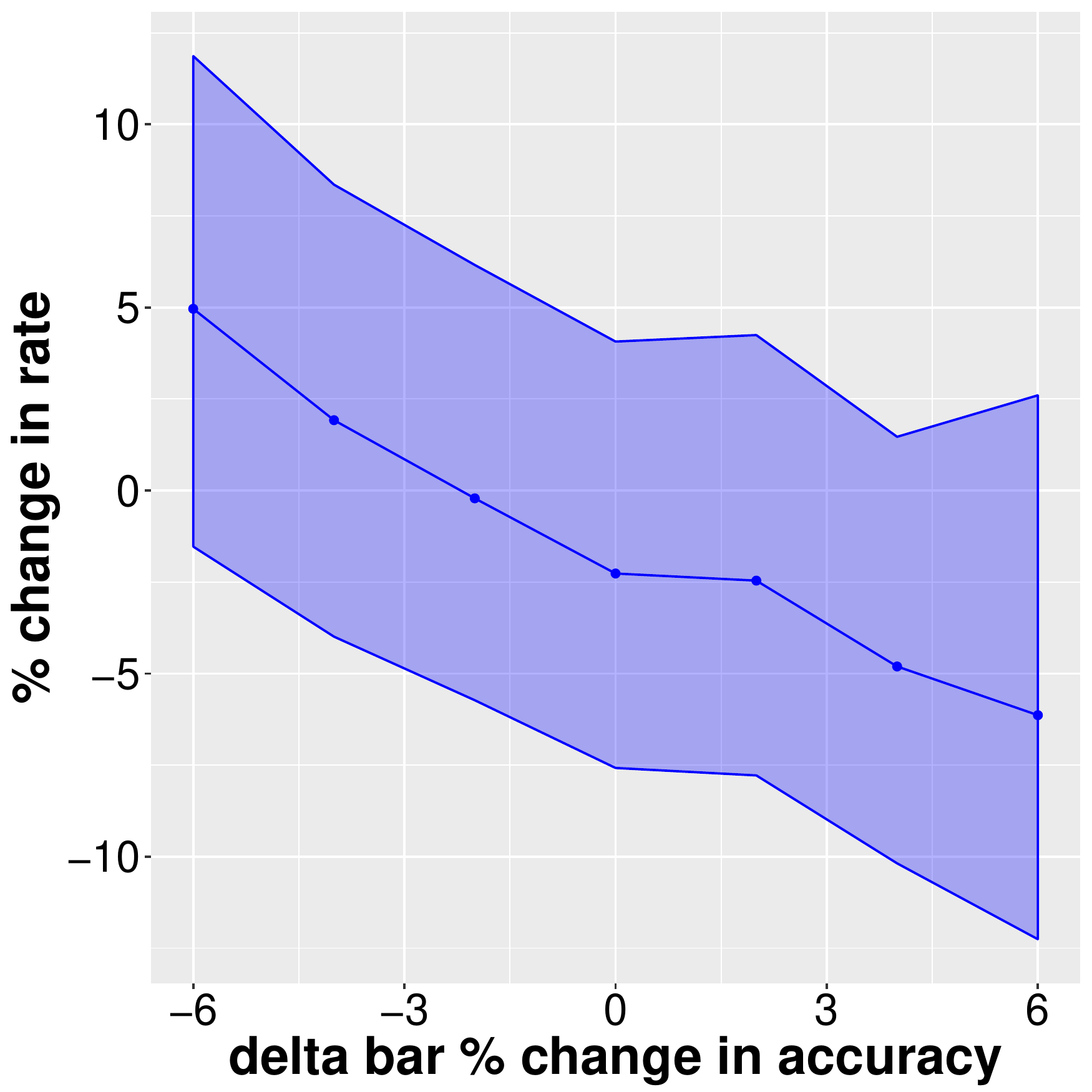}
\caption{Rate}
\end{subfigure}
\caption{Results for Denver, CO.  Top left (a): Posterior predictive distribution for homeless counts, $C_{i,1:T}^* | C_{1:25,1:T}, N_{1:25,1:T}$, in green, and the imputed total homeless population size, $H_{i,1:T}|C_{1:25,1:T}, H_{1:25,1:T}$, in blue.  The black 'x' marks correspond to the observed (raw) homeless count by year.  The count accuracy is modeled with a constant expectation.  Top right (b): Predictive distribution for total homeless population in 2017, $H_{i,2017} | C_{1:25,1:T}, N_{1:25,1:T}$.  Bottom left (c):  Posterior distribution of increase in total homeless population with increases in ZRI.  Bottom right (d): Sensitivity of the inferred increase in the homelessness rate from 2011 - 2016 to different annual changes in count accuracy.}
\label{fig:Denver_Results}
\end{figure}

\begin{figure}[ht!]
\centering
\begin{subfigure}{.4\textwidth}
  \centering
\includegraphics[width=1\textwidth]{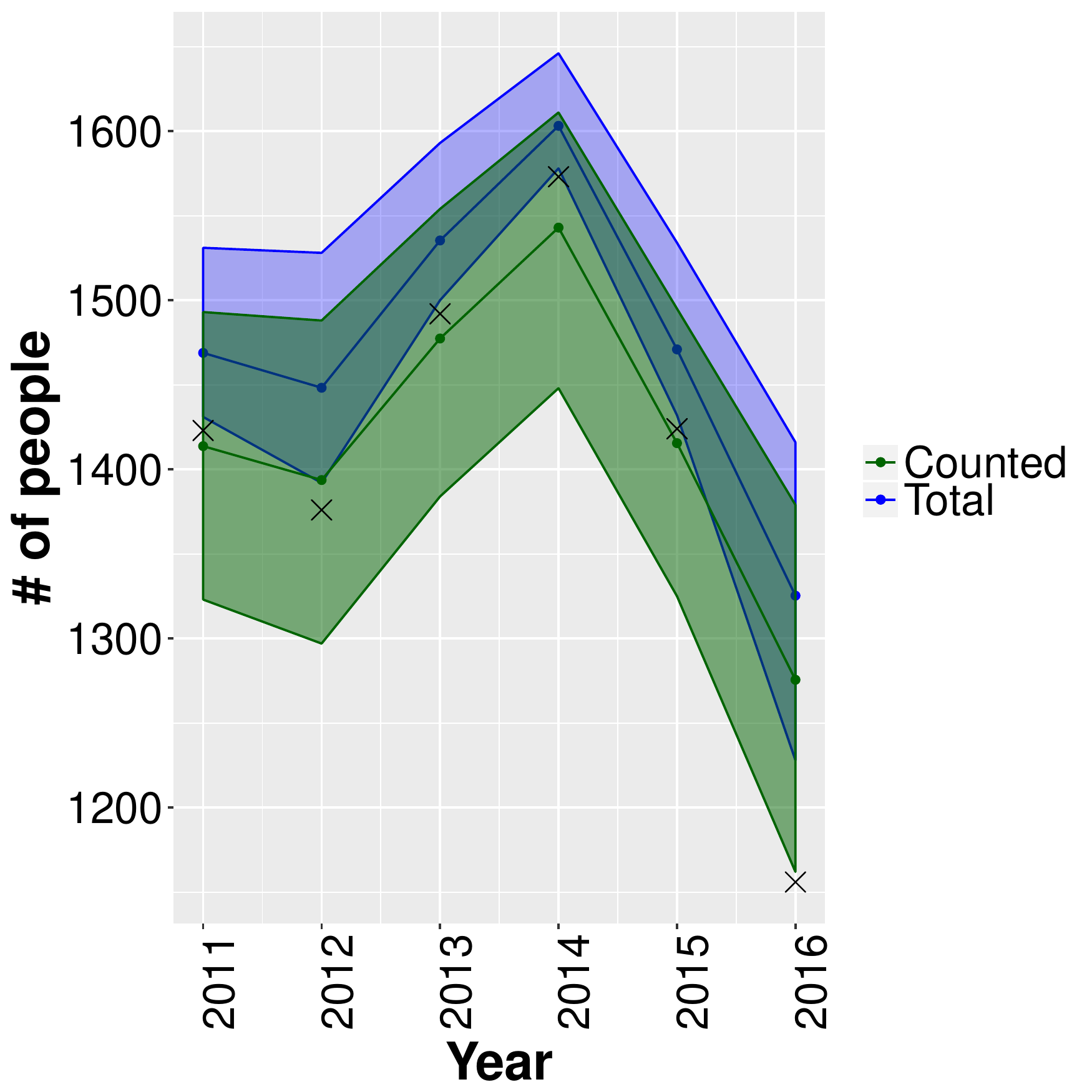}
\caption{\# of homeless}
\end{subfigure}
\begin{subfigure}{.4\textwidth}
  \centering
\includegraphics[width=1\textwidth]{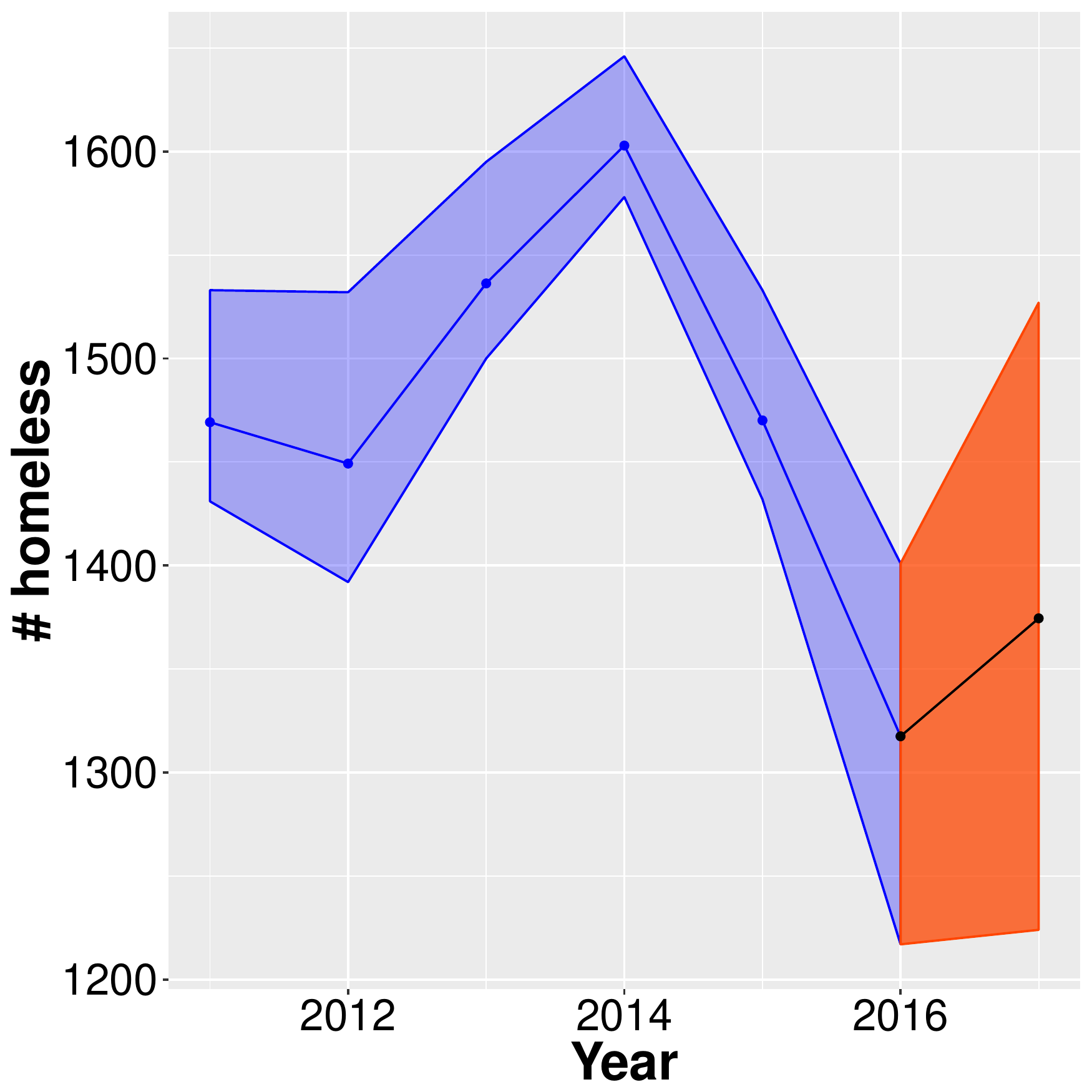}
\caption{2017 forecast}
\end{subfigure}
\begin{subfigure}{.4\textwidth}
  \centering
\includegraphics[width=1\textwidth]{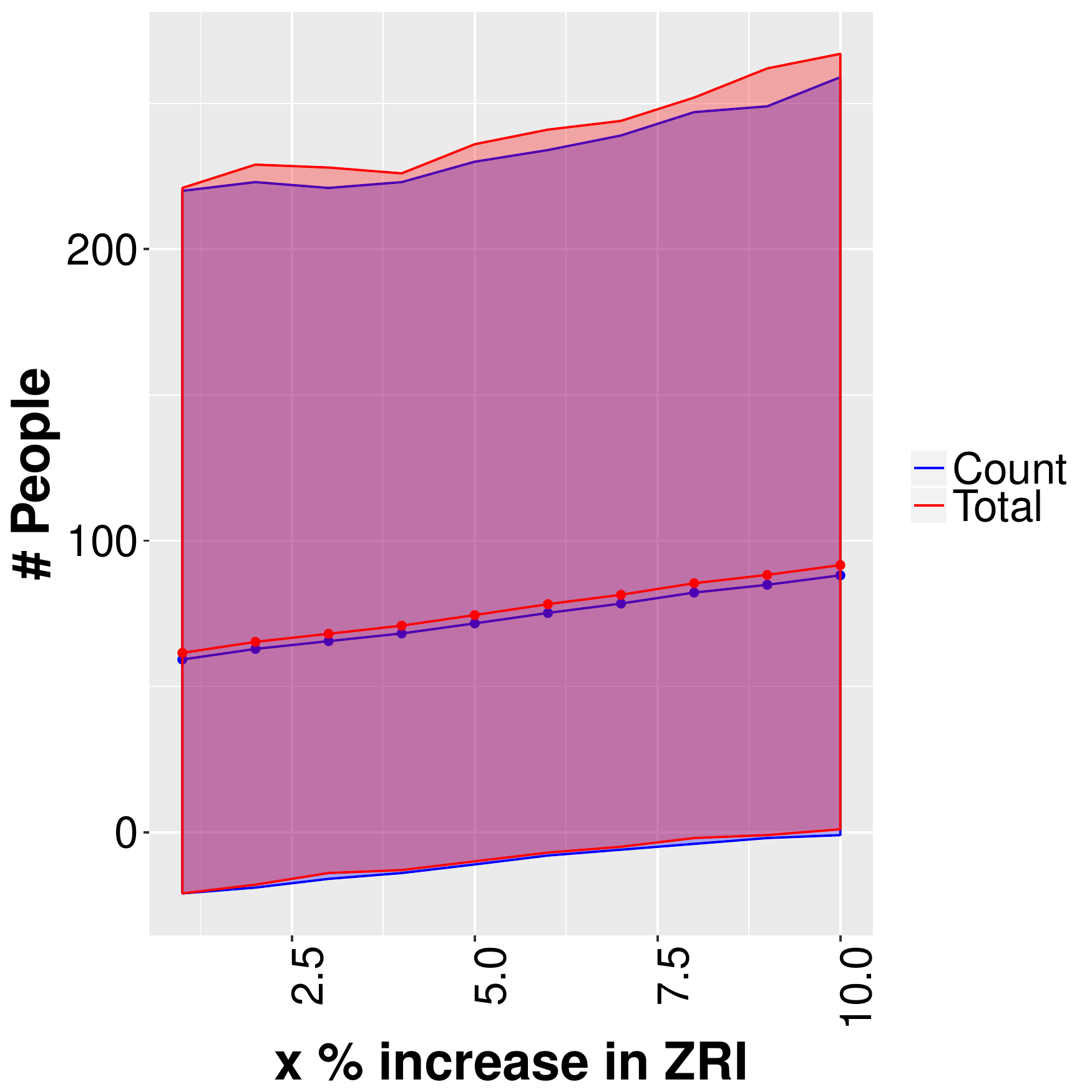}
\caption{ZRI effect}
\end{subfigure}
\begin{subfigure}{.4\textwidth}
  \centering
\includegraphics[width=1\textwidth]{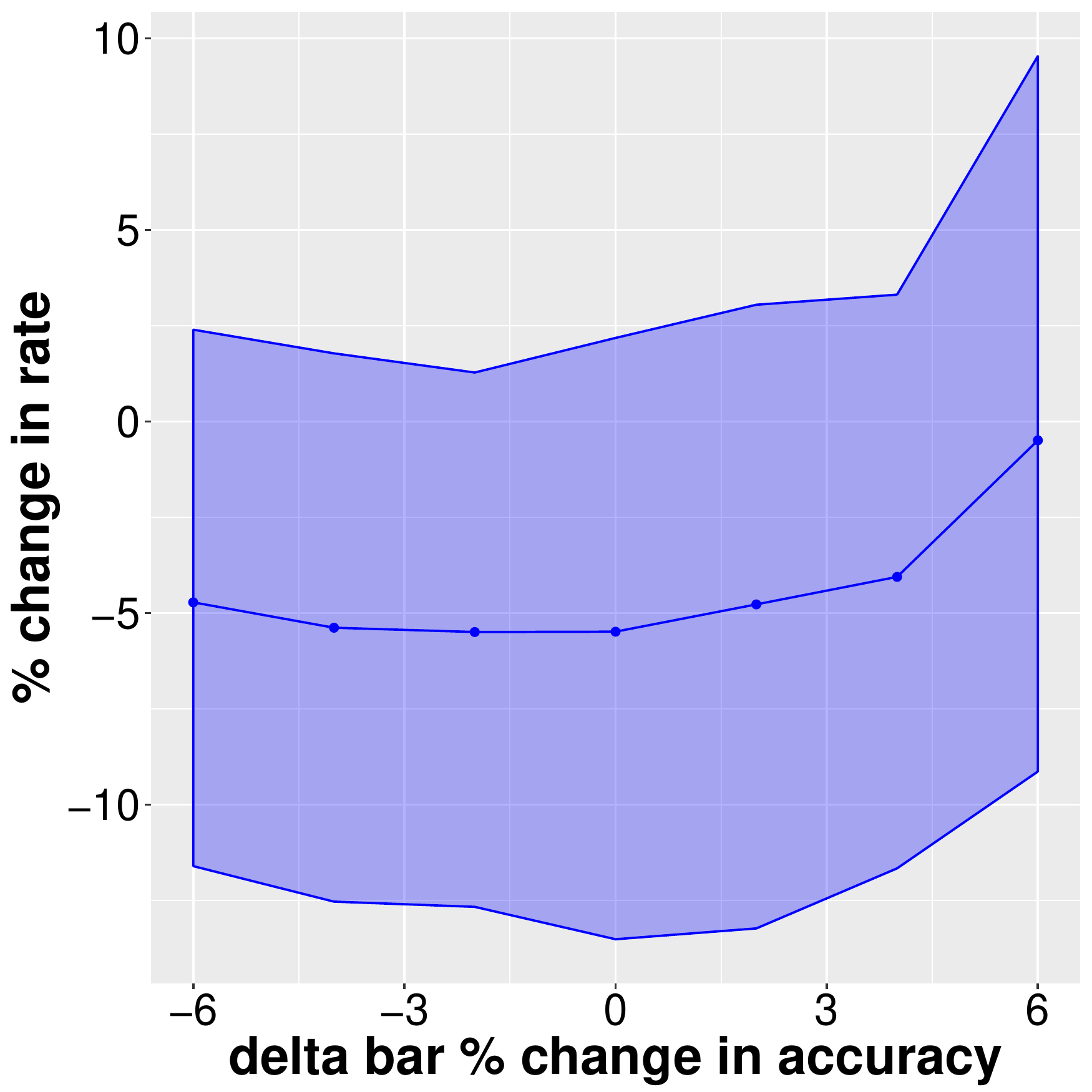}
\caption{Rate}
\end{subfigure}
\caption{Results for Pittsburgh, PA.  Top left (a): Posterior predictive distribution for homeless counts, $C_{i,1:T}^* | C_{1:25,1:T}, N_{1:25,1:T}$, in green, and the imputed total homeless population size, $H_{i,1:T}|C_{1:25,1:T}, H_{1:25,1:T}$, in blue.  The black 'x' marks correspond to the observed (raw) homeless count by year.  The count accuracy is modeled with a constant expectation.  Top right (b): Predictive distribution for total homeless population in 2017, $H_{i,2017} | C_{1:25,1:T}, N_{1:25,1:T}$.  Bottom left (c):  Posterior distribution of increase in total homeless population with increases in ZRI.  Bottom right (d): Sensitivity of the inferred increase in the homelessness rate from 2011 - 2016 to different annual changes in count accuracy.}
\label{fig:Pittsburgh_Results}
\end{figure}

\begin{figure}[ht!]
\centering
\begin{subfigure}{.4\textwidth}
  \centering
\includegraphics[width=1\textwidth]{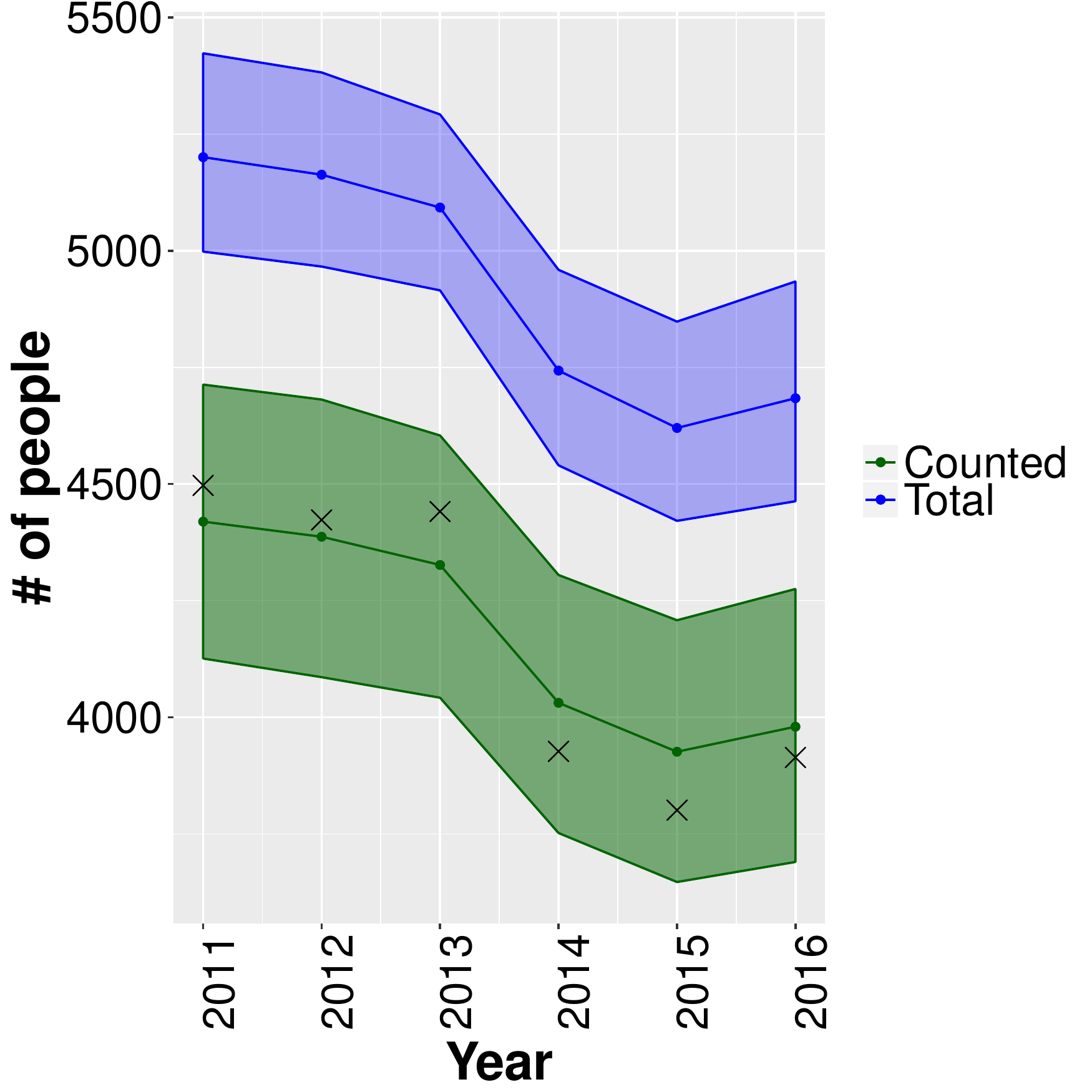}
\caption{\# of homeless}
\end{subfigure}
\begin{subfigure}{.4\textwidth}
  \centering
\includegraphics[width=1\textwidth]{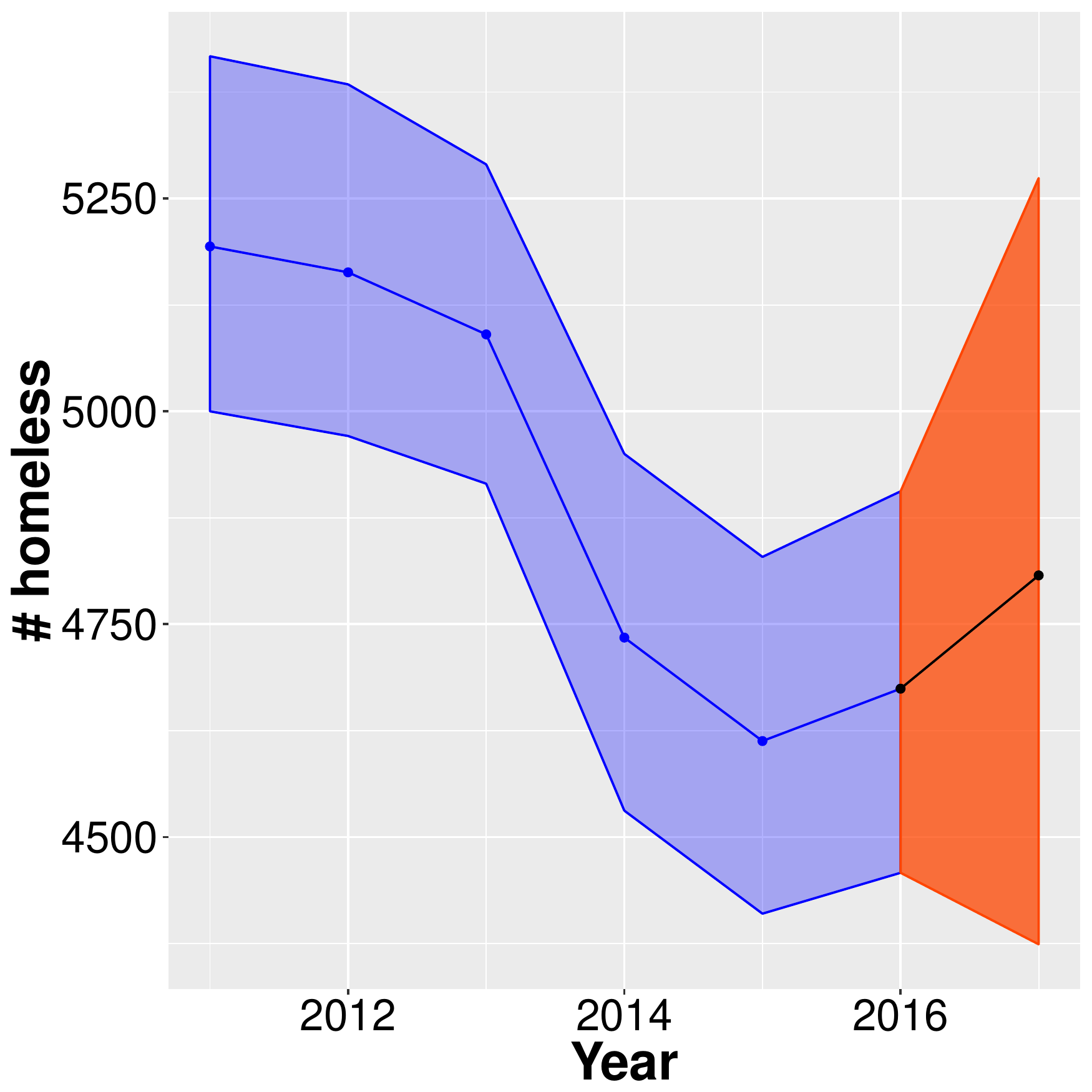}
\caption{2017 forecast}
\end{subfigure}
\begin{subfigure}{.4\textwidth}
  \centering
\includegraphics[width=1\textwidth]{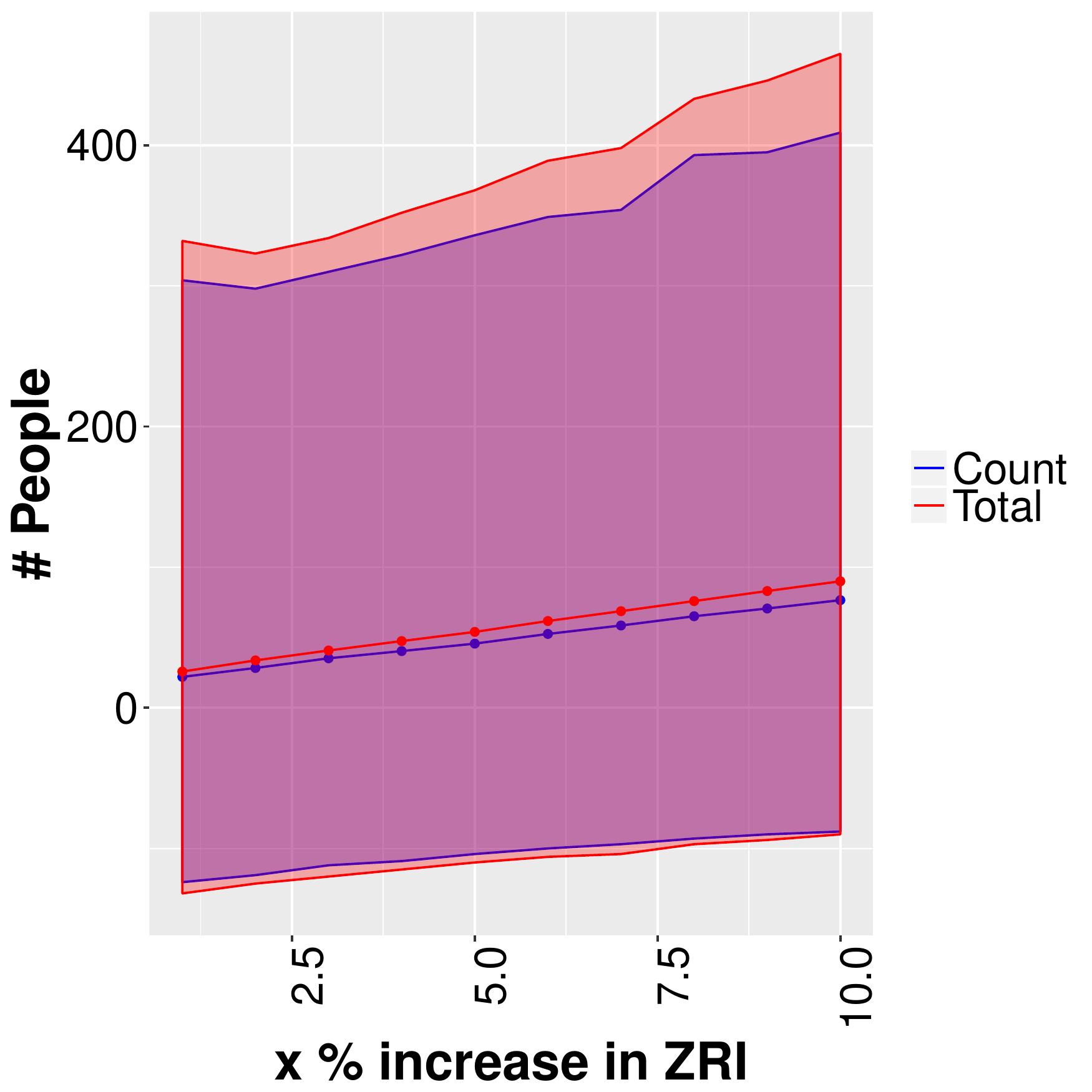}
\caption{ZRI effect}
\end{subfigure}
\begin{subfigure}{.4\textwidth}
  \centering
\includegraphics[width=1\textwidth]{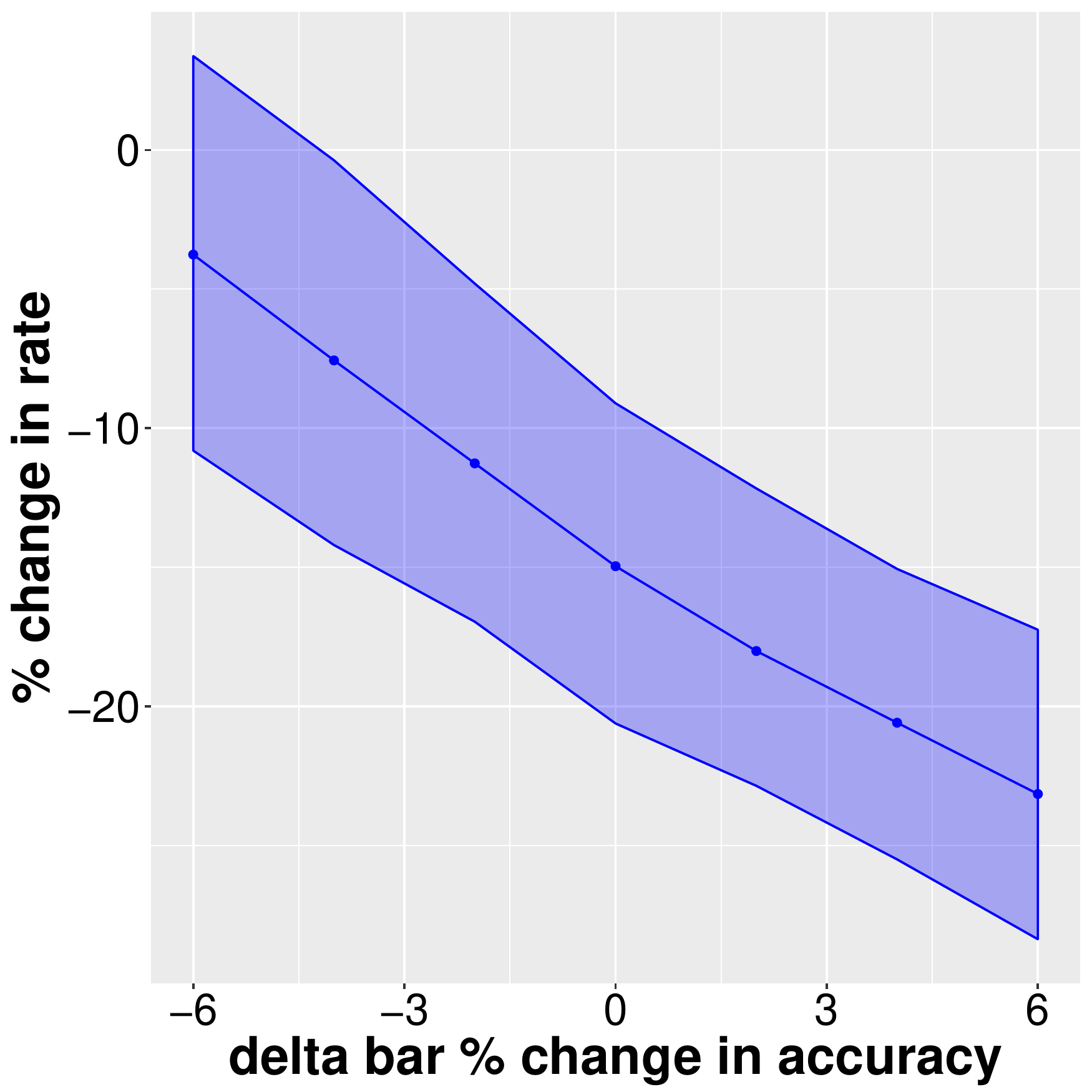}
\caption{Rate}
\end{subfigure}
\caption{Results for Portland, OR.  Top left (a): Posterior predictive distribution for homeless counts, $C_{i,1:T}^* | C_{1:25,1:T}, N_{1:25,1:T}$, in green, and the imputed total homeless population size, $H_{i,1:T}|C_{1:25,1:T}, H_{1:25,1:T}$, in blue.  The black 'x' marks correspond to the observed (raw) homeless count by year.  The count accuracy is modeled with a constant expectation.  Top right (b): Predictive distribution for total homeless population in 2017, $H_{i,2017} | C_{1:25,1:T}, N_{1:25,1:T}$.  Bottom left (c):  Posterior distribution of increase in total homeless population with increases in ZRI.  Bottom right (d): Sensitivity of the inferred increase in the homelessness rate from 2011 - 2016 to different annual changes in count accuracy.}
\label{fig:Portland_Results}
\end{figure}

\begin{figure}[ht!]
\centering
\begin{subfigure}{.4\textwidth}
  \centering
\includegraphics[width=1\textwidth]{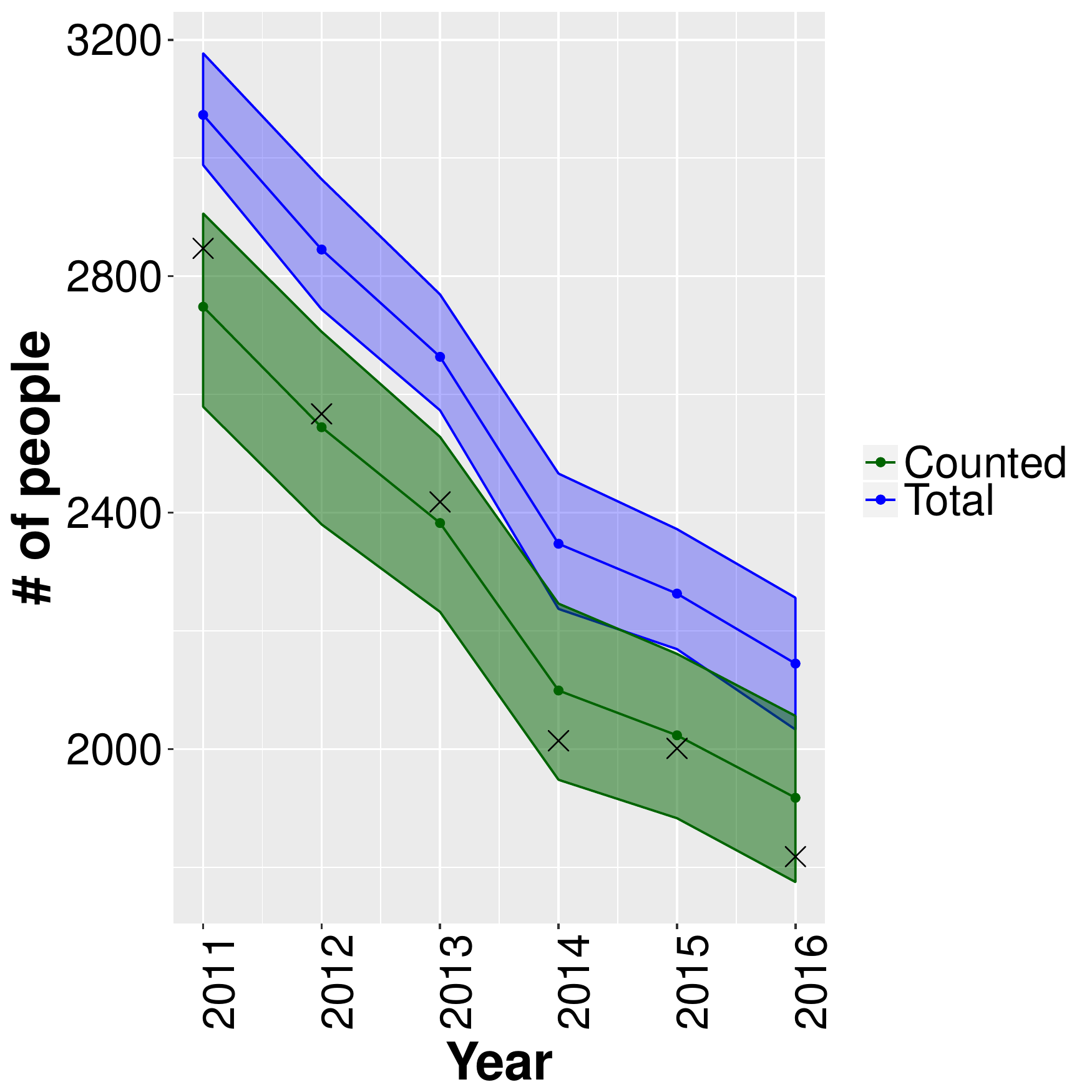}
\caption{\# of homeless}
\end{subfigure}
\begin{subfigure}{.4\textwidth}
  \centering
\includegraphics[width=1\textwidth]{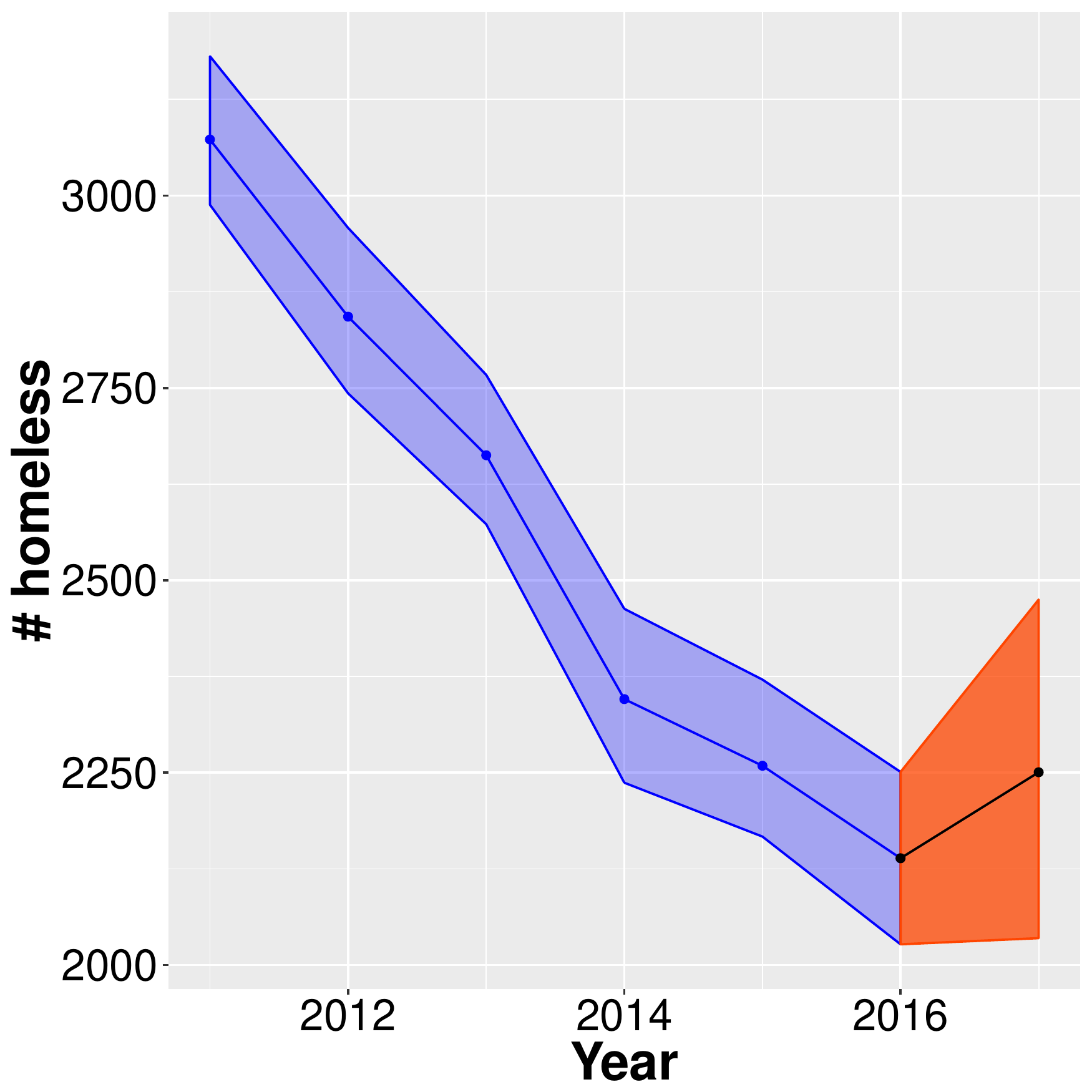}
\caption{2017 forecast}
\end{subfigure}
\begin{subfigure}{.4\textwidth}
  \centering
\includegraphics[width=1\textwidth]{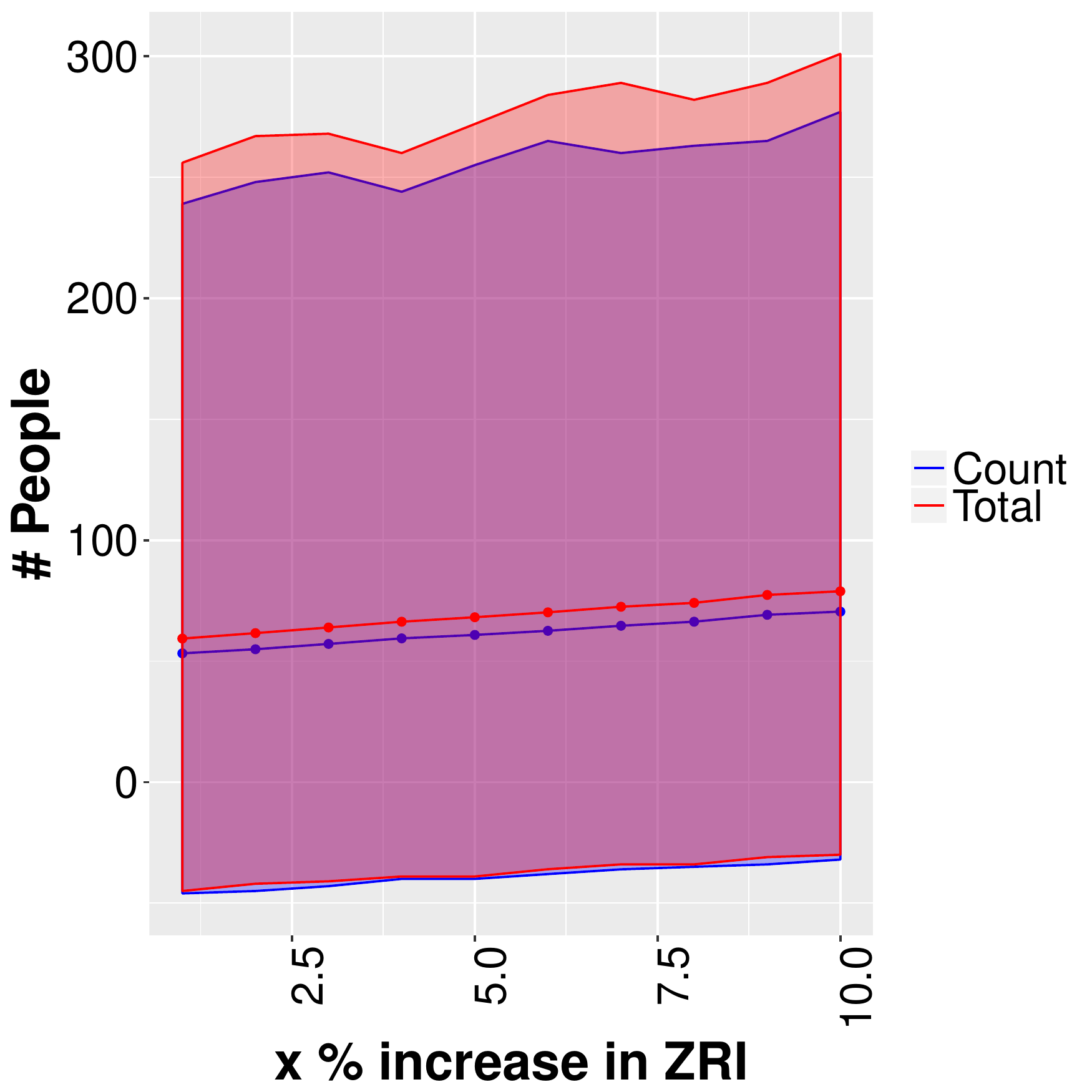}
\caption{ZRI effect}
\end{subfigure}
\begin{subfigure}{.4\textwidth}
  \centering
\includegraphics[width=1\textwidth]{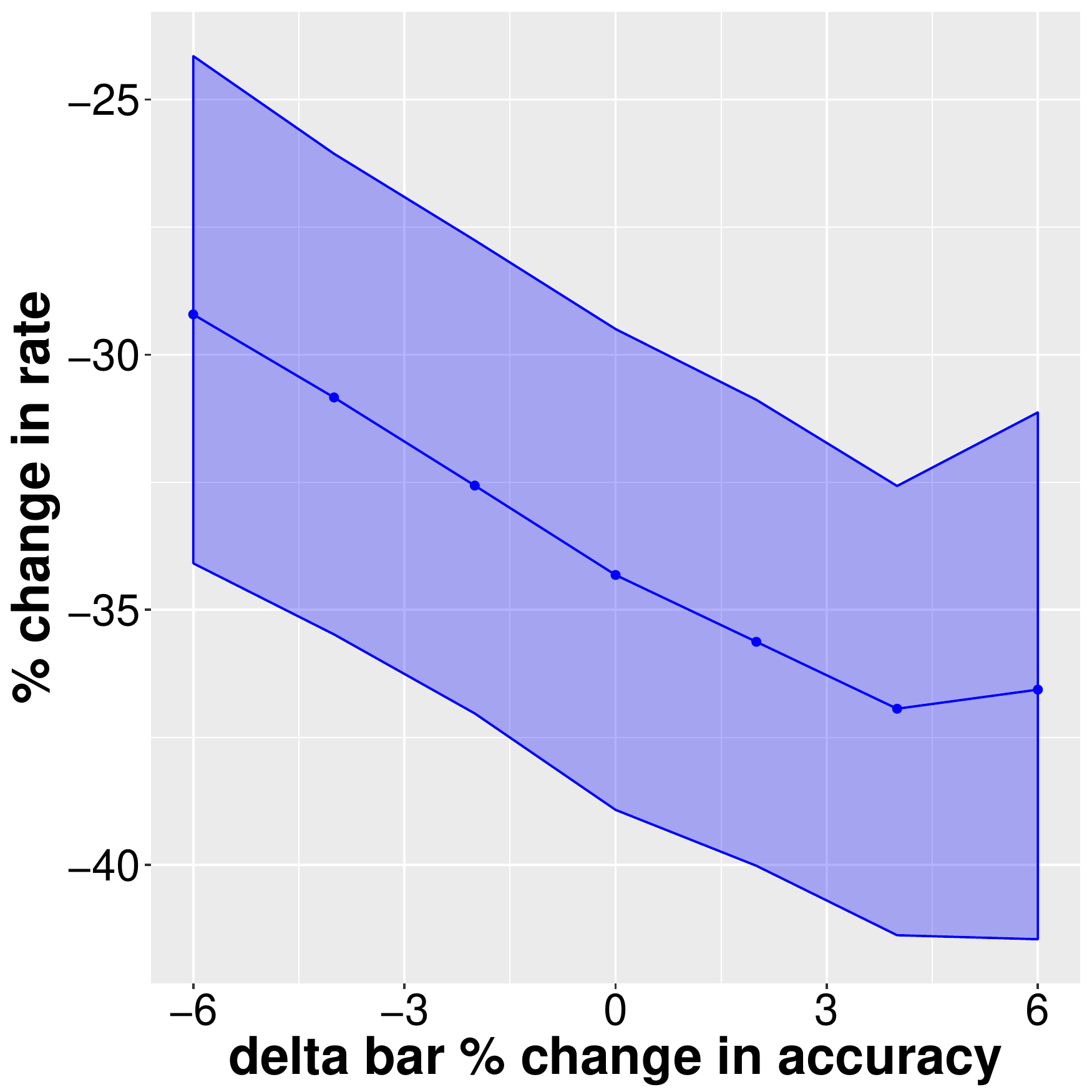}
\caption{Rate}
\end{subfigure}
\caption{Results for Charlotte, NC.  Top left (a): Posterior predictive distribution for homeless counts, $C_{i,1:T}^* | C_{1:25,1:T}, N_{1:25,1:T}$, in green, and the imputed total homeless population size, $H_{i,1:T}|C_{1:25,1:T}, H_{1:25,1:T}$, in blue.  The black 'x' marks correspond to the observed (raw) homeless count by year.  The count accuracy is modeled with a constant expectation.  Top right (b): Predictive distribution for total homeless population in 2017, $H_{i,2017} | C_{1:25,1:T}, N_{1:25,1:T}$.  Bottom left (c):  Posterior distribution of increase in total homeless population with increases in ZRI.  Bottom right (d): Sensitivity of the inferred increase in the homelessness rate from 2011 - 2016 to different annual changes in count accuracy.}
\label{fig:Charlotte_Results}
\end{figure}

\begin{figure}[ht!]
\centering
\begin{subfigure}{.4\textwidth}
  \centering
\includegraphics[width=1\textwidth]{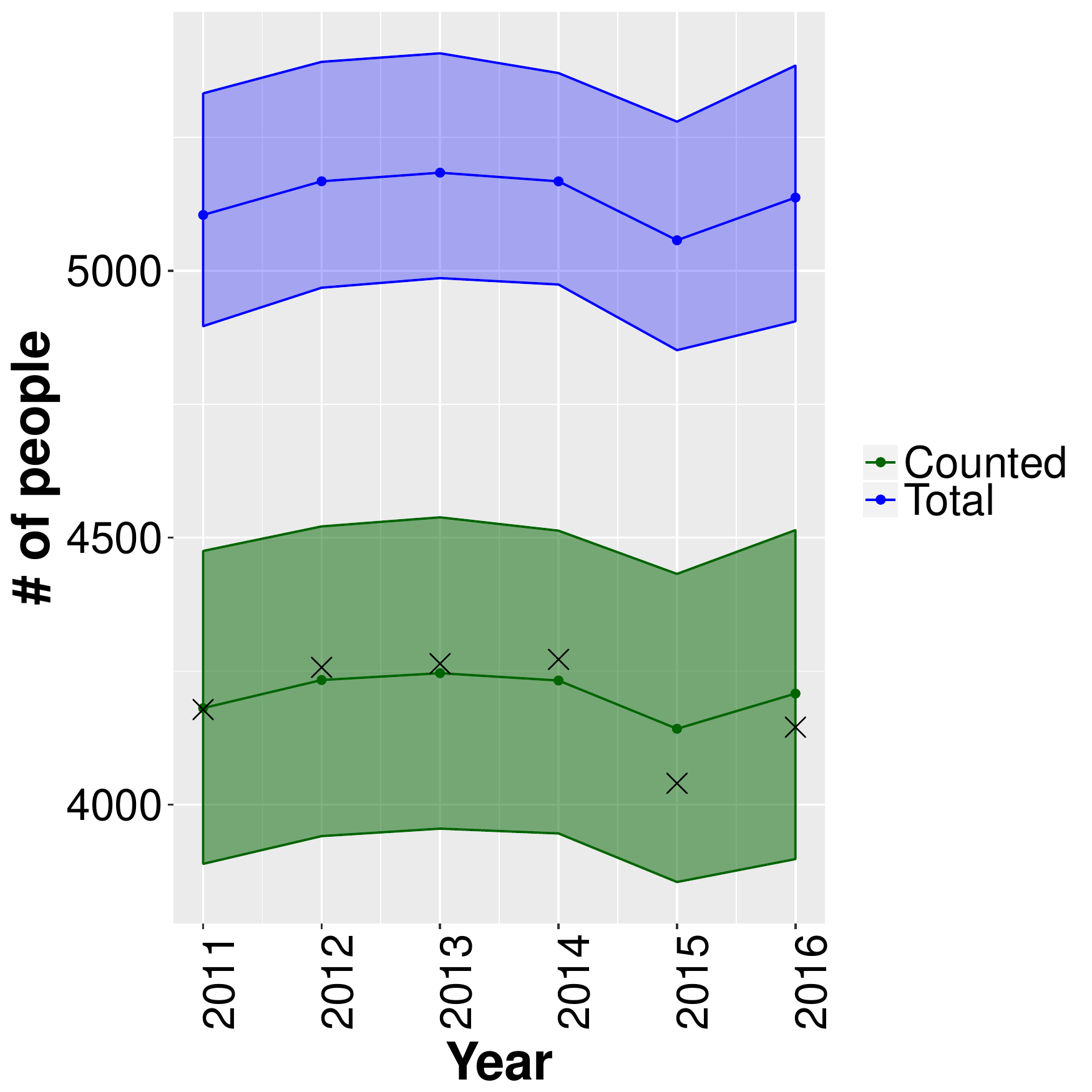}
\caption{\# of homeless}
\end{subfigure}
\begin{subfigure}{.4\textwidth}
  \centering
\includegraphics[width=1\textwidth]{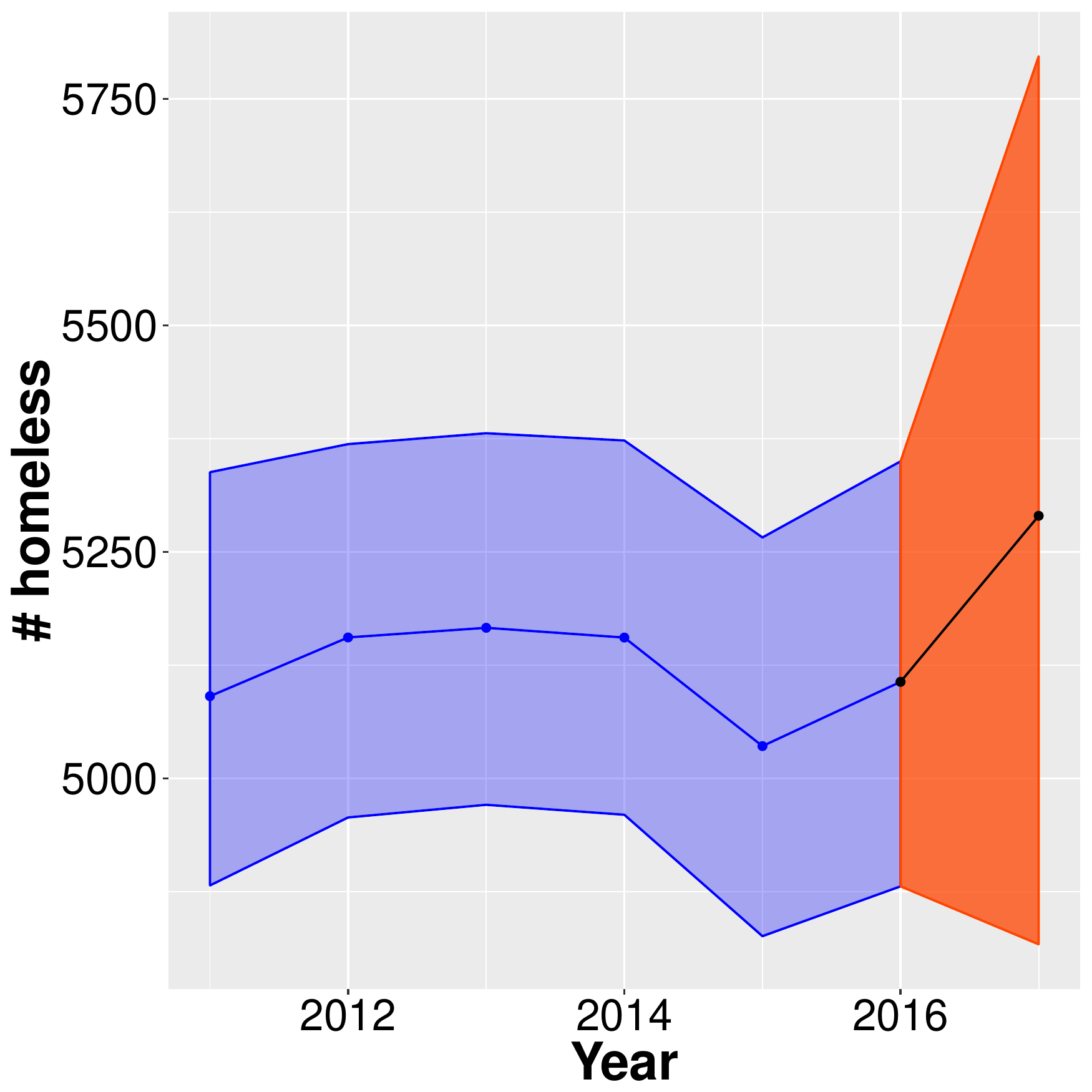}
\caption{2017 forecast}
\end{subfigure}
\begin{subfigure}{.4\textwidth}
  \centering
\includegraphics[width=1\textwidth]{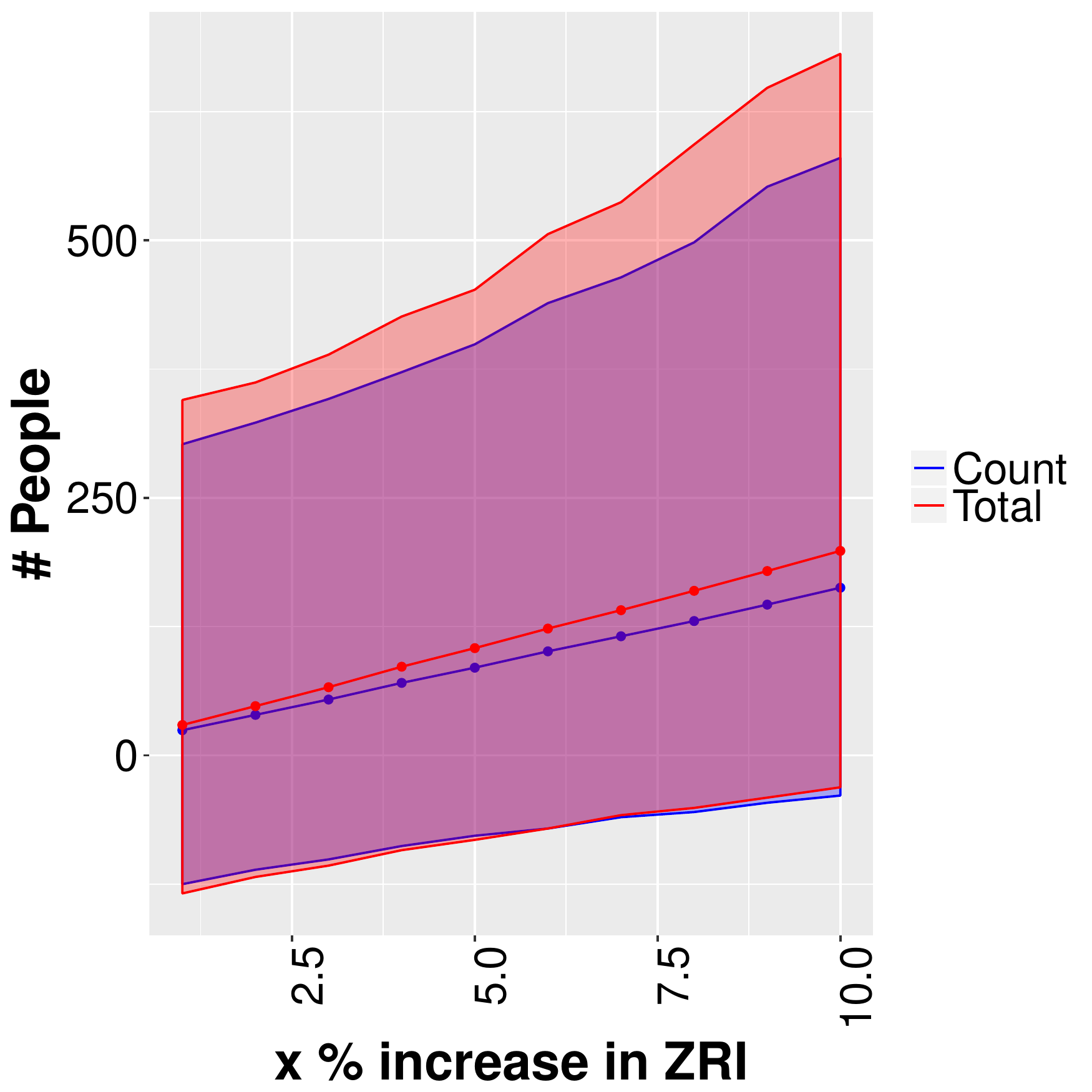}
\caption{ZRI effect}
\end{subfigure}
\begin{subfigure}{.4\textwidth}
  \centering
\includegraphics[width=1\textwidth]{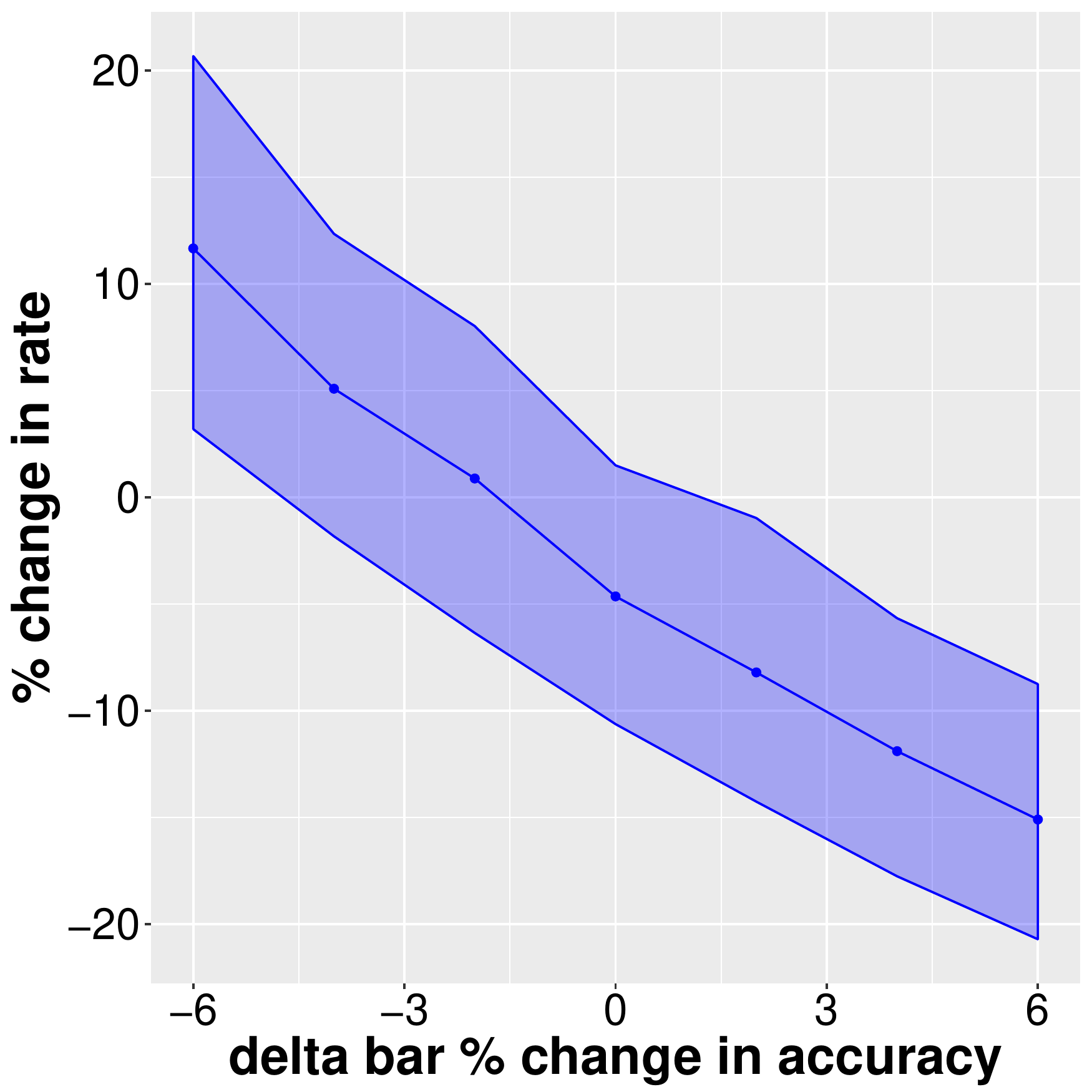}
\caption{Rate}
\end{subfigure}
\caption{Results for Sacramento, CA.  Top left (a): Posterior predictive distribution for homeless counts, $C_{i,1:T}^* | C_{1:25,1:T}, N_{1:25,1:T}$, in green, and the imputed total homeless population size, $H_{i,1:T}|C_{1:25,1:T}, H_{1:25,1:T}$, in blue.  The black 'x' marks correspond to the observed (raw) homeless count by year.  The count accuracy is modeled with a constant expectation.  Top right (b): Predictive distribution for total homeless population in 2017, $H_{i,2017} | C_{1:25,1:T}, N_{1:25,1:T}$.  Bottom left (c):  Posterior distribution of increase in total homeless population with increases in ZRI.  Bottom right (d): Sensitivity of the inferred increase in the homelessness rate from 2011 - 2016 to different annual changes in count accuracy.}
\label{fig:Sacramento_Results}
\end{figure}

\end{document}